\documentclass[twocolumn,trackchanges,twocolappendix]{aastex701}
\received{July 13, 2025}

\submitjournal{\apjs}

\usepackage{amsmath,environ}
\usepackage{listings}
\usepackage{float}
\usepackage{color}
\usepackage{mathtools}

\usepackage{rotating}

\usepackage{lineno}
\linenumbers

\newcommand{\halfwdth}{0.475\textwidth}

\newcommand{\hi}{H~{\sc i}}

\newcommand{\hneut}{H\textsuperscript{0}}

\newcommand{\ci}[1]{[C~{\sc i}]{#1}} 
\newcommand{\cii}[1]{[C~{\sc ii}]{#1}} 
\newcommand{\nii}[1]{[N~{\sc ii}]{#1}} 
\newcommand{\niii}[1]{[N~{\sc iii}]{#1}} 
\newcommand{\oi}[1]{[O~{\sc i}]{#1}} 
\newcommand{\oiii}[1]{[O~{\sc iii}]{#1}} 
\newcommand{\lcii}[1]{$L_\mathrm{[C~II]{#1}}$} 
\newcommand{\loi}[1]{$L_\mathrm{[O~I]{#1}}$} 
\newcommand{\loiii}[1]{$L_\mathrm{[O~III]{#1}}$} 
\newcommand{\oiv}[1]{[O~{\sc iv}]{#1}} 
\newcommand{\neii}[1]{[Ne~{\sc ii}]{#1}} 
\newcommand{\neiii}[1]{[Ne~{\sc iii}]{#1}} 
\newcommand{\nev}[1]{[Ne~{\sc v}]{#1}} 
\newcommand{\siii}[1]{[S~{\sc iii}]{#1}} 
\newcommand{\siv}[1]{[S~{\sc iv}]{#1}} 
\newcommand{\loiv}[1]{$L_\mathrm{[O~IV]{#1}}$} 
\newcommand{\lneii}[1]{$L_\mathrm{[Ne~II]{#1}}$} 
\newcommand{\lnev}[1]{$L_\mathrm{[Ne~V]{#1}}$} 

\newcommand{\ha}{H$\alpha$}

\newcommand{\hb}{H$\beta$}
\newcommand{\hg}{H$\gamma$}
\newcommand{\hd}{H$\delta$}
\newcommand{\niio}[1]{[N~{\sc ii}]$\lambda${#1}} 
\newcommand{\oio}[1]{[O~{\sc i}]$\lambda${#1}} 
\newcommand{\oiio}[1]{[O~{\sc ii}]$\lambda${#1}} 
\newcommand{\oiiio}[1]{[O~{\sc iii}]$\lambda${#1}} 
\newcommand{\siio}[1]{[S~{\sc ii}]$\lambda${#1}} 

\newcommand{\lhaextcorr}{$L_\mathrm{H\alpha,ext.corr.}$}

\newcommand{\lir}{$L_\mathrm{IR}$}
\newcommand{\lfir}{$L_\mathrm{FIR}$}

\newcommand{\lfuv}{$L_\mathrm{FUV}$}
\newcommand{\lline}{$L_\mathrm{line}$}
\newcommand{\vlv}{$\nu L_\nu$}

\newcommand{\cp}{C\textsuperscript{+}}
\newcommand{\op}{O\textsuperscript{+}}
\newcommand{\opp}{O\textsuperscript{2+}}

\newcommand{\npp}{N\textsuperscript{2+}}
\newcommand{\spion}{S\textsuperscript{+}}

\newcommand{\oh}{$\log$\,(O/H)}
\newcommand{\no}{$\log$\,(N/O)}

\newcommand{\edens}{$n_e$}
\newcommand{\hdens}{$n_\mathrm{H^0}$}
\newcommand{\cc}{cm\textsuperscript{-3}}
\newcommand{\td}{$T_\mathrm{dust}$}
\newcommand{\te}{$T_e$}
\newcommand{\teoiii}{$T_{e\mathrm{,[O~III]}}$}
\newcommand{\teoiiifir}{$T_{e\mathrm{,[O~III],FIR}}$}
\newcommand{\tenii}{$T_{e\mathrm{,[N~II]}}$}
\newcommand{\teniifir}{$T_{e\mathrm{,[N~II],FIR}}$}
\newcommand{\thi}{$T_\mathrm{H^0}$}
\newcommand{\md}{$M_\mathrm{dust}$}
\newcommand{\lsun}{$L_\odot$}
\newcommand{\msun}{$M_\odot$}
\newcommand{\mhi}{$M_\mathrm{H~I}$}
\newcommand{\mmol}{$M_\mathrm{mol}$}
\newcommand{\mstar}{$M_\star$}
\newcommand{\fciin}{$f_\mathrm{[C~II],neutral}$}
\newcommand{\um}{\micron}

\newcommand{\fnu}[1]{$S${#1}}
\newcommand{\tcolor}{\fnu{60}/\fnu{100}}

\newcommand{\bsf}[1]{\textbf{\textsf{#1}}}
\newcommand{\zz}{\textit{z}}

\newcommand{\hardness}{$Q_1/Q_0$}

\newcommand{\pyneb}{\texttt{PyNeb}}

\newcommand{\ppi}{Paper I}
\newcommand{\ppii}{Paper II}
\newcommand{\ppiii}{Paper III}

\begin{document}

\title{Fine-structure Line Atlas for Multi-wavelength Extragalactic Study (FLAMES) I: \\Comprehensive Low and High Redshift Catalogs and Empirical Relations for Probing Gas Conditions}

\shorttitle{FLAMES I: Low and High-\texorpdfstring{\zz{}}{z} Catalogs and Empirical Relations}
\shortauthors{Peng et al.}

\correspondingauthor{Bo Peng}
\email{bp392@cornell.edu}

\author[0000-0002-1605-0032]{Bo Peng}
\affiliation{Max-Planck-Institut für Astrophysik, Garching, D-85748, Germany}
\email{bp392@cornell.edu}

\author[0000-0003-1874-7498]{Cody Lamarche}
\affiliation{Department of Physics, Winona State University, Winona, MN 55987, USA}
\email{cody.lamarche@winona.edu}

\author[0000-0002-1895-0528]{Catie Ball}
\affiliation{Department of Astronomy, Cornell University, Ithaca, NY 14853, USA}
\email{cjb356@cornell.edu}

\author[0000-0002-4444-8929]{Amit Vishwas}
\affiliation{Cornell Center for Astrophysics and Planetary Science, Cornell University, Ithaca, NY 14853, USA}
\email{vishwas@cornell.edu}

\author{Gordon Stacey}
\affiliation{Department of Astronomy, Cornell University, Ithaca, NY 14853, USA}
\email{stacey@cornell.edu}

\author[0000-0002-8513-2971]{Christopher Rooney}
\affiliation{National Institute of Standards and Technology, Boulder, CO 80305, USA}
\email{ctr44@cornell.edu}

\author{Thomas Nikola}
\affiliation{Cornell Center for Astrophysics and Planetary Science, Cornell University, Ithaca, NY 14853, USA}
\email{tn46@cornell.edu}

\author[0000-0001-6266-0213]{Carl Ferkinhoff}
\affiliation{Department of Physics, Winona State University, Winona, MN 55987, USA}
\email{cferkinhoff@winona.edu}

\begin{abstract}

Far-infrared (FIR) and mid-infrared (MIR) fine-structure lines (FSLs) are widely used for studying galaxies nearby and faraway. 
However, interpreting these lines is complicated by factors including sample and data bias, mismatch between resolved calibrations and unresolved observations, limitations in generalizing from case studies, and unresolved issues like the origin of \cii{} emission and ``deficit.''
In this series of papers, we assemble and analyze the most comprehensive atlas of FSL data to date. 
We explore their empirical correlations (\ppi{}), compare them with photoionization models that cover multiphase gas (\ppii{}), and discuss their physical origins and the new perspectives they offer for studying physical properties (\ppiii{}). 
The first paper introduces value-added catalogs of global FSL data of low- and high-\zz{} galaxies compiled from the literature, covering most of the existing observations, supplemented with ancillary ultraviolet to FIR information. 
Our analysis focus on commonly used diagnostics, such as electron density, radiation field strength, metallicity, and electron temperature. 
We present their distributions across different galaxy samples and redshifts, and cross-validate the reliability of these diagnostics in measuring physical conditions. 
By examining empirical relations, we identify the contribution of active galactic nuclei (AGN) to \oiii{88} and \oi{63}, and reveal a bias in density measurements. 
FIR FSLs show good concordance with their optical counterparts. 
Our findings indicate FSL ratios are primarily driven by the relative abundances of emitting ions. 
Finally, we compare the FSL properties of low- and high-\zz{} galaxies, discussing both their similarities and differences. 

\end{abstract}


\section{Introduction}
\label{sec:data_intro}

The fine-structure lines (FSLs) in the mid-infrared (MIR) and far-infrared (FIR) regimes are essential tools in the study of galaxy evolution.
Several FIR FSLs, such as \oiii{52\&88\,\um{}}, \cii{158\,\um{}}, and \oi{63\,\um{}}, are among the major coolants of the interstellar medium (ISM), typically emitting at a few percent of the total infrared luminosity (\lir) \citep{tielens85}. 
Low-ionization FIR FSLs---including \cii{158\,\um{}}, \oi{63\,\um{}}, and \ci{370\&609\,\um{}}---serve as unique tracers of the transitional layers between ionized and molecular gas, known as the photo-dissociation region \citep[PDR; ][]{tielens85,stuzki88,stacey91}. 
In contrast, high-ionization MIR FSLs, such as \nev{15.5\&36\,\um{}} and \oiv{26\,\um{}}, are sensitive indicators of active galactic nucleus (AGN) activity, even in heavily obscured environments \citep{genzel98,P10}. 
Recent advances in submillimeter (sub-mm) observations, facilitated by the atmospheric window in the sub-mm regime and the sensitivity of the Atacama Large Millimeter/submillimeter Array (ALMA), have enabled the intensive use of \cii{158\,\um{}} and \oiii{88\,\um{}} lines over the past two decades to confirm and even discover galaxies at high redshift (high-\zz{}) \citep[e.g.][]{MR05,FC10,SG10,WR13,GB15,CP15,HT18,LO20,SS24,ZJ24}.

Fine-structure lines (FSLs) offer several key advantages over optical forbidden lines for studying ISM conditions and galaxy evolution. 
Due to their longer wavelengths, FSLs are far less affected by dust attenuation, providing a unique opportunity to investigate heavily obscured galaxies and regions, which often exhibit the highest star formation rates (SFR) or intense nuclear activity \citep{sanders96}.
The excitation temperatures of most FSLs are significantly lower than the ambient gas temperature (with the exception of \oi{} and \ci{} lines). 
As a result, their excitation depends solely on the density of the collisional partner, without the strong exponential temperature dependence seen in optical forbidden lines. 
This property effectively makes FSL emission a density-weighted sum over the emitting gas, with less systematic uncertainties and biases than optical lines when measuring ISM properties. 
FIR FSLs are particularly valuable for studying galaxies at $z \geq 3$. 
As the two brightest FSLs, \cii{} and \oiii{88}, are redshifted into atmospheric windows in the sub-mm regime, they remain accessible to ground-based facilities, while strong optical lines are shifted beyond $2\,\mu$m and become challenging to observe from the ground. 
Furthermore, the exceptional angular resolution provided by interferometers enables detailed studies of the ISM distribution and structures in high-\zz{} galaxies. 

Despite their diagnostic potential, our understanding of FIR FSLs remains incomplete. 
Several key questions remain regarding the physical interpretation of FIR FSLs, especially the \cii{} line. 
Notably, the long-standing \cii{}-to-infrared (IR) ``deficit'' problem \citep{kaufman99,D17} and the varying contribution of \cii{} emission from the neutral medium across different galaxies \citep{brauher08,C19} are unresolved. 
The case is further worsened by the fragmented studies of the \cii{} line. 
Progress in this area is hindered by fragmented studies of the \cii{} line. 
Depending on assumptions and sample selections, previous works have used \cii{} emission to infer a wide range of ISM properties, including the star formation rate \citep[SFR;][]{delooze14,herrera15}, atomic gas mass \citep[\mhi{};][]{heintz21}, CO-dark molecular gas mass \citep[\mmol{};][]{madden20}, total \mmol{} \citep{ZA18}, total gas mass \citep{deugenio23}, ultraviolet (UV) radiation field intensity \citep{HS10}, and the covering factor of PDRs \citep{C19}.
The widespread use of \cii{} as a probe, despite the lack of consensus, is due in part to the complex and multiphase origins of its emission. 
Additionally, \cii{} is frequently the only detectable spectral line in distant galaxies, particularly at high redshift, further motivating its extensive application. 

Although large amounts of observational data now exist for other FIR FSLs, such as \oiii{88} and \nii{205}, in both low- and high-\zz{} galaxies, combining emission from ionized and neutral gas phases within multi-line diagnostic frameworks remains a substantial challenge. 
Traditionally, theoretical studies have modeled PDR lines (e.g., \cii{} and \oi{}) and ionized gas lines (e.g., \nii{} and \oiii{}) separately, using distinct tools and methodologies. 
Only a limited number of studies have attempted to model emissions from both gas phases in a unified framework \citep[e.g.,][]{abel05,C19}.
This theoretical gap between the treatment of ionized and neutral gas emission complicates the interpretation of phenomena involving multiple lines, such as the \oiii{88}/\cii{} ratios observed in the early universe \citep{FC10,HY20}.

In addition to the challenges associated with individual emission lines, several systematic issues hinder the interpretation of FIR FSL observations. 
One prominent problem is the spatial scale gap between resolved cloud-scale studies in nearby galaxies and the integrated observations of both low-\zz{} (ultra)luminous infrared galaxies (U/LIRGs) and high-\zz{} systems. 
While early FIR FSL studies focused on galactic star-forming (SF) regions and extragalactic nuclei \citep[][and references therein]{genzel89,stacey91}, attention soon shifted to marginally resolved or galaxy-integrated measurements \citep[e.g.][]{kaufman99,SG10,F13}. 
This transition was driven by the coarse spatial resolution of FIR observatories (\textgreater 10\arcsec), which is insufficient to resolve the small angular sizes of local U/LIRGs and high-\zz{} galaxies, as well as by the limited mapping speeds of FIR/sub-mm facilities.
Consequently, methodologies and conclusions from resolved cloud-scale studies have often been directly extrapolated to unresolved or integrated observations. 
As a result, interpretations based on PDR models have become prevalent in FIR FSL studies, despite clear differences in the trends of dust and FSL luminosities at cloud versus galaxy-integrated scales, and disparities between resolved and averaged physical properties \citep{RM20a,wolfire22}. 
Observations within this spatial scale gap remain scarce \citep{K11,croxall17}, and available studies are limited in both sample size and galaxy properties.

Last but not least, another gap persists between FIR and optical studies. 
Although several spectral lines originate from the same atomic or ionic species, few studies have attempted direct comparisons between FIR FSLs and optical spectral lines. 
This is partly due to the focus of FIR research on heavily obscured regions and galaxies, and partly due to the inaccessibility of bright optical lines in high-\zz{} galaxies from the ground. 
However, the recent success of the \textit{James Webb Space Telescope} (JWST) underscores the importance of studies that combine both FIR and optical observations, which can greatly enhance our understanding of the early universe, where observational data remain limited. 

These gaps and challenges collectively highlight the need for a comprehensive study of FIR FSLs. 
Such an approach should test hypotheses across a wide range of galaxy types and properties, compare FIR FSLs to observables at other wavelengths, and leverage theoretical models that account for multiple ISM phases. 
Only through this broad perspective can we achieve robust and widely applicable results. 

As the first paper in a series, this paper compiles the majority of currently available, galaxy-integrated FSL data alongside multiwavelength ancillary data, emphasizing a wide coverage of galaxy types and properties. 
This data atlas provides a foundation for a holistic assessment of FIR FSL studies and the validity of existing conclusions. 
It also establishes a reference framework for multi-line analyses and for testing theoretical predictions. 
Using this data set, we present common FSL diagnostics and investigate empirical scaling relations between global observables. 
While such scaling relations may reflect secondary effects of more fundamental correlations, their existence (or absence) reveals the interconnections (or decoupling) of key parameters. 
Consequently, we rely on observational data and employ straightforward interpretations grounded in robust physical principles and explicit assumptions. 
Quantitative analyses based on photoionization models will be presented in \ppii{}, and some of the key questions in the study of FIR FSLs will be explored in \ppiii{}, in particular the interpretation of \cii{}, and the cause of the line ``deficit''.

The structure of the paper is as follows: Section~\ref{sec:data} describes the galaxy sample and the suite of observables compiled from the literature; Section~\ref{sec:data_results} presents the main scaling relations among global properties, including updates to known diagnostics and new correlations; Section~\ref{sec:data_discussion} discusses density diagnostics and AGN contributions, and compares low- and high-\zz{} galaxies; and Section~\ref{sec:data_summary} concludes with a summary. 
Throughout this paper, we adopt a flat $\Lambda$CDM cosmology with $H_0 = 69.3\,\mathrm{km\,s^{-1}\,Mpc^{-1}}$ and $\Omega_\mathrm{m} = 0.287$ \citep{hinshaw13}.

\section{Data}
\label{sec:data}

We assemble a comprehensive catalog of galaxy-integrated measurements, cross-matching FIR FSLs to MIR FSL and optical spectral lines, as well as key ancillary properties. 

\subsection{Collecting the Data}
\label{sec:data_fslines}

Our primary focus is the eight most commonly observed FIR FSLs originating from both ionized and neutral gas: \cii{} 158\,\um{}, \nii{} 122 and 205\,\um{}, \niii{} 57\,\um{}, \oi{} 63 and 145\,\um{}, and \oiii{} 52 and 88\,\um{}. 
These lines have been extensively studied in the literature; basic properties such as wavelength, ionization potential, and critical density are summarized in \citet{stacey11} and \citet{carilli13}. 
Although the \ci{} 370 and 609\,\um{} lines are also well studied, they are more closely related to the molecular gas and show little correlation with other FSLs in our analysis. 
We therefore exclude \ci{} lines from this work, reserving them for a future study focused on molecular gas. 
The sample is only based on the availability of FIR FSL data, making it an inclusive catalog of any galaxies with FIR FSL detections. 
However, we do not guarantee the completeness of the sample, but estimate conservatively that it contains more than 95\% of all the existing data reported in the literature. 

To further enrich our dataset, we compile a suite of MIR FSLs in low-\zz{} galaxies, including: \oiv{} 25.9\,\um{}, \neii{} 12.8\,\um{}, \neiii{} 15.6 and 36.0\,\um{}, \nev{} 14.3 and 24.3\,\um{}, \siii{} 18.7 and 33.5\,\um{}, \siv{} 10.5\,\um{}, [Si\,{\sc ii}] 34.8\,\um{}, [Ar\,{\sc ii}] 7.0\,\um{}, [Ar\,{\sc iii}] 9.0 and 21.8\,\um{}, [Ar\,{\sc v}] 13.1\,\um{}, [Cl\,{\sc ii}] 14.3\,\um{}, [P\,{\sc iii}] 17.9\,\um{}, [Fe\,{\sc ii}] 17.9, 22.9, 24.5, and 26.0\,\um{}. 
The wavelengths and transition configurations for these lines are referenced from the original data sources \citep[e.g.][]{F07,I13}.

We also collect galaxy-integrated optical spectroscopy, focusing on strong lines such as the hydrogen recombination lines (\ha{}, \hb{}, \hg{}, \hd{}) and key forbidden lines (\oiio{3727\,\AA}, \oiiio{5007\,\AA}, \oio{6300\,\AA}, \niio{6584\,\AA}, \siio{6716,6731\,\AA}). 
These lines are essential for determining elemental abundances and serve as a basis for calibrating FIR FSL diagnostics. 

In addition to spectral lines, we include photometric data where available, such as total infrared (IR), FIR, far-ultraviolet (FUV), and single-band luminosities, as well as the UV slope $\beta$ and infrared excess (IRX). 
We further compile ancillary parameters including redshift, galaxy type, luminosity distance, stellar mass (\mstar{}), atomic gas mass (where \hi{} 21\,cm data are available), lensing magnification (for strongly lensed high-\zz{} sources), elemental abundances (O/H, N/O), dust extinction, and more. 
This broad coverage provides essential context for examining galactic properties and their interrelations.

\subsection{Post Processing}
\label{sec:data_post}

Dust temperature (\td{}) is a crucial parameter, commonly estimated from the flux density ratio of FIR bands, especially the IRAS \citep{neugebauer84} 60\,\um{} to 100\,\um{} ratio (\tcolor{}). 
The conversion from color temperature to physical dust temperature is non-linear and model-dependent, often complicated by sparse or irregular photometric sampling, particularly in high-$z$ galaxies. 
To maximize data set compatibility and minimize systematics, we standardize all color temperatures to \tcolor{}, using conversions based on empirical fits and optically thin modified blackbody model (OT-MBB; see Appendix~\ref{sec:temp_cal}). 
This approach is applied to sources lacking direct \tcolor{} measurements. 
We do not further convert color temperature to physical dust temperature to avoid introducing additional model-dependent assumptions. 

Optical line fluxes are corrected for dust extinction using the Balmer decrement (\ha{}/\hb{}). 
We adopt the theoretical case B value of 2.86 \citep{osterbrock89} and apply the extinction law from \citet{calzetti00} with $R_V^\prime=4.05$ for starburst galaxies. 
The reliability of extinction correction for $\lambda > 4300$\,\AA{} is validated in Appendix~\ref{sec:data_ext} by comparison of corrected \ha{}/\hg{} ratios with theoretical expectations. 
However, at shorter wavelengths, the correction may overestimate line fluxes, as indicated by the \ha{}/\hd{} ratio in Fig.~\ref{f:data_ext}. 

For galaxies with integrated optical spectra but lacking reported O/H or N/O abundances, we compute metallicities using the S calibration of \citet{pilyugin16} (PG16S), the O3N2 and N2 calibrations of \citet{pettini04}, and N/O using the N2S2 index of \citet{perez09}, in order of descending priority and subject to data availability. 
Systematic offsets between different methods are accounted for by comparing computed abundances to literature values: the O3N2 index overestimates O/H by 0.15\,dex, and N2 index by 0.2\,dex; these offsets are applied to our results. 
The validity of abundance measurements from integrated spectra is discussed in \citet{pilyugin04}.

\subsection{Low-\texorpdfstring{\zz{}}{z} Galaxies Catalog}
\label{sec:data_nearby}

Our low-redshift sample is primarily drawn from surveys conducted with the Infrared Space Observatory (ISO; \citealt{kessler96}), the \textit{Herschel} Space Observatory \citep{pilbratt10}, and SOFIA \citep{temi18}, resulting in a value-added catalog of 1273 entries---galaxies, or galaxy systems and components, or resolved regions---at \zz{} \textless1. 
Specific treatment of galaxy pairs and resolved nuclei is detailed in Appendix~\ref{sec:flames-low_systems}. 
The catalog assembly, structure, and examples are presented in Appendix~\ref{sec:flames-low}. The complete table will be available in the online published version.

\subsection{High-\texorpdfstring{\zz{}}{z} Galaxies Catalog}
\label{sec:data_high-z}

The high-redshift value-added catalog comprises 543 galaxies or systems at \zz{} \textgreater 1, compiled from \textgreater 400 publications and selected via various methods, including sub-mm selection for dusty star-forming galaxies (DSFGs), Lyman-break or Ly$\alpha$ line selection for Lyman Break Galaxies (LBGs) and Lyman Alpha Emitters (LAEs), and rest-frame UV brightness for quasars (QSOs). 
Most cataloged galaxies have \cii{} detections, with \nii{205} and \oiii{88} being the other most commonly observed lines (37 and 64 detections). 
Catalog assembly, structure, and examples are given in Appendix~\ref{sec:flames-high}. 
The full table will be available in the online published version.

\section{Results}
\label{sec:data_results}

We present empirical correlations among line luminosities using our compiled data set. 
To minimize biases from galaxy masses (the ``bigger-things-brighter'' effect), we normalize integrated properties by common factors. 
As galaxy masses span over six orders of magnitude, even unrelated properties can show apparent correlations simply due to scaling with gas mass or energy output from star-formation. 

To address this, we focus on line ratios, which can be formally expressed as:
\begin{equation}\begin{split}
    \frac{L(\mathrm{X}^{a+}_{i\rightarrow j})}{L(\mathrm{Y}^{b+}_{k\rightarrow l})} = \frac{\mathrm{X}}{\mathrm{Y}}\ \cdot \ \frac{\mathrm{X}^{a+}/\mathrm{X}}{\mathrm{Y}^{b+}/\mathrm{Y}}\ \cdot\  \frac{\varepsilon_{\mathrm{X}^{a+},i\rightarrow j}}{\varepsilon_{\mathrm{Y}^{b+},k\rightarrow l}}
\label{equ:ratio}
\end{split}\end{equation}
Here, the line luminosity ratio reflects the elemental abundance ratio X/Y, the ionization correction factors (ICFs) X$^{a+}$/X, and the line emissivity ratio $\varepsilon_{\mathrm{X}^{a+},i\rightarrow j}/\varepsilon_{\mathrm{Y}^{b+},k\rightarrow l}$. 
By careful selection, we can isolate and then study physical conditions of interest, such as abundance, ionization state (in ICF), or electron density \edens{} and temperature \te{} (in emissivity).

\subsection{IR and UV Luminosity}
\label{sec:data_fir}

\begin{figure*}
    \centering
    \includegraphics[width=\textwidth]{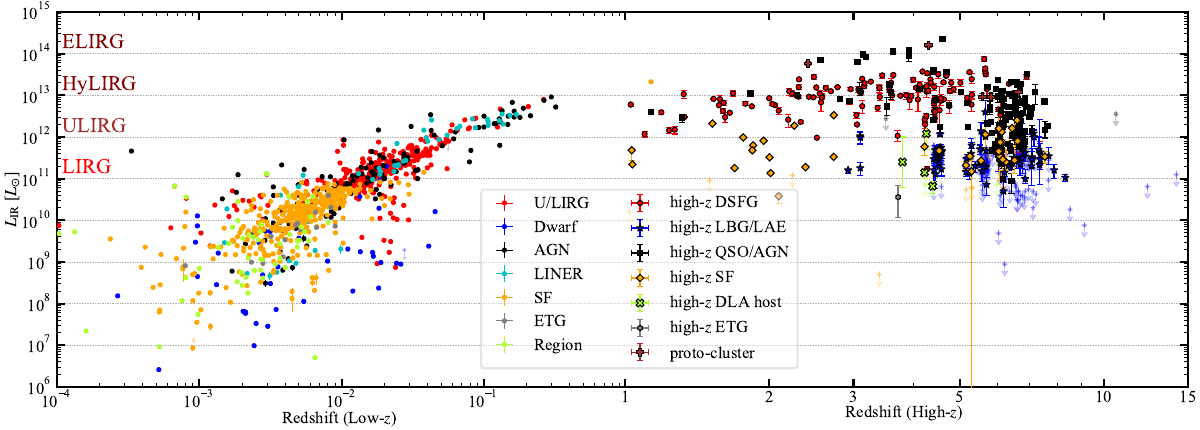}
    \caption[Redshift distribution of \lir{} of the sample.]{Redshift distribution of the IR luminosity \lir{} of the sample. The data points are divided into low-\zz{} and high-\zz{} with the redshift cut at 1. The low-\zz{} galaxies are plotted as small dots, with the color scheme: blue--dwarf galaxies, orange--star-forming (SF) galaxies, red--U/LIRGs, cyan--LINERs, black--AGNs, gray--early type galaxies (ETGs), green-yellow--regions. All high-\zz{} data points are shown as larger symbols edged in black, in the following colors and symbols: blue star--LBGs or LAEs, orange diamond--SF, red circle--DSFG, black square--QSOs or AGNs, gray hexagon--ETGs, green-yellow ``x''--damped Ly$\alpha$ systems (DLAs), brown ``+''--proto-clusters. Upper or lower limit is plotted at the 3 $\sigma$ value with the same color and a smaller symbol corresponding to its type, but is translucent and has an arrow showing the direction of the constraint.}
    \label{f:data_z-IR}
\end{figure*}

We start with an overview of the IR and UV luminosities of our sample galaxies, based on which many were selected and categorized. 

Galaxies with extremely high IR luminosities have been found in large numbers both locally (e.g., with IRAS; \citealt{sanders96}) and at high redshift through submillimeter surveys \citep[e.g.,][]{smail97,RC20}. 
In our sample, most low-\zz{} galaxies and more than half of the high-\zz{} galaxies were selected based on their IR emission. 
Fig.~\ref{f:data_z-IR} shows the distribution of IR luminosity (\lir{}) versus redshift for the entire sample. 
The redshift distribution exhibits a weak peak at $z\sim3$, consistent with the known cosmic star formation history \citep{madau14,gruppioni20}. 
However, the flattening of the upper envelope of \lir{} at $z>3$ reflects the so-called ``dust production problem'' at early epochs \citep[e.g.,][]{lesniewska19}, as well as the continued presence of highly obscured star formation out to \zz{}~$\sim$~7 \citep{gruppioni20,dayal22}. 
Because the non-uniformity in studying and reporting high-\zz{} galaxies, \lir{} and \lfir{} are both recorded only occasionally. 
In the case that \lfir{} is reported but not \lir{}, the former is multiplied by a factor of 2 to approximate \lir{} and used as \lir{} throughout this paper. 

An offset in \lir{} between low- and high-\zz{} galaxies is evident in Fig.~\ref{f:data_z-IR}. 
Almost all high-\zz{} DSFGs have \lir{} \textgreater $10^{12}$\,\lsun{}, reaching beyond $10^{14}$\,\lsun{}, qualifying as ULIRGs and hyper-luminous galaxies \citep[HyLIRGs;][]{sanders96}. 
This offset is seen for all galaxy types; UV-selected high-\zz{} galaxies (LBG/LAEs) are also much more luminous than their local ``analogs'', dwarf galaxies. 
The majority of the high-\zz{} LBG/LAEs are even dustier than the local SF galaxies, exceeding the LIRG threshold. 
These differences are primarily a result of strong selection effects in high-\zz{} galaxy searches and FIR FSL detections. 

Selection biases are introduced at multiple stages. 
Most high-\zz{} galaxies targeted for FSL observations are among the brightest detected in sub-mm or UV surveys. 
Additionally, observational sensitivity limits (reflected by the upper/lower limits in Fig.~\ref{f:data_z-IR}) mean that only the most luminous sources are detected, while fainter galaxies are missed. 
As a result, the high-\zz{} sample with sub-mm spectroscopy is dominated by exceptionally luminous galaxies, which are rare in the universe and not representative of the general galaxy population or the bulk of cosmic star formation.

\begin{figure}
    \centering
    \includegraphics[width=\halfwdth]{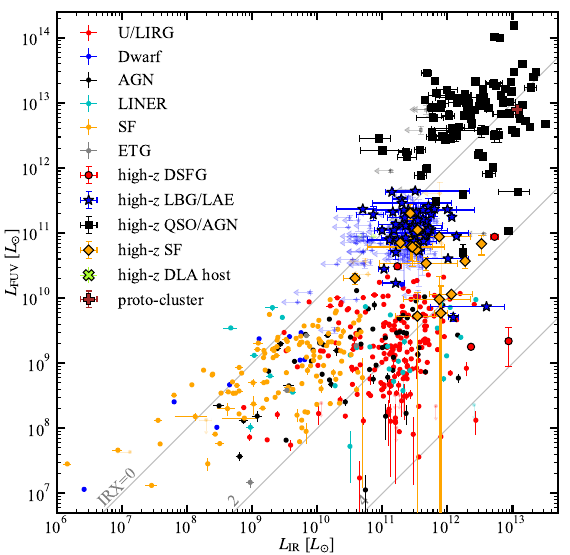}
    \caption[\lir{}-to-\lfuv{} comparison. ]{IR-to-FUV luminosity (\lir{}-to-\lfuv{}) comparison. The style of the plot is the same as that of Fig.~\ref{f:data_z-IR}. The gray lines correspond to different values of IRX.}
    \label{f:data_ir-irx}
\end{figure}

Fig.~\ref{f:data_ir-irx} compares the IR and FUV luminosities. 
The data show substantial scatter, with no tight correlation between \lir{} and \lfuv{}. 
This is reflected in the wide range of IR excess (IRX $\equiv \log$(\lir{}/\lfuv{})), which reflects the degree of dust attenuation and can vary by more than four orders of magnitude for galaxies with similar IR luminosities. 
A correlation between intrinsic luminosities and dust extinction causes clustered distribution of each galaxy-type in Fig.~\ref{f:data_ir-irx}. 
The large scatter in the IRX-\lir{} relation is also partly driven by the broad dynamic range in \lir{}, as discussed in Section~\ref{sec:data_tdust}. 

Selection effects are further illustrated when comparing FUV luminosities. 
In Fig.~\ref{f:data_ir-irx}, UV-selected high-\zz{} QSOs and LBG/LAEs populate the upper range of \lfuv{}, in contrast to high-\zz{} DSFGs, which are largely absent due to heavy dust obscuration and intrinsically redder UV spectral energy distributions (SEDs). 
Notably, the UV-selected high-\zz{} galaxies have IRX values between $-1$ and $1$, similar to local dwarfs, despite being over two orders of magnitude more luminous in both IR and FUV. 
The only high-\zz{} DSFG with available data is consistent with local U/LIRGs but is much brighter. 
Other high-\zz{} DSFGs are missing from this figure as their UV emission is heavily suppressed by dust.

\subsection{Dust Temperature}
\label{sec:data_tdust}

Dust temperature plays a pivotal role in determining the infrared luminosity of galaxies, as dictated by the Stefan-Boltzmann law. 
For example, increasing \td{} by a factor of two (e.g., from 20\,K to 40\,K, typical in U/LIRGs) leads to a sixteen-fold increase in dust emissivity. 
By contrast, doubling the electron temperature (\te{}) from 8,000\,K to 16,000\,K leads to only a five-fold increase in the emissivity of the \oiiio{5007} line. 
As a result, dust emission is more strongly weighted towards regions with the hottest dust, potentially dominating the global IR output.

\begin{figure}
    \centering
    \includegraphics[width=\halfwdth]{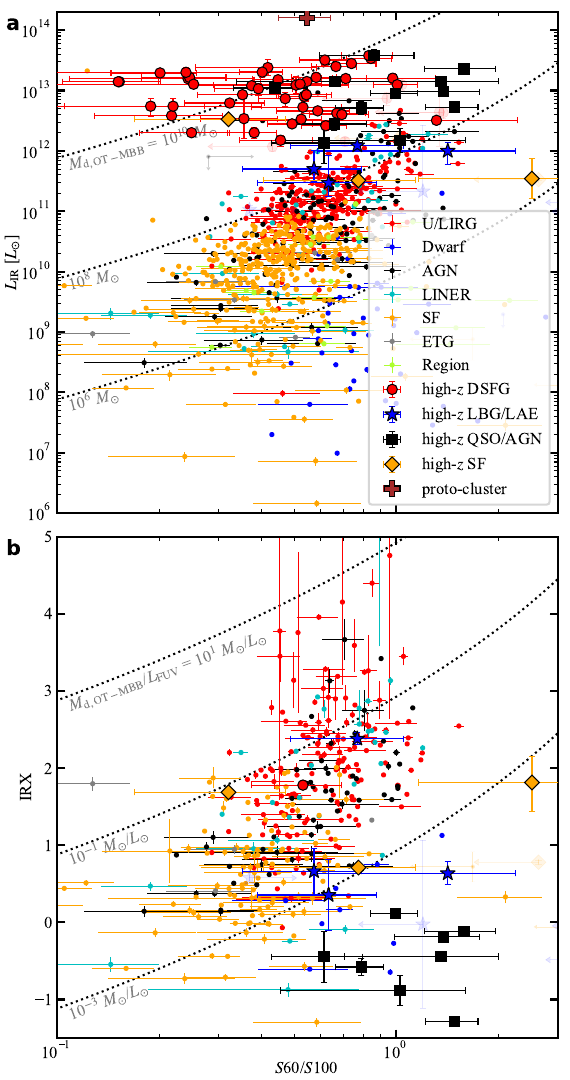}
    \caption[\tcolor{} vs. \lir{} and IRX. ]{Dust color temperature \tcolor{} vs. \bsf{a}, IR luminosity \lir{}; and \bsf{b}, IR excess IRX. The dotted lines correspond to the \tcolor{} coverted from OT-MBB models for constant dust masses \md{} or constant values of \md{}/\lfuv{}, with the values shown in the figure. The plot style is the same as in Fig.~\ref{f:data_z-IR}}
    \label{f:data_s60_100-LIR}
\end{figure}

Fig.~\ref{f:data_s60_100-LIR} compares the color temperature \tcolor{} with both \lir{} and IRX. 
Among low-\zz{} galaxies (excluding dwarfs), there is a clear positive correlation, such that galaxies with higher \lir{} systematically exhibit higher \tcolor{} and thus higher \td{}. 
The majority of low-\zz{} galaxies in Fig.~\ref{f:data_s60_100-LIR}(\bsf{a}) follow a trend from \tcolor{}~$\sim$~0.3 and \lir{}~$\sim$~10\textsuperscript{10}\,\lsun{} up to \tcolor{}~$\sim$~1 and \lir{}~$\sim$~10\textsuperscript{12}\,\lsun{}.
Comparison with the iso-\md{} model curves (dotted lines) reveals that an order-of-magnitude increase in \td{} can drive a similar increase in \lir{}, underlining the comparable contributions of dust temperature and dust mass to the total IR luminosity.

A similar pattern is observed in the \tcolor versus IRX plot (\bsf{b}). 
The model curves for constant \md{}/\lfuv{} encompass the distribution of most low-\zz{} SF and dusty galaxies, highlighting the contribution of \td{} variation.

High-\zz{} galaxies in Fig.~\ref{f:data_s60_100-LIR} generally exhibit both higher \lir{} and higher \td{} than local galaxies, extending the trend observed in low-\zz{} systems. 
This is consistent with previous findings that dust temperature does not evolve significantly up to $z\sim2$ \citep{drew22}, and indicates that higher \td{} in high-\zz{} samples may be a consequence of selection effects. 
Specifically, FIR and sub-mm surveys are biased toward detecting the most IR-luminous galaxies (\lir{}~\textgreater~10\textsuperscript{12}\,\lsun{} for DSFGs and \textgreater~10\textsuperscript{11}\,\lsun{} for UV-selected galaxies; see Fig.~\ref{f:data_z-IR}), which tend to harbor hotter dust.
It is important to note that our sample may be subject to even stronger selection effects, as only galaxies with available FSL observations are included. 
While we do not attempt a comprehensive analysis of dust properties here, we emphasize the key points: (1) the strong dependence of \lir{} on \td{}; (2) the utility of dust color as a proxy for \td{}; and (3) the strong correlation between high \lir{} and high \td{}, suggesting that the former is, at least in part, a consequence of the latter.

\subsection{Line Luminosity and ``Deficit''}
\label{sec:data_line}

\begin{figure*}
    \centering
    \includegraphics[width=\textwidth]{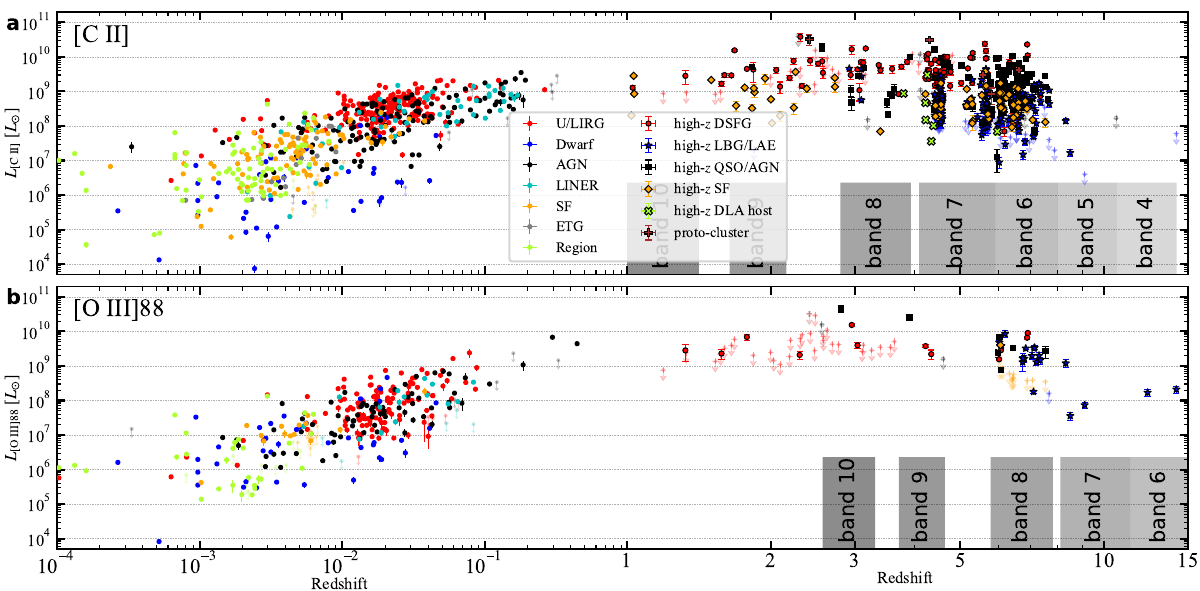}
    \caption[Redshift distribution of \lcii{} and \loiii{88}. ]{Redshift distribution of the luminosity of \bsf{a}, \cii{}; and \bsf{b}, \oiii{88}. Grey shades denote the redshift coverages of ALMA bands. The plot style is the same as in Fig.~\ref{f:data_z-IR}.}
    \label{f:data_z-line_1}
\end{figure*}

\begin{figure}
    \centering
    \includegraphics[width=\halfwdth]{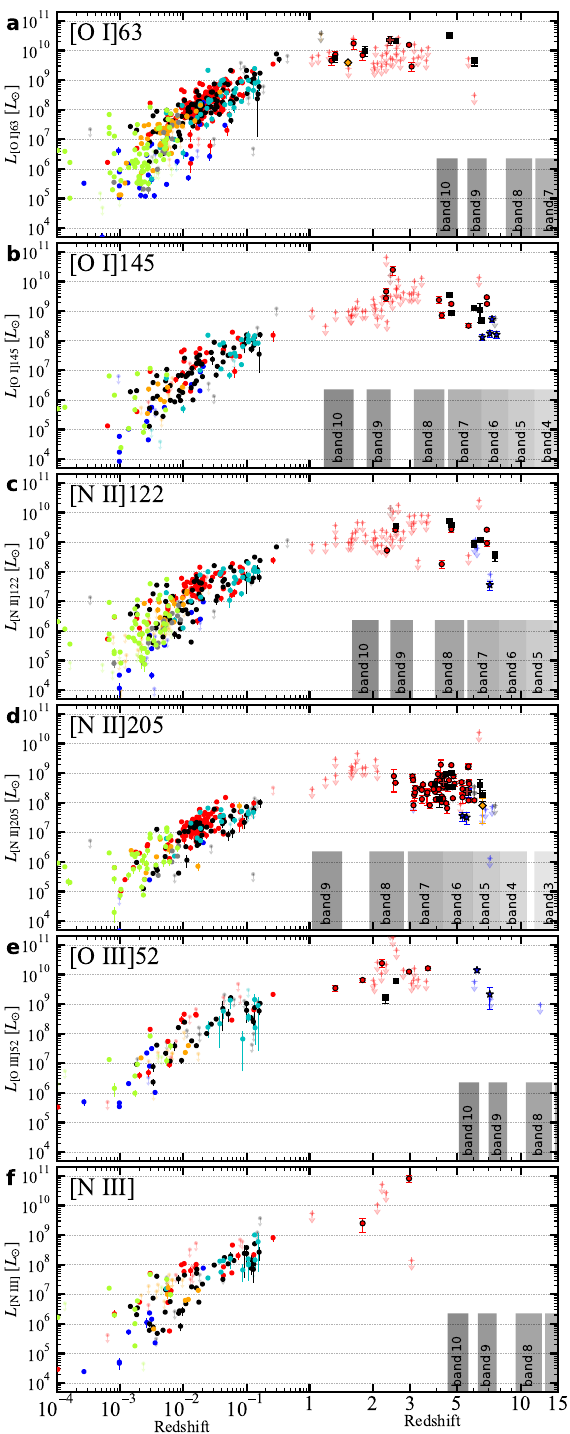}
    \caption[Redshift distribution of the other FIR FSL luminosities. ]{Redshift distribution of the other FIR FSL luminosities. continued from Fig.~\ref{f:data_z-line_1}.}
    \label{f:data_z-line_2}
\end{figure}

To provide a holistic view of the FIR FSL data across cosmic time, we present the luminosities of all the target FIR FSLs—\cii{}, \oiii{88}, \oi{63}, \oi{145}, \nii{122}, \nii{205}, \oiii{52}, and \niii{}—as a function of redshift in Fig.~\ref{f:data_z-line_1} and \ref{f:data_z-line_2}. 
As with the IR luminosity distribution, a pronounced selection effect is observed: high-\zz{} detections are systematically more luminous than their low-\zz{} counterparts. 
This bias is driven not only by galaxy properties but also by instrumental factors, including the frequency coverage and sensitivity limits of observatories such as Herschel and ALMA. 

Detections in low-\zz{} galaxies are biased in terms of the types of galaxies with available line measurements. 
The frequency coverage of the Herschel/PACS spectrometer \citep{poglitsch10} similarly limits the detection of \oi{63} and \oiii{52}, despite their high brightness, while intrinsically fainter lines such as \oi{145} and \niii{} are constrained by the instrument sensitivity.
Most of the \nii{205} detections are taken with Herschel/SPIRE that has lower sensitivity. 
Consequently, detections of these lines are mainly from bright U/LIRGs and AGNs, with far fewer data for SF and dwarf galaxies compared to \cii{} and \oiii{88} lines. 

A similar bias exists at high redshift. 
The atmospheric opacity in the FIR and high-frequency sub-mm regime restricts access to many FSLs, making Band 8 and lower frequency bands of ALMA essential for observations of lines like \cii{}, \oiii{88}, and \nii{205}. 
As a result, FIR FSL studies at high $z$ are largely biased towards bright lines in DSFGs, making any calibration potentially unrepresentative of the broader galaxy population. 
At the highest frequencies, only a few high-\zz{} detections have been made, mainly with Herschel/SPIRE \citep{griffin10}.  
Although LBGs and LAEs are representative of typical SF galaxies in the early universe, only \cii{} and \oiii{88} are commonly detected at \zz{}~\textgreater~4 due to sensitivity and atmospheric transmission. 

\begin{figure}
    \centering
    \includegraphics[width=\halfwdth]{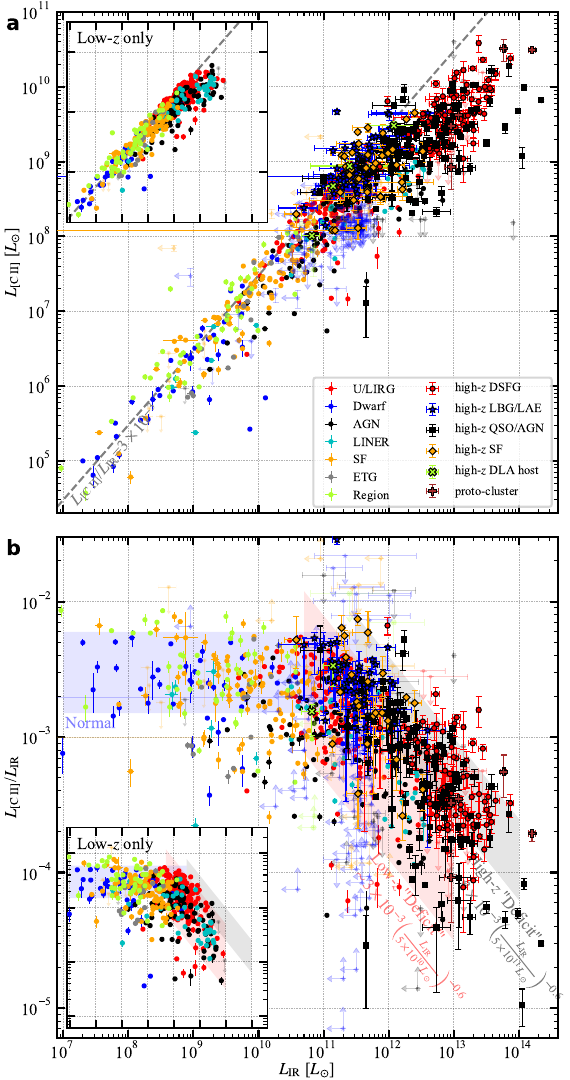}
    \caption[\lir{} vs. \lcii{} and \lcii{}/\lir{}. ]{Total IR luminosity \lir{} vs. \bsf{a}, \cii{} luminosity \lcii{}; and \bsf{b}, the residual \cii{}/IR. The dashed line in (\bsf{a}) denotes the linear relation corresponding to the non-``deficit'' trend. The blue, red, and gray shades in (\bsf{b}) highlight the non-deficit, low-\zz{} ``deficit'', and high-\zz{} ``deficit'' branches. The plot style is the same as in Fig.~\ref{f:data_z-IR}.}
    \label{f:data_IR-CII_IR}
\end{figure}

We next compare line luminosity (\lline{}) to total IR luminosity, starting with \cii{} (Fig.~\ref{f:data_IR-CII_IR}). 
Across all galaxy types, \lcii{} increases nearly linearly with \lir{}, modulo significant scatter—a manifestation of the ``bigger-things-brighter'' effect. 
However, above \lir{}~$\sim$~10\textsuperscript{11}\,\lsun{}, the correlation deviates from linearity, revealing the well-known \cii{} ``deficit'': \lcii{}/\lir{} decreases from the typical value of $\sim$0.03, following a power law of $\sim$\lir$^{-0.6}$, with considerable scatter (0.6 dex for low-\zz{} galaxies, 0.3 dex for high-\zz{} galaxies). 
Although the ``deficit'' is more pronounced in surface brightness ratios \citep{H18a}, it is clearly visible in integrated galaxy data. 

The \cii{} ``deficit'' is observed in low-\zz{} U/LIRGs and AGNs, as well as in high-\zz{} DSFGs, QSOs, and some highly luminous SF galaxies. 
In high-\zz{} galaxies, all DSFGs, QSOs, and proto-clusters occupy the ``deficit'' branch, with some reaching extreme values (\lcii{}/\lir{}~\textless~10\textsuperscript{-4}). 
Notably, the ``deficit'' branch for high-\zz{} galaxies is horizontally offset compared to the local trend, a point discussed in \ppiii{}. 
Local dwarfs and high-\zz{} LBG/LAEs generally do not exhibit the ``deficit'', with few exceptions. 

\begin{figure*}
    \centering
    \includegraphics[width=\textwidth]{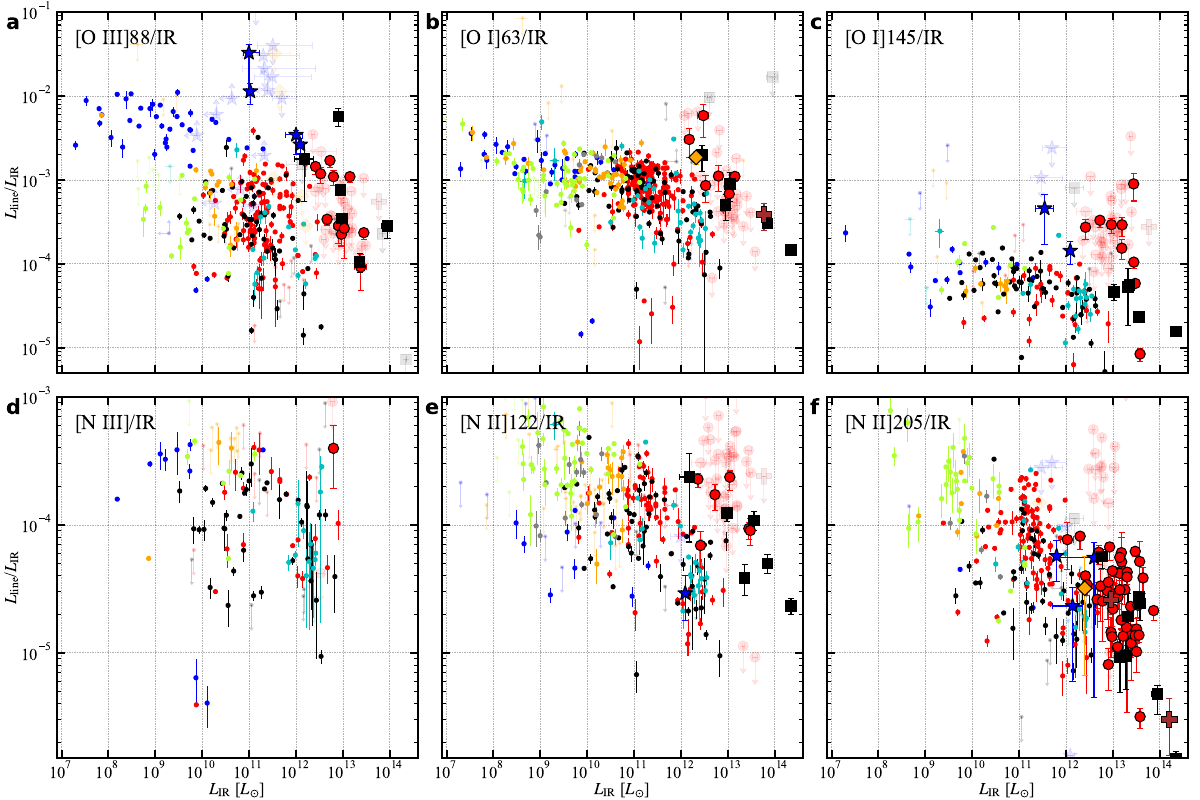}
    \caption[\lir{} vs. \lline{}/\lir{}. ]{IR luminosity \lir{} vs. \lline{}/\lir{} for \bsf{a}, \oiii{88}; \bsf{b}, \oi{63}; \bsf{c}, \oi{145}; \bsf{d}, \niii{}; \bsf{e}, \nii{122}; \bsf{f}, and \nii{205}. The plot style is the same as Fig.~\ref{f:data_z-IR}}
    \label{f:data_IR-line_IR}
\end{figure*}

The phenomenon of ``deficit'' is not unique to \cii{}. 
As previously reported \citep[e.g.,][]{D17}, all FIR FSLs (except \ci{}) show similar trends: the \lline{}/\lir{} ratio for \oiii{88}, \oi{63}, \oi{145}, \niii{}, \nii{122}, and \nii{205} declines beyond \lir{}~$\gtrsim$~10\textsuperscript{11}\,\lsun{} (Fig.~\ref{f:data_IR-line_IR}), with the magnitude and slope of the ``deficit'' varying by line—from two orders of magnitude for \oiii{88} to about one for \oi{145}. 
For \nii{} lines, dwarf galaxies fall below SF galaxies even in the non-``deficit'' regime. 
Both low- and high-\zz{} samples follow similar power-law indices, but the ``deficit'' branch is shifted to higher \lir{} for high-\zz{} galaxies.

These complex trends reflect a combination of factors, including the drivers of the \cii{} ``deficit'', variations in elemental abundance, and differences in ionization structure across galaxy types. 
A thorough exploration of the line ``deficit'' problem and its implications will be presented in \ppiii{}. 
Here, we emphasize the observational fact that all major FIR FSLs—regardless of whether they originate in neutral or ionized gas—exhibit ``deficit'' behavior at high \lir{}. 

\begin{figure}
    \centering
    \includegraphics[width=\halfwdth]{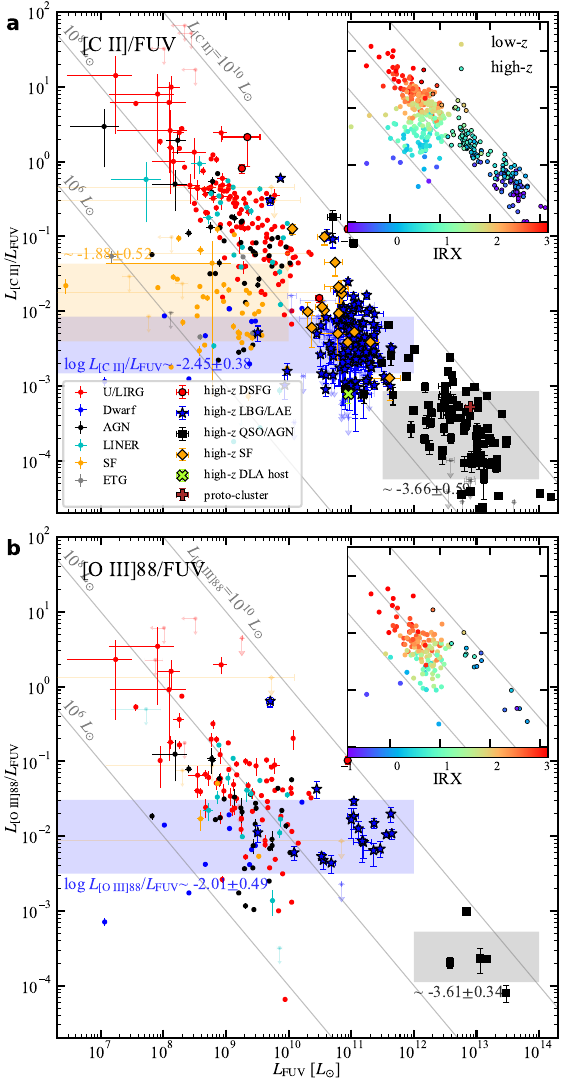}
    \caption[\lfuv{} vs. \lline{}/\lfuv{}. ]{FUV luminosity \lfuv{} vs. \bsf{a}, \cii{}/\lfuv{}; and \bsf{b}, \oiii{88}/\lfuv{}. The gray lines correspond to constant \lline{} values. The fitting results on line/FUV are shown as shades for low-\zz{} SFs (orange), dwarf \& LBG/LAEs (blue), and high-\zz{} QSOs (gray). The results and scatter are displayed beside shades. In the inset panels, IRX of each data point is plotted as color coded circles on the same coordinate, and the high-\zz{} data points are denoted with black edges.}
    \label{f:data_FUV-line_FUV}
\end{figure}

Finally, we compare the two strongest FIR FSLs, \cii{} and \oiii{88}, to \lfuv{} (Fig.~\ref{f:data_FUV-line_FUV}). 
Unlike the coherent trends seen with \lir{}, galaxies cluster by type in \lline{}/\lfuv{} versus \lfuv{} space, and galaxies with similar IRX span a broad range of \lfuv{}, except for local U/LIRGs and AGNs. 
The diagonal sequence followed by these dusty systems is a consequence of their selection and high IRX. 
The horizontal spread and clustering in other types reflect selection effects in \lfuv{} as well as the underlying \lline{}--\lfuv{} relation: a combination of UV photon-to-\lline{} correlation, and the stratified IRX distribution in non-IR selected galaxies. 
We fit \cii{}/FUV for low-\zz{} SF, low-\zz{} dwarfs plus high-\zz{} LBG/LAEs, and high-\zz{} QSOs, as well as \oiii{88}/FUV for low-\zz{} dwarfs plus high-\zz{} LBG/LAEs and high-\zz{} QSOs. 
These empirical relations may aid in galaxy classification and the prediction for designing future observations. 
The spread of LBG/LAE data points in the regimes of U/LIRGs and QSOs suggests a diverse population of these galaxies.

\subsection{Equivalent Width}
\label{sec:data_ew}

\begin{figure*}
    \centering
    \includegraphics[width=\textwidth]{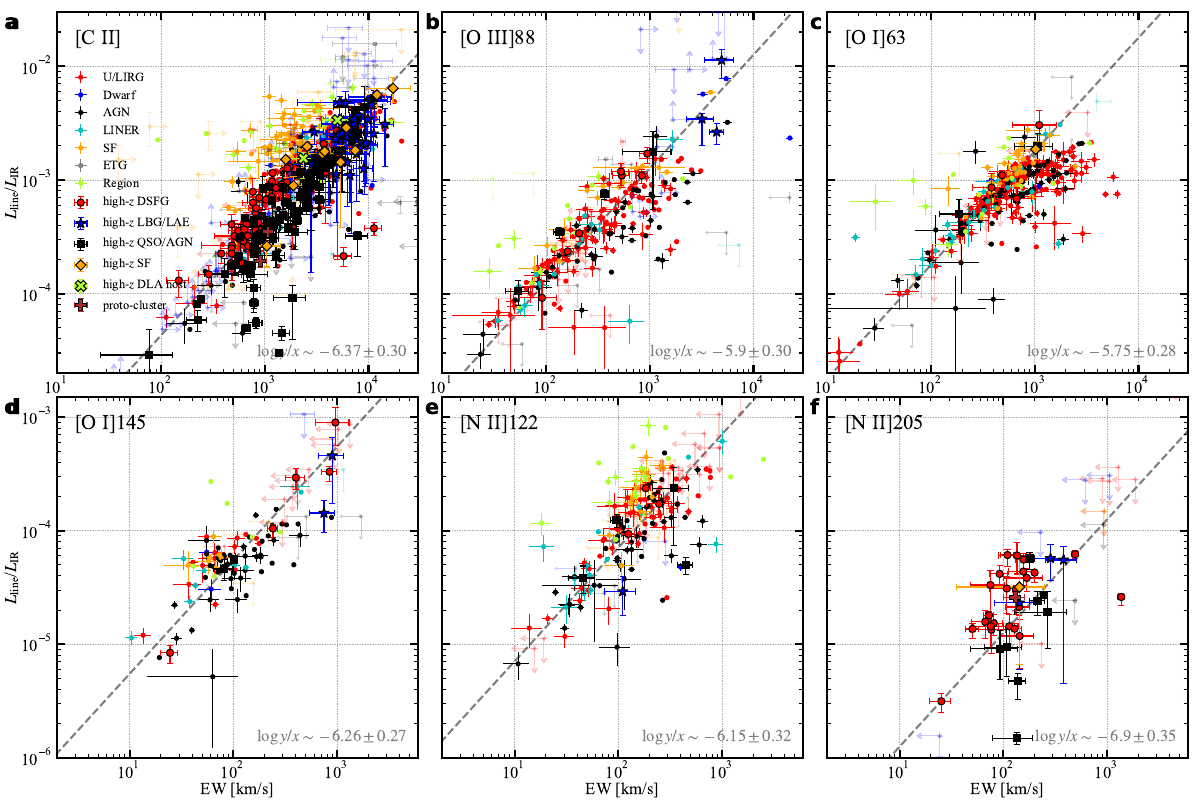}
    \caption[EW vs. \lline{}/\lir. ]{Line equivalent width vs. \lline{}/\lir{} for \bsf{a}, \cii{}; \bsf{b}, \oiii{88}; \bsf{c}, \oi{63}; \bsf{d}, \oi{145}; \bsf{e}, \nii{122}; \bsf{f}, \nii{205}. The dashed line in each panel corresponds to the linear fit, the fitting result and scatter are displayed at the lower right corner. The plot style is the same as Fig.~\ref{f:data_z-IR}.}
    \label{f:data_ew}
\end{figure*}

Given the significant time investment required for high-\zz{} FIR FSL observations, many galaxies only have a single spectral line measured. 
To maximize the scientific return from such observations, we examine the utility of the line equivalent width (EW) as a diagnostic. 
We define the rest-frame EW as $\mathrm{EW} = F_\mathrm{line}/F_{\lambda,\mathrm{cont}}\times c/\lambda_\mathrm{obs}$. 
This is mathematically equivalent to \lline{}/\vlv{}, and given the strong correlation between the monochromatic FIR continuum and \lir{}, the FSL EW serves as a first-order approximation for \lline{}/\lir{}. 

Fig.~\ref{f:data_ew} compares EW and \lline{}/\lir{} for the major FIR FSLs. 
Note that for \nii{205}, local galaxies often lack continuum measurements since most data were obtained with Herschel/SPIRE. 

The relation between EW and \lline{}/\lir{} exhibits scatter, typically $\gtrsim$0.3 dex, largely due to variations in \vlv{}/\lir{} driven by dust temperature. 
Additional scatter, especially affecting the \oi{63} EW,  arises from continuum measurement uncertainties with PACS spectrometer. 
Nevertheless, at wavelengths near the SED peak (e.g., 88\,\um{}), the correlation between EW and \lline{}/\lir{} holds well over more than two orders of magnitude. 
This demonstrates that in some cases, EW can provide a model-independent estimate of \lline{}/\lir{}, which is useful both for physical interpretation and for planning future observations. 

\begin{figure}
    \centering
    \includegraphics[width=\halfwdth]{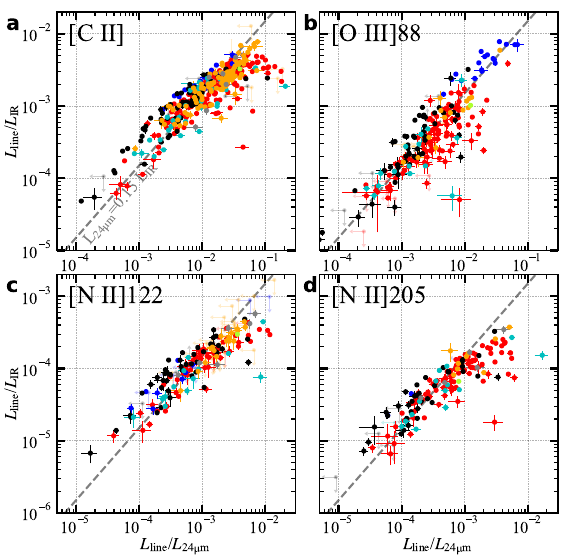}
    \caption[\lline/$L_\mathrm{24\mu m}$ vs. \lline{}/\lir{}. ]{\lline{}/24\,\um{} luminosity vs. \lline{}/\lir{} for \bsf{a}, \cii{}; \bsf{b}, \oiii{88}; \bsf{c} and \bsf{d}, \nii{} doublets. The dashed lines represent the median $L_\mathrm{24\mu m}$/\lir{} value of 0.15.}
    \label{f:data_L24}
\end{figure}

In addition to EW, another practical line-to-continuum diagnostic is the ratio of line luminosity to 24\,\um{} continuum ($L_\mathrm{24\mu m}$). 
The 24\,$\mu$m emission, associated with warm dust, is a well-established tracer of star formation \citep{kennicutt09}, and previous studies have found resolved relations between \nii{205} and 24\,\um{} emission \citep{hughes16}. 
Fig.~\ref{f:data_L24} shows \lline{}/$L_\mathrm{24\mu m}$ for \cii{}, \oiii{88}, and the \nii{} doublets as a function of \lline{}/\lir{}.

All lines display a strong linear proportionality (gray dashed line) over nearly three orders of magnitude, indicating that the integrated \lline{}/$L_\mathrm{24\mu m}$ ratio is primarily set by \lline{}/\lir{}. 
This suggests that the resolved \nii{205}--24\,\um{} relation does not hold globally, likely because the variation in $L_\mathrm{24\mu m}$/\lir{} ($\sim$0.3 dex; see Fig.~\ref{f:data_s60_100-Lcont_LIR}) is small compared to the $\sim$2 orders of magnitude of the \nii{205} ``deficit'' (Fig.~\ref{f:data_IR-line_IR}). 
The slopes for \cii{} and the \nii{} doublets are slightly shallower than unity, reflecting the impact of dust temperature on both the line ``deficit'' and the 24\,\um{}/IR ratio.

\subsection{Density Diagnostics}
\label{sec:data_density}

\begin{figure}
    \centering
    \includegraphics[width=\halfwdth]{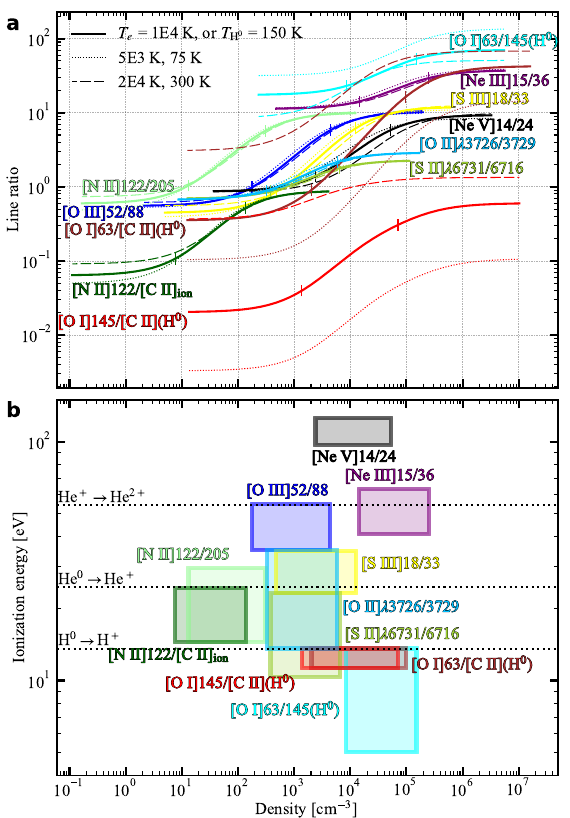}
    \caption[Density diagnostics. ]{\bsf{a}, density diagnostic curves of \nev{14/24} (black), \neiii{15/16} (purple), \oiii{52/88} (blue), \siii{18/33} (yellow), \nii{122/205} (pale green), \nii{122}/\cii{}\textsubscript{ion} (dark green), \siio{6731/6716} (yellow green), \oiio{3726/3729} (sky blue) \oi{63}/\cii{}(\hneut{}) (brown), \oi{145}/\cii{}(\hneut{}) (red), and \oi{63/145}(\hneut{}) (cyan). The theoretical line ratios are plotted against the density of the collisional companion. The solid lines correspond to line ratios computed at \te{} = 10\textsuperscript{4} K (or \thi{} = 150 K if excited by \hneut{}), the dashed and dotted lines are the theoretical curves at twice higher or lower temperatures. The two vertical ticks on each curve mark the applicable density range, or the near-linear part, of each diagnostic, approximated as 20 and 80\% in logarithm space between the low- and high-density limits. \bsf{b}, boxes and shades manifest the applicable density range on x-axis, creation energy and ionization/dissociation potential of the ions on y-axis, in the same color as (\bsf{a}). The ionization potential of \hneut{}, He\textsuperscript{0} and He\textsuperscript{+} are denoted by the horizontal dotted lines. }
    \label{f:data_ne}
\end{figure}

\begin{figure*}
    \centering
    \includegraphics[width=\halfwdth]{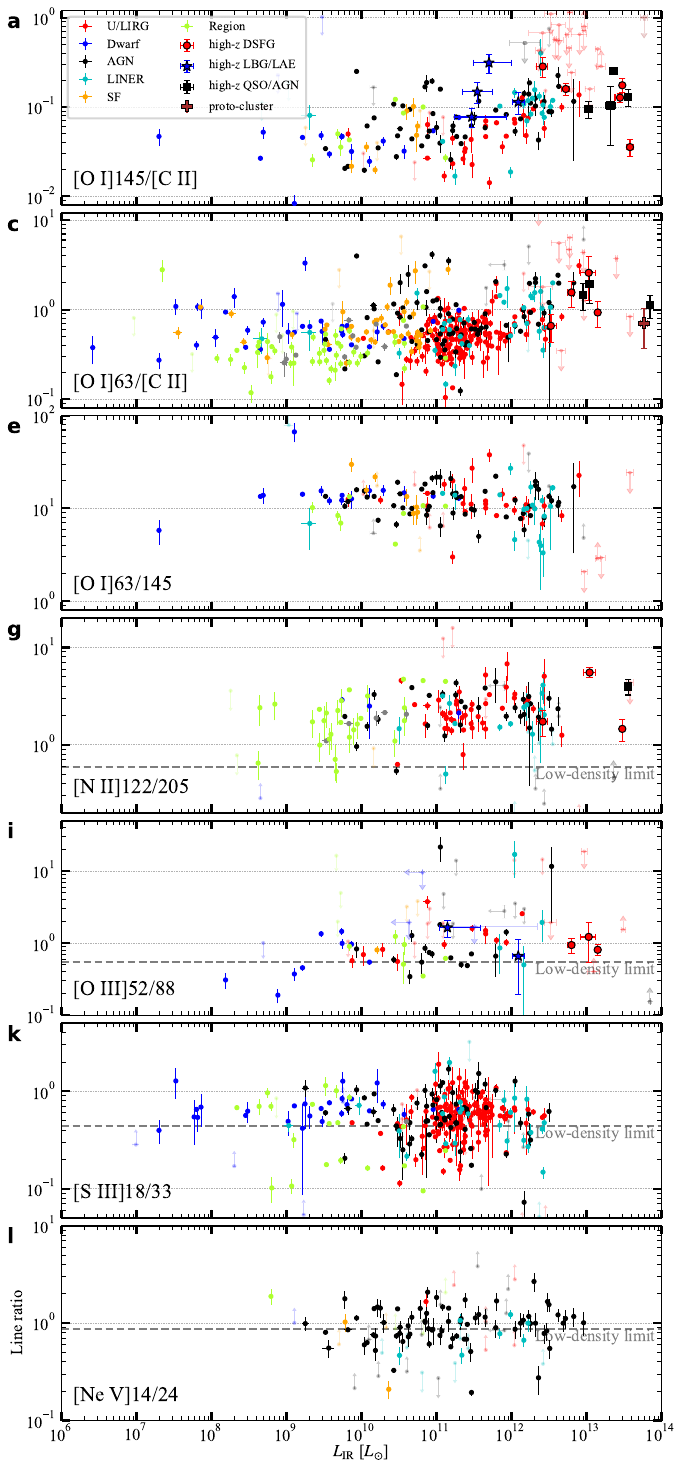}
    \includegraphics[width=0.442\textwidth]{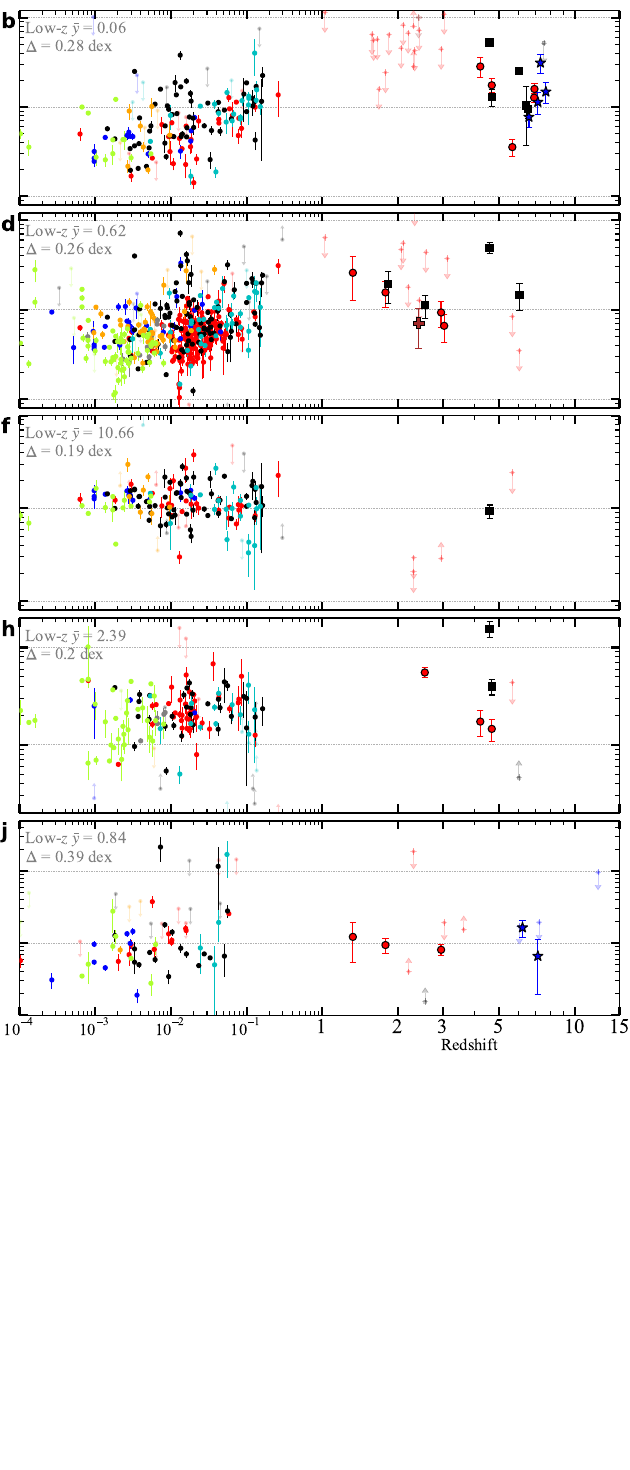}
    \caption[Demographics of density tracers. ]{Demographics of all the density related line ratios displayed along with \lir{} (left column) and redshift (right column). The median value fit on low-\zz{} galaxies is printed at the upper left corner of panels in the right column.}
    \label{f:data_IR-n}
\end{figure*}

Forbidden lines are sensitive to the density of their collisional partners due to the nature of collisional excitation. 
This property makes them valuable probes of the ISM density, as demonstrated by numerous studies \citep[c.f.][]{simpson75,S15,herrera16}. 
Density diagnostics commonly employ pairs of spectral lines from transitions of the same ion that share a common energy level, minimizing the influence of abundance, ionization, and temperature (see Eq.~\ref{equ:ratio}). 
Most such ratios trace the electron density, since electrons are typically the dominant collisional partners for ions. 
In Fig.~\ref{f:data_ne}(\bsf{a}), we plot diagnostic curves for several commonly used density-sensitive line ratios, including FIR, MIR, and optical lines.

Theoretical emissivity ratios in these curves are calculated at fixed temperature but varying density. 
Emissivities are computed using \pyneb{} \citep{Luridiana15} for electron excitation, or by solving two- or three-level atomic detailed balancing matrices for excitation by atomic hydrogen (\hneut{}) in the case of \cii{} and \oi{}. 
For \hneut{} excitation, rate coefficients are adopted from \citet{barinovs05} and \citet{abrahamsson07}, with fitting functions from \citet{draine11}. 
Several caveats apply to the diagnostic curves: (1) The \nii{}/\cii{}\textsubscript{ion} ratio is included for completeness, where \cii{}\textsubscript{ion} represents the component of \cii{} emission from ionized gas excited by electrons; (2) \oi{63/145} and \oi{}/\cii{}(\hneut{}) are computed assuming atomic hydrogen as the collisional companion and considering only the neutral gas emission of \cii{}; (3) Density diagnostics involving \cii{} depend on elemental abundances, for which solar C/O and N/O values from \citet{asplund09} are adopted; however, we caution that N/O can vary significantly among galaxies. 

Among the available density tracers, some are more effective than others. 
A key figure of merit is the dynamic range of the line ratio as a function of density. 
For example, ratios such as \siio{6716/6731} and \neiii{15/36} vary by less than a factor of 3 over their full density range. 
Temperature sensitivity is also important: most \edens{} diagnostics are relatively insensitive to \te{}, but neutral density diagnostics involving \oi{} are rather sensitive to neutral gas temperature (\thi{}), since the excitation potentials of the \oi{} lines \citep[see][table 1]{stacey11} are close to typical ISM \thi{}. 
For such ratios, a factor of 4 change in \thi{} can mimic a two-orders-of-magnitude change in \hdens{}, making the ratio effectively an estimate of pressure instead of just a density probe \citep{osterbrock89}. 

Another consideration is the position of the ions in the multiphase ISM structure, which, in turn is bounded by the ionization or dissociation energies of the relevant atomic species. 
The shaded regions in Fig.~\ref{f:data_ne}(\bsf{b}) illustrate where these diagnostics apply within the ISM as a function of ionization energy and gas density. 
For example, \nev{} is only present in highly ionized regions, while \oi{} ratios probe neutral gas. 

\begin{figure}
    \centering
    \includegraphics[width=\halfwdth]{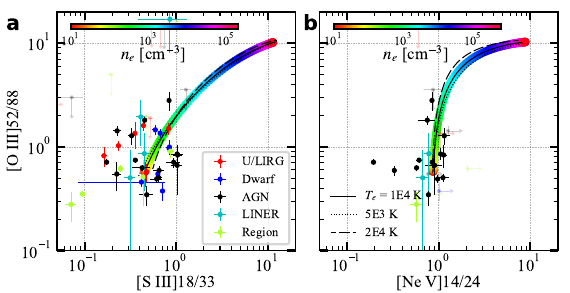}
    \includegraphics[width=\halfwdth]{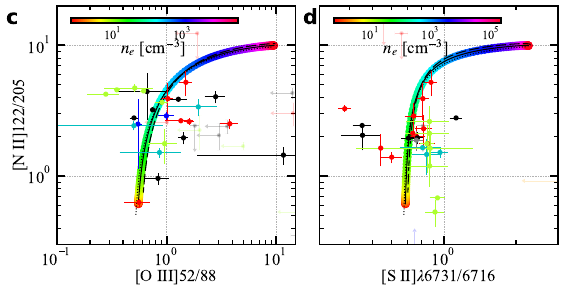}
    \includegraphics[width=\halfwdth]{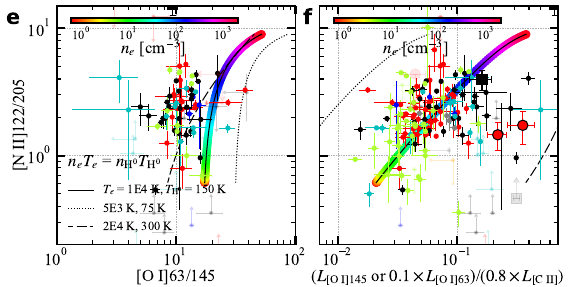}
    \caption[Comparison between different density diagnostics. ]{Comparison between different density diagnostics. Top: \oiii{52/88} vs. high-ionization diagnostics \bsf{a}, \siii{18/33}; and \bsf{b}, \nev{14/24}; middle: \nii{122/205} vs. \bsf{c}, \oiii{52/88}; and \bsf{d}, \siio{6731/6716}; bottom: \nii{122/205} vs. \bsf{e}, \oi{63/145}; and \bsf{f}, \oi{145}/\cii{}(\hneut{}), where 0.1$\times$\loi{63} is used to approximate \loi{145} if the latter is not available, and a typical \fciin{}~=~0.8. For the \edens{} tracers in the upper two rows, the solid, dashed, and dotted theoretical curves are plotted assuming \te{} is indicated in the upper right panel, with a colorbar representing \edens{} of each point along the solid line. For \edens{}-to-\hdens{} comparison in the bottom row, \thi{} values are assumed as noted in the legend in the lower left panel.}
    \label{f:data_ne-ne}
\end{figure}

The range of densities probed by a given diagnostic also affects its utility. 
For example, only \nii{} ratios can effectively probe low-density ionized gas. 
In Fig.~\ref{f:data_ne} (\bsf{b}), we approximate the usable part of each curve as 20 and 80\% in the logarithmic space of the low- and high-density limits, and mark them as vertical ticks in (\bsf{a}), and horizontal spans of shades in (\bsf{b}). 
To assess the reliability and applicability of these diagnostics, we compare high-ionization diagnostics against \oiii{52/88}, low-ionization diagnostics against \nii{122/205}, and \nii{122/205} versus \oiii{52/88} in Fig.~\ref{f:data_ne-ne}. 
Theoretical curves, assuming the same density for both ions, are overlaid for reference; deviations from these curves may indicate density stratification between different ionization zones. 
The characteristic “turnover” in the theoretical curves reflects mismatched sensitivity ranges for the two diagnostics: the horizontal branch corresponds to the regime where the x-axis tracer is most sensitive to density, while the vertical branch indicates greater sensitivity in the y-axis tracer.

Several caveats apply to these comparisons. 
In the lowest row of Fig.~\ref{f:data_ne-ne}, four degrees of freedom are involved (\edens{}, \te{}, \hdens{}, \thi{}), which we reduce by fixing \te{}/\thi{}~=~66.6 (so that for \te{}~=~10\textsuperscript{4}\,K, \thi{} = 150\,K; see \citealt{S15,osterbrock89}) and by assuming thermal equilibrium (\edens{}\te{} = \hdens{}\thi{}). 
For \oi{}/\cii{}(\hneut{}), we use \loi{145} where available (as it is less affected by self-absorption), or else \loi{63} multiplied by the median \oi{145/63} value of 0.1. 
We also assume solar C/N abundance and a \cii{} neutral fraction (\fciin{}) of 80\%. 
Further discussion of \fciin{} is provided in \ppiii{}. 
We do not compare \neiii{15/36} due to the scarcity of \neiii{36} data, which lies at the edge of Spitzer/IRS coverage \citep{houck04}.

A key diagnostic comparison is between \oiii{52/88} and \nii{122/205}. 
We observe substantial scatter, with a slight tendency for data to lie to the right of the theoretical curve, possibly suggesting $n_{e,\mathrm{O^{2+}}} > n_{e,\mathrm{N^{+}}}$. 
However, we caution that uncertainties in \oiii{52/88} are large, and most points are close to the low-density limit of \oiii{52/88}.

For high-density tracers (\nev{14/24}, \siii{18/33}, \siio{6731/6716}), no clear trends emerge with \oiii{52/88}. 
Most values cluster near the low-density limits, and the scatter is larger than expected from theory alone. 
The implications of this behavior—particularly the lack of correspondence between high-ionization tracers and others, and the tendency to cluster near the low-density limit—are discussed in more detail in Sec.~\ref{sec:data_density_bias}.

For neutral gas, \oi{63/145} does not correlate with \nii{}, likely reflecting its insensitivity to both \hdens{} and \thi{} (see Fig.~\ref{f:data_ne}). 
In contrast, \oi{145}/\cii{} shows a positive correlation with \nii{122/205}. 
However, our theoretical calculations cannot simultaneously reproduce both the observed \oi{63/145} and \oi{145}/\cii{} ratios. 
Although the higher \thi{} (long-dashed curve) is consistent with the low observed \oi{63/145} (median $\sim$10), it significantly overestimates \oi{145}/\cii{}. 
Two factors complicate the interpretation: (1) \oi{63} is often affected by self-absorption \citep{stacey83,boreiko96}, and optical depths $\tau \sim 1$ can bring observed \oi{63/145} in line with the \thi{}~=~150\,K model; (2) the strong \thi{} dependence of these ratios means our simplified assumption relating \thi{} and \te{} may not hold. 
The observed trend may indicate that the neutral gas is hotter when ionized gas is denser (i.e., an \edens{}--\thi{} correlation), but current data do not allow us to break the \hdens{}--\thi{} degeneracy. 
We defer further discussion to the comparison with the photoionization models in \ppii{}.

For high-\zz{} galaxies, only three have both \nii{} doublet and \oi{}/\cii{} measurements. 
Two of these are offset from the local galaxy trend. 
While this small sample precludes a robust conclusion regarding redshift evolution in the \nii{122/205}--\oi{}/\cii{} plane, there is an indication that \oi{}/\cii{} is systematically higher in high-\zz{} galaxies, which we explore further in Sec.~\ref{sec:data_sim} and \ppiii{}.

\subsection{Radiation Field and AGN Decomposition}
\label{sec:data_radiation}

\begin{figure*}
    \centering
    \includegraphics[width=\halfwdth]{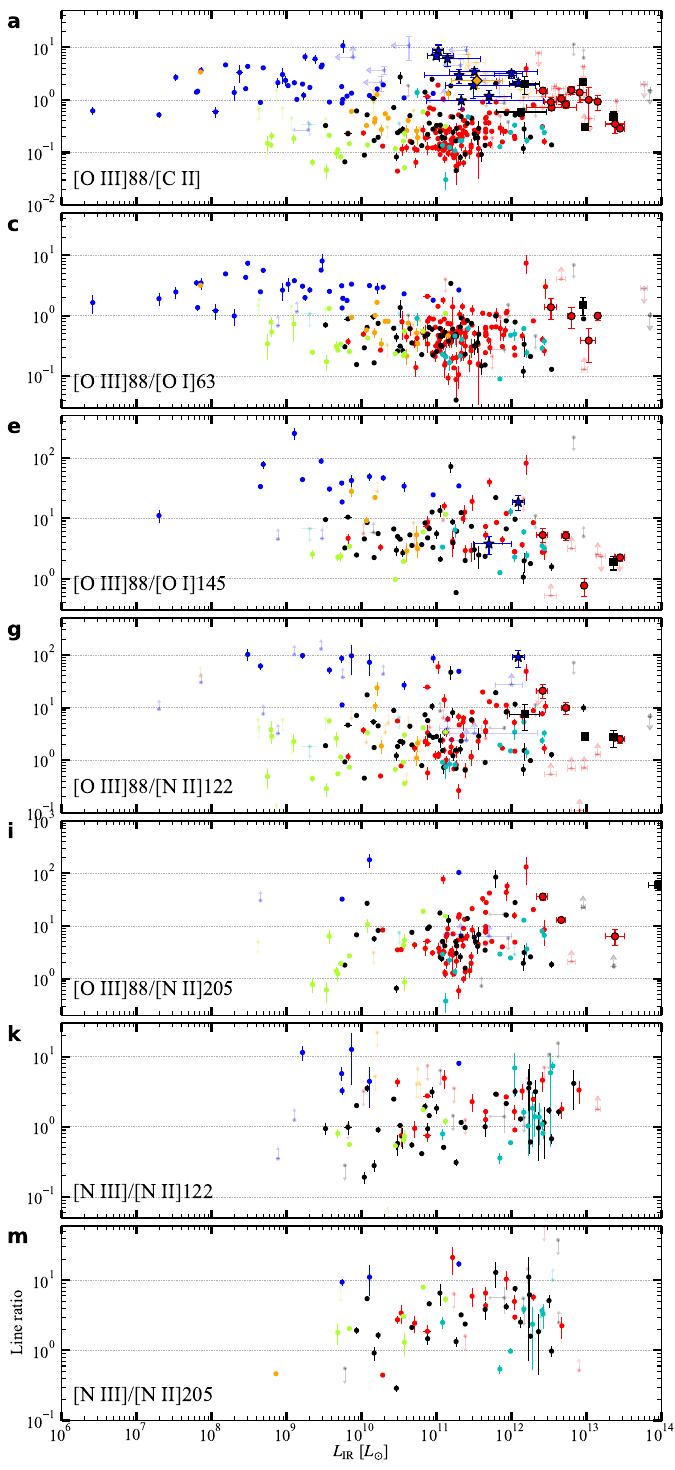}
    \includegraphics[width=0.442\textwidth]{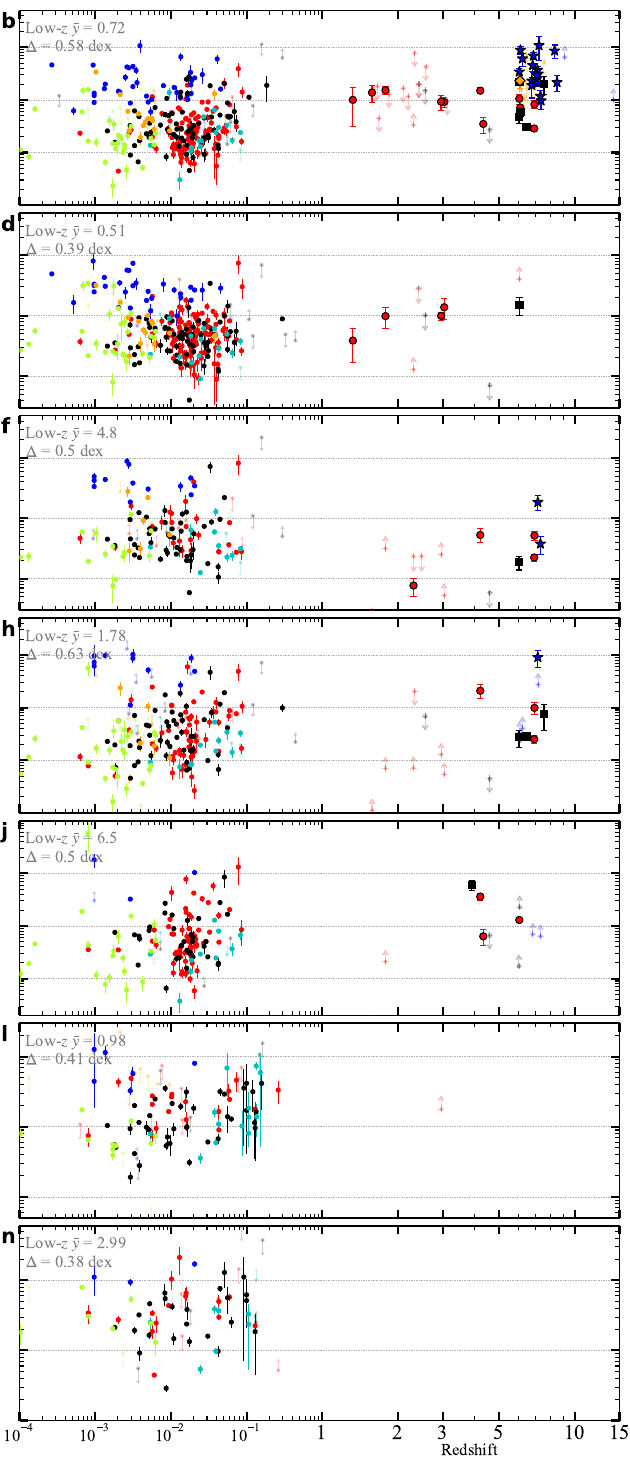}
    \caption[Demographics of FIR radiation field diagnostics.]{Demographics of FIR radiation field diagnostics.}
    \label{f:data_IR-U}
\end{figure*}

\begin{figure}
    \centering
    \includegraphics[width=\halfwdth]{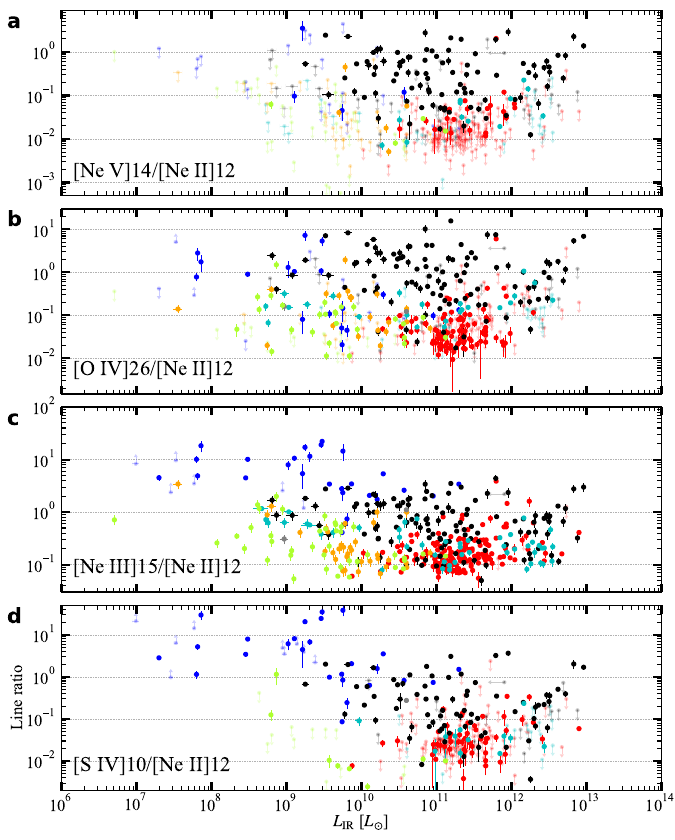}
    \caption[Demographics of MIR radiation field diagnostics.]{Demographics of MIR radiation field diagnostics.}
    \label{f:data_IR-U_MIR}
\end{figure}

Comparisons of emission from different ionization states have been extensively used to probe the radiation field conditions in the ISM. 
This is because the ICF term in Eq.~\ref{equ:ratio} reflects the ionization structure of the emitting gas. 
More specifically, ICF is influenced by two main factors: (1) the hardness of the radiation field, often quantified as the fraction of He-ionizing photons among all ionizing photons (\hardness{}); and (2) the photoionization parameter $U$, defined as the ratio of ionizing photon density to electron density, which sets the balance between ion creation (via photoionization) and destruction (via recombination). 

MIR FSLs are particularly valuable for studying the radiation field, as some emitting ions, such as O$^{3+}$ and Ne$^{4+}$, require photons with $h\nu > 50\,\mathrm{eV}$, making them sensitive tracers of radiation field hardness. 
FIR FSLs and optical lines typically trace lower ionization states, but FIR FSLs provide less temperature-biased measurements of $U$ compared to the latter. 
In this section, we demonstrate the behavior of common radiation field diagnostics across our sample, focusing on the photoionization condition typical to star-forming regions. 
We will discuss the impact of AGN on FSLs in Sec.~\ref{sec:data_agn} 

\begin{figure}
    \centering
    \includegraphics[width=\halfwdth]{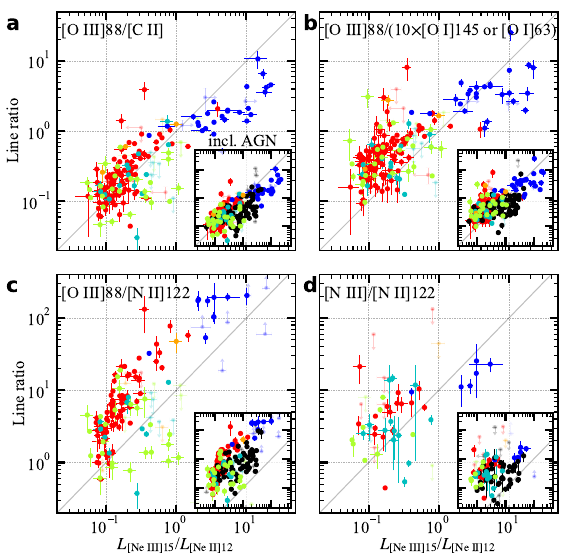}
    \caption[\neiii{15}/\neii{12} vs. radiation field diagnostics. ]{\bsf{a}, \neiii{15}/\neii{12} vs. \oiii{88}/\cii{}; \bsf{b}, \oiii{88}/(10$\times$\oi{145} or \oi{63}); \bsf{c}, \oiii{88}/(\nii{205} or 0.5$\times$\nii{122}); \bsf{d}, \niii{}/(\nii{205} or 0.5$\times$\nii{122}). The diagonal dashed line represent the slope of unity as a visual aid. The \neiii{15} and \oiii{88} luminosities plotted in the main panels have been corrected for AGN contribution, and the same figure without applying AGN correction can be found in the inset panels.}
    \label{f:data_hardness-U}
\end{figure}

We consider the utility of \oiii{88} and \niii{} as radiation field diagnostics. 
Ratios explored include \oiii{88}/\cii{}, \oiii{88}/\oi{}, \oiii{88}/\nii{}, and \niii{}/\nii{}. 
For \oi{} and \nii{}, we combine the doublets as previously described to maximize the sample size. 
We note that \oiii{88}/\nii{} depends on both the ionization structure and the N/O abundance. 

Fig.~\ref{f:data_hardness-U} compares these FIR ratios to the high-ionization MIR ratio \neiii{15}/\neii{12}, a common tracer of radiation field hardness. 
Positive correlations across two orders of magnitude empirically confirm that \oiii{88}/\cii{} is primarily correlated with the ionizing radiation field properties. 
Galaxies identified as AGN-host are excluded from this comparison for the AGN contamination to \neiii{} and \oiii{} (see Sec.~\ref{sec:data_agn} for more details), while the inset panels show the result with AGNs, showing AGNs cluster offset from the main correlation.

The FIR diagnostics \oiii{88}/\cii{} and \oiii{88}/\oi{} exhibit slopes less than unity compared to \neiii{15}/\neii{12}, as expected: \neiii{15} requires harder photons and is thus more sensitive to radiation hardness than \oiii{88}. 
The \oiii{88}/\nii{} ratio shows a nearly linear relation, but we caution that a lower N/O abundance in dwarfs leads to higher \oiii{88}/\nii{} values and a steeper slope. 
The \niii{}/\nii{} comparison is limited by data availability and large uncertainty, but generally shows weaker dependence on hardness.

\begin{figure}
    \centering
    \includegraphics[width=\halfwdth]{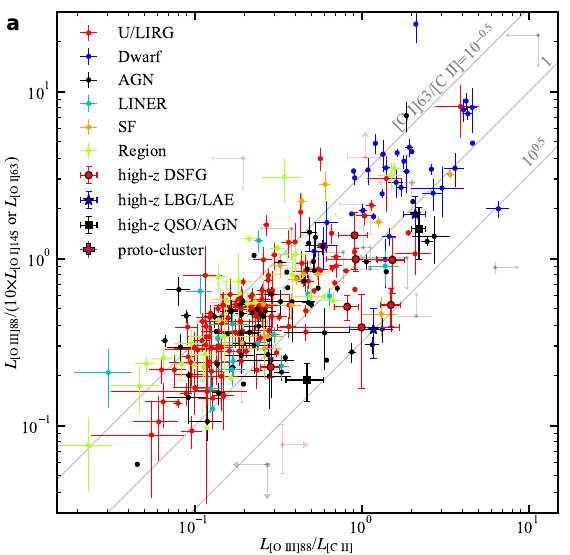}
    \includegraphics[width=\halfwdth]{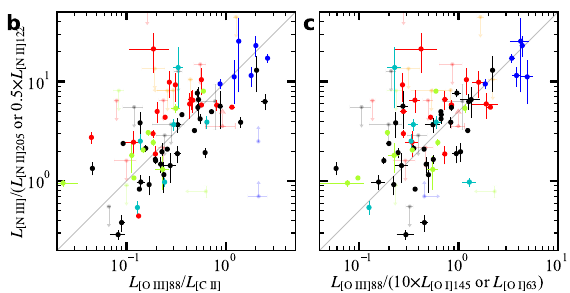}
    \caption[Comparison between radiation field diagnostics. ]{\bsf{a}, \oiii{88}/\oi{} vs. \oiii{88}/\cii{}. In lower row, \niii{}/\nii{} vs. \bsf{b}, \oiii{88}/\cii{}; and \bsf{c}, \oiii{88}/\oi{}. The diagonal dashed lines have slope of unity, and are shown as visual aids.}
    \label{f:data_U-U}
\end{figure}

We further compare these FIR ratios among themselves in Fig.~\ref{f:data_U-U}. 
\oiii{88}/\cii{} shows a tight relation to \oiii{88}/\oi{}, which partly reflects the small variation in \oi{}/\cii{} across the sample. 
High-\zz{} galaxies are offset, caused by elevated \oi{}/\cii{}. 
The \niii{}/\nii{} ratio also correlates with both \oiii{88}/\cii{} and \oiii{88}/\oi{}, supporting the use of \oiii{88}/\cii{} as a probe for the radiation field. 
Although the scatter is larger compared to \oiii{88}/\oi{}, the agreement between \niii{}/\nii{} and other ionized-to-neutral gas diagnostics suggests a similarity between the ionized gas structure and the ionized-to-neutral gas structure. 
This implies a strong correlation among low-ionization and neutral gas emission, which is explored further in \ppiii{}.

In summary, our comprehensive analysis demonstrates that FIR fine-structure line ratios—especially \oiii{88}/\cii{}—are robust empirical tracers of radiation field properties, with strong connections to both ionization structure and, potentially, the link between ionized and neutral gas phases.

\subsection{Abundance}
\label{sec:data_abundance}

\begin{figure*}
    \centering
    \includegraphics[width=\halfwdth]{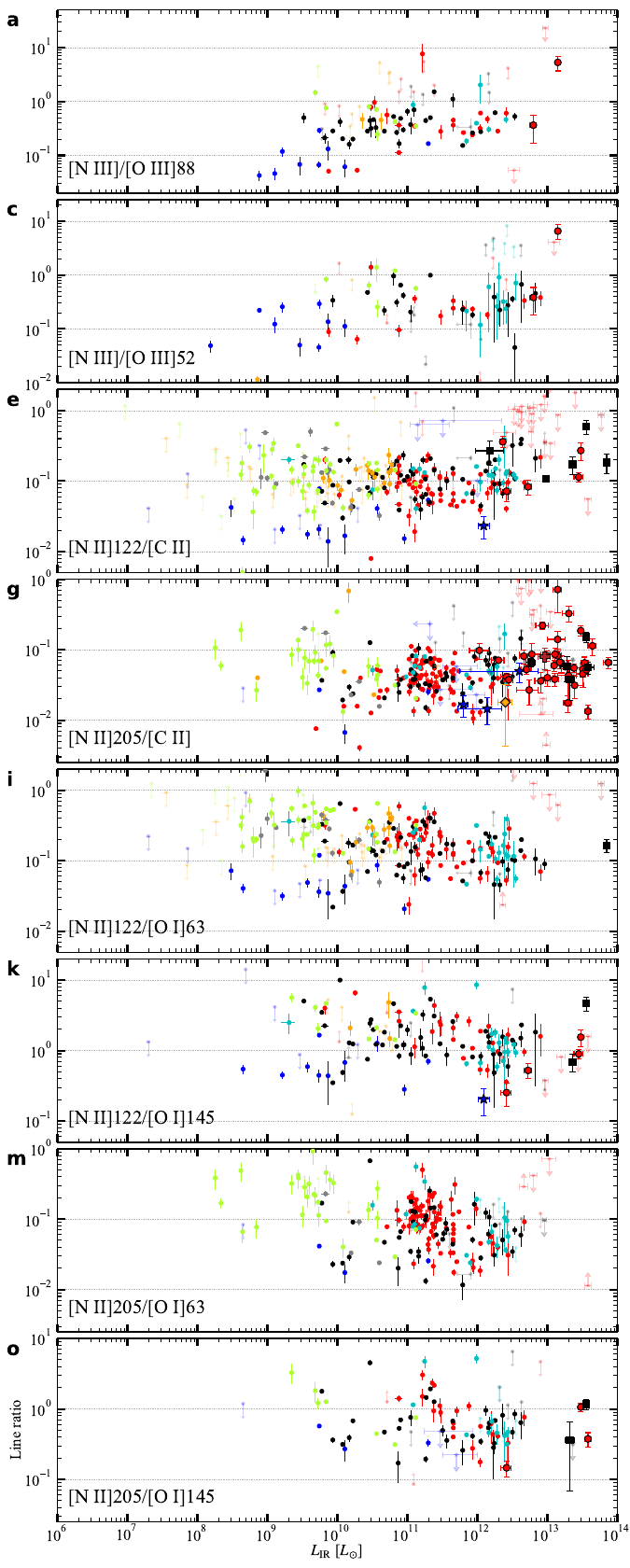}
    \includegraphics[width=0.442\textwidth]{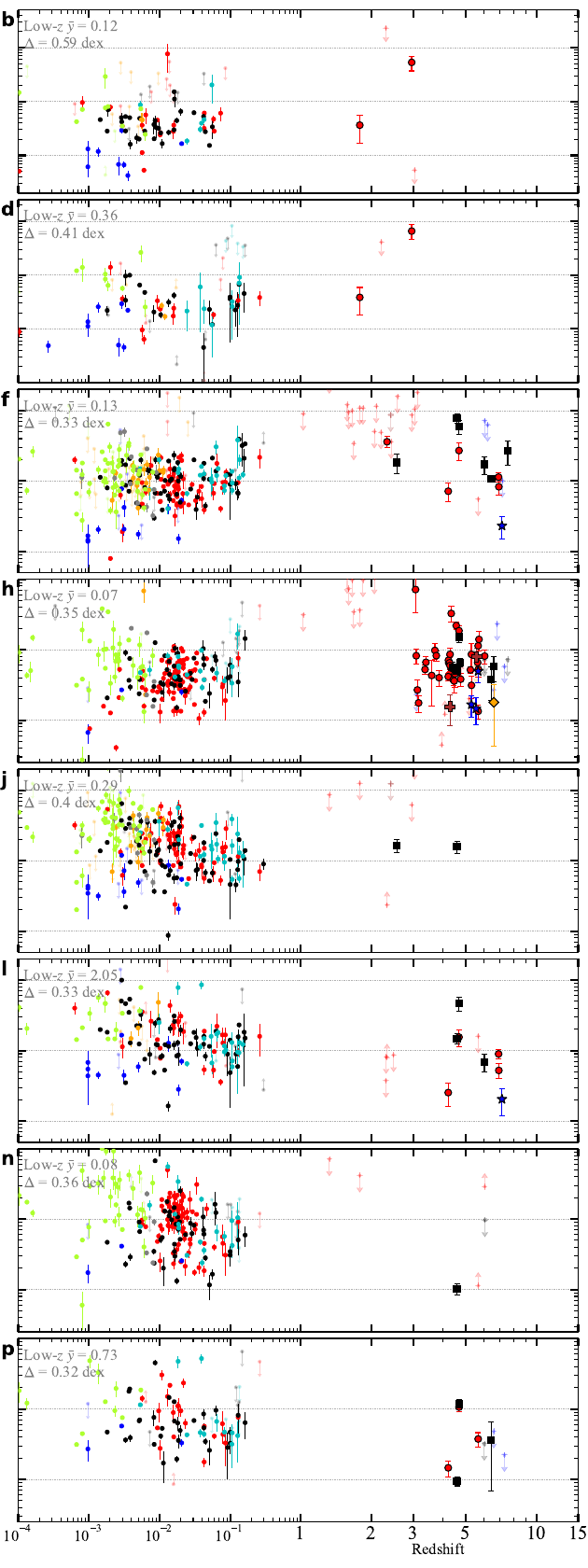}
    \caption[Demographics of all abundance diagnostics. ]{Demographics of all abundance diagnostics. }
    \label{f:data_IR-metallicity}
\end{figure*}

\begin{figure}
    \centering
    \includegraphics[width=\halfwdth]{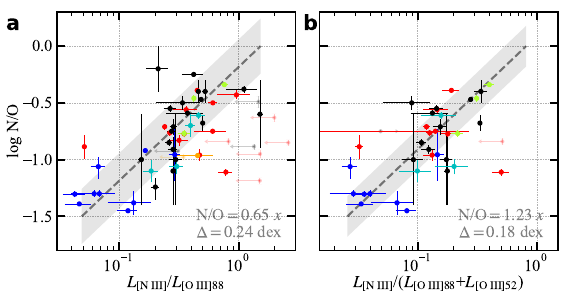}
    \caption[N/O vs. \niii{}/\oiii{}. ]{N/O abundance ratio vs. \bsf{a}, \niii{}/\oiii{88}; and \bsf{b}, \niii{}/(\oiii{88}+\oiii{52}. The gray dashed lines and shades are the linear fits, which are described at the lower right corners. The plot style is the same as Fig.~\ref{f:data_z-IR}}
    \label{f:data_N_O-NIII_OIII}
\end{figure}

Abundance has a direct impact on the observed line ratios and plays a central role in regulating cooling in the ISM. 
Traditionally, elemental abundances have been studied via optical lines, either through empirical correlations between metallicity and strong-line indices \citep[e.g., N2, S2, N2S2, O3N2;][]{denicolo02,kewley02,pettini04} or, where feasible, by directly comparing the abundances of oxygen ions to hydrogen \citep[e.g., via the $R_{23}$ method][]{mcgaugh91}, provided the electron temperature can be measured. 
However, these methods face significant challenges in metal-rich systems, including strong sensitivity to \te{} (R23, N2O2), the need for extinction correction, reliance on proxy elements (S2, N2S2), and the assumption of a tight N/O correlation (N2, O3N2).
Given the relatively subtle density variations found in Sec.~\ref{sec:data_density}, FIR FSLs experience little change in emissivity with changing physical conditions, and are therefore good tracers of the emitting ion's abundance. 
In this subsection, we explore the impact of abundance on FSLs and demonstrate that many of these lines are so strongly correlated with elemental abundances that they can serve as direct metallicity diagnostics.

It is important to note that the abundances used here are predominantly derived from optical strong line methods, which themselves have systematic uncertainties of 0.1-0.2 dex \citep{perez09,D19,stasinska19}. 
These systematic errors are not included in the reported uncertainties. 
Furthermore, many metallicity values collected in this work lack error bars in the literature. 
Therefore, a minimum scatter of $\sim$0.1 dex is expected in any abundance-related comparison. 

Although FIR lines avoid strong temperature dependence and bias toward hot regions, they are limited by the lack of a direct hydrogen proxy in the FIR or submillimeter. 
As a result, abundance determinations require either additional data of recombination lines (e.g., H$\alpha$, Pa$\alpha$, Br$\alpha$), which are still affected by dust attenuation, or free-free emission, which is extinction-free but difficult to disentangle from dust and synchrotron emission. 
Alternatively, N/O can be measured directly in the FIR, as in optical N2S2, N2, and O3N2 indices, and then used either to study nitrogen enrichment or infer O/H via an N/O–O/H relation.

The FIR \niii{}/\oiii{} line ratio (N3O3) is an excellent N/O diagnostic \citep{lester83,P21}, benefiting from the co-spatial origin of \npp{} and \opp{} and from direct density constraints via \oiii{52/88}. 
We show the correlation of \niii{}/\oiii{88} with N/O in Fig.~\ref{f:data_N_O-NIII_OIII}. 
The \niii{}/\oiii{88} ratio correlates tightly with N/O (scatter $\sim$0.25 dex). 
The best-fit linear relation, N/O~=~0.66~$\times$~\niii{}/\oiii{88}, is fully consistent with the theoretical emissivity ratio $\varepsilon_{\mathrm{[N~III]}}/\varepsilon_{\mathrm{[O~III]88}} = 1.47$ computed at \te{}~=~10\textsuperscript{4}\,K and \edens~=~50\,\cc{} using \pyneb{}.

Density corrections using the \oiii{52} line have led to alternative N3O3 indices that employ linear combinations of \oiii{88} and \oiii{52} in the denominator \citep[e.g.,][]{S22,C22}. 
We show one such form, \niii{}/(\oiii{88}+\oiii{52}), in Fig.~\ref{f:data_N_O-NIII_OIII}(\bsf{b}). 
The scaling factor of 1.23 again matches the theoretical emissivity ratio calculated to be 1.14. 
The scatter appears slightly reduced (0.19 dex), but this improvement arises only from sample selection (limited to those with \oiii{52} detections). 
When using the same sample, the scatter for the \niii{}/\oiii{88} calibration also drops to 0.18 dex.

Although density effects are non-negligible in theory, for example, $\varepsilon_{\mathrm{[N~III]}}/\varepsilon_{\mathrm{[O~III]88}}$ increases from 1.39 to 1.78 as \edens{} increases from 10 to 250\,\cc{}, the inclusion of \oiii{52} does not practically lead to tighter relations in practice, and introduces additional observational uncertainties. 
Furthermore, as discussed in Sec.~\ref{sec:data_density}, \oiii{52/88} is a poor density diagnostic in ionized ISM, and the true variation in \edens{} is small. 
We therefore recommend the simpler and more accessible \niii{}/\oiii{88} ratio for N/O measurements in extragalactic systems.

\begin{figure}
    \centering
    \includegraphics[width=\halfwdth]{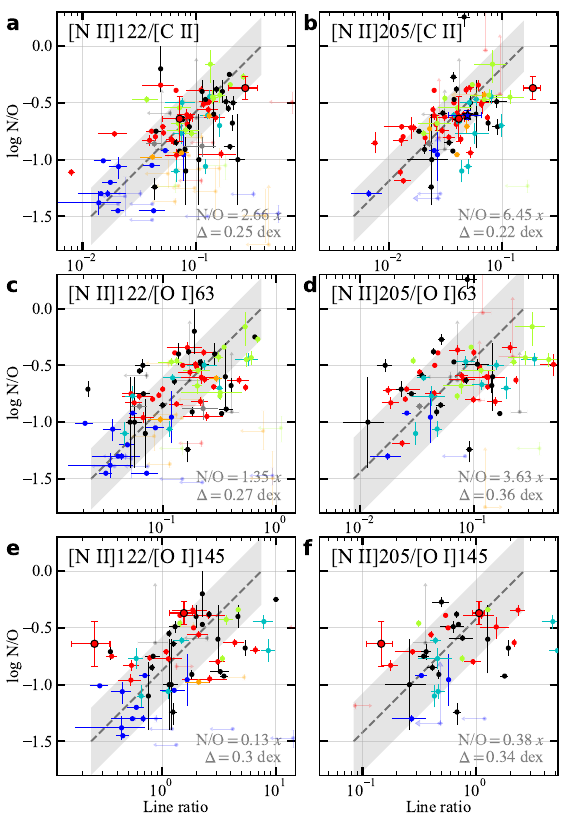}
    \caption[N/O vs. \nii{} based diagnostics. ]{N/O abundance ratio vs. \bsf{a} \& \bsf{b}, \nii{}/\cii{}; \bsf{c} \& \bsf{d}, \nii{}/\oi{63}; \bsf{e} \& \bsf{f}, \nii{}/\oi{145}. \nii{122} is compared in the left column, and \nii{205} in the right column. The gray dashed lines and shades are the linear fit result and residual scatter, which are also reported at the lower right corner. The plot style is the same as Fig.~\ref{f:data_z-IR}}
    \label{f:data_N_O-NII_CII}
\end{figure}

N/O abundance also affects the ratios between \nii{} and other low-ionization lines. 
In Fig.~\ref{f:data_N_O-NII_CII}, we show N/O versus \nii{}/\cii{} and \nii{}/\oi{} doublets, along with linear fits. 
All show clear positive correlations with N/O, with small scatters (0.25–0.3 dex), demonstrating the dominant role of abundance in determining \nii{} luminosities. 
Traditionally, \cii{}/\nii{} has been used to estimate the fraction of \cii{} emission arising from neutral gas (\fciin{}), but previous extragalactic work has often ignored the variation in nitrogen abundance. 
We argue in \ppiii{} that \cii{}/\nii{} is primarily a measure of N/O, with only a secondary dependence on \fciin{}. 

Correlations using \nii{122} are generally tighter than those using \nii{205}, for several reasons: (1) dwarf galaxies (which provide the largest range in \no{}, down to $\sim$-1.5) only have \nii{122} data; (2) \nii{122} data are typically higher quality due to better PACS sensitivity; (3) SPIRE, used for observing \nii{205}, has a smaller beam than PACS, introducing extra uncertainty. 
Among low-ionization ratios, \nii{}/\cii{} is slightly tighter than \nii{122}/\oi{}, likely due to intrinsic variation in \cii{}/\oi{} (see Sec.~\ref{sec:data_density}) and the lower quality of \oi{145} data compared to \oi{63}.

\begin{figure}
    \centering
    \includegraphics[width=\halfwdth]{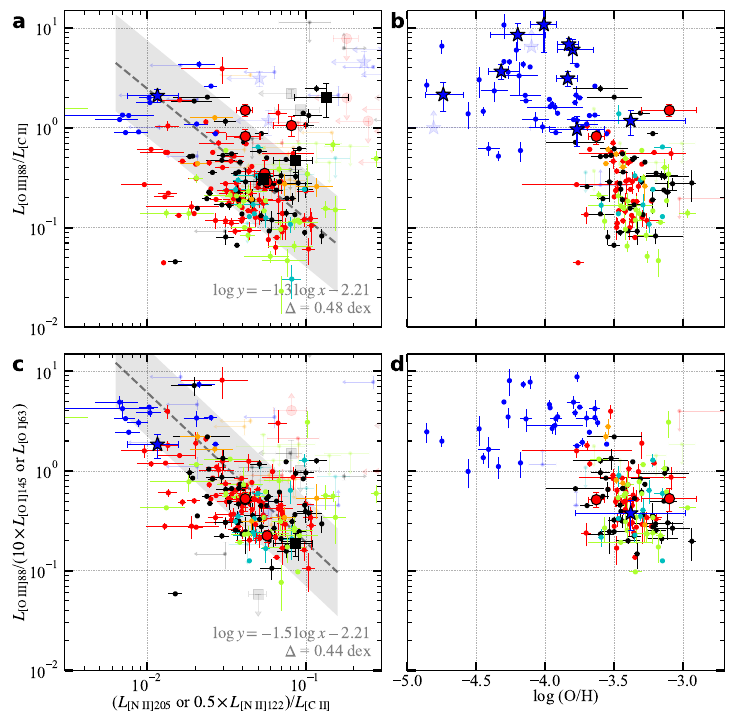}
    \caption[$U$ vs. metallicity. ]{Radiation field strength $U$--elemental abundance correlation. \oiii{88}/\cii{} versus \bsf{a}, \nii{}/\cii{}; and \bsf{b}, \oh{}. \oiii{88}/\oi{} versus \bsf{c}, \nii{}/\cii{} and \bsf{d}, \oh{}. The gray dashed line and shade correspond to the linear fit on the low-\zz{} galaxy data, which is printed at the lower right corner. }
    \label{f:data_U-N_O}
\end{figure}

It is well established, especially from optical studies, that metallicity affects ionization structure by influencing cooling rates \citep{draine11}. 
This is seen in the tight star-forming locus of the BPT diagram: from upper left to lower right, a decrease in $U$ is coupled with an increase in metallicity, and hence N/O \citep[see review by][]{kewley19}, though the detailed physics remain debated. 
Inspired by the BPT diagram and the $U$--metallicity correlation, Fig.~\ref{f:data_U-N_O} shows the correlation between the radiation field tracer \oiii{88}/\cii{} and N/O, as indicated by \nii{}/\cii{} and \niii{}/\oiii{88}.

In both panels, the diagonal distribution reflects a correlation between the \opp{} ICF and N/O. 
We fit a power law to the low-$z$ sample (left panel), and the fit parameters are given in the figure. 
However, without the additional order-of-magnitude variation in optical forbidden line emissivities caused by \te{} (see next section), the dynamic range in both axes of Fig.~\ref{f:data_U-N_O} is much smaller than in the optical BPT diagram, making the latter a more sensitive probe of such trend.

\subsection{Electron Temperature}
\label{sec:data_te}

\begin{figure*}
    \centering
    \includegraphics[width=\halfwdth]{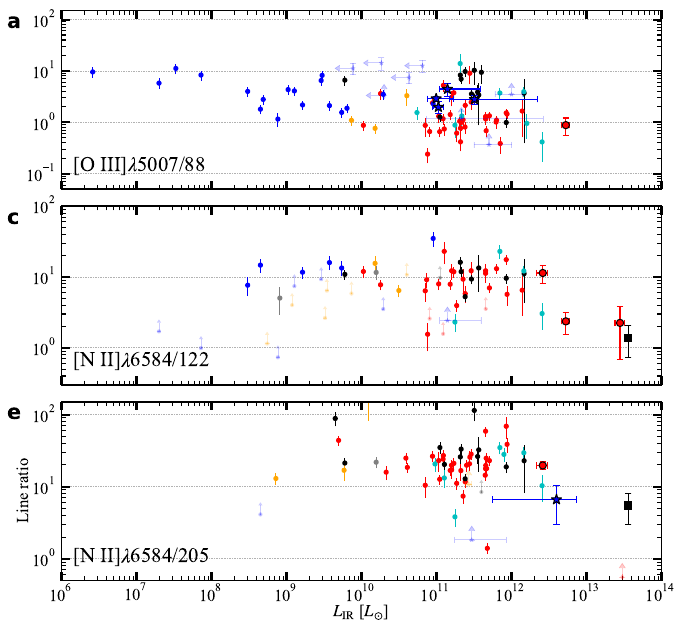}
    \includegraphics[width=0.442\textwidth]{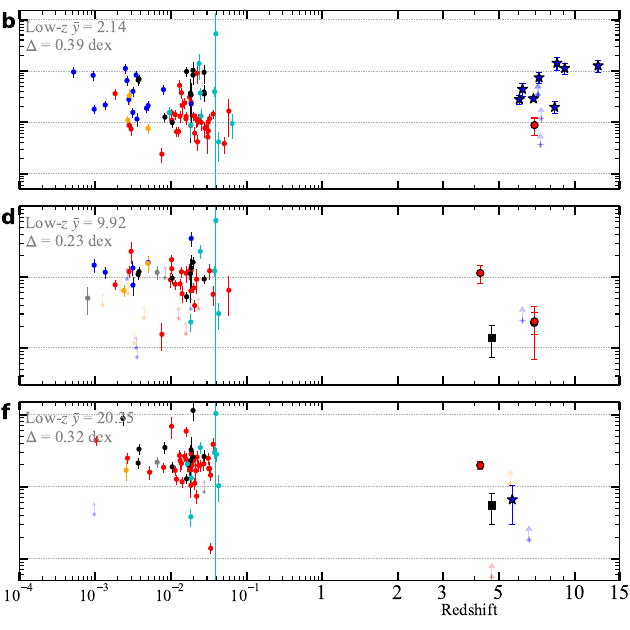}
    \caption[Demographics of electron temperature diagnostics. ]{Demographics of electron temperature diagnostics. }
    \label{f:data_IR-te}
\end{figure*}

Because optical strong line emissivities depend exponentially on the electron temperature (\te{}), their ratios with FIR fine-structure lines (FSLs) of the same ion (and with common energy levels) serve as sensitive probes of \te{}. 
Among all, the ratios \oiiio{5007}/\oiii{88} (and/or \oiii{52}) and \niio{6584}/\nii{122} (and/or \nii{205}) are widely used. 
Even greater sensitivity can be achieved by using more energetic shorter wavelength transitions, such as \oiio{3727}. 
However, these diagnostics become increasingly sensitive to extinction corrections, and only optical line fluxes at wavelengths $\gtrsim$5000\,\AA{} are used in this work. 
The accuracy of our extinction corrections is discussed in Appendix~\ref{sec:data_ext}, which verifies the reliability of line fluxes above 4400\,\AA.

\begin{figure}
    \centering
    \includegraphics[width=\halfwdth]{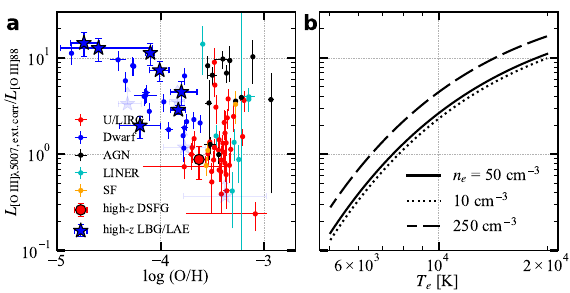}
    \caption[\teoiiifir{} vs. O/H. ]{Extinction corrected \oiiio{5007}-to-\oiii{88} ratio vs. \bsf{a}, metallicity ; and \bsf{b}, electron temperature diagnostic curve. The theoretical emissivity ratios are computed and shown assuming different electron densities \edens{}~=~50 (solid), 10 (dotted), and 250\,\cc{} (long dashed).}
    \label{f:data_te}
\end{figure}

In Fig.~\ref{f:data_te}, we show the extinction-corrected \oiiio{5007}/\oiii{88} ratio as a function of metallicity and convert this to \te{} using theoretical curves calculated with \pyneb{}. 
The distribution from dwarf galaxies to U/LIRGs demonstrates the expected "thermostatic" effect: the average \te{} decreases in more metal-rich galaxies. 
The FIR-based \teoiiifir{}-metallicity relation shows a steep decline, dropping from $\sim$20000\,K at \oh{}~=~--4.5, to 11000\,K at --4, and further to $\sim$7000\,K at --3.5. 
In comparison, the optical \teoiii{}-metallicity relation from \citet{nicholls14} is shallower, decreasing from $\sim$19000\,K at --4.5 to $\sim$13000\,K at --4 and 9000\,K at --3.5. 

This discrepancy is not due to extinction correction, as over-correcting \oiiio{5007} would lead to an overestimate of \teoiiifir{}; besides, the corrections for dwarf galaxies are modest. 
Moreover, resolved regions show good agreement between FIR-based and optical \teoiii{} estimates \citep[e.g.,][]{lamarche22}. 
We therefore propose that the difference reflects selection biases: the optical diagnostics are biased towards hotter regions, while the FIR line in the denominator of \oiiio{5007}/\oiii{88} is more representative of the total \opp{} in irradiated gas, thus lowering the observed ratio.

Another branch is apparent in the figure, rising vertically in \oiiio{5007}/\oiii{88} at an invariant \oh{}~$\sim$~--3.5, dominated by AGNs. 
This is a direct result of enhanced metal line emission in AGNs, best illustrated by the AGN branch in BPT diagram. 
The higher values of \oiiio{5007}/\oiii{88} are consistent with higher \te{} (\textgreater 12000 K) found in NLRs \citet[e.g.][]{tadhunter89}. 
Whereas we are unable to distinguish between a universally hot gas and localized heating \citet[e.g., by shock;][]{sutherland93}, and AGN photoionization models including FIR FSLs are not yet fully explored. 
Moreover, since AGN contributions to \oiii{88} are typically small (Sec.~\ref{sec:data_radiation}), our FIR-based \teoiiifir{} represents a lower limit for the \te{} in NLR. 

\begin{figure}
    \centering
    \includegraphics[width=\halfwdth]{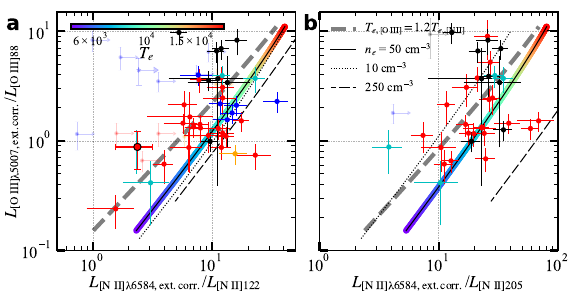}
    \caption[\teoiiifir{} vs. \teniifir{}. ]{Extinction corrected \oiiio{5007}-to-\oiii{88} ratio vs. extinction corrected \niio{6584} to \bsf{a}, \nii{122}; or \bsf{b}, \nii{205}. The thin solid, dotted, and dashed lines correspond to the theoretical curves using \edens{} = 50, 10, and 250 \cc{}, and along the solid line, each point is colored by \te{}. The thick gray dashed line is a theoretical curve assuming \teoiiifir{} = 1.2 \teniifir{} and \edens{} = 50 \cc{}. The plot style is the same as Fig.~\ref{f:data_te}.}
    \label{f:data_te_OIII-te_NII}
\end{figure}

We compare \teoiiifir{} and \teniifir{} in Fig.~\ref{f:data_te_OIII-te_NII}. 
The theoretical curve for \edens~=~50\,\cc{} (chosen as the median from \nii{122/205}) best matches the data for both \nii{122} and \nii{205}. 
In both panels, most data lie to the left of the theoretical prediction, indicating \teoiiifir{}~\textgreater~\teniifir{}. 
A relation of \teoiiifir{}~=~1.2$\times$\teniifir{} (thick dashed line) provides a better fit. 
This phenomenon, where \oiii{}-derived temperatures are higher than those from \nii{}, has been reported previously \citep[e.g.,][]{curti17}, though other studies find either agreement, the opposite trend, or more complex temperature structures regarding the two temperatures \citep[e.g.,][]{berg20,arellano20,mendez23}.

\section{Discussion}
\label{sec:data_discussion}

\subsection{Bias and Applicability of Density Diagnostics}
\label{sec:data_density_bias}

As described in Sec.~\ref{sec:data_density}, high-ionization density tracers---including \nev{14/24}, \siii{18/33}, \siio{6731/6716}, and \oiii{52/88}---do not show agreement with other density tracers. 
Instead, their observed ratios cluster near the theoretical low-density limit, and the scatter in Fig.~\ref{f:data_ne-ne} often exceeds what is expected from the 1$\sigma$ uncertainties. 
This inconsistency is exacerbated by the fact that a significant fraction of data points fall below the theoretical low-density limit.

The intensity of a collisionally excited line is linearly proportional to the density of the collisional partner up to the critical density. 
As a result, the density inferred from integrated FSL ratios is a weighted average of the electron density \edens{}, weighted by both the electron density and the emitting ion mass. 
This can be approximated as
\begin{equation}\begin{split}
	\log n_{e\mathrm{,avg}} \sim (\sum^i n_{e,i} \cdot M_i \cdot \log n_{e,i})/(\sum^i n_{e,i} \cdot M_i)
\label{equ:dens}
\end{split}\end{equation}
where $n_{e,i}$ and $M_i$ are the electron density and the mass of the emitting ion in each cloud. 
Note that Eq.~\ref{equ:dens} is only accurate (within $\sim$30\%) if all \edens{} values are below the critical density of the lines involved. 

For example, consider two clouds with the same mass of \opp{} ions but \edens{}~=~50 and 500\,\cc{}. 
Their individual \oiii{52/88} values are 0.676 and 1.68 (at \te{}~=~10\textsuperscript{4}\,K). 
Combining the two clouds, the ratio of summed fluxes is 1.48, corresponding to \edens{}~=~400\,\cc{}—consistent with Eq.~\ref{equ:dens}.

Therefore, the low observed values of high-ionization density tracers indicate that the bulk of emission originates from environments at or below the lower end of the diagnostic's applicable range. 
Empirically, these ratios are poor tracers of the true electron density in their emitting regions. 

Even for roughly half of the data points that lie above the low-density limit in Fig.~\ref{f:data_ne-ne}, the inferred \edens{} should be treated with caution. 
The existence of 1/3 to 1/2 of points below the theoretical limit is physically implausible, prompting an examination of possible systematic effects. 
Common radiative transfer effects (e.g., dust extinction) are insufficient, since optical doublets (\siio{6731,6716}) are equally affected, and the extinction difference in the MIR/FIR is too small (given known extinction curves) to explain these deviations. 
Self-absorption is not responsible either, because the denominator lines (e.g., \nev{24}, \oiii{88}) are ground-state transitions with higher optical depth; self-absorption would increase, not decrease, the observed ratios. 
While differences in the Spitzer/IRS FoV for \nev{14/24} may play a role, this does not affect other doublets observed within the same instrument module. 

Having ruled out these factors, we attribute the very low ratios primarily to underestimated uncertainties in flux calibration, extraction, or measurement. 
Since these are ratios, random errors can increase and decrease the measured value, leading to a roughly symmetric distribution about the median for ratios such as \oiii{52/88}, \nev{14/24}, and \siio{6731/6716}. 
This also naturally explains the larger observed scatter relative to theory. 
Therefore, using individual values for density measurement requires taking into account these missing uncertainties, especially systematic errors that are often not accounted for in the literature. 
Besides, when the values are close to the low- or high- density limits, ordinary error propagation fails to capture the nonlinear dependence of the diagnostics, and sophisticated statistical treatment like Bayesian inference is necessary. 

Blindly trusting \edens{} measurements from these tracers risks misleading conclusions. 
Due to atomic structure and excitation physics, the lower density limit for each tracer increases with ionization state—--progressing from \nii{}, to \oiii{} and \siii{}, to \nev{} (Fig.~\ref{f:data_ne}(\bsf{b})). 
Thus, random observational uncertainties near the low-density limit can spuriously assign high \edens{} values to a subset of sources (positive bias), creating the false impression of a correlation between \edens{} and ionization state. 
We therefore refrain from supporting the density stratification suggested by \citet{S15} based on \nii{122/205}-to-\oiii{52/88} (or similar) comparisons. 

In summary, we caution that high-critical-density lines are not reliable density diagnostics for their emission regions without careful statistical treatment. 
In contrast, the \nii{} doublets are unique in providing sensitivity to the electron densities typical of ionized gas in galaxies, with a median \edens{}~$\sim$~50\,\cc{} and remarkably little variation across all galaxies in Fig.~\ref{f:data_IR-n}.

\subsection{AGN Contribution to FSLs}
\label{sec:data_agn}

\begin{figure}
    \centering
    \includegraphics[width=\halfwdth]{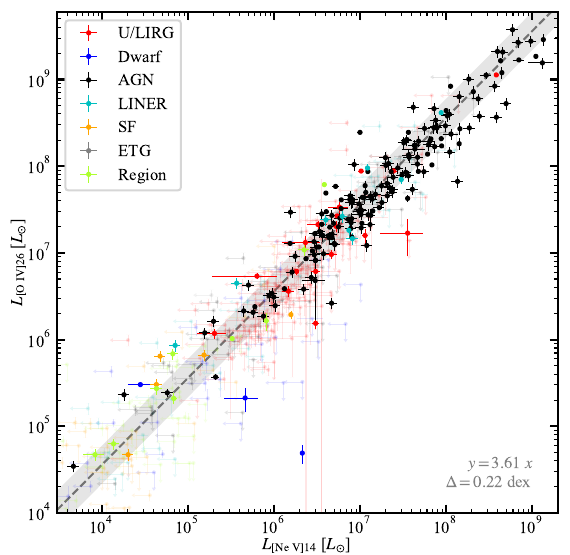}
    \caption[\nev{14}--\loiv{26}. ]{\nev{14}-to-\oiv{26} luminosity comparison. The gray dashed line and shade denotes the linear fit and scatter, which is also printed at the lower right corner.}
    \label{f:data_NeV14-OIV26}
\end{figure}

In this section, we will demonstrate the impact of AGN on FSLs, ranging from the highest ionization line \nev{} to the neutral gas line \oi{}. 

The \nev{14} line is a widely used AGN indicator, as the ionization potential for creating Ne\textsuperscript{4+} is 97.12\,eV—energies typically produced only by AGN accretion disks or strong shocks \citep{sutherland93,genzel89}. 
The \oiv{26} line is another popular AGN tracer (creation energy 54.9\,eV), with observational advantages: it is $\sim$3 times brighter than \nev{14}, and the MIR continuum at 26\,\um{} is less affected by molecular or dust features. 
However, the lower ionization potential of \oiv{26} means it can also arise in very intense or low-metallicity (hard) star-forming environments, particularly in dwarf galaxies \citep{P10}.

Fig.~\ref{f:data_NeV14-OIV26} shows the luminosity correlation between \nev{14} and \oiv{26}. 
Most data points represent known AGNs or resolved nuclear regions. 
A non-detection of these lines does not rule out an AGN, as both can be heavily attenuated by dust in extreme environments \citep{perez11}. 
We fit a linear relation, finding \loiv{26}~=~3.49~$\times$~\lnev{14} with a standard deviation of 0.22 dex.

\begin{figure}
    \centering
    \includegraphics[width=\halfwdth]{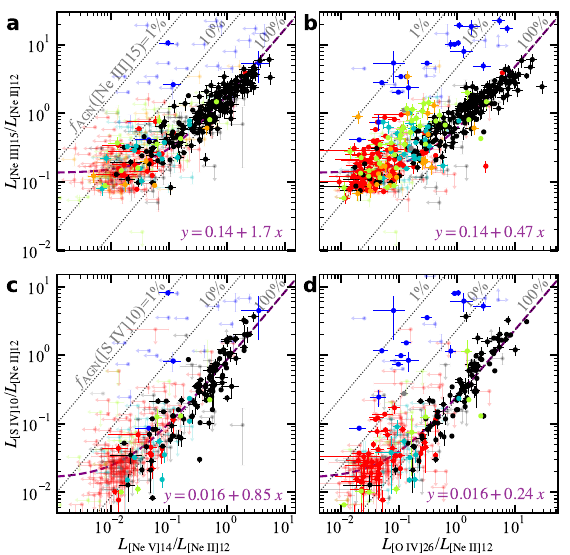}
    \caption[High-ionization lines vs. AGN lines and decomposition. ]{High-ionization lines \neiii{15} (first row) and \siv{10} (second row) vs. AGN lines \nev{14} (first column) and \oiv{26} (second column). All luminosities are normalized by \lneii{12}. The dotted lines have slope of unity, corresponding to f$_\mathrm{AGN}$ = 1\%, 10\% and 100\%. The purple dashed line is the first order polynomial fit, described at the lower right corner.}
    \label{f:data_agn-hardness}
\end{figure}

Excluding AGN contamination is important, as previous studies \citep[e.g.,][]{gorjian07} have shown that high-ionization FSLs like \neiii{15} can have significant AGN contributions. 
Utilizing AGN tracers \nev{14} and \oiv{26}, we calibrate AGN contributions to other highly ionized FSLs, starting with MIR lines \neiii{15} and \siv{10}. 
In Fig.~\ref{f:data_agn-hardness}, we plot \neiii{15} and \siv{10} against the AGN lines \nev{14} and \oiv{26}, normalizing all lines by \neii{12}. 
This normalization (creation energy 21.6\,eV) makes \neii{12} representative of the irradiated ISM, and its wavelength proximity to \neiii{15} reduces differential extinction effects. 
Although different elements are involved, all are $\alpha$-elements, so abundance variations are modest compared to the order-of-magnitude changes driven by the ionization structure.

With a rich dataset, we can disentangle AGN and SF contributions, instead of simply fitting a power law. 
We assume the luminosity of \neiii{15} and \siv{10} as the sum of star formation and AGN components: $L_\mathrm{line} = L_\mathrm{SF} + L_\mathrm{AGN}$. 
In AGN-dominated systems, high-ionization lines scale linearly with AGN tracers; in SF-dominated systems, they stay in a constant ratio to \neii{12}, typical for star-forming regions. 
Thus, we expect a first-order polynomial relation $y = a x + b$ between the AGN line ratio $x$ and the high-ionization line ratio $y$, where $a$ and $b$ are the ratios to AGN lines and low-ionization \neii{12} lines in the case of AGN-dominated and SF-dominated scenarios, respectively. 

The dotted lines in Fig.~\ref{f:data_agn-hardness} represent constant AGN fractions, while the data show a clear trend that converges towards a unity slope for AGN-dominated systems. 
We fit the first-order polynomial in logarithmic space, combining both \nev{14} and \oiv{26} to improve reliability at low AGN fractions. 
Dwarf galaxies are offset in the figure because, even though their extreme radiation fields produce energetic photons to make Ne\textsuperscript{2+} and S\textsuperscript{3+}, it still drops rapidly at the high-energy end \textgreater 50\,eV. 
Therefore, they are not included in the sample for fitting. 
Resolved regions are excluded in the fitting, either, because of possible aperture effects such that the Spitzer slit may not cover the full narrow-line region (NLR) in very nearby AGNs, and different beam sizes at different wavelengths may bias ratios. 
The fit is shown as a purple dashed line in Fig.~\ref{f:data_agn-hardness}. 

The AGN-dominated \neiii{15}/\nev{14} scaling matches the literature value \neiii{15}/\nev{14} = 1.7 \citep{spoon22}. 
Moreover, the distribution of galaxy types is consistent with the AGN fraction in \neiii{}, with $f_\mathrm{AGN,[Ne~III]} \sim 50\%$ separating AGNs from SF-dominated systems. 
However, we acknowledge that assuming a single value of \neiii{15}/\neii{12} for SF systems is a over-simplification, as indicated by the scatter.

A notable finding is that, within the Spitzer FoV, \siv{10} emission is almost always AGN-dominated over more than two orders of magnitude in \siv{10}/\neii{12}. 
This conclusion is robust to the exact fit, as the distribution remains close to linear scaling except at the lowest values. 
This is surprising, as S$^{3+}$ has a creation potential of only 34.8\,eV---energies readily produced by young massive stars. 
The lack of detailed photoionization modeling for MIR AGN lines, particularly \siv{10}, limits further interpretation. 
Regardless, this makes \siv{10} a promising short-wavelength AGN indicator, relevant for JWST, which has limited coverage and sensitivity at long-wavelength .

\begin{figure}
    \centering
    \includegraphics[width=\halfwdth]{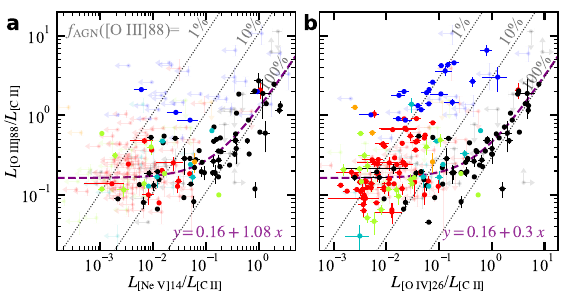}
	\includegraphics[width=\halfwdth]{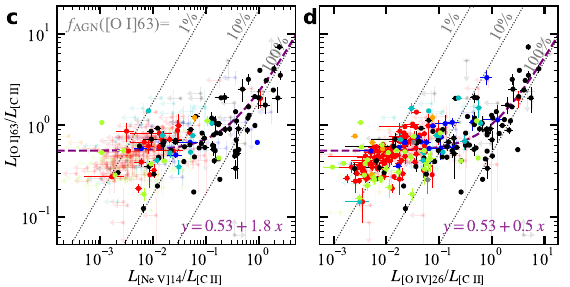}
    \caption[\oiii{88} and \oi{63} vs. AGN lines and decomposition. ]{\oiii{88} (top) and \oi{63} vs. AGN lines \nev{14} (left column) and \oiv{26} (right column). The luminosities are normalized by \lcii{}. The plot style is the same as Fig.~\ref{f:data_agn-hardness}.}
    \label{f:data_agn-OIII88}
\end{figure}

The same methodology can be applied to FIR FSLs \oiii{88} and \oi{} to assess their AGN contributions. 
The choice of \oiii{88} line is inspired by the elevated optical \oiiio{4959,5007} emission in AGN NLRs (e.g., in the BPT diagram; \citealt{baldwin81}), and that they are emitted by the same ions as FIR \oiii{} lines. 
And \oi{} lines are modeled to be enhanced in x ray-dominated regions near AGN \citep[e.g., ][and references therein]{meijerink07,wolfire22}.

In Fig.~\ref{f:data_agn-OIII88}, we compare \oiii{88} with \nev{14} and \oiv{26}, normalizing all ratios by \cii{}. 
The use of \cii{} as a denominator is motivated by its low ionization potential, making it representative of the bulk ISM, and by its $\alpha$-element origin, minimizing abundance effects.

Although the AGN lines observed by Spitzer are typically measured in much smaller apertures than the FIR FSL observations, this is acceptable for our purposes: we expect, and require, that \nev{14} and \oiv{26} emission originate from only the nuclear regions. 
The figure shows a clear upward trend, similar to that seen for the MIR high-ionization lines, directly indicating the influence of AGN activity on \oiii{88} emission. 

We performed the same decomposition and fitting procedure as in previous sections, with the results indicated in the figure. 
Compared to \neiii{15} and \siv{10}, the AGN contribution to \oiii{88} is generally modest: for most AGNs, $f_\mathrm{AGN,[O\,III]} < 50\%$ (or equivalently, \oiii{88}/\cii{}~\textless~2$\times$0.16 in the AGN-dominated branch). 
However, this difference may partly reflect the use of integrated galaxy measurements for \oiii{88}, versus nuclear-region measurements for \neiii{15} and \siv{10}, due to the different FoV of MIR and FIR observatories.

The presence of an AGN-powered component in \oiii{88} emission is not unexpected, as optical forbidden lines \oiiio{4959,5007} in AGN hosts are often dominated by AGN NLR emission. 
Nevertheless, these results highlight the incompleteness of previous treatments that attributed all \oiii{88} luminosity exclusively to star formation. 
This also offers an alternative interpretation for the strong \oiii{88} lines seen in high-\zz{} galaxies, discussed in \ppiii{}.

The same decomposition also confirms AGN contributions to \oi{} lines. 
Other high-ionization lines, such as \niii{}, may also be affected by AGN activity; however, the limited number of detections, low data quality, and the challenge of correcting for nitrogen abundance preclude a robust analysis in those cases.

\subsection{FIR-Optical Concordance}
\label{sec:data_optical}

\begin{figure*}
    \centering
    \includegraphics[width=0.75\textwidth]{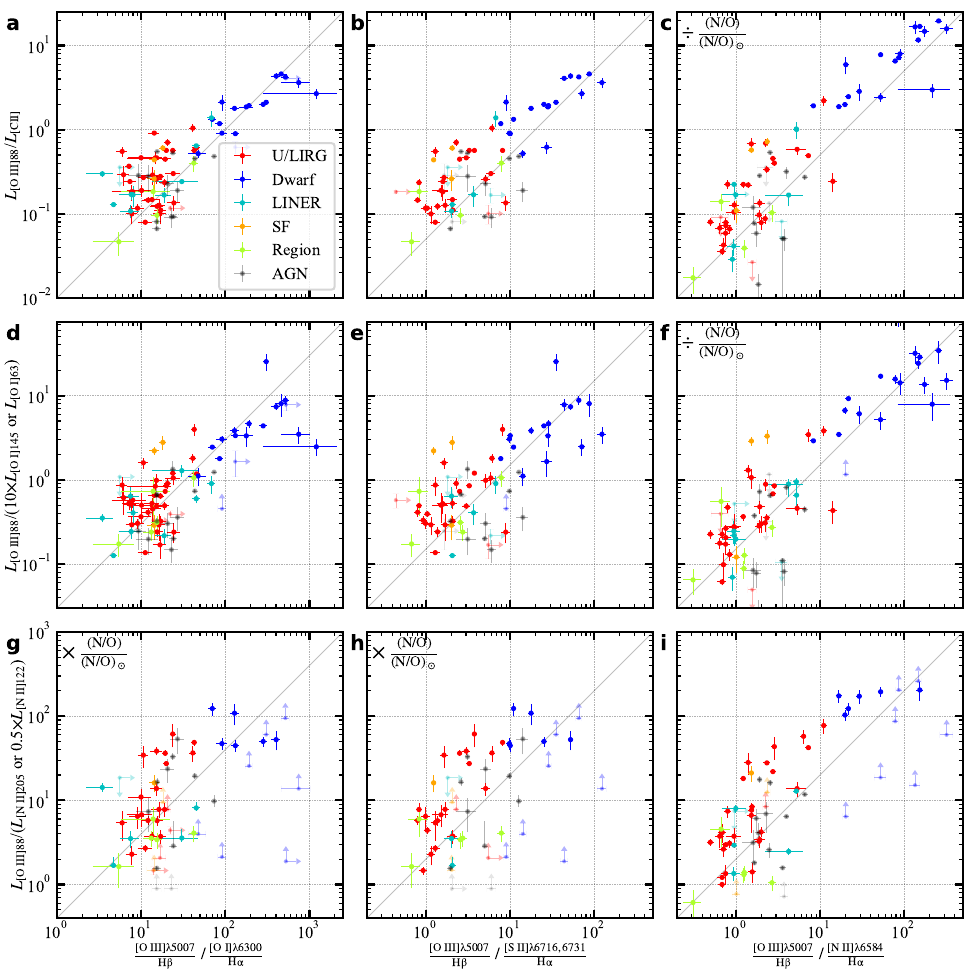}
    \caption[FIR vs. optical radiation field diagnostics.]{FIR line ratios \oiii{88} to \cii{} (first row), (10$\times$\oi{145} or \oi{63}) (second row), (\nii{205} or 0.5$\times$\nii{122}) (third row), vs. optical line ratios \oiiio{5007} to \oio{6300} (first column), \siio{6716\&6731} (second column), \niio{6584} (third column). The FIR line ratios are corrected by (N/O)/(N/O)$_\odot$ if there is mismatch in Nitrogen abundance with the optical line ratios. The diagonal gray lines are visual aids with a slope of unity.}
    \label{f:data_fir-opt_U}
\end{figure*}

One of the goals of this study is to bridge the gap between FIR and optical spectroscopic studies. 
Agreements between the two regimes have already been demonstrated in our analyses of abundance and electron temperature: the O/H and N/O values used here are derived from optical strong lines, and the electron temperature calibrations employ optical lines in the numerators. 
Here, we further demonstrate that FIR line ratios can be directly compared to their optical counterparts.

Many important ions, such as \op{}, \npp{}, and \opp{}, emit both optical forbidden lines and FIR FSLs, allowing for direct comparison of lines from the same ion. 
Furthermore, \spion{} and \cp{} ions have similar ionization potentials (10.36 and 11.26\,eV, respectively), both below the hydrogen ionization threshold (13.6 eV), motivating a comparison between the optical \siio{6716,6731} doublet and the FIR \cii{} line.

Optical line ratios of \oiiio{5007} to low-ionization lines are well-established probes of the ionization parameter $U$---for example, in BPT diagrams \citep{dopita00,kewley06,kewley19} and in comparison with models in \ppii{}. 
In Fig.~\ref{f:data_fir-opt_U}, we compare these ratios (\oiiio{5007} divided by \oio{6300}, \siio{6716,6731}, and \niio{6584}) to their FIR counterparts: \oiii{88} over \oi{}, \cii{}, and \nii{}, respectively. 
The observed ratios of the optical lines are plotted, but \oiiio{5007} is normalized by \hb{}, and all other optical lines around 6500 \AA are normalized by \ha{}, to mitigate the effect of dust extinction. 
As the relative abundance ratios of the $\alpha$-elements are less variable than N/O, we apply an N/O correction to the y-axis ratios to match the x-axis and minimize the impact of variations in nitrogen abundance. 

In all panels of Fig.~\ref{f:data_fir-opt_U}, the FIR \oiii{88} ratios track the trends of the optical \oiiio{5007} ratios over two orders of magnitude. 
The best agreement is seen for \oiiio{5007}/\niio{6584} and its FIR counterpart, followed by the \oiiio{5007}/\oio{6300} and \oiiio{5007}/\siio{6716,6731} comparisons. 
The agreement between optical and FIR \oiii{}-to-low-ionization line ratios supports our argument in Sec.~\ref{sec:data_radiation} that the large variations in \oiii{88}/low-ion ratios are primarily driven by changes in the ionization parameter $U$. 
Variations in neutral-to-ionized gas fractions, FIR line emissivities, and the optical lines' temperature dependence play secondary roles. 
This correspondence also enables the extension of established optical scaling relations and diagnostics to FIR FSLs. 

In all panels, especially \oiii{}/\nii{}, the distribution of data points follows an ``arc'' shape: dwarf galaxies, with higher $U$, populate the upper right; the trend extends down and left through star-forming galaxies and U/LIRGs with a slope shallower than linear; and at lower left, the distribution bends vertically, with AGN, LINERs, and nuclear regions becoming more prevalent. 
This arc arises because the optical ratios are sensitive not only to ionization structure but also to electron temperature. 
The \te{} diagnostics using \oiiio{5007}/\oiii{88} and \niio{6584}/\nii{205} in Sec.~\ref{sec:data_te} explain the arc-shaped distribution. 
In Fig.~\ref{f:data_te_OIII-te_NII}, the diagonal corresponds to a constant (\oiiio{5007}/\oiii{88})/(\niio{6584}/\nii{205}); however, the data and theoretical predictions are steeper, so high-\te{} galaxies have higher values of this ratio. 
In Fig.~\ref{f:data_fir-opt_U}, moving parallel to the left of the linear scaling (dotted line) corresponds to decreasing this ratio. 
Thus, as we move to lower optical \oiii{}/\nii{} and $U$, \te{} also declines, decreasing the ratio and resulting in a slope shallower than linear. 
At the low end, populated by U/LIRGs and AGNs (which have high \te{}), a second branch extends downward, producing the observed arc. 
The high \te{} at both high and low $U$ ends produces the characteristic shape.

Our investigations in this paper demonstrate a strong concordance between FIR and optical diagnostics in many aspects. 
The agreement of FIR and optical radiation field diagnostics, as well as the correlation of FIR line ratios with abundances, underscores that the amount of emitting ions primarily determines the line luminosities, and both sets of lines probe the same underlying physical properties. 
We further demonstrated the value of combining FIR and optical lines for measuring electron temperature. 
While such cross-regime ratios have traditionally been avoided beacause of concerns over extinction correction uncertainties, our results indicate that the dynamic range of the physical parameters probed is much larger than the uncertainties introduced by extinction corrections. 
Thus, the physical trends remain clear.

Combining FIR and optical lines provides complementary powerful tools for studying gas physical conditions in galaxies. 
Redundant measurements allow mutual calibration of optical and FIR diagnostics and bridge parameter space across different galaxy types: optical lines are particularly effective for metal-poor and normal environments, while FIR FSLs excel in high-SFR and dust-rich galaxies. 
Such joint diagnostics are especially valuable for high-redshift studies, where data are often sparse.

\subsection{Similarity and Difference of ISM in High-\texorpdfstring{\zz{}}{z} Galaxies}
\label{sec:data_sim}

A major motivation for high-\zz{} FIR FSL observations is to study the evolution of galaxies across cosmic time. 
Throughout this paper, we present figures showing the distribution of key line ratios as a function of $L_{\mathrm{IR}}$ and redshift (see Fig.~\ref{f:data_IR-CII_IR}, \ref{f:data_IR-line_IR}, \ref{f:data_IR-n}, \ref{f:data_IR-U}, \ref{f:data_IR-metallicity}, \ref{f:data_IR-te}), enabling direct comparison between low- and high-redshift galaxies. 

However, practical challenges limit our ability to robustly infer evolutionary trends. 
The biggest limitation is the scarcity of observations for many spectral lines---particularly at high redshift, but also at low redshift for certain transitions. 
High-\zz{} detections, especially of weak lines such as \oi{145} and \niii{}, typically have marginal reliability (S/N~$\sim$~3–5). Additionally, sample biases exist; for example, \nii{} lines are rarely detected in dwarf galaxies. 
As a result, any inferred trends are based on small and often biased samples.

It is important to distinguish between galaxy global properties (e.g., mass, SFR), typically probed by absolute luminosities, and ISM gas properties (e.g., metallicity, density), which are best traced by line ratios. 
Our focus in this work is on line diagnostics of ISM properties. 
Thus, the similarities and differences discussed here pertain to the physical conditions of the ISM at different cosmic epochs, not to galaxy-wide properties. 
The galaxy properties are known to evolve dramatically over time and receive prime focus in galaxy evolution studies, but due to observational selection effects---particularly the faintness and difficulty of FIR FSL observations---measured \lline{} values are dominated by selection rather than intrinsic evolution (see Sec.~\ref{sec:data_line}). 
We therefore refrain from discussing global galaxy evolution in this context. 

Among the empirical ISM line ratio diagnostics explored here, only a few show reliable differences between low- and high-\zz{} samples. 
In particular, the "deficit" trend for all FIR FSLs is offset by $\sim$1 dex in \lir{} between low-\zz{} and high-\zz{} galaxies (Sec.~\ref{sec:data_line}; Figs.~\ref{f:data_IR-CII_IR}, \ref{f:data_IR-line_IR}). 
The threshold for the onset of the ``deficit'' trend shifts from \lir{}~$\sim$~10\textsuperscript{11.5}\,\lsun{} at low \zz{} to 10\textsuperscript{12.5}\,\lsun{} at high \zz{}. 
High-\zz{} dusty galaxies of comparable \lir{} often show elevated line luminosities and higher \lline{}/\lir{}. 
These galaxies also display differences in their radiation field properties (Sec.~\ref{sec:data_radiation}, Fig.~\ref{f:data_IR-U}), with \oiii{}/\cii{} and \oiii{}/\nii{} ratios systematically higher. 
This difference is not seen in metal-poor systems (LBG/LAEs, dwarfs). 
Furthermore, \oi{}/\cii{} ratios are enhanced at high \zz{}, suggesting a denser and/or warmer neutral gas in high-\zz{} galaxies. 
These possible evolutionary trends in gas properties will be discussed in detail in \ppiii{}.

Other diagnostics do not show significant evolution. 
Electron densities, as probed by \nii{122/205} and \oiii{52/88}, occupy similar ranges at low and high redshifts, indicating little or no evolution in the ionized gas density. 
The only exception is W2246-0526, which shows an unusually high \nii{122/205} ratio \citep[\textgreater10;][]{FR24}, likely due to its extreme AGN-driven environment. 
For the radiation field, high-\zz{} LBG/LAEs have \oiii{}/\cii{} ratios similar to those of low-\zz{} dwarfs, in contrast to the claims in previous studies of elevated ratios at high \zz{} \citep{LN19,HY20}. 
\oiii{}/\oi{} ratios agree between low- and high-\zz{} dusty galaxies, reflecting elevated luminosities for both lines. 
We will revisit this in \ppiii{}. 
For abundance, only \nii{}/\cii{} ratios have sufficient data for comparison; no systematic evolution is observed when comparing high-\zz{} metal-enriched galaxies to their low-\zz{} counterparts, or high-\zz{} metal-poor galaxies to low-\zz{} dwarfs. 
For electron temperature, available data are sparse, but most measurements fall within the same range at both redshifts, with metallicity dependence persisting (albeit with very limited statistics).

In summary, the gas properties---especially those of ionized gas and of galaxies that are not IR-bright---show little evidence of evolution with redshift. 
This is not unexpected for two reasons: (1) comparisons are drawn within the metal-rich and metal-poor populations, while metallicity is the main driver of ISM physical property variations; and (2) many of the properties discussed here, such as density and temperature, are primarily governed by local physical processes and equilibrium conditions, rather than by large-scale cosmic evolution. 
Thus, the general lack of observed evolution suggests that similar physical processes likely regulate FIR FSL-emitting gas in galaxies throughout cosmic history.

\section{Summary}
\label{sec:data_summary}

In this paper, we present the most comprehensive catalog to date of galaxy-integrated FIR FSL data and use it to investigate a wide range of observational relations.

\begin{itemize} 
	\item We calibrate the dust color temperature \tcolor{}. 
	\item We show that the equivalent width of FIR FSL reflects the line-to-infrared luminosity ratio (\lline{}/\lir{}). No significant relation is found between \nii{} and $L_\mathrm{24\mu m}$, except for the observed \nii{} ``deficit''. 
	\item We systematically examine density diagnostics, demonstrating that \nii{122/205} and \oi{}/\cii{} are correlated, while other diagnostics are not. 
	\item We present a comprehensive analysis of radiation field diagnostics, spanning AGN tracers, hardness tracers, and ionization parameter tracers. 
	\item We explore abundance diagnostics for N/O. Both \nii{} and \niii{} show a strong correlation with N/O.  
	\item We demonstrate methods for measuring electron temperature using FIR and optical lines. 
\end{itemize}

Based on these observational relations, we draw the following conclusions.

\begin{itemize} 
	\item The FSL sample is strongly biased towards higher \lir{} and \lline{} values due to observational selection effects. 
	\item With the exception of \nii{122/205} and \oi{}/\cii{}, most density diagnostics are found to be near the low-density limit and should not be used for galaxy-integrated measurements without careful statistical treatment. 
	\item The galaxy-averaged electron density shows a median value of $\sim$50\,cm$^{-3}$, with little variation between galaxies and cross redshifts. 
	\item We identify and quantify AGN contributions to MIR (\neiii{15}, \siv{10}) and FIR (\oiii{88}, \oi{63}) emission, which can be decomposed using MIR AGN tracers; \siv{10} emission is always AGN-dominated. 
	\item \oiii{}/\cii{} and \oiii{}/\oi{} probe the ionizing radiation field strength. 
	\item The \cii{}/\nii{} ratio is empirically correlated with N/O. 
    \item FIR line ratios show strong concordance with the optical line ratios in probing the metallicity and radiation field. 
    \item The abundance of emitting ions primarily sets the FIR FSL ratios. 
	\item The electron temperature derived from \oiii{} (\teoiii{}) is systematically higher than that from \nii{} (\tenii{}). 
    \item Most of the line ratios do not show reliable differences across redshifts, with the exception of line/IR and \oiii{88}\cii{} in dusty galaxies. 
	\item The similarity suggests similar ISM properties in low- and high-\zz{} galaxies as a result of regulation by the same physical processes. The differences will be explored in detail in \ppiii{}. 
\end{itemize}


\begin{acknowledgments}
We thank Henrik Spoon for knowledge on MIR FSLs and enlightening discussions. 
B.P. acknowledges the support of NRAO SOS 1519126. 
Support for this work was also provided in part by NASA grant NNX17AF37G, NSF grant AST-1716229, NASA/SOFIA grants NAS-97001 (SOF-04-0179, SOF-05-0111), NNA17BF53C (SOF-06-0225, SOF07-0209, SOF08-0165), and the NASA New York Space Grant Consortium. 
Herschel is an ESA space observatory with science instruments provided by European-led Principal Investigator consortia and with important participation from NASA. 
This work is based in part on archival data obtained with the Spitzer Space Telescope, which was operated by the Jet Propulsion Laboratory, California Institute of Technology under a contract with NASA.

\facility{Herschel, Spitzer}

\software{Astropy \citep{astropy13,astropy18}, \pyneb{} \citep{Luridiana15}}
\end{acknowledgments}



\appendix

\section{Low-\texorpdfstring{\zz{}}{z} Galaxy FSL Catalog}
\label{sec:flames-low}

\subsection{Sample Selection}
\label{nearby_caveat_sample}

In order to construct a comprehensive table of low-\zz{} FIR FSL data, we include all the galaxies that have any published survey observations by any of the following FIR observatories: ISO \citep{kessler96}, Herschel \citep{pilbratt10}, and SOFIA \citep{temi18}. 
We also include a large amount of galaxies in Spitzer \citep{werner04} surveys for MIR FSL data. 
In addition to line data, we further include all the sources in the Revised Bright Galaxy Survey catalog \citep[RBGS;][]{S03} and the Great Observatories All-Sky LIRG Survey \citep[GOALS;][]{A09}, for a complete sample of all the LIRGs in the low-\zz{} universe.
This adds up to a total of 1273 entries. 
Being inclusive in sample selection, only 886 of the entries have FSL data, but the rest are still valuable for the population study in terms of IR or UV luminosity studies and are potential targets for future FSL surveys.

\subsection{Galaxy Systems}
\label{sec:flames-low_systems}

151 galaxy systems are included in the low-\zz{} catalog. 
These systems have at least two entries in the catalog, one referring to the system and the others being the individual members. 
Among them, 65 are mergers, the rest being galaxies with resolved nuclear or extranuclear regions. 
Because of the high fraction of mergers among U/LIRGs, all galaxies in the GOALS sample were manually checked, and the corresponding data points are assigned to either the whole system or individual members based on the spatial resolution of the observation facility. 
The resolved regions are included because the limited FoV of the PACS instrument on Herschel is only able to capture the regions in very nearby SF galaxies.

\subsection{FIR FSL Data}
\label{sec:flames-low_fir}

The FIR FSL data are taken from \citet{B08} and \citet{F14} with ISO/LWS \citep{clegg96}; SHINING \citep{H18a}, DGS \citep{M13,C15,C19}, HERCULES \citep{R15}, HERUS \citep{F13}, \citet{D17}, \citet{F16}, \citet{S15}, \citet{D15}, \citet{L17b}, KINGFISH \citep{K11,D12}, \citet{S19} with Herschel/PACS \citep{poglitsch10}; HERCULES, \citep{R15}, \citet{K16}, \citet{S15}, \citet{L17a}, \citet{L17b} with Herschel/SPIRE \citet{griffin10}; \citet{P21}, \citet{S22}, \citet{C22} with SOFIA/FIFI-LS \citep{fischer18}. 
In addition to the published data, a collection of Herschel/PACS data was also independently reduced and included. 

Among 1273 entries, 698 have FIR FSL observational data. 

The optimally extracted line fluxes in each source are used, typically recovering most of the flux for compact sources, and covering the whole FoV for extended emitter. 
The uncertainties reported in \citet{C22} are modified to represent the statistical error of the FIFI-LS data.

\subsection{MIR FSL Data}
\label{sec:flames-low_mir}

The MIR FSL data are taken from DGS \citep{M13,C15,C19}, \citet{F07}, \citet{P10}, \citet{I13}, \citet{H10}, \citet{S15}, \citet{D15}, all observed with Spitzer/IRS \citep{houck04}. 
A collection of IRS data was also reduced independently and included. 

677 of 1273 entries have MIR FSL data. 

\subsection{Optical line Data}
\label{sec:flames-low_optical}

All optical spectral line data are drawn from drift-scan observations in \citet{M06}, integrated over the whole galaxy/system. 
Information on the extinction correction can be found in Sec.~\ref{sec:data_post} and Appendix~\ref{sec:data_ext}. 

More metallicity data are also computed using the optical lines, as described in Sec.~\ref{sec:data_post}.

\subsection{\texorpdfstring{\hi{}}{H I} Data}
\label{sec:flames-low_hi}

\hi{} 21-cm line data are taken from ALFALFA \citep{H18b,D20}, \citet{S05}, and HIPASS \citep{K04}, with various facilities and beam sizes typically \textgreater3\arcmin.

\subsection{Photometry Data}
\label{sec:flames-low_photometry}

The MIR and FIR photometry data is collected from various papers, although RGBS \citep{S03}, GOALS \citep{A09}, KINGFISH \citep{K11,D12}, and DustPedia \citep{C18,D19} contribute a significant fraction. 
These photometry data include the Herschel/PACS \citep{poglitsch10} spectrometer continuum measurements at line wavelengths 52, 57, 63, 88, 122, 145, and 158 \um{}, as well as single band imaging measurements in 12, 25, 60, 100 \um{} bands of IRAS \citep{neugebauer84}; 24, 70, 100, 160 \um{} bands of Spitzer/MIPS \citep{rieke04}; 70, 100, 160 \um{} bands of Herschel/PACS \citep{poglitsch10} camera; and 250, 350, 500 \um{} bands of Herschel/SPIRE \citep{griffin10}. 

The dust color temperature \tcolor{} is computed using the recorded FIR photometries. 
See Sec.~\ref{sec:data_post} and Appendix~\ref{sec:temp_cal} for more details. 
The Herschel/PACS spectrometry photometry ratios are converted to \tcolor{} using empirical fitting results, and other imaging band ratios are converted to \tcolor{} using OT-MBB model. 

For sources with IRAS flux density reported but not \lir{} or \lfir{}, we compute both using the widely used approximation detailed in Table 1 in \citet{sanders96}. 

Additional ultraviolet to near-IR photometry data are also collected from the aforementioned sources, including the FUV and NUV bands on GALEX \citep{morrissey07}; 3.6, 4.5, 5.8, 8.0 \um{} bands on Spitzer/IRAC \citep{fazio04}; W1-4 bands on WISE \citep{wright10}.

\subsection{Ancillary Data}
\label{sec:flames-low_ancillary}

Because different sources report a varying level of information, some key quantities can be missing in the literature and have to be computed in post-processing of the catalog. 

The galaxy types based on various definitions and classifications are all recorded, while only one type is used in making plots in this paper, based on the following descending priority: region, dwarf, LINER, AGN, ETG, U/LIRG, SF. 

Note that members of U/LIRG system are also treated as U/LIRGs, even if \lir{} of individual source may not reach the threshold of 10\textsuperscript{11} \lsun{}. 

The comoving distance is computed in the framework of a flat $\Lambda$CDM cosmology with H$_0$=69.3 km/s and $\Omega_\mathrm{m}$=0.287 \citep{hinshaw13}, if the value is not reported in any data source. 
Otherwise, the value in the latest publication is used.

\subsection{Table Structure}
\label{sec:flames-low_structure}

The table of the low-\zz{} catalog\footnote{FLAMES-low. The full machine-readable table will be available for download in the online published version. Prior to that, the catalog can be obtained by contacting the correspondence author. } is organized in the following format.

\begin{enumerate}
    \item Column 1. Name, the unique identifier of the region, galaxy, or system.
    \item Column 2. Alternative name(s) of the object used in the literature. 
    \item Column 3-4. R.A. and Dec. (J2000) of the object. 
    \item Column 5-7. Merger, region, and host flag, whether the object is part of a merger, or a region of a resolved galaxy, and if the entry represents the host of the system.
    \item Column 8. System name, if the object is a member of a merger or group, or the object is a region of a resolved galaxy. 
    \item Column 9. Redshift, the redshift of the object.
    \item Column 10. Distance, the comoving distance $D$ of the object in the literature, in the unit Mpc. 
    \item Column 11. Type, the type assigned to the object based on the literature classification and \lir{}. 
    \item Column 12-27. Total IR luminosity (log \lir{}, 8-1000 \um{}), FIR luminosity (log \lfir{}, 40-500 \um{}), FUV luminosity (log \lfuv{}, 1600 \AA), and 24 \um{} monochromatic luminosity (log $L_\mathrm{24\mu m}$), sets of four columns: (1) value, (2) uncertainty, (3) limit flag, and (4) reference/method; in the unit 10\textsuperscript{8} \lsun{}.
    \item Column 28-35. Stellar mass log \mstar{} and neutral hydrogen mass log \mhi{}, same sets of four columns, in the unit log \msun{}. 
    \item Column 36-39. Dust color temperature of the IRAS photometry ratio \tcolor{}, the same sets of four columns. 
    \item Column 40-49. Absolute metallicity \oh{} and Nitrogen-to-Oxygen abundance ratio \no{}, sets of five columns: (1) value, (2) uncertainty, (3) lower uncertainties (if different from uncertainty), (4) limit flag, and (5) reference/method. 
    \item Column 40-57. UV slope $\beta$ and IR excess IRX, same sets of four columns. 
    \item Column 58-113. FIR FSLs: \oiii{52}, \niii{}, \oi{63}, \oiii{88}, \nii{122}, \oi{145}, \cii{}, \nii{205}; sets of seven columns: (1) line flux ($F_\mathrm{line}$), (2) flux uncertainty, (3) flux limit flag, (4) line luminosity (\lline{}), (5) luminosity uncertainty, (6) luminosity flag, (7) reference; fluxes are in the unit of 10\textsuperscript{-16} W m\textsuperscript{-2}, luminosities are in 10\textsuperscript{8} \lsun{}. 
    \item Column 114-253. MIR FSLs and Hydrogen recombination line: \siv{10}, \neii{12}, \nev{14}, \neiii{15}, \siii{18}, \nev{24}, \oiv{26}, \siii{33}, [Si\,{\sc ii}]34, \neiii{36}, [Ar\,{\sc ii}]7, [Ar\,{\sc iii}]9, [Ar\,{\sc v}]13, [Cl\,{\sc ii}]14, [P\,{\sc iii}]17, [Fe\,{\sc ii}]17, [Ar\,{\sc iii}]21, [Fe\,{\sc iii}]23, [Fe\,{\sc ii}]24, [Fe\,{\sc ii}]25; same sets of 7 columns and units. 
    \item Column 254-383. Galaxy integrated optical strong lines: \oiio{3727}, \hd{}, \hg{}, \hb{}, \oiiio{5007}, \oio{6300}, \ha{}, \niio{6584}, \siio{6716}, \siio{6731}; a set of 13 columns: (1-3) observed line flux $F_\mathrm{line}$, uncertainty, and limit flag, (4-6) extinction corrected line flux $F_\mathrm{line,ext.corr.}$, uncertainty, and limit flag, (7-9) extinction corrected line luminosity $L_\mathrm{line,ext.corr.}$, uncertainty, and limit flag, (10-12) line equivalent width EW value, uncertainty, and limit flag, (13) reference; units are the same as FIR and MIR data, EW are in the unit \AA.
    \item Column 384-389. Extinction $A_\mathrm{V}$ and $E_\mathrm{B-V}$; sets of three columns: value, uncertainty, and limit flag. 
    \item Column 390-513. Photometry in the bands: GALEX FUV and NUV, WISE W1 to W4, Spitzer/IRAC 3.6, 4.5, 5.8, 8.0 \um{}, IRAS 12, 25, 60, 100 \um{}, Spitzer/MIPS 24, 70, 100, 160 \um{}, Herschel/PACS 70, 100, 160 \um{}, Herschel/SPIRE 250, 350, 500 \um{}, local continuums accompanying FIR FSLs at 52, 57, 63, 88, 122, 145, 158 \um; sets of four columns: (1) flux density $S$, (2) uncertainty, (3) limit flag, and (4) reference; flux density is in the unit Jy. 
\end{enumerate}

The reference codes in the catalog correspond to LCT: this work, SHINING: \citet{H18a}, DGS: \citet{M13,C15,C19}, HERCULES: \citet{R15}, HERUS: \citet{F13}, D17: \citet{D17}, F16: \citet{F16}, Z16: \citet{Z16}, B08: \citet{B08}, F14: \citet{F14}, F07: \citet{F07}, P10: \citet{P10}, I13: \citet{I13}, RBGS: \citet{S03}, GOALS: \citet{A09}, H10: \citet{H10}, K16: \citet{K16}, P21: \citet{P21}, S21: \citet{S22}, C22: \citet{C22}, S15: \citet{S15}, D15: \citet{D15}, L17: \citet{L17a}, LR17: \citet{L17b}, I15: \citet{I15}, KINGFISH: \citet{K11,D12}, S19: \citet{S19}, M06: \citet{M06}, DustPedia: \citet{D19,C18}, S16: \citet{S16}, ALFALFA: \citet{H18b,D20}, S05: \citet{S05}, HIPASS: \citet{K04}, H07: \citet{H07}. 

A small fraction of the catalog with either \cii{} or \oiii{88} is shown in Table~\ref{tab:flames-low_example} as an example. 

\begin{rotatetable*}
\begin{deluxetable*}{lcccccrcrccrcrccr}
\centerwidetable
    \tabletypesize{\scriptsize}
    \tablecaption{Example of the low-\zz{} FSL catalog. \label{tab:flames-low_example}}
    \tablehead{
    \colhead{Name} & \colhead{Redshift} & \colhead{Type} & \colhead{...} & 
    \colhead{\lir{}} & \colhead{\tcolor{}} & \colhead{ref.} & 
    \colhead{\oh{}} & \colhead{ref.} & \colhead{...} & 
    \colhead{\loiii{88}} & 
    \colhead{ref.} & \colhead{\lcii{}} & \colhead{ref.} & 
    \colhead{...} &
    \colhead{\lhaextcorr{}} & \colhead{ref.} \\
     &  &  & & 
    \colhead{10\textsuperscript{8}\,\lsun{}} &  &  &  &  & &
    \colhead{10\textsuperscript{8}\,\lsun{}} &  & 
    \colhead{10\textsuperscript{8}\,\lsun{}} &  & &
    \colhead{10\textsuperscript{8}\,\lsun{}} & }
    \startdata
    w
NGC 4228-S               &                             & Dwarf         &     & 73.60                                 &             &             & -3.64$\pm$0.04          & D15        &     & 0.0035                                & D15        & 0.0026                                & D15     &     &                                       & \\
NGC 4236-Nuclear         & -0.0000030                  & Region, SF    &     &                                       &             &             &                         &            &     &                                       &            & $<$0.0015                             & B08     &     &                                       & \\
NGC 6503                 & -0.000060                   & SF            &     & 5.96                                  & 0.34        & 60/100      &                         &            &     &                                       &            & 0.010                                 & B08     &     &                                       & \\
M 81-Nuclei              & -0.00011                    & Region, LINER &     &                                       & $<$0.22     & 63/122      & -3.42$\pm$0.02          & F16        &     & 0.00087                               & F16        & 0.0018                                & F16     &     &                                       & \\
NGC 6822-Extranuclear    & -0.00019                    & Region, SF    &     &                                       &             &             &                         &            &     &                                       &            & 0.000095                              & B08     &     &                                       & \\
UGC 6456                 & -0.00034                    & Dwarf         &     & 0.20                                  & 0.43        & DGS         & -4.34$\pm$0.01          & DGS        &     & 0.00052                               & DGS        & 0.00099                               & DGS     &     & 0.0045                                & M06 \\
NGC 1569                 & -0.00035                    & Dwarf, AGN    &     & 12.74                                 & 0.98        & 60/100      & -4.15$^{+0.11}_{-0.11}$ & DustPedia  &     & 0.099                                 & DGS        & 0.047                                 & DGS     &     & 0.24                                  & M06 \\
M 98-Nuclear             & -0.00047                    & Region, SF    &     &                                       & 0.39        & 63/158      &                         &            &     &                                       &            & 0.11                                  & B08     &     &                                       & \\
M 33-Nuclei              & -0.00060                    & Region, SF    &     &                                       &             &             &                         &            &     & 0.00032                               & B08        & 0.00083                               & B08     &     &                                       & \\
NGC 185                  & -0.00067                    & AGN           &     & 0.0081                                & 0.24        & B08         &                         &            &     &                                       &            & 0.000011                              & F16     &     &                                       & \\
M 90-Nuclear             & -0.00078                    & Region, LINER &     & 46.17                                 & 5.48        & 88/158      & -3.42                   & F16        &     & 0.014                                 & F16        & 0.15                                  & S19     &     &                                       & M06 \\
...                      & \multicolumn{16}{c}{...} \\
NGC 3690                 & 0.010                       & U/LIRG        &     & 6313.53                               & 1.01        & 60/100      & -3.44$\pm$0.02          & PG16S      &     & 4.09                                  & SHINING    & 7.22                                  & SHINING &     & 12.51                                 & M06 \\
NGC 3557                 & 0.010                       & SF            &     & 28.39                                 & 0.34        & B08         &                         &            &     & $<$0.29                               & B08        & $<$0.088                              & B08     &     &                                       & \\
NGC 6753                 & 0.010                       & U/LIRG        &     & 761.40                                & 0.36        & 60/100      &                         &            &     & 0.19                                  & B08        & 1.52                                  & B08     &     &                                       & \\
NVSS J153412+571655      & 0.011                       & Dwarf         &     & 20.48                                 & 0.63        & DGS         & -3.95$\pm$0.01          & DGS        &     & 0.081                                 & DGS        & 0.047                                 & DGS     &     &                                       & \\
IC 4595                  & 0.011                       & SF            &     & 585.52                                & 0.38        & 60/100      &                         &            &     &                                       &            & 1.50                                  & B08     &     &                                       & \\
ESO 320-30               & 0.011                       & U/LIRG        & ... & 1645.86                               & 0.74        & 60/100      &                         &            & ... & 0.14                                  & D17        & 1.53                                  & D17     & ... &                                       & \\
IC 5063                  & 0.011                       & AGN           &     & 738.74                                & 1.38        & 60/100      &                         &            &     &                                       &            & 0.35                                  & F16     &     &                                       & \\
IC 3639                  & 0.011                       & U/LIRG, AGN   &     & 762.49                                & 0.65        & 60/100      &                         &            &     & 0.29                                  & S15        & 1.31                                  & S15     &     &                                       & \\
NGC 2369                 & 0.011                       & U/LIRG        &     & 1346.66                               & 0.53        & 60/100      &                         &            &     & 0.32                                  & D17        & 2.30                                  & D17     &     &                                       & \\
IC 5179                  & 0.011                       & U/LIRG        &     & 1523.65                               & 0.52        & 60/100      & -3.44$\pm$0.02          & PG16S      &     & 0.62                                  & D17        & 5.29                                  & D17     &     & 9.02                                  & M06 \\
NGC 6156                 & 0.011                       & U/LIRG        &     & 1385.22                               & 0.54        & 60/100      &                         &            &     & 0.52                                  & D17        & 3.80                                  & D17     &     &                                       & \\
NGC 6824                 & 0.011                       & SF            &     & 545.10                                & 0.38        & 60/100      &                         &            &     &                                       &            & 1.35                                  & B08     &     &                                       & \\
...                      & \multicolumn{16}{c}{...} \\
IRAS 07598+6508          & 0.15                        & U/LIRG, AGN   &     & 31622.78                              & 0.40        & 63/158      &                         &            &     &                                       &            & 3.24                                  & HERUS   &     &                                       & \\
LEDA 90196               & 0.16                        & U/LIRG, AGN   &     & 30199.52                              &             &             &                         &            &     &                                       &            & 20.02                                 & F16     &     &                                       & \\
3C 273                   & 0.16                        & U/LIRG, AGN   &     & 67608.30                              & 0.73        & 63/158      &                         &            &     & $<$20.06                              & F16        & 2.62                                  & HERUS   &     &                                       & \\
Mrk 1014                 & 0.16                        & U/LIRG, AGN   &     & 41686.94                              & 1.02        & 63/158      &                         &            &     &                                       &            & 6.10                                  & LCT     &     &                                       & \\
IRAS 06361-6217          & 0.16                        & LINER         &     &                                       &             &             &                         &            &     &                                       &            & 5.86                                  & F16     &     &                                       & \\
SDSS J120424.54+192509.7 & 0.17                        & U/LIRG, LINER &     & 34673.69                              &             &             &                         &            &     &                                       &            & 6.65                                  & F16     &     &                                       & \\
LEDA 70560               & 0.17                        & AGN           &     &                                       &             &             &                         &            &     &                                       &            & 6.91                                  & F16     &     &                                       & \\
IRAS 13342+3932          & 0.18                        & U/LIRG, AGN   &     & 29512.09                              &             &             &                         &            &     &                                       &            & 29.96                                 & F16     &     &                                       & \\
3C 234.0                 & 0.18                        & AGN           &     &                                       & $>$0.58     & 63/158      &                         &            &     & 9.08                                  & F16        & 4.80                                  & F16     &     &                                       & \\
IRAS 20037-1547          & 0.19                        & AGN           &     &                                       &             &             &                         &            &     &                                       &            & 19.05                                 & F16     &     &                                       & \\
6dFGS gJ004215.5-125604  & 0.26                        & U/LIRG        &     & 79432.82                              &             &             &                         &            &     &                                       &            & 8.89                                  & HERUS   &     &                                       & \\
IRAS 17002+5153          & 0.29                        & AGN           &     &                                       &             &             &                         &            &     &                                       &            & $<$12.25                              & F16     &     &                                       & \\

    \enddata
\end{deluxetable*}
\end{rotatetable*}

\section{High-\texorpdfstring{\zz{}}{z} Galaxy FSL Catalog}
\label{sec:flames-high}

\subsection{Sample Selection}
\label{sec:flames-high_sample}

In order to obtain a comprehensive coverage of the high-\zz{} FIR FSL data, high-\zz{} (\zz{} \textgreater 1) galaxies studied in more than 400 papers with FIR FSL observations were added in the catalog, with a focus on the redshift 4 and beyond where the atmospheric condition is good for \cii{} observation. 
The sample only include galaxies that have robust redshift measurements, either through FIR FSL or other spectral lines. 
The catalog also prioritize detections than limits. 
Being inclusive in sample selection, the catalog contains a conservative estimate of \textgreater 90\% of high-\zz{} FIR FSL detections of confirmed extragalactic objects in the literature. 
However, this catalog does not guarantee the completeness of high-\zz{} FIR FSL data, as the old data are scattered around the literature and could be missed in our search, and the new observations are constantly being carried out. 
Some low signal-to-noise sources or data may also be rejected, or falsified, or identified as multiple objects in future studies, as discovery and confirmation of high-\zz{} galaxies is still a tough work. 
A total of 543 sources are included in the current version of the catalog.

\subsection{FIR FSL Data}
\label{sec:flames-high_fir}

The FIR FSL data of these galaxies were observed by a variety of facilities, including Herschel/SPIRE \citep{griffin10}, APEX \citep{gusten06}, CSO, PdBI or NOEMA, ALMA, etc. 
Different from the low-\zz{} catalog, we list all the references that report unique observations or reduction results, and choose the value to record in the catalog based on the data quality and coverage of the system. 

The flux values in the catalog are observed fluxes without magnification corrections. 
In addition to the observed fluxes, the intrinsic line luminosities are also recorded, or computed if not in the literature, by correcting for the luminosity distance and gravitational lensing if present.

\subsection{Optical Data}
\label{sec:flames-high_optical}

Prior to JWST, only a handful of rest-frame optical spectra of FSL target galaxies were obtained with Keck or VLT, targeting mostly lensed galaxies \citep[e.g.][]{HK09,RJ11b}. 
With the launch of JWST, rest-frame optical spectra of high-\zz{} galaxies are rapidly accumulating with NIRSpec \citep{jakobsen22}. 
The optical line fluxes are only included if the aperture correction has been applied. 
High-\zz{} sources are often resolved into multiple components in high-resolution optical data; thus we choose the reported fluxes that best match with the beam size and velocity range of the FIR observation. 

For consistency, only gas phase extinction is included in the catalog, either extracted from the references or by applying the same procedure as in the low-\zz{} catalog if both \ha{} and \hb{} fluxes are available. 
When extinction is not available, a placeholder lower limit of 0 is added to $E$(B-V) and $A_\mathrm{V}$, and is used to compute the lower limits of extinction-corrected luminosities.

\subsection{Photometry Data}
\label{sec:flames-high_photometry}

The local continuums of FIR FSL data are recorded as photometries. 
Additional imaging photometry data at wavelength $\pm$6\% from the line wavelengths, especially with Herschel/SPIRE \cite{griffin10}, are also included in the catalog. 

\lir{} and \lfir{} are recorded whenever they are available in the literature. 
But, as noted in Sec.~\ref{sec:data_fir}, 2$\times$\lfir{} is used in this paper as \lir{} when the latter is not available. 
It is also worth noting that the IR and FIR luminosities of high-\zz{} galaxies are often estimated using photometry in the Rayleigh–Jeans tail, so the results are highly uncertain and the reported uncertainties could be severely underestimated. 

Some references perform SED fitting to derive dust properties. 
The dust mass and temperature are recorded in the catalog whenever available. 
However, they are derived with different SED models and a priori and cannot be compared directly. 

For sources with absolute FUV magnitudes available, they are recorded and converted to \lfuv at a generic wavelength 1500\,\AA{}. 

\subsection{Ancillary Data}
\label{sec:flames-high_ancillary}

The high-\zz{} galaxy types are strongly correlated to the source selection method, and some types are likely the high-\zz{} analogs for low-\zz{} types. 
Based on selections and luminosities, all relevent types are assigned and listed for each galaxy, but only one is used in the paper with the following descending priority: proto-cluster, DLA host, ETG, LAE/LBG, QSO (including AGN host), DSFG.

Many of the galaxies are gravitationally lensed. 
We record the lensing magnification factors along with their references in the catalog. 
Note that for many DSFGs, they are believed to be lensed but no lens model is available. 
A placeholder value of 6---typical magnification at \zz{}~$\sim$~4---is listed and used. 

If not available in the literautre, O/H and/or N/O are computed with the optical strong line methods described in Sec.~\ref{sec:data_post} that best suits the data availability, limited by the lack of rest-frame optical data. 

Some rest-frame optical or UV studies perform SED fitting. 
We include the stellar mass and SFR of the SED fitting results in the catalog, and list the corresponding references in either optical or UV reference columns. 
Because these SED fittings are performed with different model and a priori, extra caution must be taken if these data are used.

\subsection{Table Structure}
\label{sec:flames-high_structure}
The table of the high-\zz{} catalog\footnote{FLAMES-high. The full machine-readable table will be available for download in the online published version. Prior to that, the catalog can be obtained by contacting the correspondence author. } is organized in the following format

\begin{enumerate}
    \item Column 1. Name of the galaxy or system.
    \item Column 2. Alternative name(s) of the galaxy or system.
    \item Column 3-4. R.A. and Dec. (J2000) of the object. 
    \item Column 5. Redshift, the redshift of the object, based on FSL redshift if available. 
    \item Column 6. Type(s), the type(s) assigned to the object based on the selection method and properties. 
    \item Column 7-9. Gravitational lensing magnification factor $\mu$, uncertainty, and rerference(s), empty if not lensed.  
    \item Column 10-29. \cii{} and \oiii{88} data, sets of 11 columns: (1) observed line flux ($S\Delta v_\mathrm{line}$), (2) flux uncertainty, (3) flux limit flag, (4) intrinsic (gravitational lensing corrected) luminosity (\lline{}), (5) luminosity uncertainty, (6) luminosity flag, (7) line fullwidth half maximum (FWHM), (8) FWHM uncertainty, (9) reported line refshift, (10) reference(s) for the line data; fluxes are in the observer's unit of Jy km s\textsuperscript{-1}, luminosities are in 10\textsuperscript{8} \lsun{}, FWHM in km  s\textsuperscript{-1}. 
    \item Column 30-71. FIR FSL data of \oiii{52}, \niii{}, \oi{63}, \nii{122}, \oi{145}, \nii{205}; sets of seven columns: (1) observed line flux ($\Delta v_\mathrm{line}$), (2) flux uncertainty, (3) flux limit flag, (4) line luminosity (\lline{}), (5) luminosity uncertainty, (6) luminosity flag, (7) reference(s) for the line data; fluxes are in the observer's unit of Jy km s\textsuperscript{-1}, luminosities are in the unit 10\textsuperscript{8} \lsun{}. 
    \item Column 72-127. Photometry data in the bands: 52, 57, 63, 88, 122, 145, 158, 205 \um{}; data within $\pm 6\%$ wavelength range are also considered to be in the band; sets of seven columns: (1) flux density (S$_\lambda$), (2) uncertainty, (3) limit flag, and (4) single-band luminosity (\vlv{}), (5) luminosity uncertainty, (6) luminosity flag, (7) reference(s); flux density is in the unit mJy, luminosities are in 10\textsuperscript{8} \lsun{}. 
    \item Column 128-144. Reported total IR luminosity \lir{}, FIR luminosity \lfir{}, dust mass \md{}, and dust temperature \td{}; sets of four columns: (1) value, (2) uncertainty, (3) lower uncertainty (if different from uncertainty), (4) limit flag, one extra column records the reference(s) for any of the dust properties in column 128-143; luminosity in the unit of 10\textsuperscript{8} \lsun{}, mass in the unit of 10\textsuperscript{8} \msun{}, and temperature in the unit K. 
    \item Column 145-149. Dust color temperature \tcolor{}, computed based on other band ratios; the same sets of four columns, one extra column records the photometry bands used to compute the dust color temperature; in the unit K.
    \item Column 150-166. Absolute FUV magnitude $M_\mathrm{FUV}$, FUV luminosity \lfuv{}, UV continuum slope $\beta_\mathrm{UV}$, and infrared excess index IRX; the same sets of four columns, one extra column records the reference(s) for any of the UV properties in column 150-165, as well as the SED fitting properties in column 167-174 in some cases; magnitude in the AB magnitude unit, luminosity in the unit 10\textsuperscript{8} \lsun{}. 
    \item Column 167-174. Reported SED fitted stellar mass log $M_\mathrm{*,SED}$ and star formation rate SFR\textsubscript{SED}; the same sets of four columns; log $M_\mathrm{*,SED}$ in the unit log \msun{}, SFR in the unit \msun{}\,yr\textsuperscript{-1}. 
    \item Column 175-183. Absolute metallicity \oh{} and Nitrogen-to-Oxygen abundance ratio \no{}; the same sets of four columns, one extra column records the reference to the metallicity data. 
    \item Column 184-202. Optical line data of \ha{}, \oiiio{5007}, and \niio{6584}; sets of 6 columns: (1-3) observed line flux $F_\mathrm{obs}$, uncertainty, and limit flag, (4-6) intrinsic extinction corrected line luminosity $L_\mathrm{line,ext.corr.}$, uncertainty, and limit flag, one extra column records the reference(s) for any of the optical properties in column 184-201, as well as the SED fitting properties in column 167-174 in some cases; flux in the unit 10\textsuperscript{-18}\,erg\,s\textsuperscript{-1}\,cm\textsuperscript{-2} (10\textsuperscript{-21}\,W\,m\textsuperscript{-2}), luminosity in the unit 10\textsuperscript{8} \lsun{}. 
    \item Column 203-211. Extinction $A_\mathrm{V}$ and $E_\mathrm{B-V}$; the same sets of four columns, one extra column records the reference(s). 

\end{enumerate}

All intrinsic properties, including luminosities and masses, are corrected for gravitational lensing. 
While observe properties, like line fluxes and continuum flux densities, are not corrected. 

The reference codes in the catalog are AA25: \citet{AA25}, AH22: \citet{AH22}, AH24a: \citet{AH24a}, AH24b: \citet{AH24b}, AH25: \citet{AH25}, AJ23: \citet{AJ23}, AJ24: \citet{AJ24}, AM08: \citet{AM08}, AM16: \citet{AM16}, AS07: \citet{AS07}, AS13: \citet{AS13}, AS24: \citet{AS24}, AY08: \citet{AY08}, BA06: \citet{BA06}, BA23: \citet{BA23}, BA24: \citet{BA24}, BC06: \citet{BC06}, BC21: \citet{BC21}, BE15: \citet{BE15}, BE16: \citet{BE16}, BE18a: \citet{BE18a}, BE18b: \citet{BE18b}, BE21: \citet{BE21}, BE24: \citet{BE24}, BE25: \citet{BE25}, BI17: \citet{BI17}, B08: \citet{B08}, BJ23: \citet{BJ23}, BL08: \citet{BL08}, BL12: \citet{BL12}, BM16: \citet{BM16}, BM17: \citet{BM17}, BM18: \citet{BM18}, BM20: \citet{BM20}, BM24a: \citet{BM24a}, BM24b: \citet{BM24b}, BM25: \citet{BM25}, BP23: \citet{BP23}, BP98: \citet{BP98}, BR13: \citet{BR13}, BR15: \citet{BR15}, BR18: \citet{BR18}, BR22a: \citet{BR22a}, BR22b: \citet{BR22b}, BR22c: \citet{BR22c}, BR92: \citet{BR92}, BS24: \citet{BS24}, BT20: \citet{BT20}, BT21: \citet{BT21}, BT23: \citet{BT23}, BT24a: \citet{BT24a}, BT24b: \citet{BT24b}, BV07: \citet{BV07}, CA11: \citet{CA11}, CA23: \citet{CA23}, CA24: \citet{CA24}, CC03: \citet{CC03}, CC10: \citet{CC10}, CC13: \citet{CC13}, CC15: \citet{CC15}, CC20: \citet{CC20}, CD20: \citet{CD20}, CE09: \citet{CE09}, CF12: \citet{CF12}, CI20: \citet{CI20}, CJ14: \citet{CJ14}, CJ24: \citet{CJ24}, CK12: \citet{CK12}, CL11: \citet{CL11}, CM24: \citet{CM24}, CP02: \citet{CP02}, CP11a: \citet{CP11a}, CP11b: \citet{CP11b}, CP15: \citet{CP15}, CS13: \citet{CS13}, CS17: \citet{CS17}, CS18a: \citet{CS18a}, CS18b: \citet{CS18b}, CS24: \citet{CS24}, DC04: \citet{DC04}, DC10: \citet{DC10}, DC11: \citet{DC11}, DC14: \citet{DC14}, DC19: \citet{DC19}, DD99a: \citet{DD99a}, DD99b: \citet{DD99b}, DE09: \citet{DE09}, DG14: \citet{DG14}, DG16: \citet{DG16}, DJ24: \citet{DJ24}, DK23: \citet{DK23}, DM15: \citet{DM15}, DM20: \citet{DM20}, DR12: \citet{DR12}, DR13: \citet{DR13}, DR14: \citet{DR14}, DR17: \citet{DR17}, DR18: \citet{DR18}, DR22: \citet{DR22}, DR23: \citet{DR23}, DT16: \citet{DT16}, DT18: \citet{DT18}, DT21: \citet{DT21}, EA06: \citet{EA06}, EA07: \citet{EA07}, EA20: \citet{EA20}, EA23: \citet{EA23}, ER23: \citet{ER23}, FA20: \citet{FA20}, FC10: \citet{FC10}, FC11: \citet{FC11}, FC15: \citet{FC15}, FD98: \citet{FD98}, FD99: \citet{FD99}, FF21: \citet{FF21}, FH12: \citet{FH12}, FH13: \citet{FH13}, FN09: \citet{FN09}, FR10: \citet{FR10}, FR24: \citet{FR24}, FS09: \citet{FS09}, FS19: \citet{FS19}, FS21: \citet{FS21}, FS22: \citet{FS22}, FS24a: \citet{FS24a}, FS24b: \citet{FS24b}, FS25: \citet{FS25}, FT19: \citet{FT19}, FY21: \citet{FY21}, FY24a: \citet{FY24a}, FY24b: \citet{FY24b}, FY25: \citet{FY25}, GB15: \citet{GB15}, GB18: \citet{GB18}, GC18: \citet{GC18}, GC24: \citet{GC24}, GG23: \citet{GG23}, GJ16: \citet{GJ16}, GJ17: \citet{GJ17}, GK24: \citet{GK24}, GM20: \citet{GM20}, GM23: \citet{GM23}, GR11: \citet{GR11}, GR13: \citet{GR13}, GS12: \citet{GS12}, HA12: \citet{HA12}, HA23: \citet{HA23}, HA24: \citet{HA24}, HB15: \citet{HB15}, HK09: \citet{HK09}, HK16: \citet{HK16}, HK19: \citet{HK19}, HK23: \citet{HK23}, HR11: \citet{HR11}, HR21: \citet{HR21}, HR22: \citet{HR22}, HS10: \citet{HS10}, HT18: \citet{HT18}, HT19a: \citet{HT19a}, HT19b: \citet{HT19b}, HT23: \citet{HT23}, HT25: \citet{HT25}, HY20: \citet{HY20}, HY25: \citet{HY25}, IA16: \citet{IA16}, IA20: \citet{IA20}, ID06: \citet{ID06}, ID16: \citet{ID16}, IE21: \citet{IE21}, IH22: \citet{IH22}, IK02: \citet{IK02}, IN25: \citet{IN25}, IR10a: \citet{IR10a}, IR10b: \citet{IR10b}, IR13: \citet{IR13}, IR16: \citet{IR16}, IR25: \citet{IR25}, IT18: \citet{IT18}, IT19: \citet{IT19}, IT21a: \citet{IT21a}, IT21b: \citet{IT21b}, JE20: \citet{JE20}, JG17: \citet{JG17}, JG24a: \citet{JG24a}, JG24b: \citet{JG24b}, JG25: \citet{JG25}, JT10a: \citet{JT10a}, JT10b: \citet{JT10b}, JX24: \citet{JX24}, KA15: \citet{KA15}, KB10: \citet{KB10}, KD23: \citet{KD23}, KJ93: \citet{KJ93}, KK16: \citet{KK16}, KK17: \citet{KK17}, KK23: \citet{KK23}, KM15: \citet{KM15}, KM23: \citet{KM23}, KN13: \citet{KN13}, KN15: \citet{KN15}, KN20: \citet{KN20}, KS12: \citet{KS12}, KS25: \citet{KS25}, KT25: \citet{KT25}, KY22: \citet{KY22}, LC18: \citet{LC18}, LC19: \citet{LC19}, LD07: \citet{LD07}, LD19: \citet{LD19}, LF21: \citet{LF21}, LF24: \citet{LF24}, LH09: \citet{LH09}, LJ10: \citet{LJ10}, LJ11: \citet{LJ11}, LJ20: \citet{LJ20}, LJ22: \citet{LJ22}, LJ24: \citet{LJ24}, LK19: \citet{LK19}, LK22: \citet{LK22}, LK23: \citet{LK23}, LK24: \citet{LK24}, LM19: \citet{LM19}, LM21: \citet{LM21}, LN17a: \citet{LN17a}, LN17b: \citet{LN17b}, LN18: \citet{LN18}, LN19: \citet{LN19}, LN21a: \citet{LN21a}, LN21b: \citet{LN21b}, LO20: \citet{LO20}, LR15: \citet{LR15}, LT23: \citet{LT23}, LW24: \citet{LW24}, MC17: \citet{MC17}, MC24a: \citet{MC24a}, MC24b: \citet{MC24b}, MC25: \citet{MC25}, MD18: \citet{MD18}, MG11: \citet{MG11}, MG12: \citet{MG12}, MG14: \citet{MG14}, MH14: \citet{MH14}, MI21: \citet{MI21}, MI24: \citet{MI24}, MI25: \citet{MI25}, MJ15: \citet{MJ15}, MJ17: \citet{MJ17}, MJ19: \citet{MJ19}, MM23: \citet{MM23}, MM24: \citet{MM24}, MR05: \citet{MR05}, MR09: \citet{MR09}, MR12: \citet{MR12}, MR15: \citet{MR15}, MR22: \citet{MR22}, MR25: \citet{MR25}, MS17: \citet{MS17}, MS22: \citet{MS22}, MT18: \citet{MT18}, MT20: \citet{MT20}, MT23: \citet{MT23}, MV22: \citet{MV22}, MY16: \citet{MY16}, MY18a: \citet{MY18a}, MY18b: \citet{MY18b}, MY19: \citet{MY19}, MY91: \citet{MY91}, MZ24: \citet{MZ24}, NK23: \citet{NK23}, NM10: \citet{NM10}, NM17: \citet{NM17}, NM19a: \citet{NM19a}, NM19b: \citet{NM19b}, NM19c: \citet{NM19c}, NM20: \citet{NM20}, NR14: \citet{NR14}, NT12: \citet{NT12}, OA01: \citet{OA01}, OI16: \citet{OI16}, OK14: \citet{OK14}, OM13: \citet{OM13}, OP16: \citet{OP16}, OS09: \citet{OS09}, OS12: \citet{OS12}, OY12: \citet{OY12}, PA06: \citet{PA06}, PA17: \citet{PA17}, PA21: \citet{PA21}, PA23a: \citet{PA23a}, PA23b: \citet{PA23b}, PA24: \citet{PA24}, PB23: \citet{PB23}, PE24: \citet{PE24}, PE25: \citet{PE25}, PG23: \citet{PG23}, PJ04: \citet{PJ04}, PL11: \citet{PL11}, PL16: \citet{PL16}, PP15: \citet{PP15}, PR16: \citet{PR16}, PR18: \citet{PR18}, PR19: \citet{PR19}, RC20: \citet{RC20}, RC25: \citet{RC25}, RD08: \citet{RD08}, RD09: \citet{RD09}, RD13: \citet{RD13}, RD14: \citet{RD14}, RD18: \citet{RD18}, RD20: \citet{RD20}, RF20: \citet{RF20}, RF21: \citet{RF21}, RF23: \citet{RF23}, RJ08: \citet{RJ08}, RJ11a: \citet{RJ11a}, RJ11b: \citet{RJ11b}, RL24: \citet{RL24}, RL25: \citet{RL25}, RM15a: \citet{RM15a}, RM15b: \citet{RM15b}, RM19: \citet{RM19}, RM20a: \citet{RM20a}, RM20b: \citet{RM20b}, RM21: \citet{RM21}, RM23: \citet{RM23}, RN24: \citet{RN24}, RS21: \citet{RS21}, RT14: \citet{RT14}, RW12: \citet{RW12}, RY23: \citet{RY23}, SA10: \citet{SA10}, SA11: \citet{SA11}, SA12: \citet{SA12}, SA13: \citet{SA13}, SA24: \citet{SA24}, SC18: \citet{SC18}, SC19: \citet{SC19}, SC21: \citet{SC21}, SD15a: \citet{SD15a}, SD15b: \citet{SD15b}, SD15c: \citet{SD15c}, SD24: \citet{SD24}, SE10: \citet{SE10}, SF24: \citet{SF24}, SF25: \citet{SF25}, SG10: \citet{SG10}, SH18: \citet{SH18}, SI05: \citet{SI05}, SI23: \citet{SI23}, SJ16: \citet{SJ16}, SJ20: \citet{SJ20}, SJ25: \citet{SJ25}, SL22: \citet{SL22}, SM12: \citet{SM12}, SM16: \citet{SM16}, SM17: \citet{SM17}, SM23: \citet{SM23}, SM24: \citet{SM24}, SM25a: \citet{SM25a}, SM25b: \citet{SM25b}, SP18: \citet{SP18}, SR18: \citet{SR18}, SS22: \citet{SS22}, SS23: \citet{SS23}, SS24: \citet{SS24}, SS25: \citet{SS25}, SS98: \citet{SS98}, SV12: \citet{SV12}, SV15: \citet{SV15}, SY17: \citet{SY17}, SY19: \citet{SY19}, SY21: \citet{SY21}, SY22: \citet{SY22}, SY25: \citet{SY25}, TA12: \citet{TA12}, TH00: \citet{TH00}, TK19: \citet{TK19}, TK22: \citet{TK22}, TK25: \citet{TK25}, TR22: \citet{TR22}, TR23a: \citet{TR23a}, TR23b: \citet{TR23b}, TR24: \citet{TR24}, TS23: \citet{TS23}, TT22: \citet{TT22}, TT23: \citet{TT23}, TY20: \citet{TY20}, TY23: \citet{TY23}, UB16: \citet{UB16}, UH15: \citet{UH15}, UH17: \citet{UH17}, UH21: \citet{UH21}, UH23: \citet{UH23}, UH24a: \citet{UH24a}, UH24b: \citet{UH24b}, UH25: \citet{UH25}, VB12: \citet{VB12}, VB16: \citet{VB16}, VB17a: \citet{VB17a}, VB17b: \citet{VB17b}, VB19: \citet{VB19}, VB20: \citet{VB20}, VE11: \citet{VE11}, VF22: \citet{VF22}, VG24: \citet{VG24}, VI11: \citet{VI11}, VS03: \citet{VS03}, VV24: \citet{VV24}, WA03: \citet{WA03}, WA07: \citet{WA07}, WA13: \citet{WA13}, WB22: \citet{WB22}, WB23: \citet{WB23}, WC07: \citet{WC07}, WC13: \citet{WC13}, WC15a: \citet{WC15a}, WC15b: \citet{WC15b}, WC17: \citet{WC17}, WD15: \citet{WD15}, WE12a: \citet{WE12a}, WE12b: \citet{WE12b}, WF09: \citet{WF09}, WF12: \citet{WF12}, WF18: \citet{WF18}, WF19a: \citet{WF19a}, WF19b: \citet{WF19b}, WF21a: \citet{WF21a}, WF21b: \citet{WF21b}, WF22: \citet{WF22}, WF24a: \citet{WF24a}, WF24b: \citet{WF24b}, WG21: \citet{WG21}, WJ10: \citet{WJ10}, WJ12: \citet{WJ12}, WJ13: \citet{WJ13}, WJ17: \citet{WJ17}, WJ20: \citet{WJ20}, WJ22: \citet{WJ22}, WR13: \citet{WR13}, WR16: \citet{WR16}, WR19: \citet{WR19}, WW24: \citet{WW24}, WW25: \citet{WW25}, WY21: \citet{WY21}, WY22a: \citet{WY22a}, WY22b: \citet{WY22b}, XM24: \citet{XM24}, YJ17: \citet{YJ17}, YJ19a: \citet{YJ19a}, YJ19b: \citet{YJ19b}, YJ20: \citet{YJ20}, YJ21: \citet{YJ21}, YM15: \citet{YM15}, YM21: \citet{YM21}, YM24: \citet{YM24}, YW14: \citet{YW14}, ZA18: \citet{ZA18}, ZA24: \citet{ZA24}, ZJ15: \citet{ZJ15}, ZJ18: \citet{ZJ18}, ZJ24: \citet{ZJ24}, ZJ25: \citet{ZJ25}, ZS24: \citet{ZS24}, ZY24: \citet{ZY24}, ZZ18: \citet{ZZ18}, vI24: \citet{vI24}.

A small fraction of the catalog with either \cii{} or \oiii{88} is shown in Table~\ref{tab:flames-high_example} as an example. 

\begin{rotatetable*}
\begin{deluxetable*}{lcccccrcrccrcr}
\centerwidetable
    \tabletypesize{\scriptsize}
    \tablecaption{Example of the high-\zz{} FSL catalog. \label{tab:flames-high_example}}
    \tablehead{
    \colhead{Name} & \colhead{Redshift} & \colhead{Type} & \colhead{...} & 
    \colhead{\lir{}} & \colhead{\tcolor{}} & \colhead{ref.} & 
    \colhead{\oh{}} & \colhead{ref.} & \colhead{...} & 
    \colhead{\loiii{88}} & 
    \colhead{ref.} & \colhead{\lcii{}} & \colhead{ref.} \\
     &  &  & & 
    \colhead{10\textsuperscript{8}\,\lsun{}} &  &  &  &  & &
    \colhead{10\textsuperscript{8}\,\lsun{}} &  & 
    \colhead{10\textsuperscript{8}\,\lsun{}} &  }
    \startdata
SDSS J120924+264052      & 1.021                       & SF         &     &                                       &                  &             & -3.40$\pm$0.07          & WE12a           &     &                                       &                 & 2.01$\pm$0.28                         & MS17                \\
G15v2.19                 & 1.027                       & SMG        &     & 60000$\pm$5000                        &                  &             &                         &                 &     &                                       &                 & 12.80$\pm$1.62                        & ZZ18                \\
Abell 2218b              & 1.032                       & SF         &     & 4800$^{+500}_{-2000}$                 &                  &             & -3.47$\pm$0.10          & RW12            &     &                                       &                 & 28.37$\pm$3.52                        & MS17                \\
Abell 2667a              & 1.035                       & SF         &     &                                       &                  &             & -3.27$\pm$0.10          & RW12            &     &                                       &                 & 8.53$\pm$1.09                         & MS17                \\
SD.v1.70                 & 1.19                        & SMG        &     & 39200$\pm$4670                        &                  &             &                         &                 &     & $<$7.69                               & ZZ18            & $<$8.74                               & ZZ18                \\
HBootes03                & 1.325                       & SMG        &     & 107000$\pm$23500                      & 0.39$\pm$0.10    & 63/158      &                         &                 &     & 28.30$\pm$14.60                       & ZZ18            & 28.30$\pm$12.80                       & HS10,ZZ18,BC06      \\
SA.v1.44                 & 1.33                        & SMG        &     & 30300$\pm$1500                        &                  &             &                         &                 &     & $<$24.40                              & ZZ18            & $<$9.21                               & ZZ18                \\
SDSS J091538+382658      & 1.501                       & SF         &     &                                       &                  &             & -3.58$^{+0.28}_{-0.30}$ & WE12a           &     &                                       &                 & 21.44$\pm$0.93                        & MS17                \\
SG.v1.77                 & 1.53                        & SMG        &     & 82300$\pm$2830                        &                  &             &                         &                 &     & $<$13.70                              & ZZ18            & $<$8.46                               & ZZ18                \\
SF.v1.88                 & 1.57                        & SMG        &     & 58000$\pm$2330                        &                  &             &                         &                 &     & $<$21.50                              & ZZ18            & $<$9.71                               & ZZ18                \\
SDP.9                    & 1.577                       & SMG        &     & 81000$\pm$19000                       &                  &             &                         &                 &     & 23$\pm$7.17                           & ZZ18            & 16.70$\pm$2.87                        & ZZ18                \\
...                      & \multicolumn{13}{c}{...} \\
DEIMOS\_COSMOS\_454608   & 4.578                       & LBG        &     & $<$4030                               &                  &             &                         &                 &     &                                       &                 & 6.44$\pm$1.08                         & LO20                \\
vuds\_cosmos\_5100994794 & 4.5783                      & LBG, LAE   &     & 1700$\pm$540                          &                  &             &                         &                 &     &                                       &                 & 6.03$\pm$0.58                         & LO20,IR25           \\
DEIMOS\_COSMOS\_761315   & 4.58                        & LBG, LAE   &     & $<$4100                               &                  &             &                         &                 &     &                                       &                 & $<$1.35                               & LO20                \\
DEIMOS\_COSMOS\_665626   & 4.583                       & LBG, LAE   &     & $<$3890                               &                  &             &                         &                 &     &                                       &                 & 1.61$\pm$0.43                         & LO20                \\
DEIMOS\_COSMOS\_814483   & 4.584                       & LBG, LAE   &     & $<$5570                               &                  &             &                         &                 &     &                                       &                 & 7.95$\pm$1.84                         & LO20                \\
DEIMOS\_COSMOS\_881725   & 4.585                       & LBG, LAE   & ... & 4730$\pm$1220                         &                  &             &                         &                 & ... &                                       &                 & 6.87$\pm$0.63                         & LO20                \\
vuds\_cosmos\_5100969402 & 4.5869                      & LBG, LAE   &     & 4440$\pm$1340                         &                  &             &                         &                 &     &                                       &                 & 5.23$\pm$0.55                         & LO20                \\
W2246-0526               & 4.6009                      & QSO        &     & 2210000                               &                  &             &                         &                 &     & $<$16.10                              & FR24            & 66$\pm$2.60                           & DT21,DT18,DT16,FR24 \\
AS2COS00034.1            & 4.615                       & SMG        &     &                                       &                  &             &                         &                 &     &                                       &                 & 35$\pm$7                              & MI21                \\
AS2COS0006.1             & 4.62                        & SMG        &     &                                       &                  &             &                         &                 &     &                                       &                 & 101$\pm$8                             & MI21                \\
AS2COS0034.2             & 4.621                       & SMG        &     &                                       &                  &             &                         &                 &     &                                       &                 & 60$\pm$17                             & MI21                \\
AS2COS0001.1             & 4.625                       & SMG        &     & 206000$\pm$9290                       &                  &             &                         &                 &     &                                       &                 & 44$\pm$8                              & SJ20,JE20,MI21      \\
...                      & \multicolumn{13}{c}{...} \\
ULAS J1342+0928          & 7.54                        & QSO, group &     & 15000$^{+8770}_{-5540}$               & 1.03$\pm$0.56    & 88/158      &                         &                 &     & 26.50$\pm$9.70                        & NM19c           & 13.20$\pm$0.90                        & VB17b,VB20,NM19c    \\
J0313-1806               & 7.6422                      & QSO        &     & 18300$\pm$1500                        &                  &             &                         &                 &     &                                       &                 & 6$\pm$1.30                            & WF21a,WF24a         \\
z7 GSD 3811              & 7.664                       & LAE        &     & $<$920                                &                  &             &                         &                 &     & $<$1.60                               & BC21            & $<$0.40                               & BC21                \\
REBELS-18                & 7.675                       & LBG        &     & 3500$^{+2000}_{-1300}$                &                  &             & -3.50$\pm$0.20          & RL25            &     &                                       &                 & 10.80$\pm$0.70                        & IH22,SL22,FY25      \\
REBELS-36                & 7.6772                      & LBG, LAE   &     & $<$2700                               &                  &             & -3.85$^{+0.21}_{-0.18}$ & VF22            &     &                                       &                 & 4.41$\pm$0.90                         & IH22,VF22           \\
A2744-YD4                & 7.8758                      & LAE        &     & 1600$\pm$600                          & $>$3.67$\pm$0.77 & $>$0.016    & -3.81$^{+0.09}_{-0.13}$ & MT23,VG24       &     &                                       &                 & $<$0.20                               & LN19                \\
MACS0416-Y1              & 8.312                       & LBG, group &     & 1060$\pm$188                          & $>$3.67$\pm$0.77 & $>$0.46     & -4.20$\pm$0.20          & MZ24,HA24       &     & 12$\pm$3                              & TY20,TY23       & 1.40$\pm$0.20                         & BT20                \\
S04590                   & 8.496                       & LBG        &     & $<$184                                &                  &             & -4.74$^{+0.15}_{-0.13}$ & HK23,FS24a,NK23 &     & 0.36$\pm$0.09                         & FS24a           & 0.17$\pm$0.04                         & FS24a,HK23          \\
MACS1149-JD1             & 9.1092                      & LAE        &     & $<$77                                 &                  &             & -4.10$^{+0.04}_{-0.05}$ & SM23,MC24a      &     & 0.74$\pm$0.16                         & HT18,LN21b,TT22 & $<$0.040                              & LN19                \\
GN-z11                   & 10.6034                     & LBG, AGN   &     & $<$36000                              &                  &             & -4.23$^{+0.06}_{-0.05}$ & CA23,BA23       &     &                                       &                 & $<$1.70                               & FY24b               \\
GL-z12-1                 & 12.342                      & LBG        &     & $<$650                                &                  &             & -4.60$^{+0.52}_{-0.37}$ & CM24,ZJ25       &     & 1.70$\pm$0.40                         & ZJ24,BT23,PG23  &                                       & \\
JADES-GS-z14-0           & 14.1793                     & LBG        &     & $<$1260                               &                  &             & -4.81$^{+0.70}_{-0.40}$ & CS24            &     & 2.10$\pm$0.50                         & SS24            & $<$0.60                               & SS25                \\

    \enddata
\end{deluxetable*}
\end{rotatetable*}

\section{FIR Color Temperature Calibration}
\label{sec:temp_cal}

\begin{figure}
    \centering
    \includegraphics[width=\halfwdth]{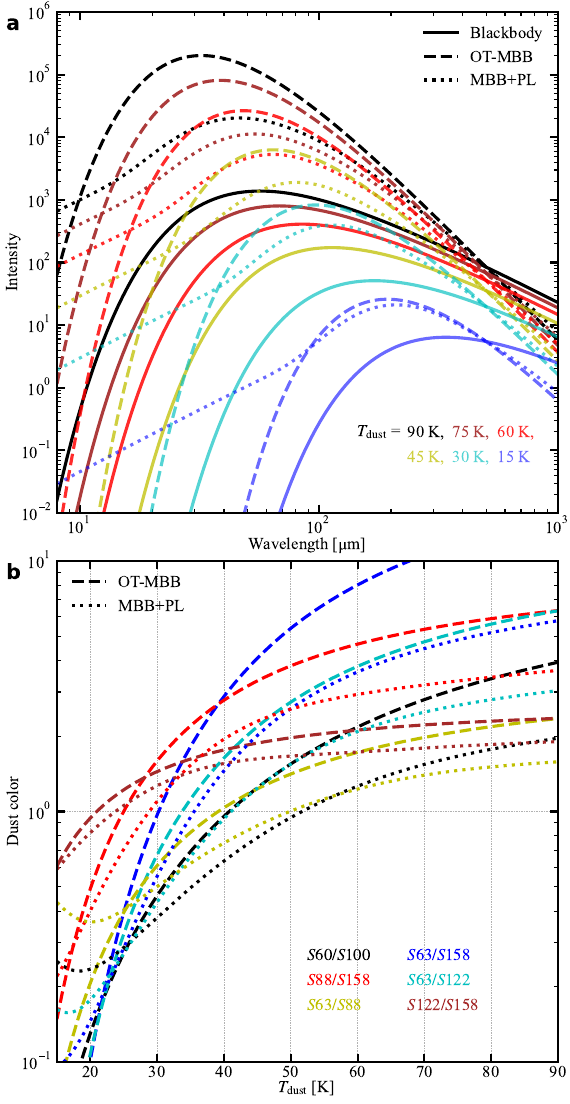}
    \caption[SED models and color-dust temperature conversion. ]{\bsf{a}, spectral profiles of the two dust SEDs considered in the paper (OT-MBB as dashed line, MBB+PL as dotted line) compared to a blackbody curve (solid line), in the wavelength range 8-to-1000 \um{}. The lines are color-coded according to the dust temperature noted in the legend. \bsf{b}, conversion between \td{} and dust color based on the two SEDs. The OT-MBB based relation is plotted in dashed line, and MBB+PL in dotted line. Curves of different colors correspond to different bands, as noted in the figure. }
    \label{f:data_sed}
\end{figure}

There is a large quantity of literature studying the optimal form of the dust SED, and no single best SED template exists. 
We therefore consider two commonly used SEDs: (1) a simple optically thin modified blackbody (OT-MBB) $S_\nu\sim \nu^\beta B_\nu(T_\mathrm{dust})$; (2) optically thick modified blackbody with MIR approximated by a truncated power law (MBB+PL) \citep[see][for definition]{casey12}. 
OT-MBB has only one free parameter $\beta$ that typically ranges between 1.5 and 2. 
$\beta=2$ is used for simplicity and has little impact on the short wavelength (\textless 160 \um{}) regime discussed in this work.
MBB+PL, however, takes three parameters, plus a normalization between the MBB and the power-law components. 
We use the values and normalizations listed in Table 1 in \citet{casey12}, which is representative of local U/LIRGs, except that $\lambda_0$ is fixed at 100 \um{}.
The SED shapes are shown in the first panel of Fig.~\ref{f:data_sed}.
The SED intensity is then normalized to the blackbody intensity at 500 \um{}, so that the absorption coefficient $\kappa_{500} = 0.051 \,\mathrm{m^2~kg^{-1}}$ \citep{clark16} can be used to convert the dust temperature \td{} and mass \md{} to luminosity \lir{}. 
Note that OT-MBB does not include the MIR component, so when computing \lir{}, a factor of 1.4 is applied on the OT-MBB integrated luminosity as compensation.

To convert the dust color to \td{}, we take the SED model flux density ratio at the corresponding wavelengths, which is plotted against the input SED \td{} in the lower panel of Fig.~\ref{f:data_sed}. 
The only exception is \tcolor{}. 
Unlike the local continuum accompanying the spectral line, S60 and S100 are measured with the broad-band filters onboard IRAS \citep{helou86}.
SO we convolve the dust SED with the filter transmission curve to derive a correct \tcolor{}-to-\td{} conversion. 

The OT-MBB and MBB+PL models produce different dust color-\td{} relations as shown in Fig.~\ref{f:data_sed}. 
At the same color, OT-MBB often yields a lower \td{}. 
One caveat for \td{} is that the optically thick MBB has a degeneracy between the unity optical depth wavelength $\lambda_0$ and \td{}. 
An optically thick MBB with a lower $\lambda_0$ value and a low \td{} is similar to that of high $\lambda_0$ and high \td{}. 
So, the high \td{}-color relation based on MBB-PL is in part due to the choice of a commonly used but high value of $\lambda_0$ = 100 \um{}. 
Therefore, we prefer the simpler form of OT-MBB and use it as the default color-\td{} conversion. 

Among the different colors of the dust, some show a more rapid change against \td{}, making them more sensitive proxies for the latter. 
S63/S158 is the best \td{} tracer in the range between 20 and 60 K, followed by S63/S122 and \tcolor{}. 
Although S63/S100 and S88/S158 also vary by more than 8 in the considered temperature range, the relation flattens markedly beyond 40 K, making them insensitive at high \td{}. 
S122/S158 is the worst \td{} measurement, varying by less than 1.2 from 40 K to 90 K.
However, it is plotted here because of the relatively common availability of \nii{122} and \cii{} continuum data, although we do not recommend using this for inferring \td{}. 

\begin{figure}
    \centering
    \includegraphics[width=\halfwdth]{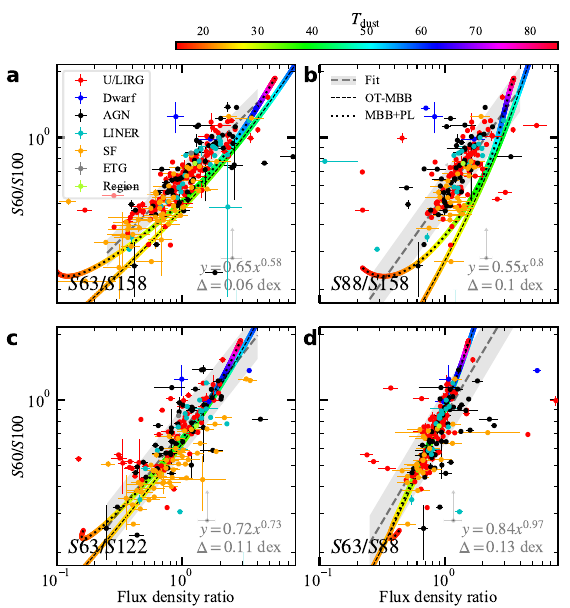}
    \caption[\tcolor{} vs. other band colors. ]{Conversion between \tcolor{} and the line continuum colors \bsf{a}, S63/S158; \bsf{b}, S88/S158; \bsf{c}, S63/S122; and \bsf{d}, S63/S88. Low-z data are shown as circles with error bars in the same plotting style as Fig.~\ref{f:data_z-IR}. The power-law fit and the 1 $\sigma$ range are shown as gray dashed lines and gray shade, with the exact form noted in the lower right corner of each panel. The SED model conversions are plotted as thin black dashed and dotted lines for OT-MBB and MBB+PL, respectively, below which curves are shown with the color manifesting the SED model \td{} at each point according to the colorbar displayed at the upper right corner.}
    \label{f:data_s60_100-color}
\end{figure}

We also calibrate and covert the line continuum color to \tcolor{}, which is more commonly available for high-\zz{} galaxies.
In Fig.~\ref{f:data_s60_100-color}, we show the continuum color-to-\td{} relation for both the local galaxies using the Herschel/PACS spectrometer continuum flux density, and the two dust SED models. 
One caveat of using the continuum measured in PACS spectrometer data is that the data often suffer from large uncertainty in the zero-point calibration especially in the case of high background, and the continuum flux density can be off by a factor \textgreater 2.  
Such a discrepancy is most apparent in the monochromatic luminosity comparison discussed below. 
Despite the uncertainty in the line continuum, a relatively tight color-to-\tcolor{} relation is present in every panel of Fig.~\ref{f:data_s60_100-color}, suggesting a systematic zero-point offset in the spectral data for consecutive observations on the same target. 
We fit a power law to each of the line continuum color, and the residual 1 $\sigma$ scatter is small within the range of 0.06 to 0.13 dex.

However, the observed color-to-\tcolor{} relations exhibit large offsets from the SED model conversion, most obvious in the two colors involving S158. 
With the calibration uncertainty in mind, the offset is likely a result of the systematic zero-point offsets rather than a real discrepancy between the dust SEDs and the observational data. 
We take a comprise, such that the low-\zz{} galaxies use the fitted empirical relations to convert line continuum colors to \tcolor{} when the latter is not available, while the high-\zz{} galaxy line continuum colors are transformed to \tcolor{} based on the OT-MBB SED. 

\begin{figure*}
    \centering
    \includegraphics[width=\textwidth]{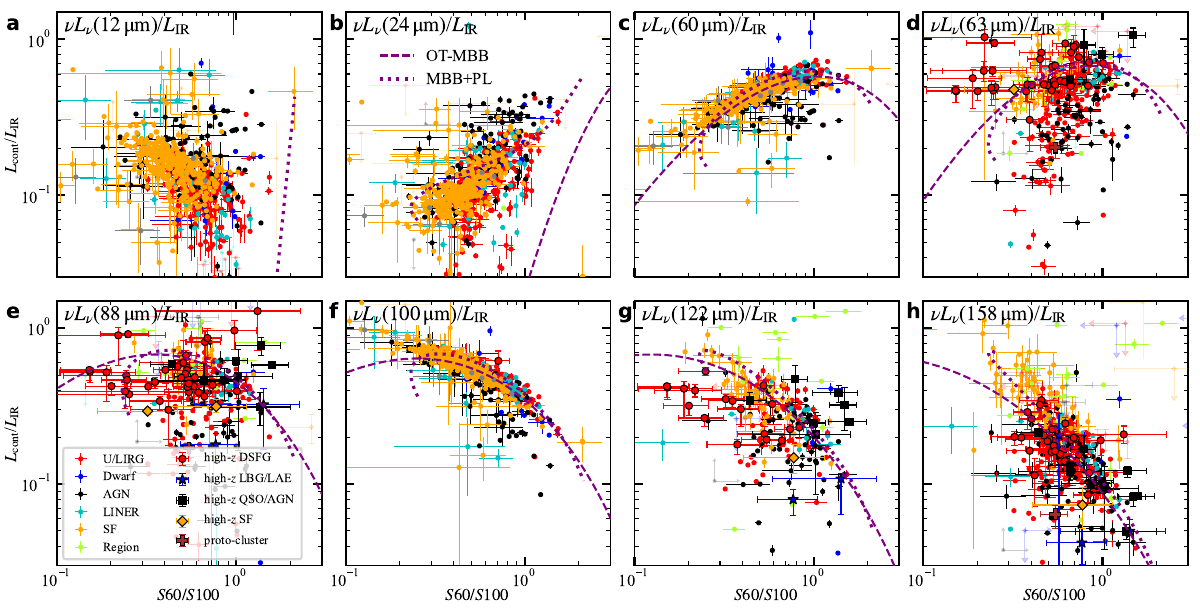}
    \caption[\vlv{}/\lir{} vs. \tcolor{}]{Monochromatic-to-total IR luminosity ratio \vlv{}/\lir{} vs. dust color \tcolor{}, for \bsf{a}-\bsf{d} 12, 24, 60, 63 \um{}; and \bsf{e}-\bsf{h} 88, 100, 122, and 158 \um{}. The plot style is the same as in Fig.~\ref{f:data_z-IR}. \vlv{}/\lir{} of the two SEDs are also shown as black dashed (OT-MBB) and dotted (MBB+PL) lines.}
    \label{f:data_s60_100-Lcont_LIR}
\end{figure*}

It is of great interest to estimate \lir{} with a few continuum flux density data points, because it is currently not possible to complete the IR wavelength coverage without 30--300 \um{} accessibility. 
Comparison of monochromatic luminosities against \lir{} also unveils the ensemble SED profile of the observation, which can be used to test the accuracy of the dust SED models used in this work. 
The monochromatic luminosity is calculated as $\nu L_\nu$, where $\nu$ is the representative frequency of the band, and $L_\nu$ is the luminosity density computed by multiplying the flux density by $4 \pi d_L^2$. 
To supplement our data, we use IRAS 25 \um{} data to approximate $L_\mathrm{24\mu m}$ using a relation $\nu L_{\nu,24} = 0.86 \nu L_{\nu,25}$ fitted to our data, and using Herschel 70 \um{} data for IRAS 60 \um{} with $\nu L_{\nu,60} = 0.81 \nu L_{\nu,70}$.
The comparison is shown in Fig.~\ref{f:data_s60_100-Lcont_LIR} against the color \tcolor{}, along with the dust SED model predictions. 

The $\nu L_\nu$/\lir{} of 12, 24, 60 and 100 \um{} show clear correlations to \tcolor{}, while the line continuum comparisons manifest large scatter, exemplified by significant underestimation at short wavelengths 63 and 88 \um{}. 
The differences again demonstrate the large zero-point calibration uncertainty for the PACS spectral continuum. 
The relationship between MIR flux density and \lir{} was estimated in several previous studies, including \citet{elbaz11,evans22,inami22}. 
In our comparison, $L_\mathrm{24\mu m}$/\lir{} shows a positive dependence on \td{}, and the SF galaxies (orange) appear to follow a flatter relation than that of IR bright galaxies, namely U/LIRGs (red) and AGNs (black). 
Besides, AGN hosts dominate the high $L_\mathrm{24\mu m}$/\lir{} parameter space because of the hot dust heated by AGN \citep{miley84,miley85}.
$L_\mathrm{24\mu m}$/\lir{} of the former varies from 0.2 to 0.07 when S60 / S100 drops from 1 to 0.3, while the U/LIRGs and AGNs have $L_\mathrm{24\mu m}$/\lir{} decreasing rapidly by more than an order of magnitude. 
Although the $L_\mathrm{24\mu m}$ to \lir{} correlation shows $\sim$ 0.3 dex of scatter for SF galaxies, making $L_\mathrm{24\mu m}$ a reasonable proxy for the latter, we caution against such use on IR bright galaxies demonstrated by the large variation and strong dependence on \td{} in Fig.~\ref{f:data_s60_100-Lcont_LIR}. 

For dust SED models, OT-MBB shows satisfactory agreement with the 60 and 100 \um{} \vlv{} correlations, but the model lacks the MIR component, making it incapable for 12 and 24 \um{} comparisons.  
MBB+PL performs slightly worse in the 60 and 100 \um{} comparison by overestimating \vlv{}/\lir{}.
At 24 \um{}, it follows a flat trend similar to that of the SF galaxies, but it still predicts too high a $\nu L_\nu$/\lir{} at low \td{}. 
At 12 \um{}, the simple power law used in MBB+PL is unable to reproduce the hot small grain emission that creates an inverse $L_{12}$/\lir{}-to-\td{} relation. 
The inferior performance of MBB+PL in matching the ensemble SED profile justifies our choice of OT-MBB for the color-to-\td{} conversion. 
Attempts to change $\beta$ in OT-MBB or the various parameters in MBB+PL, or the adoption of other dust SEDs such as a piecewise function \citep{drew22}, do not provide a consistent fit to the color-color relation and/or 24 \um{} \vlv{}. 
It is worth noting that the high-\zz{} galaxies fall along the \tcolor{}-\vlv{}/\lir{} trend that is marked by the dust SED models and populated by the local galaxy data points, validating the color-to-\tcolor{} color calibration.

\section{Verification of Extinction Correction for the Local Galaxy Optical Integrated Spectroscopic Data}
\label{sec:data_ext}

\begin{figure}
    \centering
    \includegraphics[width=\halfwdth]{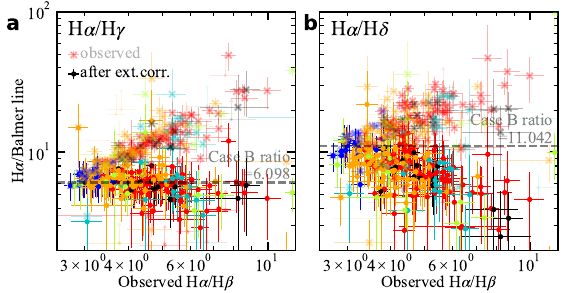}
    \caption[Verifying extinction correction. ]{Observed Balmer decrement \ha{}/\hb{} vs. \bsf{a}, Balmer line ratios \ha{}/\hg{}; and \bsf{b}, \ha{}/\hd{}. The observed line ratios for the y-axis are plotted as semi-transparent ``x'' symbols, and the ratios after dust extinction correction are plotted as opaque circles. The horizonal dashed lines indicate the theoretical line ratios based on the case B recombination rates \citep{draine11}.}
    \label{f:data_ext}
\end{figure}

Both the observed and extinction corrected line ratios of \ha{} to the two short wavelength Balmer lines \hg{} and \hd{}, are plotted against the observed Balmer decrement \ha{}/\hb{} that is used as the anchor for the host galaxy dust attenuation correction. 
The corrected \ha{}/\hg{} agree loosely with the prediction, showing the effectiveness of the correction. 
But the corrected \ha{}/\hd{} data points appear systematically lower than the case-B recombination prediction, and the downward trend is correlated with the observed \ha{}/\hb{}, meaning the \hd{} fluxes are over-corrected. 
The deviation in \ha{}/\hg{} correction is rather small, within a factor of 1.3, but it is up to a factor of 3 for \hd{}. 
Using a different extinction law such as the THEMIS attenuation law \citep{jones17} reproduces the expected \hg{}/\ha{}, but also over-estimates \hd{} and \oiio{3727} fluxes. 

The likely reason for the over-correction is that \citet{calzetti00} curve is used, but the U/LIRG attenuation curve slope may be even shallower than normal SF galaxies \citep{salim20}, thus \citet{calzetti00} or other Milky Way like attenuation curves tend to over-estimate $A_\lambda$ at short wavelengths. 
We therefore conclude the extinction correction for long wavelength optical data (\textgreater 4300 \AA) is satisfactory, but the use of the shorter wavelength data, including the telltale \oiio{3727} line, should be avoided.



\bibliographystyle{aasjournal}
\bibliography{main,low-z_ppi,high-z_ppi}{}

\begin{thebibliography}{}
\expandafter\ifx\csname natexlab\endcsname\relax\def\natexlab#1{#1}\fi
\providecommand{\url}[1]{\href{#1}{#1}}
\providecommand{\dodoi}[1]{doi:~\href{http://doi.org/#1}{\nolinkurl{#1}}}
\providecommand{\doeprint}[1]{\href{http://ascl.net/#1}{\nolinkurl{http://ascl.net/#1}}}
\providecommand{\doarXiv}[1]{\href{https://arxiv.org/abs/#1}{\nolinkurl{https://arxiv.org/abs/#1}}}

\bibitem[{{Abel} {et~al.}(2005){Abel}, {Ferland}, {Shaw}, \& {van
  Hoof}}]{abel05}
{Abel}, N.~P., {Ferland}, G.~J., {Shaw}, G., \& {van Hoof}, P.~A.~M. 2005,
  \apjs, 161, 65, \dodoi{10.1086/432913}

\bibitem[{{Abrahamsson} {et~al.}(2007){Abrahamsson}, {Krems}, \&
  {Dalgarno}}]{abrahamsson07}
{Abrahamsson}, E., {Krems}, R.~V., \& {Dalgarno}, A. 2007, \apj, 654, 1171,
  \dodoi{10.1086/509631}

\bibitem[{{Akins} {et~al.}(2022){Akins}, {Fujimoto}, {Finlator}, {Watson},
  {Knudsen}, {Richard}, {Bakx}, {Hashimoto}, {Inoue}, {Matsuo},
  {Micha{\l}owski}, \& {Tamura}}]{AH22}
{Akins}, H.~B., {Fujimoto}, S., {Finlator}, K., {et~al.} 2022, \apj, 934, 64,
  \dodoi{10.3847/1538-4357/ac795b}

\bibitem[{{Alaghband-Zadeh} {et~al.}(2013){Alaghband-Zadeh}, {Chapman},
  {Swinbank}, {Smail}, {Danielson}, {Decarli}, {Ivison}, {Meijerink}, {Weiss},
  \& {van der Werf}}]{AS13}
{Alaghband-Zadeh}, S., {Chapman}, S.~C., {Swinbank}, A.~M., {et~al.} 2013,
  \mnras, 435, 1493, \dodoi{10.1093/mnras/stt1390}

\bibitem[{{Algera} {et~al.}(2025){Algera}, {Rowland}, {Stefanon}, {Palla},
  {Sommovigo}, {Inami}, {Bouwens}, {Aravena}, {Bowler}, {Dayal}, {De Looze},
  {Ferrara}, {Fisher}, {Graziani}, {Gulis}, {Heintz}, {Hodge}, {van Leeuwen},
  {Pallottini}, {Phillips}, {Schouws}, {Smit}, {Stark}, \& {van der
  Werf}}]{AH25}
{Algera}, H., {Rowland}, L., {Stefanon}, M., {et~al.} 2025, arXiv e-prints,
  arXiv:2501.10508, \dodoi{10.48550/arXiv.2501.10508}

\bibitem[{{Algera} {et~al.}(2024{\natexlab{a}}){Algera}, {Inami}, {Sommovigo},
  {Fudamoto}, {Schneider}, {Graziani}, {Dayal}, {Bouwens}, {Aravena}, {da
  Cunha}, {Ferrara}, {Hygate}, {van Leeuwen}, {De Looze}, {Palla},
  {Pallottini}, {Smit}, {Stefanon}, {Topping}, \& {van der Werf}}]{AH24a}
{Algera}, H. S.~B., {Inami}, H., {Sommovigo}, L., {et~al.} 2024{\natexlab{a}},
  \mnras, 527, 6867, \dodoi{10.1093/mnras/stad3111}

\bibitem[{{Algera} {et~al.}(2024{\natexlab{b}}){Algera}, {Inami}, {De Looze},
  {Ferrara}, {Hirashita}, {Aravena}, {Bakx}, {Bouwens}, {Bowler}, {Da Cunha},
  {Dayal}, {Fudamoto}, {Hodge}, {Hygate}, {van Leeuwen}, {Nanayakkara},
  {Palla}, {Pallottini}, {Rowland}, {Smit}, {Sommovigo}, {Stefanon}, {Vijayan},
  \& {van der Werf}}]{AH24b}
{Algera}, H. S.~B., {Inami}, H., {De Looze}, I., {et~al.} 2024{\natexlab{b}},
  \mnras, 533, 3098, \dodoi{10.1093/mnras/stae1994}

\bibitem[{{Allam} {et~al.}(2007){Allam}, {Tucker}, {Lin}, {Diehl}, {Annis},
  {Buckley-Geer}, \& {Frieman}}]{AS07}
{Allam}, S.~S., {Tucker}, D.~L., {Lin}, H., {et~al.} 2007, \apjl, 662, L51,
  \dodoi{10.1086/519520}

\bibitem[{{{\'A}lvarez-M{\'a}rquez} {et~al.}(2023){{\'A}lvarez-M{\'a}rquez},
  {Crespo G{\'o}mez}, {Colina}, {Neeleman}, {Walter}, {Labiano},
  {P{\'e}rez-Gonz{\'a}lez}, {Bik}, {Noorgaard-Nielsen}, {Ostlin}, {Wright},
  {Alonso-Herrero}, {Azollini}, {Caputi}, {Eckart}, {Le F{\`e}vre},
  {Garc{\'\i}a-Mar{\'\i}n}, {Greve}, {Hjorth}, {Ilbert}, {Kendrew}, {Pye},
  {Tikkanen}, {Topinka}, {van der Werf}, {Ward}, {van Dishoeck}, {G{\"u}del},
  {Henning}, {Lagage}, {Ray}, \& {Waelkens}}]{AJ23}
{{\'A}lvarez-M{\'a}rquez}, J., {Crespo G{\'o}mez}, A., {Colina}, L., {et~al.}
  2023, \aap, 671, A105, \dodoi{10.1051/0004-6361/202245400}

\bibitem[{{{\'A}lvarez-M{\'a}rquez} {et~al.}(2024){{\'A}lvarez-M{\'a}rquez},
  {Colina}, {Crespo G{\'o}mez}, {Rinaldi}, {Melinder}, {{\"O}stlin},
  {Annunziatella}, {Labiano}, {Bik}, {Bosman}, {Greve}, {Wright},
  {Alonso-Herrero}, {Boogaard}, {Azollini}, {Caputi}, {Costantin}, {Eckart},
  {Garc{\'\i}a-Mar{\'\i}n}, {Gillman}, {Hjorth}, {Iani}, {Ilbert}, {Jermann},
  {Langeroodi}, {Meyer}, {Pei{\ss}ker}, {P{\'e}rez-Gonz{\'a}lez}, {Pye},
  {Tikkanen}, {Topinka}, {van der Werf}, {Walter}, {Henning}, \& {Ray}}]{AJ24}
{{\'A}lvarez-M{\'a}rquez}, J., {Colina}, L., {Crespo G{\'o}mez}, A., {et~al.}
  2024, \aap, 686, A85, \dodoi{10.1051/0004-6361/202347946}

\bibitem[{{Amvrosiadis} {et~al.}(2025){Amvrosiadis}, {Lange}, {Nightingale},
  {He}, {Frenk}, {Oman}, {Smail}, {Swinbank}, {Fragkoudi}, {Gadotti}, {Cole},
  {Borsato}, {Robertson}, {Massey}, {Cao}, \& {Li}}]{AA25}
{Amvrosiadis}, A., {Lange}, S., {Nightingale}, J.~W., {et~al.} 2025, \mnras,
  537, 1163, \dodoi{10.1093/mnras/staf048}

\bibitem[{{Ao} {et~al.}(2008){Ao}, {Wei{\ss}}, {Downes}, {Walter}, {Henkel}, \&
  {Menten}}]{AY08}
{Ao}, Y., {Wei{\ss}}, A., {Downes}, D., {et~al.} 2008, \aap, 491, 747,
  \dodoi{10.1051/0004-6361:200810482}

\bibitem[{{Aravena} {et~al.}(2008){Aravena}, {Bertoldi}, {Schinnerer}, {Weiss},
  {Jahnke}, {Carilli}, {Frayer}, {Henkel}, {Brusa}, {Menten}, {Salvato}, \&
  {Smolcic}}]{AM08}
{Aravena}, M., {Bertoldi}, F., {Schinnerer}, E., {et~al.} 2008, \aap, 491, 173,
  \dodoi{10.1051/0004-6361:200810628}

\bibitem[{{Aravena} {et~al.}(2016){Aravena}, {Spilker}, {Bethermin},
  {Bothwell}, {Chapman}, {de Breuck}, {Furstenau}, {G{\'o}nzalez-L{\'o}pez},
  {Greve}, {Litke}, {Ma}, {Malkan}, {Marrone}, {Murphy}, {Stark}, {Strandet},
  {Vieira}, {Weiss}, {Welikala}, {Wong}, \& {Collier}}]{AM16}
{Aravena}, M., {Spilker}, J.~S., {Bethermin}, M., {et~al.} 2016, \mnras, 457,
  4406, \dodoi{10.1093/mnras/stw275}

\bibitem[{{Arellano-C{\'o}rdova} \& {Rodr{\'\i}guez}(2020)}]{arellano20}
{Arellano-C{\'o}rdova}, K.~Z., \& {Rodr{\'\i}guez}, M. 2020, \mnras, 497, 672,
  \dodoi{10.1093/mnras/staa1759}

\bibitem[{{Armus} {et~al.}(2009){Armus}, {Mazzarella}, {Evans}, {Surace},
  {Sanders}, {Iwasawa}, {Frayer}, {Howell}, {Chan}, {Petric}, {Vavilkin},
  {Kim}, {Haan}, {Inami}, {Murphy}, {Appleton}, {Barnes}, {Bothun}, {Bridge},
  {Charmandaris}, {Jensen}, {Kewley}, {Lord}, {Madore}, {Marshall},
  {Melbourne}, {Rich}, {Satyapal}, {Schulz}, {Spoon}, {Sturm}, {U}, {Veilleux},
  \& {Xu}}]{A09}
{Armus}, L., {Mazzarella}, J.~M., {Evans}, A.~S., {et~al.} 2009, \pasp, 121,
  559, \dodoi{10.1086/600092}

\bibitem[{{Arribas} {et~al.}(2024){Arribas}, {Perna}, {Rodr{\'\i}guez Del
  Pino}, {Lamperti}, {D'Eugenio}, {P{\'e}rez-Gonz{\'a}lez}, {Jones}, {Crespo
  G{\'o}mez}, {Curti}, {Lim}, {{\'A}lvarez-M{\'a}rquez}, {Bunker}, {Carniani},
  {Charlot}, {Jakobsen}, {Maiolino}, {{\"U}bler}, {Willott}, {B{\"o}ker},
  {Chevallard}, {Circosta}, {Cresci}, {Kumari}, {Parlanti}, {Scholtz},
  {Venturi}, \& {Witstok}}]{AS24}
{Arribas}, S., {Perna}, M., {Rodr{\'\i}guez Del Pino}, B., {et~al.} 2024, \aap,
  688, A146, \dodoi{10.1051/0004-6361/202348824}

\bibitem[{{Asplund} {et~al.}(2009){Asplund}, {Grevesse}, {Sauval}, \&
  {Scott}}]{asplund09}
{Asplund}, M., {Grevesse}, N., {Sauval}, A.~J., \& {Scott}, P. 2009, \araa, 47,
  481, \dodoi{10.1146/annurev.astro.46.060407.145222}

\bibitem[{{Astropy Collaboration} {et~al.}(2013){Astropy Collaboration},
  {Robitaille}, {Tollerud}, {Greenfield}, {Droettboom}, {Bray}, {Aldcroft},
  {Davis}, {Ginsburg}, {Price-Whelan}, {Kerzendorf}, {Conley}, {Crighton},
  {Barbary}, {Muna}, {Ferguson}, {Grollier}, {Parikh}, {Nair}, {Unther},
  {Deil}, {Woillez}, {Conseil}, {Kramer}, {Turner}, {Singer}, {Fox}, {Weaver},
  {Zabalza}, {Edwards}, {Azalee Bostroem}, {Burke}, {Casey}, {Crawford},
  {Dencheva}, {Ely}, {Jenness}, {Labrie}, {Lim}, {Pierfederici}, {Pontzen},
  {Ptak}, {Refsdal}, {Servillat}, \& {Streicher}}]{astropy13}
{Astropy Collaboration}, {Robitaille}, T.~P., {Tollerud}, E.~J., {et~al.} 2013,
  \aap, 558, A33, \dodoi{10.1051/0004-6361/201322068}

\bibitem[{{Astropy Collaboration} {et~al.}(2018){Astropy Collaboration},
  {Price-Whelan}, {Sip{\H{o}}cz}, {G{\"u}nther}, {Lim}, {Crawford}, {Conseil},
  {Shupe}, {Craig}, {Dencheva}, {Ginsburg}, {VanderPlas}, {Bradley},
  {P{\'e}rez-Su{\'a}rez}, {de Val-Borro}, {Aldcroft}, {Cruz}, {Robitaille},
  {Tollerud}, {Ardelean}, {Babej}, {Bach}, {Bachetti}, {Bakanov}, {Bamford},
  {Barentsen}, {Barmby}, {Baumbach}, {Berry}, {Biscani}, {Boquien}, {Bostroem},
  {Bouma}, {Brammer}, {Bray}, {Breytenbach}, {Buddelmeijer}, {Burke},
  {Calderone}, {Cano Rodr{\'\i}guez}, {Cara}, {Cardoso}, {Cheedella}, {Copin},
  {Corrales}, {Crichton}, {D'Avella}, {Deil}, {Depagne}, {Dietrich}, {Donath},
  {Droettboom}, {Earl}, {Erben}, {Fabbro}, {Ferreira}, {Finethy}, {Fox},
  {Garrison}, {Gibbons}, {Goldstein}, {Gommers}, {Greco}, {Greenfield},
  {Groener}, {Grollier}, {Hagen}, {Hirst}, {Homeier}, {Horton}, {Hosseinzadeh},
  {Hu}, {Hunkeler}, {Ivezi{\'c}}, {Jain}, {Jenness}, {Kanarek}, {Kendrew},
  {Kern}, {Kerzendorf}, {Khvalko}, {King}, {Kirkby}, {Kulkarni}, {Kumar},
  {Lee}, {Lenz}, {Littlefair}, {Ma}, {Macleod}, {Mastropietro}, {McCully},
  {Montagnac}, {Morris}, {Mueller}, {Mumford}, {Muna}, {Murphy}, {Nelson},
  {Nguyen}, {Ninan}, {N{\"o}the}, {Ogaz}, {Oh}, {Parejko}, {Parley}, {Pascual},
  {Patil}, {Patil}, {Plunkett}, {Prochaska}, {Rastogi}, {Reddy Janga},
  {Sabater}, {Sakurikar}, {Seifert}, {Sherbert}, {Sherwood-Taylor}, {Shih},
  {Sick}, {Silbiger}, {Singanamalla}, {Singer}, {Sladen}, {Sooley},
  {Sornarajah}, {Streicher}, {Teuben}, {Thomas}, {Tremblay}, {Turner},
  {Terr{\'o}n}, {van Kerkwijk}, {de la Vega}, {Watkins}, {Weaver}, {Whitmore},
  {Woillez}, {Zabalza}, \& {Astropy Contributors}}]{astropy18}
{Astropy Collaboration}, {Price-Whelan}, A.~M., {Sip{\H{o}}cz}, B.~M., {et~al.}
  2018, \aj, 156, 123, \dodoi{10.3847/1538-3881/aabc4f}

\bibitem[{{Ba{\~n}ados} {et~al.}(2018{\natexlab{a}}){Ba{\~n}ados}, {Carilli},
  {Walter}, {Momjian}, {Decarli}, {Farina}, {Mazzucchelli}, \&
  {Venemans}}]{BE18b}
{Ba{\~n}ados}, E., {Carilli}, C., {Walter}, F., {et~al.} 2018{\natexlab{a}},
  \apjl, 861, L14, \dodoi{10.3847/2041-8213/aac511}

\bibitem[{{Ba{\~n}ados} {et~al.}(2015){Ba{\~n}ados}, {Decarli}, {Walter},
  {Venemans}, {Farina}, \& {Fan}}]{BE15}
{Ba{\~n}ados}, E., {Decarli}, R., {Walter}, F., {et~al.} 2015, \apjl, 805, L8,
  \dodoi{10.1088/2041-8205/805/1/L8}

\bibitem[{{Ba{\~n}ados} {et~al.}(2016){Ba{\~n}ados}, {Venemans}, {Decarli},
  {Farina}, {Mazzucchelli}, {Walter}, {Fan}, {Stern}, {Schlafly}, {Chambers},
  {Rix}, {Jiang}, {McGreer}, {Simcoe}, {Wang}, {Yang}, {Morganson}, {De Rosa},
  {Greiner}, {Balokovi{\'c}}, {Burgett}, {Cooper}, {Draper}, {Flewelling},
  {Hodapp}, {Jun}, {Kaiser}, {Kudritzki}, {Magnier}, {Metcalfe}, {Miller},
  {Schindler}, {Tonry}, {Wainscoat}, {Waters}, \& {Yang}}]{BE16}
{Ba{\~n}ados}, E., {Venemans}, B.~P., {Decarli}, R., {et~al.} 2016, \apjs, 227,
  11, \dodoi{10.3847/0067-0049/227/1/11}

\bibitem[{{Ba{\~n}ados} {et~al.}(2018{\natexlab{b}}){Ba{\~n}ados}, {Venemans},
  {Mazzucchelli}, {Farina}, {Walter}, {Wang}, {Decarli}, {Stern}, {Fan},
  {Davies}, {Hennawi}, {Simcoe}, {Turner}, {Rix}, {Yang}, {Kelson}, {Rudie}, \&
  {Winters}}]{BE18a}
{Ba{\~n}ados}, E., {Venemans}, B.~P., {Mazzucchelli}, C., {et~al.}
  2018{\natexlab{b}}, \nat, 553, 473, \dodoi{10.1038/nature25180}

\bibitem[{{Ba{\~n}ados} {et~al.}(2021){Ba{\~n}ados}, {Mazzucchelli}, {Momjian},
  {Eilers}, {Wang}, {Schindler}, {Connor}, {Andika}, {Barth}, {Carilli},
  {Davies}, {Decarli}, {Fan}, {Farina}, {Hennawi}, {Pensabene}, {Stern},
  {Venemans}, {Wenzl}, \& {Yang}}]{BE21}
{Ba{\~n}ados}, E., {Mazzucchelli}, C., {Momjian}, E., {et~al.} 2021, \apj, 909,
  80, \dodoi{10.3847/1538-4357/abe239}

\bibitem[{{Ba{\~n}ados} {et~al.}(2025){Ba{\~n}ados}, {Momjian}, {Connor},
  {Belladitta}, {Decarli}, {Mazzucchelli}, {Venemans}, {Walter}, {Wang}, {Xie},
  {Barth}, {Eilers}, {Fan}, {Khusanova}, {Schindler}, {Stern}, {Yang},
  {Andika}, {Carilli}, {Farina}, {Fabian}, {Hennawi}, {Pensabene}, \&
  {Rojas-Ruiz}}]{BE25}
{Ba{\~n}ados}, E., {Momjian}, E., {Connor}, T., {et~al.} 2025, Nature
  Astronomy, 9, 293, \dodoi{10.1038/s41550-024-02431-4}

\bibitem[{{Baier-Soto} {et~al.}(2022){Baier-Soto}, {Herrera-Camus},
  {F{\"o}rster Schreiber}, {Contursi}, {Genzel}, {Lutz}, \& {Tacconi}}]{BR22a}
{Baier-Soto}, R., {Herrera-Camus}, R., {F{\"o}rster Schreiber}, N.~M., {et~al.}
  2022, \aap, 664, L5, \dodoi{10.1051/0004-6361/202243642}

\bibitem[{{Bakx} {et~al.}(2020){Bakx}, {Tamura}, {Hashimoto}, {Inoue}, {Lee},
  {Mawatari}, {Ota}, {Umehata}, {Zackrisson}, {Hatsukade}, {Kohno}, {Matsuda},
  {Matsuo}, {Okamoto}, {Shibuya}, {Shimizu}, {Taniguchi}, \& {Yoshida}}]{BT20}
{Bakx}, T. J.~L.~C., {Tamura}, Y., {Hashimoto}, T., {et~al.} 2020, \mnras, 493,
  4294, \dodoi{10.1093/mnras/staa509}

\bibitem[{{Bakx} {et~al.}(2021){Bakx}, {Sommovigo}, {Carniani}, {Ferrara},
  {Akins}, {Fujimoto}, {Hagimoto}, {Knudsen}, {Pallottini}, {Tamura}, \&
  {Watson}}]{BT21}
{Bakx}, T. J.~L.~C., {Sommovigo}, L., {Carniani}, S., {et~al.} 2021, \mnras,
  508, L58, \dodoi{10.1093/mnrasl/slab104}

\bibitem[{{Bakx} {et~al.}(2023){Bakx}, {Zavala}, {Mitsuhashi}, {Treu},
  {Fontana}, {Tadaki}, {Casey}, {Castellano}, {Glazebrook}, {Hagimoto},
  {Ikeda}, {Jones}, {Leethochawalit}, {Mason}, {Morishita}, {Nanayakkara},
  {Pentericci}, {Roberts-Borsani}, {Santini}, {Serjeant}, {Tamura}, {Trenti},
  \& {Vanzella}}]{BT23}
{Bakx}, T. J.~L.~C., {Zavala}, J.~A., {Mitsuhashi}, I., {et~al.} 2023, \mnras,
  519, 5076, \dodoi{10.1093/mnras/stac3723}

\bibitem[{{Bakx} {et~al.}(2024{\natexlab{a}}){Bakx}, {Algera}, {Venemans},
  {Sommovigo}, {Fujimoto}, {Carniani}, {Hagimoto}, {Hashimoto}, {Inoue},
  {Salak}, {Serjeant}, {Vallini}, {Eales}, {Ferrara}, {Fudamoto}, {Imamura},
  {Inoue}, {Knudsen}, {Matsuo}, {Sugahara}, {Tamura}, {Taniguchi}, \&
  {Yamanaka}}]{BT24a}
{Bakx}, T. J.~L.~C., {Algera}, H. S.~B., {Venemans}, B., {et~al.}
  2024{\natexlab{a}}, \mnras, 532, 2270, \dodoi{10.1093/mnras/stae1613}

\bibitem[{{Bakx} {et~al.}(2024{\natexlab{b}}){Bakx}, {Amvrosiadis}, {Bendo},
  {Algera}, {Serjeant}, {Bonavera}, {Borsato}, {Chen}, {Cox},
  {Gonz{\'a}lez-Nuevo}, {Hagimoto}, {Harrington}, {Ivison}, {Kamieneski},
  {Marchetti}, {Riechers}, {Tsukui}, {van der Werf}, {Yang}, {Zavala},
  {Andreani}, {Berta}, {Cooray}, {De Zotti}, {Eales}, {Ikeda}, {Knudsen},
  {Mitsuhashi}, {Negrello}, {Neri}, {Omont}, {Scott}, {Tamura}, {Temi}, \&
  {Urquhart}}]{BT24b}
{Bakx}, T.~J.~L.~C., {Amvrosiadis}, A., {Bendo}, G.~J., {et~al.}
  2024{\natexlab{b}}, \mnras, 535, 1533, \dodoi{10.1093/mnras/stae2409}

\bibitem[{{Baldwin} {et~al.}(1981){Baldwin}, {Phillips}, \&
  {Terlevich}}]{baldwin81}
{Baldwin}, J.~A., {Phillips}, M.~M., \& {Terlevich}, R. 1981, \pasp, 93, 5,
  \dodoi{10.1086/130766}

\bibitem[{{Banados} {et~al.}(2024){Banados}, {Khusanova}, {Decarli}, {Momjian},
  {Walter}, {Connor}, {Carilli}, {Mazzucchelli}, {Rojas-Ruiz}, \&
  {Venemans}}]{BE24}
{Banados}, E., {Khusanova}, Y., {Decarli}, R., {et~al.} 2024, arXiv e-prints,
  arXiv:2408.12299, \dodoi{10.48550/arXiv.2408.12299}

\bibitem[{{Barinovs} {et~al.}(2005){Barinovs}, {van Hemert}, {Krems}, \&
  {Dalgarno}}]{barinovs05}
{Barinovs}, {\u{G}}., {van Hemert}, M.~C., {Krems}, R., \& {Dalgarno}, A. 2005,
  \apj, 620, 537, \dodoi{10.1086/426860}

\bibitem[{{Barisic} {et~al.}(2017){Barisic}, {Faisst}, {Capak}, {Pavesi},
  {Riechers}, {Scoville}, {Cooke}, {Kartaltepe}, {Casey}, \& {Smolcic}}]{BI17}
{Barisic}, I., {Faisst}, A.~L., {Capak}, P.~L., {et~al.} 2017, \apj, 845, 41,
  \dodoi{10.3847/1538-4357/aa7eda}

\bibitem[{{Barvainis} {et~al.}(1992){Barvainis}, {Antonucci}, \&
  {Coleman}}]{BR92}
{Barvainis}, R., {Antonucci}, R., \& {Coleman}, P. 1992, \apjl, 399, L19,
  \dodoi{10.1086/186596}

\bibitem[{{Beelen} {et~al.}(2006){Beelen}, {Cox}, {Benford}, {Dowell},
  {Kov{\'a}cs}, {Bertoldi}, {Omont}, \& {Carilli}}]{BA06}
{Beelen}, A., {Cox}, P., {Benford}, D.~J., {et~al.} 2006, \apj, 642, 694,
  \dodoi{10.1086/500636}

\bibitem[{{Belokurov} {et~al.}(2007){Belokurov}, {Evans}, {Moiseev}, {King},
  {Hewett}, {Pettini}, {Wyrzykowski}, {McMahon}, {Smith}, {Gilmore}, {Sanchez},
  {Udalski}, {Koposov}, {Zucker}, \& {Walcher}}]{BV07}
{Belokurov}, V., {Evans}, N.~W., {Moiseev}, A., {et~al.} 2007, \apjl, 671, L9,
  \dodoi{10.1086/524948}

\bibitem[{{Berg} {et~al.}(2020){Berg}, {Pogge}, {Skillman}, {Croxall},
  {Moustakas}, {Rogers}, \& {Sun}}]{berg20}
{Berg}, D.~A., {Pogge}, R.~W., {Skillman}, E.~D., {et~al.} 2020, \apj, 893, 96,
  \dodoi{10.3847/1538-4357/ab7eab}

\bibitem[{{Bergamini} {et~al.}(2023){Bergamini}, {Acebron}, {Grillo}, {Rosati},
  {Caminha}, {Mercurio}, {Vanzella}, {Mason}, {Treu}, {Angora}, {Brammer},
  {Meneghetti}, {Nonino}, {Boyett}, {Brada{\v{c}}}, {Castellano}, {Fontana},
  {Morishita}, {Paris}, {Prieto-Lyon}, {Roberts-Borsani}, {Roy}, {Santini},
  {Vulcani}, {Wang}, \& {Yang}}]{BP23}
{Bergamini}, P., {Acebron}, A., {Grillo}, C., {et~al.} 2023, \apj, 952, 84,
  \dodoi{10.3847/1538-4357/acd643}

\bibitem[{{Best} {et~al.}(1998){Best}, {Longair}, \& {Roettgering}}]{BP98}
{Best}, P.~N., {Longair}, M.~S., \& {Roettgering}, H.~J.~A. 1998, \mnras, 295,
  549, \dodoi{10.1046/j.1365-8711.1998.01245.x}

\bibitem[{{B{\'e}thermin} {et~al.}(2016){B{\'e}thermin}, {De Breuck},
  {Gullberg}, {Aravena}, {Bothwell}, {Chapman}, {Gonzalez}, {Greve}, {Litke},
  {Ma}, {Malkan}, {Marrone}, {Murphy}, {Spilker}, {Stark}, {Strandet},
  {Vieira}, {Wei{\ss}}, \& {Welikala}}]{BM16}
{B{\'e}thermin}, M., {De Breuck}, C., {Gullberg}, B., {et~al.} 2016, \aap, 586,
  L7, \dodoi{10.1051/0004-6361/201527739}

\bibitem[{{B{\'e}thermin} {et~al.}(2020){B{\'e}thermin}, {Fudamoto}, {Ginolfi},
  {Loiacono}, {Khusanova}, {Capak}, {Cassata}, {Faisst}, {Le F{\`e}vre},
  {Schaerer}, {Silverman}, {Yan}, {Amorin}, {Bardelli}, {Boquien}, {Cimatti},
  {Davidzon}, {Dessauges-Zavadsky}, {Fujimoto}, {Gruppioni}, {Hathi}, {Ibar},
  {Jones}, {Koekemoer}, {Lagache}, {Lemaux}, {Moreau}, {Oesch}, {Pozzi},
  {Riechers}, {Talia}, {Toft}, {Vallini}, {Vergani}, {Zamorani}, \&
  {Zucca}}]{BM20}
{B{\'e}thermin}, M., {Fudamoto}, Y., {Ginolfi}, M., {et~al.} 2020, \aap, 643,
  A2, \dodoi{10.1051/0004-6361/202037649}

\bibitem[{{Bik} {et~al.}(2024){Bik}, {{\'A}lvarez-M{\'a}rquez}, {Colina},
  {Crespo G{\'o}mez}, {Peissker}, {Walter}, {Boogaard}, {{\"O}stlin}, {Greve},
  {Wright}, {Alonso-Herrero}, {Caputi}, {Costantin}, {Eckart}, {Gillman},
  {Hjorth}, {Iani}, {Jermann}, {Labiano}, {Langeroodi}, {Melinder},
  {P{\'e}rez-Gonz{\'a}lez}, {Pye}, {Rinaldi}, {Tikkanen}, {van der Werf},
  {G{\"u}del}, {Henning}, {Lagage}, {Ray}, \& {van Dishoeck}}]{BA24}
{Bik}, A., {{\'A}lvarez-M{\'a}rquez}, J., {Colina}, L., {et~al.} 2024, \aap,
  686, A3, \dodoi{10.1051/0004-6361/202348845}

\bibitem[{{Binggeli} {et~al.}(2021){Binggeli}, {Inoue}, {Hashimoto}, {Toribio},
  {Zackrisson}, {Ramstedt}, {Mawatari}, {Harikane}, {Matsuo}, {Okamoto}, {Ota},
  {Shimizu}, {Tamura}, {Taniguchi}, \& {Umehata}}]{BC21}
{Binggeli}, C., {Inoue}, A.~K., {Hashimoto}, T., {et~al.} 2021, \aap, 646, A26,
  \dodoi{10.1051/0004-6361/202038180}

\bibitem[{{Birkin} {et~al.}(2023){Birkin}, {Hutchison}, {Welch}, {Spilker},
  {Aravena}, {Bayliss}, {Cathey}, {Chapman}, {Gonzalez}, {Gururajan},
  {Hayward}, {Khullar}, {Kim}, {Mahler}, {Malkan}, {Narayanan}, {Olivier},
  {Phadke}, {Reuter}, {Rigby}, {Smith}, {Solimano}, {Sulzenauer}, {Vieira},
  {Vizgan}, \& {Weiss}}]{BJ23}
{Birkin}, J.~E., {Hutchison}, T.~A., {Welch}, B., {et~al.} 2023, \apj, 958, 64,
  \dodoi{10.3847/1538-4357/acf712}

\bibitem[{{Bischetti} {et~al.}(2025){Bischetti}, {Feruglio}, {Carniani},
  {D'Odorico}, {Salvestrini}, \& {Fiore}}]{BM25}
{Bischetti}, M., {Feruglio}, C., {Carniani}, S., {et~al.} 2025, arXiv e-prints,
  arXiv:2504.15357, \dodoi{10.48550/arXiv.2504.15357}

\bibitem[{{Bischetti} {et~al.}(2018){Bischetti}, {Piconcelli}, {Feruglio},
  {Duras}, {Bongiorno}, {Carniani}, {Marconi}, {Pappalardo}, {Schneider},
  {Travascio}, {Valiante}, {Vietri}, {Zappacosta}, \& {Fiore}}]{BM18}
{Bischetti}, M., {Piconcelli}, E., {Feruglio}, C., {et~al.} 2018, \aap, 617,
  A82, \dodoi{10.1051/0004-6361/201833249}

\bibitem[{{Bischetti} {et~al.}(2024){Bischetti}, {Choi}, {Fiore}, {Feruglio},
  {Carniani}, {D'Odorico}, {Ba{\~n}ados}, {Chen}, {Decarli}, {Gallerani},
  {Hlavacek-Larrondo}, {Lai}, {Leighly}, {Mazzucchelli}, {Perreault-Levasseur},
  {Tripodi}, {Walter}, {Wang}, {Yang}, {Zanchettin}, \& {Zhu}}]{BM24a}
{Bischetti}, M., {Choi}, H., {Fiore}, F., {et~al.} 2024, \apj, 970, 9,
  \dodoi{10.3847/1538-4357/ad4a77}

\bibitem[{{Boreiko} \& {Betz}(1996)}]{boreiko96}
{Boreiko}, R.~T., \& {Betz}, A.~L. 1996, \apjl, 464, L83,
  \dodoi{10.1086/310094}

\bibitem[{{Borys} {et~al.}(2006){Borys}, {Blain}, {Dey}, {Le Floc'h},
  {Jannuzi}, {Barnard}, {Bian}, {Brodwin}, {Men{\'e}ndez-Delmestre},
  {Thompson}, {Brand}, {Brown}, {Dowell}, {Eisenhardt}, {Farrah}, {Frayer},
  {Higdon}, {Higdon}, {Phillips}, {Soifer}, {Stern}, \& {Weedman}}]{BC06}
{Borys}, C., {Blain}, A.~W., {Dey}, A., {et~al.} 2006, \apj, 636, 134,
  \dodoi{10.1086/497983}

\bibitem[{{Bosman} {et~al.}(2024){Bosman}, {{\'A}lvarez-M{\'a}rquez}, {Colina},
  {Walter}, {Alonso-Herrero}, {Ward}, {{\~A}-stlin}, {Greve}, {Wright}, {Bik},
  {Boogaard}, {Caputi}, {Costantin}, {Eckart}, {Garc{\'\i}a-Mar{\'\i}n},
  {Gillman}, {Hjorth}, {Iani}, {Ilbert}, {Jermann}, {Labiano}, {Langeroodi},
  {Pei{\ss}ker}, {Rinaldi}, {Topinka}, {van der Werf}, {G{\"u}del}, {Henning},
  {Lagage}, {Ray}, {van Dishoeck}, \& {Vandenbussche}}]{BS24}
{Bosman}, S. E.~I., {{\'A}lvarez-M{\'a}rquez}, J., {Colina}, L., {et~al.} 2024,
  Nature Astronomy, 8, 1054, \dodoi{10.1038/s41550-024-02273-0}

\bibitem[{{Bouwens} {et~al.}(2022){Bouwens}, {Smit}, {Schouws}, {Stefanon},
  {Bowler}, {Endsley}, {Gonzalez}, {Inami}, {Stark}, {Oesch}, {Hodge},
  {Aravena}, {da Cunha}, {Dayal}, {de Looze}, {Ferrara}, {Fudamoto},
  {Graziani}, {Li}, {Nanayakkara}, {Pallottini}, {Schneider}, {Sommovigo},
  {Topping}, {van der Werf}, {Algera}, {Barrufet}, {Hygate}, {Labb{\'e}},
  {Riechers}, \& {Witstok}}]{BR22c}
{Bouwens}, R.~J., {Smit}, R., {Schouws}, S., {et~al.} 2022, \apj, 931, 160,
  \dodoi{10.3847/1538-4357/ac5a4a}

\bibitem[{{Bowler} {et~al.}(2018){Bowler}, {Bourne}, {Dunlop}, {McLure}, \&
  {McLeod}}]{BR18}
{Bowler}, R.~A.~A., {Bourne}, N., {Dunlop}, J.~S., {McLure}, R.~J., \&
  {McLeod}, D.~J. 2018, \mnras, 481, 1631, \dodoi{10.1093/mnras/sty2368}

\bibitem[{{Bowler} {et~al.}(2022){Bowler}, {Cullen}, {McLure}, {Dunlop}, \&
  {Avison}}]{BR22b}
{Bowler}, R.~A.~A., {Cullen}, F., {McLure}, R.~J., {Dunlop}, J.~S., \&
  {Avison}, A. 2022, \mnras, 510, 5088, \dodoi{10.1093/mnras/stab3744}

\bibitem[{{Brada{\v{c}}} {et~al.}(2017){Brada{\v{c}}}, {Garcia-Appadoo},
  {Huang}, {Vallini}, {Quinn Finney}, {Hoag}, {Lemaux}, {Borello Schmidt},
  {Treu}, {Carilli}, {Dijkstra}, {Ferrara}, {Fontana}, {Jones}, {Ryan}, {Wagg},
  \& {Gonzalez}}]{BM17}
{Brada{\v{c}}}, M., {Garcia-Appadoo}, D., {Huang}, K.-H., {et~al.} 2017, \apjl,
  836, L2, \dodoi{10.3847/2041-8213/836/1/L2}

\bibitem[{{Brada{\v{c}}} {et~al.}(2024){Brada{\v{c}}}, {Strait}, {Mowla},
  {Iyer}, {Noirot}, {Willott}, {Brammer}, {Abraham}, {Asada}, {Desprez},
  {Estrada-Carpenter}, {Harshan}, {Martis}, {Matharu}, {Muzzin},
  {Rihtar{\v{s}}i{\v{c}}}, {Sarrouh}, \& {Sawicki}}]{BM24b}
{Brada{\v{c}}}, M., {Strait}, V., {Mowla}, L., {et~al.} 2024, \apjl, 961, L21,
  \dodoi{10.3847/2041-8213/ad0e73}

\bibitem[{{Bradley} {et~al.}(2008){Bradley}, {Bouwens}, {Ford}, {Illingworth},
  {Jee}, {Ben{\'\i}tez}, {Broadhurst}, {Franx}, {Frye}, {Infante}, {Motta},
  {Rosati}, {White}, \& {Zheng}}]{BL08}
{Bradley}, L.~D., {Bouwens}, R.~J., {Ford}, H.~C., {et~al.} 2008, \apj, 678,
  647, \dodoi{10.1086/533519}

\bibitem[{{Bradley} {et~al.}(2012){Bradley}, {Bouwens}, {Zitrin}, {Smit},
  {Coe}, {Ford}, {Zheng}, {Illingworth}, {Ben{\'\i}tez}, \&
  {Broadhurst}}]{BL12}
{Bradley}, L.~D., {Bouwens}, R.~J., {Zitrin}, A., {et~al.} 2012, \apj, 747, 3,
  \dodoi{10.1088/0004-637X/747/1/3}

\bibitem[{{Brauher} {et~al.}(2008{\natexlab{a}}){Brauher}, {Dale}, \&
  {Helou}}]{brauher08}
{Brauher}, J.~R., {Dale}, D.~A., \& {Helou}, G. 2008{\natexlab{a}}, \apjs, 178,
  280, \dodoi{10.1086/590249}

\bibitem[{{Brauher} {et~al.}(2008{\natexlab{b}}){Brauher}, {Dale}, \&
  {Helou}}]{B08}
---. 2008{\natexlab{b}}, \apjs, 178, 280, \dodoi{10.1086/590249}

\bibitem[{{Bunker} {et~al.}(2023){Bunker}, {Saxena}, {Cameron}, {Willott},
  {Curtis-Lake}, {Jakobsen}, {Carniani}, {Smit}, {Maiolino}, {Witstok},
  {Curti}, {D'Eugenio}, {Jones}, {Ferruit}, {Arribas}, {Charlot}, {Chevallard},
  {Giardino}, {de Graaff}, {Looser}, {L{\"u}tzgendorf}, {Maseda}, {Rawle},
  {Rix}, {Del Pino}, {Alberts}, {Egami}, {Eisenstein}, {Endsley}, {Hainline},
  {Hausen}, {Johnson}, {Rieke}, {Rieke}, {Robertson}, {Shivaei}, {Stark},
  {Sun}, {Tacchella}, {Tang}, {Williams}, {Willmer}, {Baker}, {Baum},
  {Bhatawdekar}, {Bowler}, {Boyett}, {Chen}, {Circosta}, {Helton}, {Ji},
  {Kumari}, {Lyu}, {Nelson}, {Parlanti}, {Perna}, {Sandles}, {Scholtz},
  {Suess}, {Topping}, {{\"U}bler}, {Wallace}, \& {Whitler}}]{BA23}
{Bunker}, A.~J., {Saxena}, A., {Cameron}, A.~J., {et~al.} 2023, \aap, 677, A88,
  \dodoi{10.1051/0004-6361/202346159}

\bibitem[{{Bussmann} {et~al.}(2013){Bussmann}, {P{\'e}rez-Fournon}, {Amber},
  {Calanog}, {Gurwell}, {Dannerbauer}, {De Bernardis}, {Fu}, {Harris}, {Krips},
  {Lapi}, {Maiolino}, {Omont}, {Riechers}, {Wardlow}, {Baker}, {Birkinshaw},
  {Bock}, {Bourne}, {Clements}, {Cooray}, {De Zotti}, {Dunne}, {Dye}, {Eales},
  {Farrah}, {Gavazzi}, {Gonz{\'a}lez Nuevo}, {Hopwood}, {Ibar}, {Ivison},
  {Laporte}, {Maddox}, {Mart{\'\i}nez-Navajas}, {Michalowski}, {Negrello},
  {Oliver}, {Roseboom}, {Scott}, {Serjeant}, {Smith}, {Smith}, {Streblyanska},
  {Valiante}, {van der Werf}, {Verma}, {Vieira}, {Wang}, \& {Wilner}}]{BR13}
{Bussmann}, R.~S., {P{\'e}rez-Fournon}, I., {Amber}, S., {et~al.} 2013, \apj,
  779, 25, \dodoi{10.1088/0004-637X/779/1/25}

\bibitem[{{Bussmann} {et~al.}(2015){Bussmann}, {Riechers}, {Fialkov},
  {Scudder}, {Hayward}, {Cowley}, {Bock}, {Calanog}, {Chapman}, {Cooray}, {De
  Bernardis}, {Farrah}, {Fu}, {Gavazzi}, {Hopwood}, {Ivison}, {Jarvis},
  {Lacey}, {Loeb}, {Oliver}, {P{\'e}rez-Fournon}, {Rigopoulou}, {Roseboom},
  {Scott}, {Smith}, {Vieira}, {Wang}, \& {Wardlow}}]{BR15}
{Bussmann}, R.~S., {Riechers}, D., {Fialkov}, A., {et~al.} 2015, \apj, 812, 43,
  \dodoi{10.1088/0004-637X/812/1/43}

\bibitem[{{Calanog} {et~al.}(2014){Calanog}, {Fu}, {Cooray}, {Wardlow}, {Ma},
  {Amber}, {Baker}, {Baes}, {Bock}, {Bourne}, {Bussmann}, {Casey}, {Chapman},
  {Clements}, {Conley}, {Dannerbauer}, {De Zotti}, {Dunne}, {Dye}, {Eales},
  {Farrah}, {Furlanetto}, {Harris}, {Ivison}, {Kim}, {Maddox}, {Magdis},
  {Messias}, {Micha{\l}owski}, {Negrello}, {Nightingale}, {O'Bryan}, {Oliver},
  {Riechers}, {Scott}, {Serjeant}, {Simpson}, {Smith}, {Timmons}, {Thacker},
  {Valiante}, \& {Vieira}}]{CJ14}
{Calanog}, J.~A., {Fu}, H., {Cooray}, A., {et~al.} 2014, \apj, 797, 138,
  \dodoi{10.1088/0004-637X/797/2/138}

\bibitem[{{Calzetti} {et~al.}(2000){Calzetti}, {Armus}, {Bohlin}, {Kinney},
  {Koornneef}, \& {Storchi-Bergmann}}]{calzetti00}
{Calzetti}, D., {Armus}, L., {Bohlin}, R.~C., {et~al.} 2000, \apj, 533, 682,
  \dodoi{10.1086/308692}

\bibitem[{{Cameron} {et~al.}(2023){Cameron}, {Katz}, {Rey}, \& {Saxena}}]{CA23}
{Cameron}, A.~J., {Katz}, H., {Rey}, M.~P., \& {Saxena}, A. 2023, \mnras, 523,
  3516, \dodoi{10.1093/mnras/stad1579}

\bibitem[{{Capak} {et~al.}(2011){Capak}, {Riechers}, {Scoville}, {Carilli},
  {Cox}, {Neri}, {Robertson}, {Salvato}, {Schinnerer}, {Yan}, {Wilson}, {Yun},
  {Civano}, {Elvis}, {Karim}, {Mobasher}, \& {Staguhn}}]{CP11b}
{Capak}, P.~L., {Riechers}, D., {Scoville}, N.~Z., {et~al.} 2011, \nat, 470,
  233, \dodoi{10.1038/nature09681}

\bibitem[{{Capak} {et~al.}(2015){Capak}, {Carilli}, {Jones}, {Casey},
  {Riechers}, {Sheth}, {Carollo}, {Ilbert}, {Karim}, {Lefevre}, {Lilly},
  {Scoville}, {Smolcic}, \& {Yan}}]{CP15}
{Capak}, P.~L., {Carilli}, C., {Jones}, G., {et~al.} 2015, \nat, 522, 455,
  \dodoi{10.1038/nature14500}

\bibitem[{{Carilli} {et~al.}(2003){Carilli}, {Lewis}, {Djorgovski}, {Mahabal},
  {Cox}, {Bertoldi}, \& {Omont}}]{CC03}
{Carilli}, C.~L., {Lewis}, G.~F., {Djorgovski}, S.~G., {et~al.} 2003, Science,
  300, 773, \dodoi{10.1126/science.1082600}

\bibitem[{{Carilli} {et~al.}(2013){Carilli}, {Riechers}, {Walter}, {Maiolino},
  {Wagg}, {Lentati}, {McMahon}, \& {Wolfe}}]{CC13}
{Carilli}, C.~L., {Riechers}, D., {Walter}, F., {et~al.} 2013, \apj, 763, 120,
  \dodoi{10.1088/0004-637X/763/2/120}

\bibitem[{{Carilli} \& {Walter}(2013)}]{carilli13}
{Carilli}, C.~L., \& {Walter}, F. 2013, \araa, 51, 105,
  \dodoi{10.1146/annurev-astro-082812-140953}

\bibitem[{{Carilli} {et~al.}(2010){Carilli}, {Daddi}, {Riechers}, {Walter},
  {Weiss}, {Dannerbauer}, {Morrison}, {Wagg}, {Dav{\'e}}, {Elbaz}, {Stern},
  {Dickinson}, {Krips}, \& {Aravena}}]{CC10}
{Carilli}, C.~L., {Daddi}, E., {Riechers}, D., {et~al.} 2010, \apj, 714, 1407,
  \dodoi{10.1088/0004-637X/714/2/1407}

\bibitem[{{Carniani} {et~al.}(2018{\natexlab{a}}){Carniani}, {Maiolino},
  {Smit}, \& {Amor{\'\i}n}}]{CS18b}
{Carniani}, S., {Maiolino}, R., {Smit}, R., \& {Amor{\'\i}n}, R.
  2018{\natexlab{a}}, \apjl, 854, L7, \dodoi{10.3847/2041-8213/aaab45}

\bibitem[{{Carniani} {et~al.}(2013){Carniani}, {Marconi}, {Biggs}, {Cresci},
  {Cupani}, {D'Odorico}, {Humphreys}, {Maiolino}, {Mannucci}, {Molaro},
  {Nagao}, {Testi}, \& {Zwaan}}]{CS13}
{Carniani}, S., {Marconi}, A., {Biggs}, A., {et~al.} 2013, \aap, 559, A29,
  \dodoi{10.1051/0004-6361/201322320}

\bibitem[{{Carniani} {et~al.}(2017){Carniani}, {Maiolino}, {Pallottini},
  {Vallini}, {Pentericci}, {Ferrara}, {Castellano}, {Vanzella}, {Grazian},
  {Gallerani}, {Santini}, {Wagg}, \& {Fontana}}]{CS17}
{Carniani}, S., {Maiolino}, R., {Pallottini}, A., {et~al.} 2017, \aap, 605,
  A42, \dodoi{10.1051/0004-6361/201630366}

\bibitem[{{Carniani} {et~al.}(2018{\natexlab{b}}){Carniani}, {Maiolino},
  {Amorin}, {Pentericci}, {Pallottini}, {Ferrara}, {Willott}, {Smit},
  {Matthee}, {Sobral}, {Santini}, {Castellano}, {De Barros}, {Fontana},
  {Grazian}, \& {Guaita}}]{CS18a}
{Carniani}, S., {Maiolino}, R., {Amorin}, R., {et~al.} 2018{\natexlab{b}},
  \mnras, 478, 1170, \dodoi{10.1093/mnras/sty1088}

\bibitem[{{Carniani} {et~al.}(2024){Carniani}, {Hainline}, {D'Eugenio},
  {Eisenstein}, {Jakobsen}, {Witstok}, {Johnson}, {Chevallard}, {Maiolino},
  {Helton}, {Willott}, {Robertson}, {Alberts}, {Arribas}, {Baker},
  {Bhatawdekar}, {Boyett}, {Bunker}, {Cameron}, {Cargile}, {Charlot}, {Curti},
  {Curtis-Lake}, {Egami}, {Giardino}, {Isaak}, {Ji}, {Jones}, {Kumari},
  {Maseda}, {Parlanti}, {P{\'e}rez-Gonz{\'a}lez}, {Rawle}, {Rieke}, {Rieke},
  {Del Pino}, {Saxena}, {Scholtz}, {Smit}, {Sun}, {Tacchella}, {{\"U}bler},
  {Venturi}, {Williams}, \& {Willmer}}]{CS24}
{Carniani}, S., {Hainline}, K., {D'Eugenio}, F., {et~al.} 2024, \nat, 633, 318,
  \dodoi{10.1038/s41586-024-07860-9}

\bibitem[{{Casey}(2012)}]{casey12}
{Casey}, C.~M. 2012, \mnras, 425, 3094,
  \dodoi{10.1111/j.1365-2966.2012.21455.x}

\bibitem[{{Castellano} {et~al.}(2024){Castellano}, {Napolitano}, {Fontana},
  {Roberts-Borsani}, {Treu}, {Vanzella}, {Zavala}, {Arrabal Haro},
  {Calabr{\`o}}, {Llerena}, {Mascia}, {Merlin}, {Paris}, {Pentericci},
  {Santini}, {Bakx}, {Bergamini}, {Cupani}, {Dickinson}, {Filippenko},
  {Glazebrook}, {Grillo}, {Kelly}, {Malkan}, {Mason}, {Morishita},
  {Nanayakkara}, {Rosati}, {Sani}, {Wang}, \& {Yoon}}]{CM24}
{Castellano}, M., {Napolitano}, L., {Fontana}, A., {et~al.} 2024, \apj, 972,
  143, \dodoi{10.3847/1538-4357/ad5f88}

\bibitem[{{Cathey} {et~al.}(2024){Cathey}, {Gonzalez}, {Lower}, {Phadke},
  {Spilker}, {Aravena}, {Bayliss}, {Birkin}, {Birrer}, {Chapman}, {Dahle},
  {Hayward}, {Hezaveh}, {Hill}, {Hutchison}, {Kim}, {Mahler}, {Marrone},
  {Narayanan}, {Navarre}, {Reuter}, {Rigby}, {Sharon}, {Solimano},
  {Sulzenauer}, {Vieira}, \& {Vizgan}}]{CJ24}
{Cathey}, J., {Gonzalez}, A.~H., {Lower}, S., {et~al.} 2024, \apj, 967, 11,
  \dodoi{10.3847/1538-4357/ad33c9}

\bibitem[{{Chapin} {et~al.}(2009){Chapin}, {Pope}, {Scott}, {Aretxaga},
  {Austermann}, {Chary}, {Coppin}, {Halpern}, {Hughes}, {Lowenthal},
  {Morrison}, {Perera}, {Scott}, {Wilson}, \& {Yun}}]{CE09}
{Chapin}, E.~L., {Pope}, A., {Scott}, D., {et~al.} 2009, \mnras, 398, 1793,
  \dodoi{10.1111/j.1365-2966.2009.15267.x}

\bibitem[{{Chartab} {et~al.}(2022){Chartab}, {Cooray}, {Ma}, {Nayyeri},
  {Zilliot}, {Lopez}, {Fadda}, {Herrera-Camus}, {Malkan}, {Rigopoulou},
  {Sheth}, \& {Wardlow}}]{C22}
{Chartab}, N., {Cooray}, A., {Ma}, J., {et~al.} 2022, Nature Astronomy, 6, 844,
  \dodoi{10.1038/s41550-022-01679-y}

\bibitem[{{Cheng} {et~al.}(2020){Cheng}, {Cao}, {Lu}, {Li}, {Yang},
  {Rigopoulou}, {Charmandaris}, {Gao}, {Xu}, {van der Werf}, {Diaz Santos},
  {Privon}, {Zhao}, {Cao}, {Dai}, {Huang}, {Sanders}, {Wang}, {Wang}, \&
  {Zhu}}]{CC20}
{Cheng}, C., {Cao}, X., {Lu}, N., {et~al.} 2020, \apj, 898, 33,
  \dodoi{10.3847/1538-4357/ab980b}

\bibitem[{{Cicone} {et~al.}(2015){Cicone}, {Maiolino}, {Gallerani}, {Neri},
  {Ferrara}, {Sturm}, {Fiore}, {Piconcelli}, \& {Feruglio}}]{CC15}
{Cicone}, C., {Maiolino}, R., {Gallerani}, S., {et~al.} 2015, \aap, 574, A14,
  \dodoi{10.1051/0004-6361/201424980}

\bibitem[{{Clark} {et~al.}(2016){Clark}, {Schofield}, {Gomez}, \&
  {Davies}}]{clark16}
{Clark}, C. J.~R., {Schofield}, S.~P., {Gomez}, H.~L., \& {Davies}, J.~I. 2016,
  \mnras, 459, 1646, \dodoi{10.1093/mnras/stw647}

\bibitem[{{Clark} {et~al.}(2018){Clark}, {Verstocken}, {Bianchi}, {Fritz},
  {Viaene}, {Smith}, {Baes}, {Casasola}, {Cassara}, {Davies}, {De Looze}, {De
  Vis}, {Evans}, {Galametz}, {Jones}, {Lianou}, {Madden}, {Mosenkov}, \&
  {Xilouris}}]{C18}
{Clark}, C.~J.~R., {Verstocken}, S., {Bianchi}, S., {et~al.} 2018, \aap, 609,
  A37, \dodoi{10.1051/0004-6361/201731419}

\bibitem[{{Clegg} {et~al.}(1996){Clegg}, {Ade}, {Armand}, {Baluteau}, {Barlow},
  {Buckley}, {Berges}, {Burgdorf}, {Caux}, {Ceccarelli}, {Cerulli}, {Church},
  {Cotin}, {Cox}, {Cruvellier}, {Culhane}, {Davis}, {di Giorgio}, {Diplock},
  {Drummond}, {Emery}, {Ewart}, {Fischer}, {Furniss}, {Glencross},
  {Greenhouse}, {Griffin}, {Gry}, {Harwood}, {Hazell}, {Joubert}, {King},
  {Lim}, {Liseau}, {Long}, {Lorenzetti}, {Molinari}, {Murray}, {Naylor},
  {Nisini}, {Norman}, {Omont}, {Orfei}, {Patrick}, {Pequignot}, {Pouliquen},
  {Price}, {Nguyen-Q-Rieu}, {Rogers}, {Robinson}, {Saisse}, {Saraceno},
  {Serra}, {Sidher}, {Smith}, {Smith}, {Spinoglio}, {Swinyard}, {Texier},
  {Towlson}, {Trams}, {Unger}, \& {White}}]{clegg96}
{Clegg}, P.~E., {Ade}, P.~A.~R., {Armand}, C., {et~al.} 1996, \aap, 315, L38

\bibitem[{{Combes} {et~al.}(2012){Combes}, {Rex}, {Rawle}, {Egami}, {Boone},
  {Smail}, {Richard}, {Ivison}, {Gurwell}, {Casey}, {Omont}, {Berciano Alba},
  {Dessauges-Zavadsky}, {Edge}, {Fazio}, {Kneib}, {Okabe}, {Pell{\'o}},
  {P{\'e}rez-Gonz{\'a}lez}, {Schaerer}, {Smith}, {Swinbank}, \& {van der
  Werf}}]{CF12}
{Combes}, F., {Rex}, M., {Rawle}, T.~D., {et~al.} 2012, \aap, 538, L4,
  \dodoi{10.1051/0004-6361/201118750}

\bibitem[{{Conley} {et~al.}(2011){Conley}, {Cooray}, {Vieira}, {Gonz{\'a}lez
  Solares}, {Kim}, {Aguirre}, {Amblard}, {Auld}, {Baker}, {Beelen}, {Blain},
  {Blundell}, {Bock}, {Bradford}, {Bridge}, {Brisbin}, {Burgarella},
  {Carpenter}, {Chanial}, {Chapin}, {Christopher}, {Clements}, {Cox},
  {Djorgovski}, {Dowell}, {Eales}, {Earle}, {Ellsworth-Bowers}, {Farrah},
  {Franceschini}, {Frayer}, {Fu}, {Gavazzi}, {Glenn}, {Griffin}, {Gurwell},
  {Halpern}, {Ibar}, {Ivison}, {Jarvis}, {Kamenetzky}, {Krips}, {Levenson},
  {Lupu}, {Mahabal}, {Maloney}, {Maraston}, {Marchetti}, {Marsden},
  {Matsuhara}, {Mortier}, {Murphy}, {Naylor}, {Neri}, {Nguyen}, {Oliver},
  {Omont}, {Page}, {Papageorgiou}, {Pearson}, {P{\'e}rez-Fournon}, {Pohlen},
  {Rangwala}, {Rawlings}, {Raymond}, {Riechers}, {Rodighiero}, {Roseboom},
  {Rowan-Robinson}, {Schulz}, {Scott}, {Scott}, {Serra}, {Seymour}, {Shupe},
  {Smith}, {Symeonidis}, {Tugwell}, {Vaccari}, {Valiante}, {Valtchanov},
  {Verma}, {Viero}, {Vigroux}, {Wang}, {Wiebe}, {Wright}, {Xu}, {Zeimann},
  {Zemcov}, \& {Zmuidzinas}}]{CA11}
{Conley}, A., {Cooray}, A., {Vieira}, J.~D., {et~al.} 2011, \apjl, 732, L35,
  \dodoi{10.1088/2041-8205/732/2/L35}

\bibitem[{{Coppin} {et~al.}(2012){Coppin}, {Danielson}, {Geach}, {Hodge},
  {Swinbank}, {Wardlow}, {Bertoldi}, {Biggs}, {Brandt}, {Caselli}, {Chapman},
  {Dannerbauer}, {Dunlop}, {Greve}, {Hamann}, {Ivison}, {Karim}, {Knudsen},
  {Menten}, {Schinnerer}, {Smail}, {Spaans}, {Walter}, {Webb}, \& {van der
  Werf}}]{CK12}
{Coppin}, K.~E.~K., {Danielson}, A.~L.~R., {Geach}, J.~E., {et~al.} 2012,
  \mnras, 427, 520, \dodoi{10.1111/j.1365-2966.2012.21977.x}

\bibitem[{{Cormier} {et~al.}(2015){Cormier}, {Madden}, {Lebouteiller}, {Abel},
  {Hony}, {Galliano}, {R{\'e}my-Ruyer}, {Bigiel}, {Baes}, {Boselli},
  {Chevance}, {Cooray}, {De Looze}, {Doublier}, {Galametz}, {Hughes},
  {Karczewski}, {Lee}, {Lu}, \& {Spinoglio}}]{C15}
{Cormier}, D., {Madden}, S.~C., {Lebouteiller}, V., {et~al.} 2015, \aap, 578,
  A53, \dodoi{10.1051/0004-6361/201425207}

\bibitem[{{Cormier} {et~al.}(2019){Cormier}, {Abel}, {Hony}, {Lebouteiller},
  {Madden}, {Polles}, {Galliano}, {De Looze}, {Galametz}, \&
  {Lambert-Huyghe}}]{C19}
{Cormier}, D., {Abel}, N.~P., {Hony}, S., {et~al.} 2019, \aap, 626, A23,
  \dodoi{10.1051/0004-6361/201834457}

\bibitem[{{Cortzen} {et~al.}(2020){Cortzen}, {Magdis}, {Valentino}, {Daddi},
  {Liu}, {Rigopoulou}, {Sargent}, {Riechers}, {Cormier}, {Hodge}, {Walter},
  {Elbaz}, {B{\'e}thermin}, {Greve}, {Kokorev}, \& {Toft}}]{CI20}
{Cortzen}, I., {Magdis}, G.~E., {Valentino}, F., {et~al.} 2020, \aap, 634, L14,
  \dodoi{10.1051/0004-6361/201937217}

\bibitem[{{Cowie} {et~al.}(2011){Cowie}, {Hu}, \& {Songaila}}]{CL11}
{Cowie}, L.~L., {Hu}, E.~M., \& {Songaila}, A. 2011, \apjl, 735, L38,
  \dodoi{10.1088/2041-8205/735/2/L38}

\bibitem[{{Cox} {et~al.}(2002){Cox}, {Omont}, {Djorgovski}, {Bertoldi}, {Pety},
  {Carilli}, {Isaak}, {Beelen}, {McMahon}, \& {Castro}}]{CP02}
{Cox}, P., {Omont}, A., {Djorgovski}, S.~G., {et~al.} 2002, \aap, 387, 406,
  \dodoi{10.1051/0004-6361:20020382}

\bibitem[{{Cox} {et~al.}(2011){Cox}, {Krips}, {Neri}, {Omont}, {G{\"u}sten},
  {Menten}, {Wyrowski}, {Wei{\ss}}, {Beelen}, {Gurwell}, {Dannerbauer},
  {Ivison}, {Negrello}, {Aretxaga}, {Hughes}, {Auld}, {Baes}, {Blundell},
  {Buttiglione}, {Cava}, {Cooray}, {Dariush}, {Dunne}, {Dye}, {Eales},
  {Frayer}, {Fritz}, {Gavazzi}, {Hopwood}, {Ibar}, {Jarvis}, {Maddox},
  {Micha{\l}owski}, {Pascale}, {Pohlen}, {Rigby}, {Smith}, {Swinbank}, {Temi},
  {Valtchanov}, {van der Werf}, \& {de Zotti}}]{CP11a}
{Cox}, P., {Krips}, M., {Neri}, R., {et~al.} 2011, \apj, 740, 63,
  \dodoi{10.1088/0004-637X/740/2/63}

\bibitem[{{Crespo G{\'o}mez} {et~al.}(2024){Crespo G{\'o}mez}, {Colina},
  {{\'A}lvarez-M{\'a}rquez}, {Bik}, {Boogaard}, {{\"O}stlin}, {Pei{\ss}ker},
  {Walter}, {Labiano}, {P{\'e}rez-Gonz{\'a}lez}, {Greve}, {Wright},
  {Alonso-Herrero}, {Caputi}, {Costantin}, {Eckart}, {Garc{\'\i}a-Mar{\'\i}n},
  {Gillman}, {Hjorth}, {Iani}, {Langeroodi}, {Pye}, {Rinaldi}, {Tikkanen}, {van
  der Werf}, {Lagage}, \& {van Dishoeck}}]{CA24}
{Crespo G{\'o}mez}, A., {Colina}, L., {{\'A}lvarez-M{\'a}rquez}, J., {et~al.}
  2024, \aap, 691, A325, \dodoi{10.1051/0004-6361/202449750}

\bibitem[{{Croxall} {et~al.}(2017){Croxall}, {Smith}, {Pellegrini}, {Groves},
  {Bolatto}, {Herrera-Camus}, {Sandstrom}, {Draine}, {Wolfire}, {Armus},
  {Boquien}, {Brandl}, {Dale}, {Galametz}, {Hunt}, {Kennicutt}, {Kreckel},
  {Rigopoulou}, {van der Werf}, \& {Wilson}}]{croxall17}
{Croxall}, K.~V., {Smith}, J.~D., {Pellegrini}, E., {et~al.} 2017, \apj, 845,
  96, \dodoi{10.3847/1538-4357/aa8035}

\bibitem[{{Cunningham} {et~al.}(2020){Cunningham}, {Chapman}, {Aravena}, {De
  Breuck}, {B{\'e}thermin}, {Chen}, {Dong}, {Gonzalez}, {Greve}, {Litke}, {Ma},
  {Malkan}, {Marrone}, {Miller}, {Phadke}, {Reuter}, {Rotermund}, {Spilker},
  {Stark}, {Strandet}, {Vieira}, \& {Wei{\ss}}}]{CD20}
{Cunningham}, D.~J.~M., {Chapman}, S.~C., {Aravena}, M., {et~al.} 2020, \mnras,
  494, 4090, \dodoi{10.1093/mnras/staa820}

\bibitem[{{Curti} {et~al.}(2017){Curti}, {Cresci}, {Mannucci}, {Marconi},
  {Maiolino}, \& {Esposito}}]{curti17}
{Curti}, M., {Cresci}, G., {Mannucci}, F., {et~al.} 2017, \mnras, 465, 1384,
  \dodoi{10.1093/mnras/stw2766}

\bibitem[{{Daddi} {et~al.}(2009){Daddi}, {Dannerbauer}, {Stern}, {Dickinson},
  {Morrison}, {Elbaz}, {Giavalisco}, {Mancini}, {Pope}, \& {Spinrad}}]{DE09}
{Daddi}, E., {Dannerbauer}, H., {Stern}, D., {et~al.} 2009, \apj, 694, 1517,
  \dodoi{10.1088/0004-637X/694/2/1517}

\bibitem[{{Dale} {et~al.}(2012){Dale}, {Aniano}, {Engelbracht}, {Hinz},
  {Krause}, {Montiel}, {Roussel}, {Appleton}, {Armus}, {Beir{\~a}o}, {Bolatto},
  {Brandl}, {Calzetti}, {Crocker}, {Croxall}, {Draine}, {Galametz}, {Gordon},
  {Groves}, {Hao}, {Helou}, {Hunt}, {Johnson}, {Kennicutt}, {Koda}, {Leroy},
  {Li}, {Meidt}, {Miller}, {Murphy}, {Rahman}, {Rix}, {Sandstrom}, {Sauvage},
  {Schinnerer}, {Skibba}, {Smith}, {Tabatabaei}, {Walter}, {Wilson}, {Wolfire},
  \& {Zibetti}}]{D12}
{Dale}, D.~A., {Aniano}, G., {Engelbracht}, C.~W., {et~al.} 2012, \apj, 745,
  95, \dodoi{10.1088/0004-637X/745/1/95}

\bibitem[{{Dayal} {et~al.}(2022){Dayal}, {Ferrara}, {Sommovigo}, {Bouwens},
  {Oesch}, {Smit}, {Gonzalez}, {Schouws}, {Stefanon}, {Kobayashi}, {Bremer},
  {Algera}, {Aravena}, {Bowler}, {da Cunha}, {Fudamoto}, {Graziani}, {Hodge},
  {Inami}, {De Looze}, {Pallottini}, {Riechers}, {Schneider}, {Stark}, \&
  {Endsley}}]{dayal22}
{Dayal}, P., {Ferrara}, A., {Sommovigo}, L., {et~al.} 2022, \mnras, 512, 989,
  \dodoi{10.1093/mnras/stac537}

\bibitem[{{De Breuck} {et~al.}(2011){De Breuck}, {Maiolino}, {Caselli},
  {Coppin}, {Hailey-Dunsheath}, \& {Nagao}}]{DC11}
{De Breuck}, C., {Maiolino}, R., {Caselli}, P., {et~al.} 2011, \aap, 530, L8,
  \dodoi{10.1051/0004-6361/201116868}

\bibitem[{{De Breuck} {et~al.}(2004){De Breuck}, {Bertoldi}, {Carilli},
  {Omont}, {Venemans}, {R{\"o}ttgering}, {Overzier}, {Reuland}, {Miley},
  {Ivison}, \& {van Breugel}}]{DC04}
{De Breuck}, C., {Bertoldi}, F., {Carilli}, C., {et~al.} 2004, \aap, 424, 1,
  \dodoi{10.1051/0004-6361:20035885}

\bibitem[{{De Breuck} {et~al.}(2010){De Breuck}, {Seymour}, {Stern}, {Willner},
  {Eisenhardt}, {Fazio}, {Galametz}, {Lacy}, {Rettura}, {Rocca-Volmerange}, \&
  {Vernet}}]{DC10}
{De Breuck}, C., {Seymour}, N., {Stern}, D., {et~al.} 2010, \apj, 725, 36,
  \dodoi{10.1088/0004-637X/725/1/36}

\bibitem[{{De Breuck} {et~al.}(2014){De Breuck}, {Williams}, {Swinbank},
  {Caselli}, {Coppin}, {Davis}, {Maiolino}, {Nagao}, {Smail}, {Walter},
  {Wei{\ss}}, \& {Zwaan}}]{DC14}
{De Breuck}, C., {Williams}, R.~J., {Swinbank}, M., {et~al.} 2014, \aap, 565,
  A59, \dodoi{10.1051/0004-6361/201323331}

\bibitem[{{De Breuck} {et~al.}(2019){De Breuck}, {Wei{\ss}}, {B{\'e}thermin},
  {Cunningham}, {Apostolovski}, {Aravena}, {Archipley}, {Chapman}, {Chen},
  {Fu}, {Jarugula}, {Malkan}, {Mangian}, {Phadke}, {Reuter}, {Stacey},
  {Strandet}, {Vieira}, \& {Vishwas}}]{DC19}
{De Breuck}, C., {Wei{\ss}}, A., {B{\'e}thermin}, M., {et~al.} 2019, \aap, 631,
  A167, \dodoi{10.1051/0004-6361/201936169}

\bibitem[{{De Looze} {et~al.}(2014){De Looze}, {Cormier}, {Lebouteiller},
  {Madden}, {Baes}, {Bendo}, {Boquien}, {Boselli}, {Clements}, {Cortese},
  {Cooray}, {Galametz}, {Galliano}, {Graci{\'a}-Carpio}, {Isaak}, {Karczewski},
  {Parkin}, {Pellegrini}, {R{\'e}my-Ruyer}, {Spinoglio}, {Smith}, \&
  {Sturm}}]{delooze14}
{De Looze}, I., {Cormier}, D., {Lebouteiller}, V., {et~al.} 2014, \aap, 568,
  A62, \dodoi{10.1051/0004-6361/201322489}

\bibitem[{{De Vis} {et~al.}(2019){De Vis}, {Jones}, {Viaene}, {Casasola},
  {Clark}, {Baes}, {Bianchi}, {Cassara}, {Davies}, {De Looze}, {Galametz},
  {Galliano}, {Lianou}, {Madden}, {Manilla-Robles}, {Mosenkov}, {Nersesian},
  {Roychowdhury}, {Xilouris}, \& {Ysard}}]{D19}
{De Vis}, P., {Jones}, A., {Viaene}, S., {et~al.} 2019, \aap, 623, A5,
  \dodoi{10.1051/0004-6361/201834444}

\bibitem[{{Deane} {et~al.}(2013){Deane}, {Rawlings}, {Garrett}, {Heywood},
  {Jarvis}, {Kl{\"o}ckner}, {Marshall}, \& {McKean}}]{DR13}
{Deane}, R.~P., {Rawlings}, S., {Garrett}, M.~A., {et~al.} 2013, \mnras, 434,
  3322, \dodoi{10.1093/mnras/stt1241}

\bibitem[{{Decarli} {et~al.}(2012){Decarli}, {Walter}, {Neri}, {Bertoldi},
  {Carilli}, {Cox}, {Kneib}, {Lestrade}, {Maiolino}, {Omont}, {Richard},
  {Riechers}, {Thanjavur}, \& {Weiss}}]{DR12}
{Decarli}, R., {Walter}, F., {Neri}, R., {et~al.} 2012, \apj, 752, 2,
  \dodoi{10.1088/0004-637X/752/1/2}

\bibitem[{{Decarli} {et~al.}(2014){Decarli}, {Walter}, {Carilli}, {Bertoldi},
  {Cox}, {Ferkinhoff}, {Groves}, {Maiolino}, {Neri}, {Riechers}, \&
  {Weiss}}]{DR14}
{Decarli}, R., {Walter}, F., {Carilli}, C., {et~al.} 2014, \apjl, 782, L17,
  \dodoi{10.1088/2041-8205/782/2/L17}

\bibitem[{{Decarli} {et~al.}(2017){Decarli}, {Walter}, {Venemans},
  {Ba{\~n}ados}, {Bertoldi}, {Carilli}, {Fan}, {Farina}, {Mazzucchelli},
  {Riechers}, {Rix}, {Strauss}, {Wang}, \& {Yang}}]{DR17}
{Decarli}, R., {Walter}, F., {Venemans}, B.~P., {et~al.} 2017, \nat, 545, 457,
  \dodoi{10.1038/nature22358}

\bibitem[{{Decarli} {et~al.}(2018){Decarli}, {Walter}, {Venemans},
  {Ba{\~n}ados}, {Bertoldi}, {Carilli}, {Fan}, {Farina}, {Mazzucchelli},
  {Riechers}, {Rix}, {Strauss}, {Wang}, \& {Yang}}]{DR18}
---. 2018, \apj, 854, 97, \dodoi{10.3847/1538-4357/aaa5aa}

\bibitem[{{Decarli} {et~al.}(2022){Decarli}, {Pensabene}, {Venemans}, {Walter},
  {Ba{\~n}ados}, {Bertoldi}, {Carilli}, {Cox}, {Fan}, {Farina}, {Ferkinhoff},
  {Groves}, {Li}, {Mazzucchelli}, {Neri}, {Riechers}, {Uzgil}, {Wang}, {Wang},
  {Weiss}, {Winters}, \& {Yang}}]{DR22}
{Decarli}, R., {Pensabene}, A., {Venemans}, B., {et~al.} 2022, \aap, 662, A60,
  \dodoi{10.1051/0004-6361/202142871}

\bibitem[{{Decarli} {et~al.}(2023){Decarli}, {Pensabene}, {Diaz-Santos},
  {Ferkinhoff}, {Strauss}, {Venemans}, {Walter}, {Ba{\~n}ados}, {Bertoldi},
  {Fan}, {Farina}, {Riechers}, {Rix}, \& {Wang}}]{DR23}
{Decarli}, R., {Pensabene}, A., {Diaz-Santos}, T., {et~al.} 2023, \aap, 673,
  A157, \dodoi{10.1051/0004-6361/202245674}

\bibitem[{{Denicol{\'o}} {et~al.}(2002){Denicol{\'o}}, {Terlevich}, \&
  {Terlevich}}]{denicolo02}
{Denicol{\'o}}, G., {Terlevich}, R., \& {Terlevich}, E. 2002, \mnras, 330, 69,
  \dodoi{10.1046/j.1365-8711.2002.05041.x}

\bibitem[{{Dessauges-Zavadsky} {et~al.}(2015){Dessauges-Zavadsky}, {Zamojski},
  {Schaerer}, {Combes}, {Egami}, {Swinbank}, {Richard}, {Sklias}, {Rawle},
  {Rex}, {Kneib}, {Boone}, \& {Blain}}]{DM15}
{Dessauges-Zavadsky}, M., {Zamojski}, M., {Schaerer}, D., {et~al.} 2015, \aap,
  577, A50, \dodoi{10.1051/0004-6361/201424661}

\bibitem[{{D'Eugenio} {et~al.}(2023){D'Eugenio}, {Daddi}, {Liu}, \&
  {Gobat}}]{deugenio23}
{D'Eugenio}, C., {Daddi}, E., {Liu}, D., \& {Gobat}, R. 2023, \aap, 678, L9,
  \dodoi{10.1051/0004-6361/202347233}

\bibitem[{{D{\'\i}az-Santos} {et~al.}(2016){D{\'\i}az-Santos}, {Assef},
  {Blain}, {Tsai}, {Aravena}, {Eisenhardt}, {Wu}, {Stern}, \& {Bridge}}]{DT16}
{D{\'\i}az-Santos}, T., {Assef}, R.~J., {Blain}, A.~W., {et~al.} 2016, \apjl,
  816, L6, \dodoi{10.3847/2041-8205/816/1/L6}

\bibitem[{{D{\'\i}az-Santos} {et~al.}(2017){D{\'\i}az-Santos}, {Armus},
  {Charmandaris}, {Lu}, {Stierwalt}, {Stacey}, {Malhotra}, {van der Werf},
  {Howell}, {Privon}, {Mazzarella}, {Goldsmith}, {Murphy}, {Barcos-Mu{\~n}oz},
  {Linden}, {Inami}, {Larson}, {Evans}, {Appleton}, {Iwasawa}, {Lord},
  {Sanders}, \& {Surace}}]{D17}
{D{\'\i}az-Santos}, T., {Armus}, L., {Charmandaris}, V., {et~al.} 2017, \apj,
  846, 32, \dodoi{10.3847/1538-4357/aa81d7}

\bibitem[{{D{\'\i}az-Santos} {et~al.}(2018){D{\'\i}az-Santos}, {Assef},
  {Blain}, {Aravena}, {Stern}, {Tsai}, {Eisenhardt}, {Wu}, {Jun}, {Dibert},
  {Inami}, {Lansbury}, \& {Leclercq}}]{DT18}
{D{\'\i}az-Santos}, T., {Assef}, R.~J., {Blain}, A.~W., {et~al.} 2018, Science,
  362, 1034, \dodoi{10.1126/science.aap7605}

\bibitem[{{D{\'\i}az-Santos} {et~al.}(2021){D{\'\i}az-Santos}, {Assef},
  {Eisenhardt}, {Jun}, {Jones}, {Blain}, {Stern}, {Aravena}, {Tsai}, {Lake},
  {Wu}, \& {Gonz{\'a}lez-L{\'o}pez}}]{DT21}
{D{\'\i}az-Santos}, T., {Assef}, R.~J., {Eisenhardt}, P. R.~M., {et~al.} 2021,
  \aap, 654, A37, \dodoi{10.1051/0004-6361/202140455}

\bibitem[{{Diego} {et~al.}(2024){Diego}, {Adams}, {Willner}, {Harvey},
  {Broadhurst}, {Cohen}, {Jansen}, {Summers}, {Windhorst}, {D'Silva},
  {Koekemoer}, {Coe}, {Conselice}, {Driver}, {Frye}, {Grogin}, {Marshall},
  {Nonino}, {Ortiz}, {Pirzkal}, {Robotham}, {Ryan}, {Willmer}, {Yan}, {Sun},
  {Hainline}, {Berkheimer}, {Polletta}, \& {Zitrin}}]{DJ24}
{Diego}, J.~M., {Adams}, N.~J., {Willner}, S.~P., {et~al.} 2024, \aap, 690,
  A114, \dodoi{10.1051/0004-6361/202349119}

\bibitem[{{Dimaratos} {et~al.}(2015){Dimaratos}, {Cormier}, {Bigiel}, \&
  {Madden}}]{D15}
{Dimaratos}, A., {Cormier}, D., {Bigiel}, F., \& {Madden}, S.~C. 2015, \aap,
  580, A135, \dodoi{10.1051/0004-6361/201526447}

\bibitem[{{Doherty} {et~al.}(2020){Doherty}, {Geach}, {Ivison}, \&
  {Dye}}]{DM20}
{Doherty}, M.~J., {Geach}, J.~E., {Ivison}, R.~J., \& {Dye}, S. 2020, \apj,
  905, 152, \dodoi{10.3847/1538-4357/abc5b9}

\bibitem[{{Dopita} {et~al.}(2000){Dopita}, {Kewley}, {Heisler}, \&
  {Sutherland}}]{dopita00}
{Dopita}, M.~A., {Kewley}, L.~J., {Heisler}, C.~A., \& {Sutherland}, R.~S.
  2000, \apj, 542, 224, \dodoi{10.1086/309538}

\bibitem[{{Downes} {et~al.}(1999{\natexlab{a}}){Downes}, {Neri}, {Wiklind},
  {Wilner}, \& {Shaver}}]{DD99b}
{Downes}, D., {Neri}, R., {Wiklind}, T., {Wilner}, D.~J., \& {Shaver}, P.~A.
  1999{\natexlab{a}}, \apjl, 513, L1, \dodoi{10.1086/311896}

\bibitem[{{Downes} {et~al.}(1999{\natexlab{b}}){Downes}, {Neri}, {Greve},
  {Guilloteau}, {Casoli}, {Hughes}, {Lutz}, {Menten}, {Wilner}, {Andreani},
  {Bertoldi}, {Carilli}, {Dunlop}, {Genzel}, {Gueth}, {Ivison}, {Mann},
  {Mellier}, {Oliver}, {Peacock}, {Rigopoulou}, {Rowan-Robinson}, {Schilke},
  {Serjeant}, {Tacconi}, \& {Wright}}]{DD99a}
{Downes}, D., {Neri}, R., {Greve}, A., {et~al.} 1999{\natexlab{b}}, \aap, 347,
  809, \dodoi{10.48550/arXiv.astro-ph/9907139}

\bibitem[{{Draine}(2011)}]{draine11}
{Draine}, B.~T. 2011, {Physics of the Interstellar and Intergalactic Medium}

\bibitem[{{Drew} \& {Casey}(2022)}]{drew22}
{Drew}, P.~M., \& {Casey}, C.~M. 2022, \apj, 930, 142,
  \dodoi{10.3847/1538-4357/ac6270}

\bibitem[{{Drouart} {et~al.}(2016){Drouart}, {Rocca-Volmerange}, {De Breuck},
  {Fioc}, {Lehnert}, {Seymour}, {Stern}, \& {Vernet}}]{DG16}
{Drouart}, G., {Rocca-Volmerange}, B., {De Breuck}, C., {et~al.} 2016, \aap,
  593, A109, \dodoi{10.1051/0004-6361/201526880}

\bibitem[{{Drouart} {et~al.}(2014){Drouart}, {De Breuck}, {Vernet}, {Seymour},
  {Lehnert}, {Barthel}, {Bauer}, {Ibar}, {Galametz}, {Haas}, {Hatch},
  {Mullaney}, {Nesvadba}, {Rocca-Volmerange}, {R{\"o}ttgering}, {Stern}, \&
  {Wylezalek}}]{DG14}
{Drouart}, G., {De Breuck}, C., {Vernet}, J., {et~al.} 2014, \aap, 566, A53,
  \dodoi{10.1051/0004-6361/201323310}

\bibitem[{{Duncan} {et~al.}(2023){Duncan}, {Windhorst}, {Koekemoer},
  {R{\"o}ttgering}, {Cohen}, {Jansen}, {Summers}, {Tompkins}, {Hutchison},
  {Conselice}, {Driver}, {Yan}, {Adams}, {Cheng}, {Coe}, {Diego}, {Dole},
  {Frye}, {Gim}, {Grogin}, {Holwerda}, {Lim}, {Marshall}, {Nonino}, {Pirzkal},
  {Robotham}, {Ryan}, \& {Willmer}}]{DK23}
{Duncan}, K.~J., {Windhorst}, R.~A., {Koekemoer}, A.~M., {et~al.} 2023, \mnras,
  522, 4548, \dodoi{10.1093/mnras/stad1267}

\bibitem[{{Durbala} {et~al.}(2020){Durbala}, {Finn}, {Crone Odekon}, {Haynes},
  {Koopmann}, \& {O'Donoghue}}]{D20}
{Durbala}, A., {Finn}, R.~A., {Crone Odekon}, M., {et~al.} 2020, \aj, 160, 271,
  \dodoi{10.3847/1538-3881/abc018}

\bibitem[{{Eilers} {et~al.}(2020){Eilers}, {Hennawi}, {Decarli}, {Davies},
  {Venemans}, {Walter}, {Ba{\~n}ados}, {Fan}, {Farina}, {Mazzucchelli},
  {Novak}, {Schindler}, {Simcoe}, {Wang}, \& {Yang}}]{EA20}
{Eilers}, A.-C., {Hennawi}, J.~F., {Decarli}, R., {et~al.} 2020, \apj, 900, 37,
  \dodoi{10.3847/1538-4357/aba52e}

\bibitem[{{Eilers} {et~al.}(2023){Eilers}, {Simcoe}, {Yue}, {Mackenzie},
  {Matthee}, {{\v{D}}urov{\v{c}}{\'\i}kov{\'a}}, {Kashino}, {Bordoloi}, \&
  {Lilly}}]{EA23}
{Eilers}, A.-C., {Simcoe}, R.~A., {Yue}, M., {et~al.} 2023, \apj, 950, 68,
  \dodoi{10.3847/1538-4357/acd776}

\bibitem[{{Elbaz} {et~al.}(2011){Elbaz}, {Dickinson}, {Hwang},
  {D{\'\i}az-Santos}, {Magdis}, {Magnelli}, {Le Borgne}, {Galliano},
  {Pannella}, {Chanial}, {Armus}, {Charmandaris}, {Daddi}, {Aussel}, {Popesso},
  {Kartaltepe}, {Altieri}, {Valtchanov}, {Coia}, {Dannerbauer}, {Dasyra},
  {Leiton}, {Mazzarella}, {Alexander}, {Buat}, {Burgarella}, {Chary}, {Gilli},
  {Ivison}, {Juneau}, {Le Floc'h}, {Lutz}, {Morrison}, {Mullaney}, {Murphy},
  {Pope}, {Scott}, {Brodwin}, {Calzetti}, {Cesarsky}, {Charlot}, {Dole},
  {Eisenhardt}, {Ferguson}, {F{\"o}rster Schreiber}, {Frayer}, {Giavalisco},
  {Huynh}, {Koekemoer}, {Papovich}, {Reddy}, {Surace}, {Teplitz}, {Yun}, \&
  {Wilson}}]{elbaz11}
{Elbaz}, D., {Dickinson}, M., {Hwang}, H.~S., {et~al.} 2011, \aap, 533, A119,
  \dodoi{10.1051/0004-6361/201117239}

\bibitem[{{El{\'\i}asd{\'o}ttir} {et~al.}(2007){El{\'\i}asd{\'o}ttir},
  {Limousin}, {Richard}, {Hjorth}, {Kneib}, {Natarajan}, {Pedersen}, {Jullo},
  \& {Paraficz}}]{EA07}
{El{\'\i}asd{\'o}ttir}, {\'A}., {Limousin}, M., {Richard}, J., {et~al.} 2007,
  arXiv e-prints, arXiv:0710.5636, \dodoi{10.48550/arXiv.0710.5636}

\bibitem[{{Endsley} {et~al.}(2023){Endsley}, {Stark}, {Lyu}, {Wang}, {Yang},
  {Fan}, {Smit}, {Bouwens}, {Hainline}, \& {Schouws}}]{ER23}
{Endsley}, R., {Stark}, D.~P., {Lyu}, J., {et~al.} 2023, \mnras, 520, 4609,
  \dodoi{10.1093/mnras/stad266}

\bibitem[{{Evans} {et~al.}(2006){Evans}, {Solomon}, {Tacconi}, {Vavilkin}, \&
  {Downes}}]{EA06}
{Evans}, A.~S., {Solomon}, P.~M., {Tacconi}, L.~J., {Vavilkin}, T., \&
  {Downes}, D. 2006, \aj, 132, 2398, \dodoi{10.1086/508416}

\bibitem[{{Evans} {et~al.}(2022){Evans}, {Frayer}, {Charmandaris}, {Armus},
  {Inami}, {Surace}, {Linden}, {Soifer}, {Diaz-Santos}, {Larson}, {Rich},
  {Song}, {Barcos-Munoz}, {Mazzarella}, {Privon}, {U}, {Medling}, {B{\"o}ker},
  {Aalto}, {Iwasawa}, {Howell}, {van der Werf}, {Appleton}, {Bohn}, {Brown},
  {Hayward}, {Hoshioka}, {Kemper}, {Lai}, {Law}, {Malkan}, {Marshall},
  {Murphy}, {Sanders}, \& {Stierwalt}}]{evans22}
{Evans}, A.~S., {Frayer}, D.~T., {Charmandaris}, V., {et~al.} 2022, \apjl, 940,
  L8, \dodoi{10.3847/2041-8213/ac9971}

\bibitem[{{Fadely} {et~al.}(2010){Fadely}, {Allam}, {Baker}, {Lin}, {Lutz},
  {Shapley}, {Shin}, {Allyn Smith}, {Strauss}, \& {Tucker}}]{FR10}
{Fadely}, R., {Allam}, S.~S., {Baker}, A.~J., {et~al.} 2010, \apj, 723, 729,
  \dodoi{10.1088/0004-637X/723/1/729}

\bibitem[{{Faisst} {et~al.}(2020){Faisst}, {Fudamoto}, {Oesch}, {Scoville},
  {Riechers}, {Pavesi}, \& {Capak}}]{FA20}
{Faisst}, A.~L., {Fudamoto}, Y., {Oesch}, P.~A., {et~al.} 2020, \mnras, 498,
  4192, \dodoi{10.1093/mnras/staa2545}

\bibitem[{{Falkendal} {et~al.}(2019){Falkendal}, {De Breuck}, {Lehnert},
  {Drouart}, {Vernet}, {Emonts}, {Lee}, {Nesvadba}, {Seymour}, {B{\'e}thermin},
  {Kolwa}, {Gullberg}, \& {Wylezalek}}]{FT19}
{Falkendal}, T., {De Breuck}, C., {Lehnert}, M.~D., {et~al.} 2019, \aap, 621,
  A27, \dodoi{10.1051/0004-6361/201732485}

\bibitem[{{Farrah} {et~al.}(2007){Farrah}, {Bernard-Salas}, {Spoon}, {Soifer},
  {Armus}, {Brandl}, {Charmandaris}, {Desai}, {Higdon}, {Devost}, \&
  {Houck}}]{F07}
{Farrah}, D., {Bernard-Salas}, J., {Spoon}, H.~W.~W., {et~al.} 2007, \apj, 667,
  149, \dodoi{10.1086/520834}

\bibitem[{{Farrah} {et~al.}(2013){Farrah}, {Lebouteiller}, {Spoon},
  {Bernard-Salas}, {Pearson}, {Rigopoulou}, {Smith}, {Gonz{\'a}lez-Alfonso},
  {Clements}, {Efstathiou}, {Cormier}, {Afonso}, {Petty}, {Harris}, {Hurley},
  {Borys}, {Verma}, {Cooray}, \& {Salvatelli}}]{F13}
{Farrah}, D., {Lebouteiller}, V., {Spoon}, H.~W.~W., {et~al.} 2013, \apj, 776,
  38, \dodoi{10.1088/0004-637X/776/1/38}

\bibitem[{{Fazio} {et~al.}(2004){Fazio}, {Hora}, {Allen}, {Ashby}, {Barmby},
  {Deutsch}, {Huang}, {Kleiner}, {Marengo}, {Megeath}, {Melnick}, {Pahre},
  {Patten}, {Polizotti}, {Smith}, {Taylor}, {Wang}, {Willner}, {Hoffmann},
  {Pipher}, {Forrest}, {McMurty}, {McCreight}, {McKelvey}, {McMurray}, {Koch},
  {Moseley}, {Arendt}, {Mentzell}, {Marx}, {Losch}, {Mayman}, {Eichhorn},
  {Krebs}, {Jhabvala}, {Gezari}, {Fixsen}, {Flores}, {Shakoorzadeh}, {Jungo},
  {Hakun}, {Workman}, {Karpati}, {Kichak}, {Whitley}, {Mann}, {Tollestrup},
  {Eisenhardt}, {Stern}, {Gorjian}, {Bhattacharya}, {Carey}, {Nelson},
  {Glaccum}, {Lacy}, {Lowrance}, {Laine}, {Reach}, {Stauffer}, {Surace},
  {Wilson}, {Wright}, {Hoffman}, {Domingo}, \& {Cohen}}]{fazio04}
{Fazio}, G.~G., {Hora}, J.~L., {Allen}, L.~E., {et~al.} 2004, \apjs, 154, 10,
  \dodoi{10.1086/422843}

\bibitem[{{Ferkinhoff} {et~al.}(2015){Ferkinhoff}, {Brisbin}, {Nikola},
  {Stacey}, {Sheth}, {Hailey-Dunsheath}, \& {Falgarone}}]{FC15}
{Ferkinhoff}, C., {Brisbin}, D., {Nikola}, T., {et~al.} 2015, \apj, 806, 260,
  \dodoi{10.1088/0004-637X/806/2/260}

\bibitem[{{Ferkinhoff} {et~al.}(2010){Ferkinhoff}, {Hailey-Dunsheath},
  {Nikola}, {Parshley}, {Stacey}, {Benford}, \& {Staguhn}}]{FC10}
{Ferkinhoff}, C., {Hailey-Dunsheath}, S., {Nikola}, T., {et~al.} 2010, \apjl,
  714, L147, \dodoi{10.1088/2041-8205/714/1/L147}

\bibitem[{{Ferkinhoff} {et~al.}(2011){Ferkinhoff}, {Brisbin}, {Nikola},
  {Parshley}, {Stacey}, {Phillips}, {Falgarone}, {Benford}, {Staguhn}, \&
  {Tucker}}]{FC11}
{Ferkinhoff}, C., {Brisbin}, D., {Nikola}, T., {et~al.} 2011, \apjl, 740, L29,
  \dodoi{10.1088/2041-8205/740/1/L29}

\bibitem[{{Fern{\'a}ndez Aranda} {et~al.}(2024){Fern{\'a}ndez Aranda},
  {D{\'\i}az Santos}, {Hatziminaoglou}, {Assef}, {Aravena}, {Eisenhardt},
  {Ferkinhoff}, {Pensabene}, {Nikola}, {Andreani}, {Vishwas}, {Stacey},
  {Decarli}, {Blain}, {Brisbin}, {Charmandaris}, {Jun}, {Li}, {Liao}, {Martin},
  {Stern}, {Tsai}, {Wu}, \& {Zewdie}}]{FR24}
{Fern{\'a}ndez Aranda}, R., {D{\'\i}az Santos}, T., {Hatziminaoglou}, E.,
  {et~al.} 2024, \aap, 682, A166, \dodoi{10.1051/0004-6361/202347869}

\bibitem[{{Fern{\'a}ndez-Ontiveros} {et~al.}(2016){Fern{\'a}ndez-Ontiveros},
  {Spinoglio}, {Pereira-Santaella}, {Malkan}, {Andreani}, \& {Dasyra}}]{F16}
{Fern{\'a}ndez-Ontiveros}, J.~A., {Spinoglio}, L., {Pereira-Santaella}, M.,
  {et~al.} 2016, \apjs, 226, 19, \dodoi{10.3847/0067-0049/226/2/19}

\bibitem[{{Finkelstein} {et~al.}(2009){Finkelstein}, {Papovich}, {Rudnick},
  {Egami}, {Le Floc'h}, {Rieke}, {Rigby}, \& {Willmer}}]{FS09}
{Finkelstein}, S.~L., {Papovich}, C., {Rudnick}, G., {et~al.} 2009, \apj, 700,
  376, \dodoi{10.1088/0004-637X/700/1/376}

\bibitem[{{Fischer} {et~al.}(2018){Fischer}, {Beckmann}, {Bryant}, {Colditz},
  {Fumi}, {Geis}, {Hamidouche}, {Henning}, {H{\"o}nle}, {Iserlohe}, {Klein},
  {Krabbe}, {Looney}, {Poglitsch}, {Raab}, {Rebell}, {Rosenthal}, {Savage},
  {Schweitzer}, {Trinh}, \& {Vacca}}]{fischer18}
{Fischer}, C., {Beckmann}, S., {Bryant}, A., {et~al.} 2018, Journal of
  Astronomical Instrumentation, 7, 1840003, \dodoi{10.1142/S2251171718400032}

\bibitem[{{Fischer} {et~al.}(2014){Fischer}, {Abel}, {Gonz{\'a}lez-Alfonso},
  {Dudley}, {Satyapal}, \& {van Hoof}}]{F14}
{Fischer}, J., {Abel}, N.~P., {Gonz{\'a}lez-Alfonso}, E., {et~al.} 2014, \apj,
  795, 117, \dodoi{10.1088/0004-637X/795/2/117}

\bibitem[{{F{\"o}rster Schreiber} {et~al.}(2009){F{\"o}rster Schreiber},
  {Genzel}, {Bouch{\'e}}, {Cresci}, {Davies}, {Buschkamp}, {Shapiro},
  {Tacconi}, {Hicks}, {Genel}, {Shapley}, {Erb}, {Steidel}, {Lutz},
  {Eisenhauer}, {Gillessen}, {Sternberg}, {Renzini}, {Cimatti}, {Daddi},
  {Kurk}, {Lilly}, {Kong}, {Lehnert}, {Nesvadba}, {Verma}, {McCracken},
  {Arimoto}, {Mignoli}, \& {Onodera}}]{FN09}
{F{\"o}rster Schreiber}, N.~M., {Genzel}, R., {Bouch{\'e}}, N., {et~al.} 2009,
  \apj, 706, 1364, \dodoi{10.1088/0004-637X/706/2/1364}

\bibitem[{{Fraternali} {et~al.}(2021){Fraternali}, {Karim}, {Magnelli},
  {G{\'o}mez-Guijarro}, {Jim{\'e}nez-Andrade}, \& {Posses}}]{FF21}
{Fraternali}, F., {Karim}, A., {Magnelli}, B., {et~al.} 2021, \aap, 647, A194,
  \dodoi{10.1051/0004-6361/202039807}

\bibitem[{{Frayer} {et~al.}(1998){Frayer}, {Ivison}, {Scoville}, {Yun},
  {Evans}, {Smail}, {Blain}, \& {Kneib}}]{FD98}
{Frayer}, D.~T., {Ivison}, R.~J., {Scoville}, N.~Z., {et~al.} 1998, \apjl, 506,
  L7, \dodoi{10.1086/311639}

\bibitem[{{Frayer} {et~al.}(1999){Frayer}, {Ivison}, {Scoville}, {Evans},
  {Yun}, {Smail}, {Barger}, {Blain}, \& {Kneib}}]{FD99}
---. 1999, \apjl, 514, L13, \dodoi{10.1086/311940}

\bibitem[{{Fu} {et~al.}(2012){Fu}, {Jullo}, {Cooray}, {Bussmann}, {Ivison},
  {P{\'e}rez-Fournon}, {Djorgovski}, {Scoville}, {Yan}, {Riechers}, {Aguirre},
  {Auld}, {Baes}, {Baker}, {Bradford}, {Cava}, {Clements}, {Dannerbauer},
  {Dariush}, {De Zotti}, {Dole}, {Dunne}, {Dye}, {Eales}, {Frayer}, {Gavazzi},
  {Gurwell}, {Harris}, {Herranz}, {Hopwood}, {Hoyos}, {Ibar}, {Jarvis}, {Kim},
  {Leeuw}, {Lupu}, {Maddox}, {Mart{\'\i}nez-Navajas}, {Micha{\l}owski},
  {Negrello}, {Omont}, {Rosenman}, {Scott}, {Serjeant}, {Smail}, {Swinbank},
  {Valiante}, {Verma}, {Vieira}, {Wardlow}, \& {van der Werf}}]{FH12}
{Fu}, H., {Jullo}, E., {Cooray}, A., {et~al.} 2012, \apj, 753, 134,
  \dodoi{10.1088/0004-637X/753/2/134}

\bibitem[{{Fu} {et~al.}(2013){Fu}, {Cooray}, {Feruglio}, {Ivison}, {Riechers},
  {Gurwell}, {Bussmann}, {Harris}, {Altieri}, {Aussel}, {Baker}, {Bock},
  {Boylan-Kolchin}, {Bridge}, {Calanog}, {Casey}, {Cava}, {Chapman},
  {Clements}, {Conley}, {Cox}, {Farrah}, {Frayer}, {Hopwood}, {Jia}, {Magdis},
  {Marsden}, {Mart{\'\i}nez-Navajas}, {Negrello}, {Neri}, {Oliver}, {Omont},
  {Page}, {P{\'e}rez-Fournon}, {Schulz}, {Scott}, {Smith}, {Vaccari},
  {Valtchanov}, {Vieira}, {Viero}, {Wang}, {Wardlow}, \& {Zemcov}}]{FH13}
{Fu}, H., {Cooray}, A., {Feruglio}, C., {et~al.} 2013, \nat, 498, 338,
  \dodoi{10.1038/nature12184}

\bibitem[{{Fudamoto} {et~al.}(2021){Fudamoto}, {Oesch}, {Schouws}, {Stefanon},
  {Smit}, {Bouwens}, {Bowler}, {Endsley}, {Gonzalez}, {Inami}, {Labbe},
  {Stark}, {Aravena}, {Barrufet}, {da Cunha}, {Dayal}, {Ferrara}, {Graziani},
  {Hodge}, {Hutter}, {Li}, {De Looze}, {Nanayakkara}, {Pallottini}, {Riechers},
  {Schneider}, {Ucci}, {van der Werf}, \& {White}}]{FY21}
{Fudamoto}, Y., {Oesch}, P.~A., {Schouws}, S., {et~al.} 2021, \nat, 597, 489,
  \dodoi{10.1038/s41586-021-03846-z}

\bibitem[{{Fudamoto} {et~al.}(2024{\natexlab{a}}){Fudamoto}, {Inoue}, {Coe},
  {Welch}, {Acebron}, {Ricotti}, {Mandelker}, {Windhorst}, {Xu}, {Sugahara},
  {Bauer}, {Brada{\v{c}}}, {Bradley}, {Diego}, {Florian}, {Frye}, {Fujimoto},
  {Hashimoto}, {Henry}, {Mahler}, {Oesch}, {Ravindranath}, {Rigby}, {Sharon},
  {Strait}, {Tamura}, {Trenti}, {Vanzella}, {Zackrisson}, \& {Zitrin}}]{FY24a}
{Fudamoto}, Y., {Inoue}, A.~K., {Coe}, D., {et~al.} 2024{\natexlab{a}}, \apj,
  961, 71, \dodoi{10.3847/1538-4357/ad0f95}

\bibitem[{{Fudamoto} {et~al.}(2024{\natexlab{b}}){Fudamoto}, {Oesch}, {Walter},
  {Decarli}, {Carilli}, {Ferrara}, {Barrufet}, {Bouwens}, {Dessauges-Zavadsky},
  {Nelson}, {Dannerbauer}, {Illingworth}, {Inoue}, {Marques-Chaves},
  {P{\'e}rez-Fournon}, {Riechers}, {Schaerer}, {Smit}, {Sugahara}, \& {van der
  Werf}}]{FY24b}
{Fudamoto}, Y., {Oesch}, P.~A., {Walter}, F., {et~al.} 2024{\natexlab{b}},
  \mnras, 530, 340, \dodoi{10.1093/mnras/stae556}

\bibitem[{{Fudamoto} {et~al.}(2025){Fudamoto}, {Inoue}, {Bouwens}, {Inami},
  {Smit}, {Stark}, {Aravena}, {Pallottini}, {Hashimoto}, {Oguri}, {Bowler}, {da
  Cunha}, {Dayal}, {Ferrara}, {Fujimoto}, {Heintz}, {Hygate}, {van Leeuwen},
  {De Looze}, {Rowland}, {Stefanon}, {Sugahara}, {Witstok}, \& {van der
  Werf}}]{FY25}
{Fudamoto}, Y., {Inoue}, A.~K., {Bouwens}, R., {et~al.} 2025, arXiv e-prints,
  arXiv:2504.03831, \dodoi{10.48550/arXiv.2504.03831}

\bibitem[{{Fujimoto} {et~al.}(2019){Fujimoto}, {Ouchi}, {Ferrara},
  {Pallottini}, {Ivison}, {Behrens}, {Gallerani}, {Arata}, {Yajima}, \&
  {Nagamine}}]{FS19}
{Fujimoto}, S., {Ouchi}, M., {Ferrara}, A., {et~al.} 2019, \apj, 887, 107,
  \dodoi{10.3847/1538-4357/ab480f}

\bibitem[{{Fujimoto} {et~al.}(2021){Fujimoto}, {Oguri}, {Brammer}, {Yoshimura},
  {Laporte}, {Gonz{\'a}lez-L{\'o}pez}, {Caminha}, {Kohno}, {Zitrin}, {Richard},
  {Ouchi}, {Bauer}, {Smail}, {Hatsukade}, {Ono}, {Kokorev}, {Umehata},
  {Schaerer}, {Knudsen}, {Sun}, {Magdis}, {Valentino}, {Ao}, {Toft},
  {Dessauges-Zavadsky}, {Shimasaku}, {Caputi}, {Kusakabe}, {Morokuma-Matsui},
  {Shotaro}, {Egami}, {Lee}, {Rawle}, \& {Espada}}]{FS21}
{Fujimoto}, S., {Oguri}, M., {Brammer}, G., {et~al.} 2021, \apj, 911, 99,
  \dodoi{10.3847/1538-4357/abd7ec}

\bibitem[{{Fujimoto} {et~al.}(2022){Fujimoto}, {Brammer}, {Watson}, {Magdis},
  {Kokorev}, {Greve}, {Toft}, {Walter}, {Valiante}, {Ginolfi}, {Schneider},
  {Valentino}, {Colina}, {Vestergaard}, {Marques-Chaves}, {Fynbo}, {Krips},
  {Steinhardt}, {Cortzen}, {Rizzo}, \& {Oesch}}]{FS22}
{Fujimoto}, S., {Brammer}, G.~B., {Watson}, D., {et~al.} 2022, \nat, 604, 261,
  \dodoi{10.1038/s41586-022-04454-1}

\bibitem[{{Fujimoto} {et~al.}(2024{\natexlab{a}}){Fujimoto}, {Ouchi},
  {Nakajima}, {Harikane}, {Isobe}, {Brammer}, {Oguri}, {Gim{\'e}nez-Arteaga},
  {Heintz}, {Kokorev}, {Bauer}, {Ferrara}, {Kojima}, {Lagos}, {Laura},
  {Schaerer}, {Shimasaku}, {Hatsukade}, {Kohno}, {Sun}, {Valentino}, {Watson},
  {Fudamoto}, {Inoue}, {Gonz{\'a}lez-L{\'o}pez}, {Koekemoer}, {Knudsen}, {Lee},
  {Magdis}, {Richard}, {Strait}, {Sugahara}, {Tamura}, {Toft}, {Umehata}, \&
  {Walth}}]{FS24a}
{Fujimoto}, S., {Ouchi}, M., {Nakajima}, K., {et~al.} 2024{\natexlab{a}}, \apj,
  964, 146, \dodoi{10.3847/1538-4357/ad235c}

\bibitem[{{Fujimoto} {et~al.}(2024{\natexlab{b}}){Fujimoto}, {Ouchi}, {Kohno},
  {Valentino}, {Gim\textbackslash'enez-Arteaga}, {Brammer}, {Furtak},
  {Kohandel}, {Oguri}, {Pallottini}, {Richard}, {Zitrin}, {Bauer},
  {Boylan-Kolchin}, {Dessauges-Zavadsky}, {Egami}, {Finkelstein}, {Ma},
  {Smail}, {Watson}, {Hutchison}, {Rigby}, {Welch}, {Ao}, {Bradley}, {Caminha},
  {Caputi}, {Espada}, {Endsley}, {Fudamoto},
  {Gonz\textbackslash'alez-L\textbackslash'opez}, {Hatsukade}, {Koekemoer},
  {Kokorev}, {Laporte}, {Lee}, {Magdis}, {Ono}, {Rizzo}, {Shibuya},
  {Shimasaku}, {Sun}, {Toft}, {Umehata}, {Wang}, \& {Yajima}}]{FS24b}
{Fujimoto}, S., {Ouchi}, M., {Kohno}, K., {et~al.} 2024{\natexlab{b}}, arXiv
  e-prints, arXiv:2402.18543, \dodoi{10.48550/arXiv.2402.18543}

\bibitem[{{Fujimoto} {et~al.}(2025){Fujimoto}, {Bezanson}, {Labbe}, {Brammer},
  {Price}, {Wang}, {Weaver}, {Fudamoto}, {Oesch}, {Williams}, {Dayal},
  {Feldmann}, {Greene}, {Leja}, {Whitaker}, {Zitrin}, {Cutler}, {Furtak},
  {Pan}, {Chemerynska}, {Kokorev}, {Miller}, {Atek}, {van Dokkum}, {Juneau},
  {Kassin}, {Khullar}, {Marchesini}, {Maseda}, {Nelson}, {Setton}, \&
  {Smit}}]{FS25}
{Fujimoto}, S., {Bezanson}, R., {Labbe}, I., {et~al.} 2025, \apjs, 278, 45,
  \dodoi{10.3847/1538-4365/adc677}

\bibitem[{{Gallerani} {et~al.}(2012){Gallerani}, {Neri}, {Maiolino},
  {Mart{\'\i}n}, {De Breuck}, {Walter}, {Caselli}, {Krips}, {Meneghetti},
  {Nagao}, {Wagg}, \& {Walmsley}}]{GS12}
{Gallerani}, S., {Neri}, R., {Maiolino}, R., {et~al.} 2012, \aap, 543, A114,
  \dodoi{10.1051/0004-6361/201118705}

\bibitem[{{Gavazzi} {et~al.}(2011){Gavazzi}, {Cooray}, {Conley}, {Aguirre},
  {Amblard}, {Auld}, {Beelen}, {Blain}, {Blundell}, {Bock}, {Bradford},
  {Bridge}, {Brisbin}, {Burgarella}, {Chanial}, {Chapin}, {Christopher},
  {Clements}, {Cox}, {Djorgovski}, {Dowell}, {Eales}, {Earle},
  {Ellsworth-Bowers}, {Farrah}, {Franceschini}, {Fu}, {Glenn}, {Gonz{\'a}lez
  Solares}, {Griffin}, {Gurwell}, {Halpern}, {Ibar}, {Ivison}, {Jarvis},
  {Kamenetzky}, {Kim}, {Krips}, {Levenson}, {Lupu}, {Mahabal}, {Maloney},
  {Maraston}, {Marchetti}, {Marsden}, {Matsuhara}, {Mortier}, {Murphy},
  {Naylor}, {Neri}, {Nguyen}, {Oliver}, {Omont}, {Page}, {Papageorgiou},
  {Pearson}, {P{\'e}rez-Fournon}, {Pohlen}, {Rangwala}, {Rawlings}, {Raymond},
  {Riechers}, {Rodighiero}, {Roseboom}, {Rowan-Robinson}, {Schulz}, {Scott},
  {Scott}, {Serra}, {Seymour}, {Shupe}, {Smith}, {Symeonidis}, {Tugwell},
  {Vaccari}, {Valiante}, {Valtchanov}, {Verma}, {Vieira}, {Vigroux}, {Wang},
  {Wardlow}, {Wiebe}, {Wright}, {Xu}, {Zeimann}, {Zemcov}, \&
  {Zmuidzinas}}]{GR11}
{Gavazzi}, R., {Cooray}, A., {Conley}, A., {et~al.} 2011, \apj, 738, 125,
  \dodoi{10.1088/0004-637X/738/2/125}

\bibitem[{{Geach} {et~al.}(2016){Geach}, {Narayanan}, {Matsuda}, {Hayes},
  {Mas-Ribas}, {Dijkstra}, {Steidel}, {Chapman}, {Feldmann}, {Avison},
  {Agertz}, {Ao}, {Birkinshaw}, {Bremer}, {Clements}, {Dannerbauer}, {Farrah},
  {Harrison}, {Kubo}, {Micha{\l}owski}, {Scott}, {Smith}, {Spaans}, {Simpson},
  {Swinbank}, {Taniguchi}, {van der Werf}, {Verma}, \& {Yamada}}]{GJ16}
{Geach}, J.~E., {Narayanan}, D., {Matsuda}, Y., {et~al.} 2016, \apj, 832, 37,
  \dodoi{10.3847/0004-637X/832/1/37}

\bibitem[{{Genzel} {et~al.}(1989){Genzel}, {Harris}, \& {Stutzki}}]{genzel89}
{Genzel}, R., {Harris}, A.~I., \& {Stutzki}, J. 1989, in Infrared Spectroscopy
  in Astronomy, ed. E.~{B{\"o}hm-Vitense}, 115

\bibitem[{{Genzel} {et~al.}(1998){Genzel}, {Lutz}, {Sturm}, {Egami}, {Kunze},
  {Moorwood}, {Rigopoulou}, {Spoon}, {Sternberg}, {Tacconi-Garman}, {Tacconi},
  \& {Thatte}}]{genzel98}
{Genzel}, R., {Lutz}, D., {Sturm}, E., {et~al.} 1998, \apj, 498, 579,
  \dodoi{10.1086/305576}

\bibitem[{{George} {et~al.}(2013){George}, {Ivison}, {Hopwood}, {Riechers},
  {Bussmann}, {Cox}, {Dye}, {Krips}, {Negrello}, {Neri}, {Serjeant},
  {Valtchanov}, {Baes}, {Bourne}, {Clements}, {de Zotti}, {Dunne}, {Eales},
  {Ibar}, {Maddox}, {Smith}, {Valiante}, \& {van der Werf}}]{GR13}
{George}, R.~D., {Ivison}, R.~J., {Hopwood}, R., {et~al.} 2013, \mnras, 436,
  L99, \dodoi{10.1093/mnrasl/slt122}

\bibitem[{{Gim{\'e}nez-Arteaga} {et~al.}(2024){Gim{\'e}nez-Arteaga},
  {Fujimoto}, {Valentino}, {Brammer}, {Mason}, {Rizzo}, {Rusakov}, {Colina},
  {Prieto-Lyon}, {Oesch}, {Espada}, {Heintz}, {Knudsen}, {Dessauges-Zavadsky},
  {Laporte}, {Lee}, {Magdis}, {Ono}, {Ao}, {Ouchi}, {Kohno}, \&
  {Koekemoer}}]{GC24}
{Gim{\'e}nez-Arteaga}, C., {Fujimoto}, S., {Valentino}, F., {et~al.} 2024,
  \aap, 686, A63, \dodoi{10.1051/0004-6361/202349135}

\bibitem[{{Ginolfi} {et~al.}(2020){Ginolfi}, {Jones}, {B{\'e}thermin},
  {Faisst}, {Lemaux}, {Schaerer}, {Fudamoto}, {Oesch}, {Dessauges-Zavadsky},
  {Fujimoto}, {Carniani}, {Le F{\`e}vre}, {Cassata}, {Silverman}, {Capak},
  {Yan}, {Bardelli}, {Cucciati}, {Gal}, {Gruppioni}, {Hathi}, {Lubin},
  {Maiolino}, {Morselli}, {Pelliccia}, {Talia}, {Vergani}, \&
  {Zamorani}}]{GM20}
{Ginolfi}, M., {Jones}, G.~C., {B{\'e}thermin}, M., {et~al.} 2020, \aap, 643,
  A7, \dodoi{10.1051/0004-6361/202038284}

\bibitem[{{Giulietti} {et~al.}(2023){Giulietti}, {Lapi}, {Massardi}, {Behiri},
  {Torsello}, {D'Amato}, {Ronconi}, {Perrotta}, \& {Bressan}}]{GM23}
{Giulietti}, M., {Lapi}, A., {Massardi}, M., {et~al.} 2023, \apj, 943, 151,
  \dodoi{10.3847/1538-4357/aca53f}

\bibitem[{{Glazer} {et~al.}(2024){Glazer}, {Brad{\u{a}}c}, {Sanders},
  {Fujimoto}, {Bolan}, {Ferrara}, {Strait}, {Jones}, {Lemaux}, {Vallini}, \&
  {Ryan}}]{GK24}
{Glazer}, K., {Brad{\u{a}}c}, M., {Sanders}, R.~L., {et~al.} 2024, \mnras, 531,
  945, \dodoi{10.1093/mnras/stae1178}

\bibitem[{{G{\'o}mez-Guijarro} {et~al.}(2018){G{\'o}mez-Guijarro}, {Toft},
  {Karim}, {Magnelli}, {Magdis}, {Jim{\'e}nez-Andrade}, {Capak}, {Fraternali},
  {Fujimoto}, {Riechers}, {Schinnerer}, {Smol{\v{c}}i{\'c}}, {Aravena},
  {Bertoldi}, {Cortzen}, {Hasinger}, {Hu}, {Jones}, {Koekemoer}, {Lee},
  {McCracken}, {Micha{\l}owski}, {Navarrete}, {Povi{\'c}}, {Puglisi},
  {Romano-D{\'\i}az}, {Sheth}, {Silverman}, {Staguhn}, {Steinhardt},
  {Stockmann}, {Tanaka}, {Valentino}, {van Kampen}, \& {Zirm}}]{GC18}
{G{\'o}mez-Guijarro}, C., {Toft}, S., {Karim}, A., {et~al.} 2018, \apj, 856,
  121, \dodoi{10.3847/1538-4357/aab206}

\bibitem[{{Gonz{\'a}lez-L{\'o}pez} {et~al.}(2017){Gonz{\'a}lez-L{\'o}pez},
  {Bauer}, {Romero-Ca{\~n}izales}, {Kneissl}, {Villard}, {Carvajal}, {Kim},
  {Laporte}, {Anguita}, {Aravena}, {Bouwens}, {Bradley}, {Carrasco}, {Demarco},
  {Ford}, {Ibar}, {Infante}, {Messias}, {Mu{\~n}oz Arancibia}, {Nagar},
  {Padilla}, {Treister}, {Troncoso}, \& {Zitrin}}]{GJ17}
{Gonz{\'a}lez-L{\'o}pez}, J., {Bauer}, F.~E., {Romero-Ca{\~n}izales}, C.,
  {et~al.} 2017, \aap, 597, A41, \dodoi{10.1051/0004-6361/201628806}

\bibitem[{{Gorjian} {et~al.}(2007){Gorjian}, {Cleary}, {Werner}, \&
  {Lawrence}}]{gorjian07}
{Gorjian}, V., {Cleary}, K., {Werner}, M.~W., \& {Lawrence}, C.~R. 2007, \apjl,
  655, L73, \dodoi{10.1086/511975}

\bibitem[{{Griffin} {et~al.}(2010){Griffin}, {Abergel}, {Abreu}, {Ade},
  {Andr{\'e}}, {Augueres}, {Babbedge}, {Bae}, {Baillie}, {Baluteau}, {Barlow},
  {Bendo}, {Benielli}, {Bock}, {Bonhomme}, {Brisbin}, {Brockley-Blatt},
  {Caldwell}, {Cara}, {Castro-Rodriguez}, {Cerulli}, {Chanial}, {Chen},
  {Clark}, {Clements}, {Clerc}, {Coker}, {Communal}, {Conversi}, {Cox},
  {Crumb}, {Cunningham}, {Daly}, {Davis}, {de Antoni}, {Delderfield}, {Devin},
  {di Giorgio}, {Didschuns}, {Dohlen}, {Donati}, {Dowell}, {Dowell}, {Duband},
  {Dumaye}, {Emery}, {Ferlet}, {Ferrand}, {Fontignie}, {Fox}, {Franceschini},
  {Frerking}, {Fulton}, {Garcia}, {Gastaud}, {Gear}, {Glenn}, {Goizel},
  {Griffin}, {Grundy}, {Guest}, {Guillemet}, {Hargrave}, {Harwit}, {Hastings},
  {Hatziminaoglou}, {Herman}, {Hinde}, {Hristov}, {Huang}, {Imhof}, {Isaak},
  {Israelsson}, {Ivison}, {Jennings}, {Kiernan}, {King}, {Lange}, {Latter},
  {Laurent}, {Laurent}, {Leeks}, {Lellouch}, {Levenson}, {Li}, {Li},
  {Lilienthal}, {Lim}, {Liu}, {Lu}, {Madden}, {Mainetti}, {Marliani}, {McKay},
  {Mercier}, {Molinari}, {Morris}, {Moseley}, {Mulder}, {Mur}, {Naylor},
  {Nguyen}, {O'Halloran}, {Oliver}, {Olofsson}, {Olofsson}, {Orfei}, {Page},
  {Pain}, {Panuzzo}, {Papageorgiou}, {Parks}, {Parr-Burman}, {Pearce},
  {Pearson}, {P{\'e}rez-Fournon}, {Pinsard}, {Pisano}, {Podosek}, {Pohlen},
  {Polehampton}, {Pouliquen}, {Rigopoulou}, {Rizzo}, {Roseboom}, {Roussel},
  {Rowan-Robinson}, {Rownd}, {Saraceno}, {Sauvage}, {Savage}, {Savini},
  {Sawyer}, {Scharmberg}, {Schmitt}, {Schneider}, {Schulz}, {Schwartz},
  {Shafer}, {Shupe}, {Sibthorpe}, {Sidher}, {Smith}, {Smith}, {Smith},
  {Spencer}, {Stobie}, {Sudiwala}, {Sukhatme}, {Surace}, {Stevens}, {Swinyard},
  {Trichas}, {Tourette}, {Triou}, {Tseng}, {Tucker}, {Turner}, {Vaccari},
  {Valtchanov}, {Vigroux}, {Virique}, {Voellmer}, {Walker}, {Ward}, {Waskett},
  {Weilert}, {Wesson}, {White}, {Whitehouse}, {Wilson}, {Winter}, {Woodcraft},
  {Wright}, {Xu}, {Zavagno}, {Zemcov}, {Zhang}, \& {Zonca}}]{griffin10}
{Griffin}, M.~J., {Abergel}, A., {Abreu}, A., {et~al.} 2010, \aap, 518, L3,
  \dodoi{10.1051/0004-6361/201014519}

\bibitem[{{Gruppioni} {et~al.}(2020){Gruppioni}, {B{\'e}thermin}, {Loiacono},
  {Le F{\`e}vre}, {Capak}, {Cassata}, {Faisst}, {Schaerer}, {Silverman}, {Yan},
  {Bardelli}, {Boquien}, {Carraro}, {Cimatti}, {Dessauges-Zavadsky}, {Ginolfi},
  {Fujimoto}, {Hathi}, {Jones}, {Khusanova}, {Koekemoer}, {Lagache}, {Lemaux},
  {Oesch}, {Pozzi}, {Riechers}, {Rodighiero}, {Romano}, {Talia}, {Vallini},
  {Vergani}, {Zamorani}, \& {Zucca}}]{gruppioni20}
{Gruppioni}, C., {B{\'e}thermin}, M., {Loiacono}, F., {et~al.} 2020, \aap, 643,
  A8, \dodoi{10.1051/0004-6361/202038487}

\bibitem[{{Gullberg} {et~al.}(2015){Gullberg}, {De Breuck}, {Vieira},
  {Wei{\ss}}, {Aguirre}, {Aravena}, {B{\'e}thermin}, {Bradford}, {Bothwell},
  {Carlstrom}, {Chapman}, {Fassnacht}, {Gonzalez}, {Greve}, {Hezaveh},
  {Holzapfel}, {Husband}, {Ma}, {Malkan}, {Marrone}, {Menten}, {Murphy},
  {Reichardt}, {Spilker}, {Stark}, {Strandet}, \& {Welikala}}]{GB15}
{Gullberg}, B., {De Breuck}, C., {Vieira}, J.~D., {et~al.} 2015, \mnras, 449,
  2883, \dodoi{10.1093/mnras/stv372}

\bibitem[{{Gullberg} {et~al.}(2018){Gullberg}, {Swinbank}, {Smail}, {Biggs},
  {Bertoldi}, {De Breuck}, {Chapman}, {Chen}, {Cooke}, {Coppin}, {Cox},
  {Dannerbauer}, {Dunlop}, {Edge}, {Farrah}, {Geach}, {Greve}, {Hodge}, {Ibar},
  {Ivison}, {Karim}, {Schinnerer}, {Scott}, {Simpson}, {Stach}, {Thomson}, {van
  der Werf}, {Walter}, {Wardlow}, \& {Weiss}}]{GB18}
{Gullberg}, B., {Swinbank}, A.~M., {Smail}, I., {et~al.} 2018, \apj, 859, 12,
  \dodoi{10.3847/1538-4357/aabe8c}

\bibitem[{{Gururajan} {et~al.}(2023){Gururajan}, {Bethermin}, {Sulzenauer},
  {Theul{\'e}}, {Spilker}, {Aravena}, {Chapman}, {Gonzalez}, {Greve},
  {Narayanan}, {Reuter}, {Vieira}, \& {Weiss}}]{GG23}
{Gururajan}, G., {Bethermin}, M., {Sulzenauer}, N., {et~al.} 2023, \aap, 676,
  A89, \dodoi{10.1051/0004-6361/202346449}

\bibitem[{{G{\"u}sten} {et~al.}(2006){G{\"u}sten}, {Nyman}, {Schilke},
  {Menten}, {Cesarsky}, \& {Booth}}]{gusten06}
{G{\"u}sten}, R., {Nyman}, L.~{\r{A}}., {Schilke}, P., {et~al.} 2006, \aap,
  454, L13, \dodoi{10.1051/0004-6361:20065420}

\bibitem[{{Hailey-Dunsheath} {et~al.}(2010){Hailey-Dunsheath}, {Nikola},
  {Stacey}, {Oberst}, {Parshley}, {Benford}, {Staguhn}, \& {Tucker}}]{HS10}
{Hailey-Dunsheath}, S., {Nikola}, T., {Stacey}, G.~J., {et~al.} 2010, \apjl,
  714, L162, \dodoi{10.1088/2041-8205/714/1/L162}

\bibitem[{{Hainline} {et~al.}(2009){Hainline}, {Shapley}, {Kornei}, {Pettini},
  {Buckley-Geer}, {Allam}, \& {Tucker}}]{HK09}
{Hainline}, K.~N., {Shapley}, A.~E., {Kornei}, K.~A., {et~al.} 2009, \apj, 701,
  52, \dodoi{10.1088/0004-637X/701/1/52}

\bibitem[{{Harikane} {et~al.}(2020){Harikane}, {Ouchi}, {Inoue}, {Matsuoka},
  {Tamura}, {Bakx}, {Fujimoto}, {Moriwaki}, {Ono}, {Nagao}, {Tadaki}, {Kojima},
  {Shibuya}, {Egami}, {Ferrara}, {Gallerani}, {Hashimoto}, {Kohno}, {Matsuda},
  {Matsuo}, {Pallottini}, {Sugahara}, \& {Vallini}}]{HY20}
{Harikane}, Y., {Ouchi}, M., {Inoue}, A.~K., {et~al.} 2020, \apj, 896, 93,
  \dodoi{10.3847/1538-4357/ab94bd}

\bibitem[{{Harikane} {et~al.}(2025){Harikane}, {Sanders}, {Ellis}, {Jones},
  {Ouchi}, {Laporte}, {Roberts-Borsani}, {Katz}, {Nakajima}, {Ono}, \&
  {Gupta}}]{HY25}
{Harikane}, Y., {Sanders}, R.~L., {Ellis}, R., {et~al.} 2025, arXiv e-prints,
  arXiv:2505.09186, \dodoi{10.48550/arXiv.2505.09186}

\bibitem[{{Harrington} {et~al.}(2019){Harrington}, {Vishwas}, {Wei{\ss}},
  {Magnelli}, {Grassitelli}, {Zaja{\v{c}}ek}, {Jim{\'e}nez-Andrade}, {Leung},
  {Bertoldi}, {Romano-D{\'\i}az}, {Frayer}, {Kamieneski}, {Riechers}, {Stacey},
  {Yun}, \& {Wang}}]{HK19}
{Harrington}, K.~C., {Vishwas}, A., {Wei{\ss}}, A., {et~al.} 2019, \mnras, 488,
  1489, \dodoi{10.1093/mnras/stz1740}

\bibitem[{{Harris} {et~al.}(2012){Harris}, {Baker}, {Frayer}, {Smail},
  {Swinbank}, {Riechers}, {van der Werf}, {Auld}, {Baes}, {Bussmann},
  {Buttiglione}, {Cava}, {Clements}, {Cooray}, {Dannerbauer}, {Dariush}, {De
  Zotti}, {Dunne}, {Dye}, {Eales}, {Fritz}, {Gonz{\'a}lez-Nuevo}, {Hopwood},
  {Ibar}, {Ivison}, {Jarvis}, {Maddox}, {Negrello}, {Rigby}, {Smith}, {Temi},
  \& {Wardlow}}]{HA12}
{Harris}, A.~I., {Baker}, A.~J., {Frayer}, D.~T., {et~al.} 2012, \apj, 752,
  152, \dodoi{10.1088/0004-637X/752/2/152}

\bibitem[{{Harshan} {et~al.}(2024){Harshan}, {Tripodi}, {Martis},
  {Rihtar{\v{s}}i{\v{c}}}, {Brada{\v{c}}}, {Asada}, {Brammer}, {Desprez},
  {Estrada-Carpenter}, {Matharu}, {Markov}, {Muzzin}, {Mowla}, {Noirot},
  {Sarrouh}, {Sawicki}, {Strait}, \& {Willott}}]{HA24}
{Harshan}, A., {Tripodi}, R., {Martis}, N.~S., {et~al.} 2024, \apjl, 977, L36,
  \dodoi{10.3847/2041-8213/ad9741}

\bibitem[{{Hashimoto} {et~al.}(2019{\natexlab{a}}){Hashimoto}, {Inoue},
  {Tamura}, {Matsuo}, {Mawatari}, \& {Yamaguchi}}]{HT19b}
{Hashimoto}, T., {Inoue}, A.~K., {Tamura}, Y., {et~al.} 2019{\natexlab{a}},
  \pasj, 71, 109, \dodoi{10.1093/pasj/psz094}

\bibitem[{{Hashimoto} {et~al.}(2018){Hashimoto}, {Laporte}, {Mawatari},
  {Ellis}, {Inoue}, {Zackrisson}, {Roberts-Borsani}, {Zheng}, {Tamura},
  {Bauer}, {Fletcher}, {Harikane}, {Hatsukade}, {Hayatsu}, {Matsuda}, {Matsuo},
  {Okamoto}, {Ouchi}, {Pell{\'o}}, {Rydberg}, {Shimizu}, {Taniguchi},
  {Umehata}, \& {Yoshida}}]{HT18}
{Hashimoto}, T., {Laporte}, N., {Mawatari}, K., {et~al.} 2018, \nat, 557, 392,
  \dodoi{10.1038/s41586-018-0117-z}

\bibitem[{{Hashimoto} {et~al.}(2019{\natexlab{b}}){Hashimoto}, {Inoue},
  {Mawatari}, {Tamura}, {Matsuo}, {Furusawa}, {Harikane}, {Shibuya}, {Knudsen},
  {Kohno}, {Ono}, {Zackrisson}, {Okamoto}, {Kashikawa}, {Oesch}, {Ouchi},
  {Ota}, {Shimizu}, {Taniguchi}, {Umehata}, \& {Watson}}]{HT19a}
{Hashimoto}, T., {Inoue}, A.~K., {Mawatari}, K., {et~al.} 2019{\natexlab{b}},
  \pasj, 71, 71, \dodoi{10.1093/pasj/psz049}

\bibitem[{{Hashimoto} {et~al.}(2023){Hashimoto}, {{\'A}lvarez-M{\'a}rquez},
  {Fudamoto}, {Colina}, {Inoue}, {Nakazato}, {Ceverino}, {Yoshida},
  {Costantin}, {Sugahara}, {G{\'o}mez}, {Blanco-Prieto}, {Mawatari}, {Arribas},
  {Marques-Chaves}, {Pereira-Santaella}, {Bakx}, {Hagimoto}, {Hashigaya},
  {Matsuo}, {Tamura}, {Usui}, \& {Ren}}]{HT23}
{Hashimoto}, T., {{\'A}lvarez-M{\'a}rquez}, J., {Fudamoto}, Y., {et~al.} 2023,
  \apjl, 955, L2, \dodoi{10.3847/2041-8213/acf57c}

\bibitem[{{Hatsukade} {et~al.}(2015){Hatsukade}, {Tamura}, {Iono}, {Matsuda},
  {Hayashi}, \& {Oguri}}]{HB15}
{Hatsukade}, B., {Tamura}, Y., {Iono}, D., {et~al.} 2015, \pasj, 67, 93,
  \dodoi{10.1093/pasj/psv061}

\bibitem[{{Haynes} {et~al.}(2018){Haynes}, {Giovanelli}, {Kent}, {Adams},
  {Balonek}, {Craig}, {Fertig}, {Finn}, {Giovanardi}, {Hallenbeck}, {Hess},
  {Hoffman}, {Huang}, {Jones}, {Koopmann}, {Kornreich}, {Leisman}, {Miller},
  {Moorman}, {O'Connor}, {O'Donoghue}, {Papastergis}, {Troischt}, {Stark}, \&
  {Xiao}}]{H18b}
{Haynes}, M.~P., {Giovanelli}, R., {Kent}, B.~R., {et~al.} 2018, \apj, 861, 49,
  \dodoi{10.3847/1538-4357/aac956}

\bibitem[{{Heintz} {et~al.}(2021){Heintz}, {Watson}, {Oesch}, {Narayanan}, \&
  {Madden}}]{heintz21}
{Heintz}, K.~E., {Watson}, D., {Oesch}, P.~A., {Narayanan}, D., \& {Madden},
  S.~C. 2021, \apj, 922, 147, \dodoi{10.3847/1538-4357/ac2231}

\bibitem[{{Heintz} {et~al.}(2023){Heintz}, {Gim{\'e}nez-Arteaga}, {Fujimoto},
  {Brammer}, {Espada}, {Gillman}, {Gonz{\'a}lez-L{\'o}pez}, {Greve},
  {Harikane}, {Hatsukade}, {Knudsen}, {Koekemoer}, {Kohno}, {Kokorev}, {Lee},
  {Magdis}, {Nelson}, {Rizzo}, {Sanders}, {Schaerer}, {Shapley}, {Strait},
  {Toft}, {Valentino}, {van der Wel}, {Vijayan}, {Watson}, {Bauer},
  {Christiansen}, \& {Wilson}}]{HK23}
{Heintz}, K.~E., {Gim{\'e}nez-Arteaga}, C., {Fujimoto}, S., {et~al.} 2023,
  \apjl, 944, L30, \dodoi{10.3847/2041-8213/acb2cf}

\bibitem[{{Helou}(1986)}]{helou86}
{Helou}, G. 1986, \apjl, 311, L33, \dodoi{10.1086/184793}

\bibitem[{{Herard-Demanche} {et~al.}(2025){Herard-Demanche}, {Bouwens},
  {Oesch}, {Naidu}, {Decarli}, {Nelson}, {Brammer}, {Weibel}, {Xiao},
  {Stefanon}, {Walter}, {Matthee}, {Meyer}, {Wuyts}, {Reddy}, {Rowland}, {van
  Leeuwen}, {Haro}, {Dannerbauer}, {Shapley}, {Chisholm}, {van Dokkum},
  {Labbe}, {Illingworth}, {Schaerer}, \& {Shivaei}}]{HT25}
{Herard-Demanche}, T., {Bouwens}, R.~J., {Oesch}, P.~A., {et~al.} 2025, \mnras,
  537, 788, \dodoi{10.1093/mnras/staf030}

\bibitem[{{Herrera-Camus} {et~al.}(2015){Herrera-Camus}, {Bolatto}, {Wolfire},
  {Smith}, {Croxall}, {Kennicutt}, {Calzetti}, {Helou}, {Walter}, {Leroy},
  {Draine}, {Brandl}, {Armus}, {Sandstrom}, {Dale}, {Aniano}, {Meidt},
  {Boquien}, {Hunt}, {Galametz}, {Tabatabaei}, {Murphy}, {Appleton}, {Roussel},
  {Engelbracht}, \& {Beirao}}]{herrera15}
{Herrera-Camus}, R., {Bolatto}, A.~D., {Wolfire}, M.~G., {et~al.} 2015, \apj,
  800, 1, \dodoi{10.1088/0004-637X/800/1/1}

\bibitem[{{Herrera-Camus} {et~al.}(2016){Herrera-Camus}, {Bolatto}, {Smith},
  {Draine}, {Pellegrini}, {Wolfire}, {Croxall}, {de Looze}, {Calzetti},
  {Kennicutt}, {Crocker}, {Armus}, {van der Werf}, {Sandstrom}, {Galametz},
  {Brandl}, {Groves}, {Rigopoulou}, {Walter}, {Leroy}, {Boquien}, {Tabatabaei},
  \& {Beirao}}]{herrera16}
{Herrera-Camus}, R., {Bolatto}, A., {Smith}, J.~D., {et~al.} 2016, \apj, 826,
  175, \dodoi{10.3847/0004-637X/826/2/175}

\bibitem[{{Herrera-Camus} {et~al.}(2018){Herrera-Camus}, {Sturm},
  {Graci{\'a}-Carpio}, {Lutz}, {Contursi}, {Veilleux}, {Fischer},
  {Gonz{\'a}lez-Alfonso}, {Poglitsch}, {Tacconi}, {Genzel}, {Maiolino},
  {Sternberg}, {Davies}, \& {Verma}}]{H18a}
{Herrera-Camus}, R., {Sturm}, E., {Graci{\'a}-Carpio}, J., {et~al.} 2018, \apj,
  861, 94, \dodoi{10.3847/1538-4357/aac0f6}

\bibitem[{{Herrera-Camus} {et~al.}(2021){Herrera-Camus}, {F{\"o}rster
  Schreiber}, {Genzel}, {Tacconi}, {Bolatto}, {Davies}, {Fisher}, {Lutz},
  {Naab}, {Shimizu}, {Tadaki}, \& {{\"U}bler}}]{HR21}
{Herrera-Camus}, R., {F{\"o}rster Schreiber}, N., {Genzel}, R., {et~al.} 2021,
  \aap, 649, A31, \dodoi{10.1051/0004-6361/202039704}

\bibitem[{{Herrera-Camus} {et~al.}(2022){Herrera-Camus}, {F{\"o}rster
  Schreiber}, {Price}, {{\"U}bler}, {Bolatto}, {Davies}, {Fisher}, {Genzel},
  {Lutz}, {Naab}, {Nestor}, {Shimizu}, {Sternberg}, {Tacconi}, \&
  {Tadaki}}]{HR22}
{Herrera-Camus}, R., {F{\"o}rster Schreiber}, N.~M., {Price}, S.~H., {et~al.}
  2022, \aap, 665, L8, \dodoi{10.1051/0004-6361/202142562}

\bibitem[{{Hinshaw} {et~al.}(2013){Hinshaw}, {Larson}, {Komatsu}, {Spergel},
  {Bennett}, {Dunkley}, {Nolta}, {Halpern}, {Hill}, {Odegard}, {Page}, {Smith},
  {Weiland}, {Gold}, {Jarosik}, {Kogut}, {Limon}, {Meyer}, {Tucker}, {Wollack},
  \& {Wright}}]{hinshaw13}
{Hinshaw}, G., {Larson}, D., {Komatsu}, E., {et~al.} 2013, \apjs, 208, 19,
  \dodoi{10.1088/0067-0049/208/2/19}

\bibitem[{{Ho}(2007)}]{H07}
{Ho}, L.~C. 2007, \apj, 669, 821, \dodoi{10.1086/521917}

\bibitem[{{Hopwood} {et~al.}(2011){Hopwood}, {Wardlow}, {Cooray}, {Khostovan},
  {Kim}, {Negrello}, {da Cunha}, {Burgarella}, {Aretxaga}, {Auld}, {Baes},
  {Barton}, {Bertoldi}, {Bonfield}, {Blundell}, {Buttiglione}, {Cava},
  {Clements}, {Cooke}, {Dannerbauer}, {Dariush}, {de Zotti}, {Dunlop}, {Dunne},
  {Dye}, {Eales}, {Fritz}, {Frayer}, {Gurwell}, {Hughes}, {Ibar}, {Ivison},
  {Jarvis}, {Lagache}, {Leeuw}, {Maddox}, {Micha{\l}owski}, {Omont}, {Pascale},
  {Pohlen}, {Rigby}, {Rodighiero}, {Scott}, {Serjeant}, {Smail}, {Smith},
  {Temi}, {Thompson}, {Valtchanov}, {van der Werf}, {Verma}, \&
  {Vieira}}]{HR11}
{Hopwood}, R., {Wardlow}, J., {Cooray}, A., {et~al.} 2011, \apjl, 728, L4,
  \dodoi{10.1088/2041-8205/728/1/L4}

\bibitem[{{Houck} {et~al.}(2004){Houck}, {Roellig}, {Van Cleve}, {Forrest},
  {Herter}, {Lawrence}, {Matthews}, {Reitsema}, {Soifer}, {Watson}, {Weedman},
  {Huisjen}, {Troeltzsch}, {Barry}, {Bernard-Salas}, {Blacken}, {Brandl},
  {Charmandaris}, {Devost}, {Gull}, {Hall}, {Henderson}, {Higdon}, {Pirger},
  {Schoenwald}, {Sloan}, {Uchida}, {Appleton}, {Armus}, {Burgdorf},
  {Fajardo-Acosta}, {Grillmair}, {Ingalls}, {Morris}, \& {Teplitz}}]{houck04}
{Houck}, J.~R., {Roellig}, T.~L., {Van Cleve}, J., {et~al.} 2004, in Society of
  Photo-Optical Instrumentation Engineers (SPIE) Conference Series, Vol. 5487,
  Optical, Infrared, and Millimeter Space Telescopes, ed. J.~C. {Mather},
  62--76, \dodoi{10.1117/12.550517}

\bibitem[{{Howell} {et~al.}(2010){Howell}, {Armus}, {Mazzarella}, {Evans},
  {Surace}, {Sanders}, {Petric}, {Appleton}, {Bothun}, {Bridge}, {Chan},
  {Charmandaris}, {Frayer}, {Haan}, {Inami}, {Kim}, {Lord}, {Madore},
  {Melbourne}, {Schulz}, {U}, {Vavilkin}, {Veilleux}, \& {Xu}}]{H10}
{Howell}, J.~H., {Armus}, L., {Mazzarella}, J.~M., {et~al.} 2010, \apj, 715,
  572, \dodoi{10.1088/0004-637X/715/1/572}

\bibitem[{{Huang} {et~al.}(2016){Huang}, {Brada{\v{c}}}, {Lemaux}, {Ryan},
  {Hoag}, {Castellano}, {Amor{\'\i}n}, {Fontana}, {Brammer}, {Cain}, {Lubin},
  {Merlin}, {Schmidt}, {Schrabback}, {Treu}, {Gonzalez}, {von der Linden}, \&
  {Knight}}]{HK16}
{Huang}, K.-H., {Brada{\v{c}}}, M., {Lemaux}, B.~C., {et~al.} 2016, \apj, 817,
  11, \dodoi{10.3847/0004-637X/817/1/11}

\bibitem[{{Hughes} {et~al.}(2016){Hughes}, {Baes}, {Schirm}, {Parkin}, {Wu},
  {De Looze}, {Wilson}, {Viaene}, {Bendo}, {Boselli}, {Cormier}, {Ibar},
  {Karczewski}, {Lu}, \& {Spinoglio}}]{hughes16}
{Hughes}, T.~M., {Baes}, M., {Schirm}, M.~R.~P., {et~al.} 2016, \aap, 587, A45,
  \dodoi{10.1051/0004-6361/201527644}

\bibitem[{{Hygate} {et~al.}(2023){Hygate}, {Hodge}, {da Cunha}, {Rybak},
  {Schouws}, {Inami}, {Stefanon}, {Graziani}, {Schneider}, {Dayal}, {Bouwens},
  {Smit}, {Bowler}, {Endsley}, {Gonzalez}, {Oesch}, {Stark}, {Algera},
  {Aravena}, {Barrufet}, {Ferrara}, {Fudamoto}, {Hilhorst}, {De Looze},
  {Nanayakkara}, {Pallottini}, {Riechers}, {Sommovigo}, {Topping}, \& {van der
  Werf}}]{HA23}
{Hygate}, A.~P.~S., {Hodge}, J.~A., {da Cunha}, E., {et~al.} 2023, \mnras, 524,
  1775, \dodoi{10.1093/mnras/stad1212}

\bibitem[{{Iani} {et~al.}(2021){Iani}, {Zanella}, {Vernet}, {Richard},
  {Gronke}, {Harrison}, {Arrigoni-Battaia}, {Rodighiero}, {Burkert},
  {Behrendt}, {Chen}, {Emsellem}, {Fensch}, {Hibon}, {Hilker}, {Le Floc'h},
  {Mainieri}, {Swinbank}, {Valentino}, {Vanzella}, \& {Zwaan}}]{IE21}
{Iani}, E., {Zanella}, A., {Vernet}, J., {et~al.} 2021, \mnras, 507, 3830,
  \dodoi{10.1093/mnras/stab2376}

\bibitem[{{Ikeda} {et~al.}(2025){Ikeda}, {Tadaki}, {Mitsuhashi}, {Aravena}, {De
  Looze}, {F{\"o}rster Schreiber}, {Gonz{\'a}lez-L{\'o}pez}, {Herrera-Camus},
  {Spilker}, {Barcos-Mu{\~n}oz}, {Bowler}, {Calistro Rivera}, {da Cunha},
  {Davies}, {D{\'\i}az-Santos}, {Ferrara}, {Killi}, {Lee}, {Li}, {Lutz},
  {Posses}, {Smit}, {Solimano}, {Telikova}, {{\"U}bler}, {Veilleux}, \&
  {Villanueva}}]{IR25}
{Ikeda}, R., {Tadaki}, K.-i., {Mitsuhashi}, I., {et~al.} 2025, \aap, 693, A237,
  \dodoi{10.1051/0004-6361/202451811}

\bibitem[{{Inami} {et~al.}(2013){Inami}, {Armus}, {Charmandaris}, {Groves},
  {Kewley}, {Petric}, {Stierwalt}, {D{\'\i}az-Santos}, {Surace}, {Rich},
  {Haan}, {Howell}, {Evans}, {Mazzarella}, {Marshall}, {Appleton}, {Lord},
  {Spoon}, {Frayer}, {Matsuhara}, \& {Veilleux}}]{I13}
{Inami}, H., {Armus}, L., {Charmandaris}, V., {et~al.} 2013, \apj, 777, 156,
  \dodoi{10.1088/0004-637X/777/2/156}

\bibitem[{{Inami} {et~al.}(2022{\natexlab{a}}){Inami}, {Algera}, {Schouws},
  {Sommovigo}, {Bouwens}, {Smit}, {Stefanon}, {Bowler}, {Endsley}, {Ferrara},
  {Oesch}, {Stark}, {Aravena}, {Barrufet}, {da Cunha}, {Dayal}, {De Looze},
  {Fudamoto}, {Gonzalez}, {Graziani}, {Hodge}, {Hygate}, {Nanayakkara},
  {Pallottini}, {Riechers}, {Schneider}, {Topping}, \& {van der Werf}}]{IH22}
{Inami}, H., {Algera}, H. S.~B., {Schouws}, S., {et~al.} 2022{\natexlab{a}},
  \mnras, 515, 3126, \dodoi{10.1093/mnras/stac1779}

\bibitem[{{Inami} {et~al.}(2022{\natexlab{b}}){Inami}, {Surace}, {Armus},
  {Evans}, {Larson}, {Barcos-Munoz}, {Stierwalt}, {Mazzarella}, {Privon},
  {Song}, {Linden}, {Hayward}, {B{\"o}ker}, {U}, {Bohn}, {Charmandaris},
  {Diaz-Santos}, {Howell}, {Lai}, {Medling}, {Rich}, {Aalto}, {Appleton},
  {Brown}, {Hoshioka}, {Iwasawa}, {Kemper}, {Law}, {Malkan}, {Marshall},
  {Murphy}, {Sanders}, \& {van der Werf}}]{inami22}
{Inami}, H., {Surace}, J., {Armus}, L., {et~al.} 2022{\natexlab{b}}, \apjl,
  940, L6, \dodoi{10.3847/2041-8213/ac9389}

\bibitem[{{Inoue} {et~al.}(2020){Inoue}, {Hashimoto}, {Chihara}, \&
  {Koike}}]{IA20}
{Inoue}, A.~K., {Hashimoto}, T., {Chihara}, H., \& {Koike}, C. 2020, \mnras,
  495, 1577, \dodoi{10.1093/mnras/staa1203}

\bibitem[{{Inoue} {et~al.}(2016){Inoue}, {Tamura}, {Matsuo}, {Mawatari},
  {Shimizu}, {Shibuya}, {Ota}, {Yoshida}, {Zackrisson}, {Kashikawa}, {Kohno},
  {Umehata}, {Hatsukade}, {Iye}, {Matsuda}, {Okamoto}, \& {Yamaguchi}}]{IA16}
{Inoue}, A.~K., {Tamura}, Y., {Matsuo}, H., {et~al.} 2016, Science, 352, 1559,
  \dodoi{10.1126/science.aaf0714}

\bibitem[{{Iono} {et~al.}(2006){Iono}, {Yun}, {Elvis}, {Peck}, {Ho}, {Wilner},
  {Hunter}, {Matsushita}, \& {Muller}}]{ID06}
{Iono}, D., {Yun}, M.~S., {Elvis}, M., {et~al.} 2006, \apjl, 645, L97,
  \dodoi{10.1086/506344}

\bibitem[{{Iono} {et~al.}(2016){Iono}, {Yun}, {Aretxaga}, {Hatsukade},
  {Hughes}, {Ikarashi}, {Izumi}, {Kawabe}, {Kohno}, {Lee}, {Matsuda},
  {Nakanishi}, {Saito}, {Tamura}, {Ueda}, {Umehata}, {Wilson}, {Michiyama}, \&
  {Ando}}]{ID16}
{Iono}, D., {Yun}, M.~S., {Aretxaga}, I., {et~al.} 2016, \apjl, 829, L10,
  \dodoi{10.3847/2041-8205/829/1/L10}

\bibitem[{{Isaak} {et~al.}(2002){Isaak}, {Priddey}, {McMahon}, {Omont},
  {Peroux}, {Sharp}, \& {Withington}}]{IK02}
{Isaak}, K.~G., {Priddey}, R.~S., {McMahon}, R.~G., {et~al.} 2002, \mnras, 329,
  149, \dodoi{10.1046/j.1365-8711.2002.04966.x}

\bibitem[{{Ishii} {et~al.}(2025){Ishii}, {Hashimoto}, {Ferkinhoff}, {Rybak},
  {Inoue}, {Michiyama}, {Donevski}, {Fujimoto}, {Salak}, {Kuno}, {Matsuo},
  {Mawatari}, {Tamura}, {Izumi}, {Nagao}, {Nakazato}, {Osone}, {Sugahara},
  {Usui}, {Wakasugi}, {Yajima}, {Bakx}, {Fudamoto}, {Meyer}, {Walter}, \&
  {Yoshida}}]{IN25}
{Ishii}, N., {Hashimoto}, T., {Ferkinhoff}, C., {et~al.} 2025, \pasj, 77, 139,
  \dodoi{10.1093/pasj/psae105}

\bibitem[{{Israel} {et~al.}(2015){Israel}, {Rosenberg}, \& {van der
  Werf}}]{I15}
{Israel}, F.~P., {Rosenberg}, M.~J.~F., \& {van der Werf}, P. 2015, \aap, 578,
  A95, \dodoi{10.1051/0004-6361/201425175}

\bibitem[{{Ivison} {et~al.}(2010{\natexlab{a}}){Ivison}, {Smail},
  {Papadopoulos}, {Wold}, {Richard}, {Swinbank}, {Kneib}, \& {Owen}}]{IR10b}
{Ivison}, R.~J., {Smail}, I., {Papadopoulos}, P.~P., {et~al.}
  2010{\natexlab{a}}, \mnras, 404, 198,
  \dodoi{10.1111/j.1365-2966.2010.16322.x}

\bibitem[{{Ivison} {et~al.}(2010{\natexlab{b}}){Ivison}, {Swinbank},
  {Swinyard}, {Smail}, {Pearson}, {Rigopoulou}, {Polehampton}, {Baluteau},
  {Barlow}, {Blain}, {Bock}, {Clements}, {Coppin}, {Cooray}, {Danielson},
  {Dwek}, {Edge}, {Franceschini}, {Fulton}, {Glenn}, {Griffin}, {Isaak},
  {Leeks}, {Lim}, {Naylor}, {Oliver}, {Page}, {P{\'e}rez Fournon},
  {Rowan-Robinson}, {Savini}, {Scott}, {Spencer}, {Valtchanov}, {Vigroux}, \&
  {Wright}}]{IR10a}
{Ivison}, R.~J., {Swinbank}, A.~M., {Swinyard}, B., {et~al.}
  2010{\natexlab{b}}, \aap, 518, L35, \dodoi{10.1051/0004-6361/201014548}

\bibitem[{{Ivison} {et~al.}(2013){Ivison}, {Swinbank}, {Smail}, {Harris},
  {Bussmann}, {Cooray}, {Cox}, {Fu}, {Kov{\'a}cs}, {Krips}, {Narayanan},
  {Negrello}, {Neri}, {Pe{\~n}arrubia}, {Richard}, {Riechers}, {Rowlands},
  {Staguhn}, {Targett}, {Amber}, {Baker}, {Bourne}, {Bertoldi}, {Bremer},
  {Calanog}, {Clements}, {Dannerbauer}, {Dariush}, {De Zotti}, {Dunne},
  {Eales}, {Farrah}, {Fleuren}, {Franceschini}, {Geach}, {George}, {Helly},
  {Hopwood}, {Ibar}, {Jarvis}, {Kneib}, {Maddox}, {Omont}, {Scott}, {Serjeant},
  {Smith}, {Thompson}, {Valiante}, {Valtchanov}, {Vieira}, \& {van der
  Werf}}]{IR13}
{Ivison}, R.~J., {Swinbank}, A.~M., {Smail}, I., {et~al.} 2013, \apj, 772, 137,
  \dodoi{10.1088/0004-637X/772/2/137}

\bibitem[{{Ivison} {et~al.}(2016){Ivison}, {Lewis}, {Weiss}, {Arumugam},
  {Simpson}, {Holland}, {Maddox}, {Dunne}, {Valiante}, {van der Werf}, {Omont},
  {Dannerbauer}, {Smail}, {Bertoldi}, {Bremer}, {Bussmann}, {Cai}, {Clements},
  {Cooray}, {De Zotti}, {Eales}, {Fuller}, {Gonzalez-Nuevo}, {Ibar},
  {Negrello}, {Oteo}, {P{\'e}rez-Fournon}, {Riechers}, {Stevens}, {Swinbank},
  \& {Wardlow}}]{IR16}
{Ivison}, R.~J., {Lewis}, A.~J.~R., {Weiss}, A., {et~al.} 2016, \apj, 832, 78,
  \dodoi{10.3847/0004-637X/832/1/78}

\bibitem[{{Izumi} {et~al.}(2018){Izumi}, {Onoue}, {Shirakata}, {Nagao},
  {Kohno}, {Matsuoka}, {Imanishi}, {Strauss}, {Kashikawa}, {Schulze},
  {Silverman}, {Fujimoto}, {Harikane}, {Toba}, {Umehata}, {Nakanishi},
  {Greene}, {Tamura}, {Taniguchi}, {Yamaguchi}, {Goto}, {Hashimoto},
  {Ikarashi}, {Iono}, {Iwasawa}, {Lee}, {Makiya}, {Minezaki}, \& {Tang}}]{IT18}
{Izumi}, T., {Onoue}, M., {Shirakata}, H., {et~al.} 2018, \pasj, 70, 36,
  \dodoi{10.1093/pasj/psy026}

\bibitem[{{Izumi} {et~al.}(2019){Izumi}, {Onoue}, {Matsuoka}, {Nagao},
  {Strauss}, {Imanishi}, {Kashikawa}, {Fujimoto}, {Kohno}, {Toba}, {Umehata},
  {Goto}, {Ueda}, {Shirakata}, {Silverman}, {Greene}, {Harikane}, {Hashimoto},
  {Ikarashi}, {Iono}, {Iwasawa}, {Lee}, {Minezaki}, {Nakanishi}, {Tamura},
  {Tang}, \& {Taniguchi}}]{IT19}
{Izumi}, T., {Onoue}, M., {Matsuoka}, Y., {et~al.} 2019, \pasj, 71, 111,
  \dodoi{10.1093/pasj/psz096}

\bibitem[{{Izumi} {et~al.}(2021{\natexlab{a}}){Izumi}, {Onoue}, {Matsuoka},
  {Strauss}, {Fujimoto}, {Umehata}, {Imanishi}, {Kawamuro}, {Nagao}, {Toba},
  {Kohno}, {Kashikawa}, {Inayoshi}, {Kawaguchi}, {Iwasawa}, {Inoue}, {Goto},
  {Baba}, {Schramm}, {Suh}, {Harikane}, {Ueda}, {Silverman}, {Hashimoto},
  {Hashimoto}, {Ikarashi}, {Iono}, {Lee}, {Lee}, {Minezaki}, {Nakanishi},
  {Nakano}, {Tamura}, \& {Tang}}]{IT21a}
---. 2021{\natexlab{a}}, \apj, 908, 235, \dodoi{10.3847/1538-4357/abd7ef}

\bibitem[{{Izumi} {et~al.}(2021{\natexlab{b}}){Izumi}, {Matsuoka}, {Fujimoto},
  {Onoue}, {Strauss}, {Umehata}, {Imanishi}, {Kohno}, {Kawaguchi}, {Kawamuro},
  {Baba}, {Nagao}, {Toba}, {Inayoshi}, {Silverman}, {Inoue}, {Ikarashi},
  {Iwasawa}, {Kashikawa}, {Hashimoto}, {Nakanishi}, {Ueda}, {Schramm}, {Lee},
  \& {Suh}}]{IT21b}
{Izumi}, T., {Matsuoka}, Y., {Fujimoto}, S., {et~al.} 2021{\natexlab{b}}, \apj,
  914, 36, \dodoi{10.3847/1538-4357/abf6dc}

\bibitem[{{Jakobsen} {et~al.}(2022){Jakobsen}, {Ferruit}, {Alves de Oliveira},
  {Arribas}, {Bagnasco}, {Barho}, {Beck}, {Birkmann}, {B{\"o}ker}, {Bunker},
  {Charlot}, {de Jong}, {de Marchi}, {Ehrenwinkler}, {Falcolini}, {Fels},
  {Franx}, {Franz}, {Funke}, {Giardino}, {Gnata}, {Holota}, {Honnen}, {Jensen},
  {Jentsch}, {Johnson}, {Jollet}, {Karl}, {Kling}, {K{\"o}hler}, {Kolm},
  {Kumari}, {Lander}, {Lemke}, {L{\'o}pez-Caniego}, {L{\"u}tzgendorf},
  {Maiolino}, {Manjavacas}, {Marston}, {Maschmann}, {Maurer}, {Messerschmidt},
  {Moseley}, {Mosner}, {Mott}, {Muzerolle}, {Pirzkal}, {Pittet}, {Plitzke},
  {Posselt}, {Rapp}, {Rauscher}, {Rawle}, {Rix}, {R{\"o}del}, {Rumler},
  {Sabbi}, {Salvignol}, {Schmid}, {Sirianni}, {Smith}, {Strada}, {te Plate},
  {Valenti}, {Wettemann}, {Wiehe}, {Wiesmayer}, {Willott}, {Wright}, {Zeidler},
  \& {Zincke}}]{jakobsen22}
{Jakobsen}, P., {Ferruit}, P., {Alves de Oliveira}, C., {et~al.} 2022, \aap,
  661, A80, \dodoi{10.1051/0004-6361/202142663}

\bibitem[{{Ji} {et~al.}(2024){Ji}, {{\"U}bler}, {Maiolino}, {D'Eugenio},
  {Arribas}, {Bunker}, {Charlot}, {Perna}, {Rodr{\'\i}guez Del Pino},
  {B{\"o}ker}, {Cresci}, {Curti}, {Kumari}, \& {Lamperti}}]{JX24}
{Ji}, X., {{\"U}bler}, H., {Maiolino}, R., {et~al.} 2024, \mnras, 535, 881,
  \dodoi{10.1093/mnras/stae2375}

\bibitem[{{Jim{\'e}nez-Andrade} {et~al.}(2020){Jim{\'e}nez-Andrade}, {Zavala},
  {Magnelli}, {Casey}, {Liu}, {Romano-D{\'\i}az}, {Schinnerer}, {Harrington},
  {Aretxaga}, {Karim}, {Staguhn}, {Burnham}, {Monta{\~n}a},
  {Smol{\v{c}}i{\'c}}, {Yun}, {Bertoldi}, \& {Hughes}}]{JE20}
{Jim{\'e}nez-Andrade}, E.~F., {Zavala}, J.~A., {Magnelli}, B., {et~al.} 2020,
  \apj, 890, 171, \dodoi{10.3847/1538-4357/ab6dec}

\bibitem[{{Jones} {et~al.}(2017{\natexlab{a}}){Jones}, {K{\"o}hler}, {Ysard},
  {Bocchio}, \& {Verstraete}}]{jones17}
{Jones}, A.~P., {K{\"o}hler}, M., {Ysard}, N., {Bocchio}, M., \& {Verstraete},
  L. 2017{\natexlab{a}}, \aap, 602, A46, \dodoi{10.1051/0004-6361/201630225}

\bibitem[{{Jones} {et~al.}(2017{\natexlab{b}}){Jones}, {Willott}, {Carilli},
  {Ferrara}, {Wang}, \& {Wagg}}]{JG17}
{Jones}, G.~C., {Willott}, C.~J., {Carilli}, C.~L., {et~al.}
  2017{\natexlab{b}}, \apj, 845, 175, \dodoi{10.3847/1538-4357/aa7d0d}

\bibitem[{{Jones} {et~al.}(2024{\natexlab{a}}){Jones}, {{\"U}bler}, {Perna},
  {Arribas}, {Bunker}, {Carniani}, {Charlot}, {Maiolino}, {Del Pino},
  {Willott}, {Bowler}, {B{\"o}ker}, {Cameron}, {Chevallard}, {Cresci}, {Curti},
  {D'Eugenio}, {Kumari}, {Saxena}, {Scholtz}, {Venturi}, \& {Witstok}}]{JG24a}
{Jones}, G.~C., {{\"U}bler}, H., {Perna}, M., {et~al.} 2024{\natexlab{a}},
  \aap, 682, A122, \dodoi{10.1051/0004-6361/202347838}

\bibitem[{{Jones} {et~al.}(2024{\natexlab{b}}){Jones}, {Bowler}, {Bunker},
  {Arribas}, {Carniani}, {Charlot}, {Perna}, {Rodr{\'\i}guez Del Pino},
  {{\"U}bler}, {Willott}, {Chevallard}, {Cresci}, {Parlanti}, {Scholtz}, \&
  {Venturi}}]{JG24b}
{Jones}, G.~C., {Bowler}, R., {Bunker}, A.~J., {et~al.} 2024{\natexlab{b}},
  arXiv e-prints, arXiv:2412.15027, \dodoi{10.48550/arXiv.2412.15027}

\bibitem[{{Jones} {et~al.}(2025){Jones}, {Bunker}, {Telikova}, {Arribas},
  {Carniani}, {Charlot}, {D'Eugenio}, {Maiolino}, {Perna}, {Rodr{\'\i}guez Del
  Pino}, {{\"U}bler}, {Willott}, {Aravena}, {B{\"o}ker}, {Cresci}, {Curti},
  {Gonz{\'a}lez-L{\'o}pez}, {Herrera-Camus}, {Lamperti}, {Parlanti},
  {P{\'e}rez-Gonz{\'a}lez}, \& {Villanueva}}]{JG25}
{Jones}, G.~C., {Bunker}, A.~J., {Telikova}, K., {et~al.} 2025, \mnras, 540,
  3311, \dodoi{10.1093/mnras/staf899}

\bibitem[{{Jones} {et~al.}(2010{\natexlab{a}}){Jones}, {Ellis}, {Jullo}, \&
  {Richard}}]{JT10b}
{Jones}, T., {Ellis}, R., {Jullo}, E., \& {Richard}, J. 2010{\natexlab{a}},
  \apjl, 725, L176, \dodoi{10.1088/2041-8205/725/2/L176}

\bibitem[{{Jones} {et~al.}(2010{\natexlab{b}}){Jones}, {Swinbank}, {Ellis},
  {Richard}, \& {Stark}}]{JT10a}
{Jones}, T.~A., {Swinbank}, A.~M., {Ellis}, R.~S., {Richard}, J., \& {Stark},
  D.~P. 2010{\natexlab{b}}, \mnras, 404, 1247,
  \dodoi{10.1111/j.1365-2966.2010.16378.x}

\bibitem[{{Kade} {et~al.}(2023){Kade}, {Knudsen}, {Vlemmings}, {Stanley},
  {Gullberg}, \& {K{\"o}nig}}]{KK23}
{Kade}, K., {Knudsen}, K.~K., {Vlemmings}, W., {et~al.} 2023, \aap, 673, A116,
  \dodoi{10.1051/0004-6361/202141839}

\bibitem[{{Kamenetzky} {et~al.}(2016){Kamenetzky}, {Rangwala}, {Glenn},
  {Maloney}, \& {Conley}}]{K16}
{Kamenetzky}, J., {Rangwala}, N., {Glenn}, J., {Maloney}, P.~R., \& {Conley},
  A. 2016, \apj, 829, 93, \dodoi{10.3847/0004-637X/829/2/93}

\bibitem[{{Kanekar} {et~al.}(2013){Kanekar}, {Wagg}, {Chary}, \&
  {Carilli}}]{KN13}
{Kanekar}, N., {Wagg}, J., {Chary}, R.~R., \& {Carilli}, C.~L. 2013, \apjl,
  771, L20, \dodoi{10.1088/2041-8205/771/2/L20}

\bibitem[{{Kashikawa} {et~al.}(2015){Kashikawa}, {Ishizaki}, {Willott},
  {Onoue}, {Im}, {Furusawa}, {Toshikawa}, {Ishikawa}, {Niino}, {Shimasaku},
  {Ouchi}, \& {Hibon}}]{KN15}
{Kashikawa}, N., {Ishizaki}, Y., {Willott}, C.~J., {et~al.} 2015, \apj, 798,
  28, \dodoi{10.1088/0004-637X/798/1/28}

\bibitem[{{Kashino} {et~al.}(2023){Kashino}, {Lilly}, {Simcoe}, {Bordoloi},
  {Mackenzie}, {Matthee}, \& {Eilers}}]{KD23}
{Kashino}, D., {Lilly}, S.~J., {Simcoe}, R.~A., {et~al.} 2023, \nat, 617, 261,
  \dodoi{10.1038/s41586-023-05901-3}

\bibitem[{{Kato} {et~al.}(2020){Kato}, {Matsuoka}, {Onoue}, {Koyama}, {Toba},
  {Akiyama}, {Fujimoto}, {Imanishi}, {Iwasawa}, {Izumi}, {Kashikawa},
  {Kawaguchi}, {Lee}, {Minezaki}, {Nagao}, {Noboriguchi}, \& {Strauss}}]{KN20}
{Kato}, N., {Matsuoka}, Y., {Onoue}, M., {et~al.} 2020, \pasj, 72, 84,
  \dodoi{10.1093/pasj/psaa074}

\bibitem[{{Kaufman} {et~al.}(1999){Kaufman}, {Wolfire}, {Hollenbach}, \&
  {Luhman}}]{kaufman99}
{Kaufman}, M.~J., {Wolfire}, M.~G., {Hollenbach}, D.~J., \& {Luhman}, M.~L.
  1999, \apj, 527, 795, \dodoi{10.1086/308102}

\bibitem[{{Kennicutt} {et~al.}(2009){Kennicutt}, {Hao}, {Calzetti},
  {Moustakas}, {Dale}, {Bendo}, {Engelbracht}, {Johnson}, \&
  {Lee}}]{kennicutt09}
{Kennicutt}, Robert~C., J., {Hao}, C.-N., {Calzetti}, D., {et~al.} 2009, \apj,
  703, 1672, \dodoi{10.1088/0004-637X/703/2/1672}

\bibitem[{{Kennicutt} {et~al.}(2011){Kennicutt}, {Calzetti}, {Aniano},
  {Appleton}, {Armus}, {Beir{\~a}o}, {Bolatto}, {Brandl}, {Crocker}, {Croxall},
  {Dale}, {Donovan Meyer}, {Draine}, {Engelbracht}, {Galametz}, {Gordon},
  {Groves}, {Hao}, {Helou}, {Hinz}, {Hunt}, {Johnson}, {Koda}, {Krause},
  {Leroy}, {Li}, {Meidt}, {Montiel}, {Murphy}, {Rahman}, {Rix}, {Roussel},
  {Sandstrom}, {Sauvage}, {Schinnerer}, {Skibba}, {Smith}, {Srinivasan},
  {Vigroux}, {Walter}, {Wilson}, {Wolfire}, \& {Zibetti}}]{K11}
{Kennicutt}, R.~C., {Calzetti}, D., {Aniano}, G., {et~al.} 2011, \pasp, 123,
  1347, \dodoi{10.1086/663818}

\bibitem[{{Kessler} {et~al.}(1996){Kessler}, {Steinz}, {Anderegg}, {Clavel},
  {Drechsel}, {Estaria}, {Faelker}, {Riedinger}, {Robson}, {Taylor}, \&
  {Xim{\'e}nez de Ferr{\'a}n}}]{kessler96}
{Kessler}, M.~F., {Steinz}, J.~A., {Anderegg}, M.~E., {et~al.} 1996, \aap, 315,
  L27

\bibitem[{{Kewley} \& {Dopita}(2002)}]{kewley02}
{Kewley}, L.~J., \& {Dopita}, M.~A. 2002, \apjs, 142, 35,
  \dodoi{10.1086/341326}

\bibitem[{{Kewley} {et~al.}(2006){Kewley}, {Groves}, {Kauffmann}, \&
  {Heckman}}]{kewley06}
{Kewley}, L.~J., {Groves}, B., {Kauffmann}, G., \& {Heckman}, T. 2006, \mnras,
  372, 961, \dodoi{10.1111/j.1365-2966.2006.10859.x}

\bibitem[{{Kewley} {et~al.}(2019){Kewley}, {Nicholls}, \&
  {Sutherland}}]{kewley19}
{Kewley}, L.~J., {Nicholls}, D.~C., \& {Sutherland}, R.~S. 2019, \araa, 57,
  511, \dodoi{10.1146/annurev-astro-081817-051832}

\bibitem[{{Khusanova} {et~al.}(2022){Khusanova}, {Ba{\~n}ados}, {Mazzucchelli},
  {Rojas-Ruiz}, {Momjian}, {Walter}, {Decarli}, {Venemans}, {Farina}, {Meyer},
  {Wang}, \& {Yang}}]{KY22}
{Khusanova}, Y., {Ba{\~n}ados}, E., {Mazzucchelli}, C., {et~al.} 2022, \aap,
  664, A39, \dodoi{10.1051/0004-6361/202243660}

\bibitem[{{Killi} {et~al.}(2023){Killi}, {Watson}, {Fujimoto}, {Akins},
  {Knudsen}, {Richard}, {Harikane}, {Rigopoulou}, {Rizzo}, {Ginolfi},
  {Popping}, \& {Kokorev}}]{KM23}
{Killi}, M., {Watson}, D., {Fujimoto}, S., {et~al.} 2023, \mnras, 521, 2526,
  \dodoi{10.1093/mnras/stad687}

\bibitem[{{Kimball} {et~al.}(2015){Kimball}, {Lacy}, {Lonsdale}, \&
  {Macquart}}]{KA15}
{Kimball}, A.~E., {Lacy}, M., {Lonsdale}, C.~J., \& {Macquart}, J.~P. 2015,
  \mnras, 452, 88, \dodoi{10.1093/mnras/stv1160}

\bibitem[{{Kiyota} {et~al.}(2025){Kiyota}, {Ouchi}, {Xu}, {Nakazato}, {Soga},
  {Yajima}, {Fujimoto}, {Harikane}, {Nakajima}, {Ono}, {Sun}, {Kusakabe},
  {Ceverino}, {Hatsukade}, {Iono}, {Kohno}, \& {Nakanishi}}]{KT25}
{Kiyota}, T., {Ouchi}, M., {Xu}, Y., {et~al.} 2025, arXiv e-prints,
  arXiv:2504.03156, \dodoi{10.48550/arXiv.2504.03156}

\bibitem[{{Kneib} {et~al.}(1993){Kneib}, {Mellier}, {Fort}, \& {Mathez}}]{KJ93}
{Kneib}, J.~P., {Mellier}, Y., {Fort}, B., \& {Mathez}, G. 1993, \aap, 273, 367

\bibitem[{{Knudsen} {et~al.}(2016){Knudsen}, {Richard}, {Kneib}, {Jauzac},
  {Cl{\'e}ment}, {Drouart}, {Egami}, \& {Lindroos}}]{KK16}
{Knudsen}, K.~K., {Richard}, J., {Kneib}, J.-P., {et~al.} 2016, \mnras, 462,
  L6, \dodoi{10.1093/mnrasl/slw114}

\bibitem[{{Knudsen} {et~al.}(2017){Knudsen}, {Watson}, {Frayer}, {Christensen},
  {Gallazzi}, {Micha{\l}owski}, {Richard}, \& {Zavala}}]{KK17}
{Knudsen}, K.~K., {Watson}, D., {Frayer}, D., {et~al.} 2017, \mnras, 466, 138,
  \dodoi{10.1093/mnras/stw3066}

\bibitem[{{Koester} {et~al.}(2010){Koester}, {Gladders}, {Hennawi}, {Sharon},
  {Wuyts}, {Rigby}, {Bayliss}, \& {Dahle}}]{KB10}
{Koester}, B.~P., {Gladders}, M.~D., {Hennawi}, J.~F., {et~al.} 2010, \apjl,
  723, L73, \dodoi{10.1088/2041-8205/723/1/L73}

\bibitem[{{Kolupuri} {et~al.}(2025){Kolupuri}, {Decarli}, {Neri}, {Cox},
  {Ferkinhoff}, {Bertoldi}, {Weiss}, {Venemans}, {Riechers}, {Paolo Farina}, \&
  {Walter}}]{KS25}
{Kolupuri}, S., {Decarli}, R., {Neri}, R., {et~al.} 2025, \aap, 695, A201,
  \dodoi{10.1051/0004-6361/202452374}

\bibitem[{{K{\"o}nig} {et~al.}(2012){K{\"o}nig}, {Greve}, {Seymour},
  {Rawlings}, {Papadopoulos}, {Ivison}, {De Breuck}, {Stevens}, {Smail}, \&
  {Kovacs}}]{KS12}
{K{\"o}nig}, S., {Greve}, T.~R., {Seymour}, N., {et~al.} 2012, in Journal of
  Physics Conference Series, Vol. 372, Journal of Physics Conference Series
  (IOP), 012064, \dodoi{10.1088/1742-6596/372/1/012064}

\bibitem[{{Koribalski} {et~al.}(2004){Koribalski}, {Staveley-Smith}, {Kilborn},
  {Ryder}, {Kraan-Korteweg}, {Ryan-Weber}, {Ekers}, {Jerjen}, {Henning},
  {Putman}, {Zwaan}, {de Blok}, {Calabretta}, {Disney}, {Minchin}, {Bhathal},
  {Boyce}, {Drinkwater}, {Freeman}, {Gibson}, {Green}, {Haynes}, {Juraszek},
  {Kesteven}, {Knezek}, {Mader}, {Marquarding}, {Meyer}, {Mould}, {Oosterloo},
  {O'Brien}, {Price}, {Sadler}, {Schr{\"o}der}, {Stewart}, {Stootman}, {Waugh},
  {Warren}, {Webster}, \& {Wright}}]{K04}
{Koribalski}, B.~S., {Staveley-Smith}, L., {Kilborn}, V.~A., {et~al.} 2004,
  \aj, 128, 16, \dodoi{10.1086/421744}

\bibitem[{{Kubo} {et~al.}(2015){Kubo}, {Yamada}, {Ichikawa}, {Kajisawa},
  {Matsuda}, \& {Tanaka}}]{KM15}
{Kubo}, M., {Yamada}, T., {Ichikawa}, T., {et~al.} 2015, \apj, 799, 38,
  \dodoi{10.1088/0004-637X/799/1/38}

\bibitem[{{Lamarche} {et~al.}(2019){Lamarche}, {Stacey}, {Vishwas}, {Brisbin},
  {Ferkinhoff}, {Nikola}, {Higdon}, \& {Higdon}}]{LC19}
{Lamarche}, C., {Stacey}, G.~J., {Vishwas}, A., {et~al.} 2019, \apj, 882, 1,
  \dodoi{10.3847/1538-4357/ab3389}

\bibitem[{{Lamarche} {et~al.}(2018){Lamarche}, {Verma}, {Vishwas}, {Stacey},
  {Brisbin}, {Ferkinhoff}, {Nikola}, {Higdon}, {Higdon}, \& {Tecza}}]{LC18}
{Lamarche}, C., {Verma}, A., {Vishwas}, A., {et~al.} 2018, \apj, 867, 140,
  \dodoi{10.3847/1538-4357/aae394}

\bibitem[{{Lamarche} {et~al.}(2022){Lamarche}, {Smith}, {Kreckel}, {Linden},
  {Rogers}, {Skillman}, {Berg}, {Murphy}, {Pogge}, {Donnelly}, {Kennicutt},
  {Bolatto}, {Croxall}, {Groves}, \& {Ferkinhoff}}]{lamarche22}
{Lamarche}, C., {Smith}, J.~D., {Kreckel}, K., {et~al.} 2022, \apj, 925, 194,
  \dodoi{10.3847/1538-4357/ac3b4f}

\bibitem[{{Lambert} {et~al.}(2023){Lambert}, {Posses}, {Aravena},
  {G{\'o}nzalez-L{\'o}pez}, {Assef}, {D{\'\i}az-Santos}, {Brisbin}, {Decarli},
  {Herrera-Camus}, {Mej{\'\i}a}, \& {Ricci}}]{LT23}
{Lambert}, T.~S., {Posses}, A., {Aravena}, M., {et~al.} 2023, \mnras, 518,
  3183, \dodoi{10.1093/mnras/stac3016}

\bibitem[{{Lapham} {et~al.}(2017){Lapham}, {Young}, \& {Crocker}}]{L17b}
{Lapham}, R.~C., {Young}, L.~M., \& {Crocker}, A. 2017, \apj, 840, 51,
  \dodoi{10.3847/1538-4357/aa6d83}

\bibitem[{{Laporte} {et~al.}(2021{\natexlab{a}}){Laporte}, {Meyer}, {Ellis},
  {Robertson}, {Chisholm}, \& {Roberts-Borsani}}]{LN21b}
{Laporte}, N., {Meyer}, R.~A., {Ellis}, R.~S., {et~al.} 2021{\natexlab{a}},
  \mnras, 505, 3336, \dodoi{10.1093/mnras/stab1239}

\bibitem[{{Laporte} {et~al.}(2017){Laporte}, {Ellis}, {Boone}, {Bauer},
  {Qu{\'e}nard}, {Roberts-Borsani}, {Pell{\'o}}, {P{\'e}rez-Fournon}, \&
  {Streblyanska}}]{LN17b}
{Laporte}, N., {Ellis}, R.~S., {Boone}, F., {et~al.} 2017, \apjl, 837, L21,
  \dodoi{10.3847/2041-8213/aa62aa}

\bibitem[{{Laporte} {et~al.}(2019){Laporte}, {Katz}, {Ellis}, {Lagache},
  {Bauer}, {Boone}, {Inoue}, {Hashimoto}, {Matsuo}, {Mawatari}, \&
  {Tamura}}]{LN19}
{Laporte}, N., {Katz}, H., {Ellis}, R.~S., {et~al.} 2019, \mnras, 487, L81,
  \dodoi{10.1093/mnrasl/slz094}

\bibitem[{{Laporte} {et~al.}(2021{\natexlab{b}}){Laporte}, {Zitrin}, {Ellis},
  {Fujimoto}, {Brammer}, {Richard}, {Oguri}, {Caminha}, {Kohno}, {Yoshimura},
  {Ao}, {Bauer}, {Caputi}, {Egami}, {Espada}, {Gonz{\'a}lez-L{\'o}pez},
  {Hatsukade}, {Knudsen}, {Lee}, {Magdis}, {Ouchi}, {Valentino}, \&
  {Wang}}]{LN21a}
{Laporte}, N., {Zitrin}, A., {Ellis}, R.~S., {et~al.} 2021{\natexlab{b}},
  \mnras, 505, 4838, \dodoi{10.1093/mnras/stab191}

\bibitem[{{Le F{\`e}vre} {et~al.}(2020){Le F{\`e}vre}, {B{\'e}thermin},
  {Faisst}, {Jones}, {Capak}, {Cassata}, {Silverman}, {Schaerer}, {Yan},
  {Amorin}, {Bardelli}, {Boquien}, {Cimatti}, {Dessauges-Zavadsky},
  {Giavalisco}, {Hathi}, {Fudamoto}, {Fujimoto}, {Ginolfi}, {Gruppioni},
  {Hemmati}, {Ibar}, {Koekemoer}, {Khusanova}, {Lagache}, {Lemaux}, {Loiacono},
  {Maiolino}, {Mancini}, {Narayanan}, {Morselli}, {M{\'e}ndez-Hern{\`a}ndez},
  {Oesch}, {Pozzi}, {Romano}, {Riechers}, {Scoville}, {Talia}, {Tasca},
  {Thomas}, {Toft}, {Vallini}, {Vergani}, {Walter}, {Zamorani}, \&
  {Zucca}}]{LO20}
{Le F{\`e}vre}, O., {B{\'e}thermin}, M., {Faisst}, A., {et~al.} 2020, \aap,
  643, A1, \dodoi{10.1051/0004-6361/201936965}

\bibitem[{{Lee} {et~al.}(2024){Lee}, {Akiyama}, {Kohno}, {Iono}, {Imanishi},
  {Hatsukade}, {Umehata}, {Nagao}, {Toba}, {Chen}, {Egusa}, {Ichikawa},
  {Izumi}, {Matsumoto}, {Schramm}, \& {Matsuoka}}]{LK24}
{Lee}, K., {Akiyama}, M., {Kohno}, K., {et~al.} 2024, \apj, 972, 111,
  \dodoi{10.3847/1538-4357/ad5be5}

\bibitem[{{Lee} {et~al.}(2019){Lee}, {Nagao}, {De Breuck}, {Carniani},
  {Cresci}, {Hatsukade}, {Kawabe}, {Kohno}, {Maiolino}, {Mannucci}, {Marconi},
  {Nakanishi}, {Saito}, {Tamura}, {Troncoso}, {Umehata}, \& {Yun}}]{LM19}
{Lee}, M.~M., {Nagao}, T., {De Breuck}, C., {et~al.} 2019, \apjl, 883, L29,
  \dodoi{10.3847/2041-8213/ab412e}

\bibitem[{{Lee} {et~al.}(2021){Lee}, {Nagao}, {De Breuck}, {Carniani},
  {Cresci}, {Hatsukade}, {Kawabe}, {Kohno}, {Maiolino}, {Mannucci}, {Marconi},
  {Nakanishi}, {Troncoso}, \& {Umehata}}]{LM21}
---. 2021, \apj, 913, 41, \dodoi{10.3847/1538-4357/abe7ea}

\bibitem[{{Lelli} {et~al.}(2021){Lelli}, {Di Teodoro}, {Fraternali}, {Man},
  {Zhang}, {De Breuck}, {Davis}, \& {Maiolino}}]{LF21}
{Lelli}, F., {Di Teodoro}, E.~M., {Fraternali}, F., {et~al.} 2021, Science,
  371, 713, \dodoi{10.1126/science.abc1893}

\bibitem[{{Le{\'s}niewska} \& {Micha{\l}owski}(2019)}]{lesniewska19}
{Le{\'s}niewska}, A., \& {Micha{\l}owski}, M.~J. 2019, \aap, 624, L13,
  \dodoi{10.1051/0004-6361/201935149}

\bibitem[{{Lester} {et~al.}(1983){Lester}, {Dinerstein}, {Werner}, {Watson}, \&
  {Genzel}}]{lester83}
{Lester}, D.~F., {Dinerstein}, H.~L., {Werner}, M.~W., {Watson}, D.~M., \&
  {Genzel}, R.~L. 1983, \apj, 271, 618, \dodoi{10.1086/161229}

\bibitem[{{Lestrade} {et~al.}(2011){Lestrade}, {Carilli}, {Thanjavur}, {Kneib},
  {Riechers}, {Bertoldi}, {Walter}, \& {Omont}}]{LJ11}
{Lestrade}, J.-F., {Carilli}, C.~L., {Thanjavur}, K., {et~al.} 2011, \apjl,
  739, L30, \dodoi{10.1088/2041-8205/739/1/L30}

\bibitem[{{Lestrade} {et~al.}(2010){Lestrade}, {Combes}, {Salom{\'e}}, {Omont},
  {Bertoldi}, {Andr{\'e}}, \& {Schneider}}]{LJ10}
{Lestrade}, J.~F., {Combes}, F., {Salom{\'e}}, P., {et~al.} 2010, \aap, 522,
  L4, \dodoi{10.1051/0004-6361/201015673}

\bibitem[{{Li} {et~al.}(2022){Li}, {Venemans}, {Walter}, {Decarli}, {Wang}, \&
  {Cai}}]{LJ22}
{Li}, J., {Venemans}, B.~P., {Walter}, F., {et~al.} 2022, \apj, 930, 27,
  \dodoi{10.3847/1538-4357/ac61d7}

\bibitem[{{Li} {et~al.}(2020){Li}, {Wang}, {Cox}, {Gao}, {Walter}, {Wagg},
  {Menten}, {Bertoldi}, {Shao}, {Venemans}, {Decarli}, {Riechers}, {Neri},
  {Fan}, {Omont}, \& {Narayanan}}]{LJ20}
{Li}, J., {Wang}, R., {Cox}, P., {et~al.} 2020, \apj, 900, 131,
  \dodoi{10.3847/1538-4357/ababac}

\bibitem[{{Li} {et~al.}(2024){Li}, {Da Cunha}, {Gonz{\'a}lez-L{\'o}pez},
  {Aravena}, {De Looze}, {F{\"o}rster Schreiber}, {Herrera-Camus}, {Spilker},
  {Tadaki}, {Barcos-Munoz}, {Battisti}, {Birkin}, {Bowler}, {Davies},
  {D{\'\i}az-Santos}, {Ferrara}, {Fisher}, {Hodge}, {Ikeda}, {Killi}, {Lee},
  {Liu}, {Lutz}, {Mitsuhashi}, {Naab}, {Posses}, {Rela{\~n}o}, {Solimano},
  {{\"U}bler}, {van der Giessen}, \& {Villanueva}}]{LJ24}
{Li}, J., {Da Cunha}, E., {Gonz{\'a}lez-L{\'o}pez}, J., {et~al.} 2024, \apj,
  976, 70, \dodoi{10.3847/1538-4357/ad7fee}

\bibitem[{{Lin} {et~al.}(2009){Lin}, {Buckley-Geer}, {Allam}, {Tucker},
  {Diehl}, {Kubik}, {Kubo}, {Annis}, {Frieman}, {Oguri}, \& {Inada}}]{LH09}
{Lin}, H., {Buckley-Geer}, E., {Allam}, S.~S., {et~al.} 2009, \apj, 699, 1242,
  \dodoi{10.1088/0004-637X/699/2/1242}

\bibitem[{{Litke} {et~al.}(2019){Litke}, {Marrone}, {Spilker}, {Aravena},
  {B{\'e}thermin}, {Chapman}, {Chen}, {de Breuck}, {Dong}, {Gonzalez}, {Greve},
  {Hayward}, {Hezaveh}, {Jarugula}, {Ma}, {Morningstar}, {Narayanan}, {Phadke},
  {Reuter}, {Vieira}, \& {Weiss}}]{LK19}
{Litke}, K.~C., {Marrone}, D.~P., {Spilker}, J.~S., {et~al.} 2019, \apj, 870,
  80, \dodoi{10.3847/1538-4357/aaf057}

\bibitem[{{Litke} {et~al.}(2022){Litke}, {Marrone}, {Aravena}, {B{\'e}thermin},
  {Chapman}, {Dong}, {Hayward}, {Hill}, {Jarugula}, {Malkan}, {Narayanan},
  {Reuter}, {Spilker}, {Sulzenauer}, {Vieira}, \& {Wei{\ss}}}]{LK22}
{Litke}, K.~C., {Marrone}, D.~P., {Aravena}, M., {et~al.} 2022, \apj, 928, 179,
  \dodoi{10.3847/1538-4357/ac58f9}

\bibitem[{{Litke} {et~al.}(2023){Litke}, {Marrone}, {Aravena}, {Archipley},
  {B{\'e}thermin}, {Burgoyne}, {Cathey}, {Chapman}, {Gonzalez}, {Greve},
  {Gururajan}, {Hayward}, {Malkan}, {Phadke}, {Reuter}, {Rotermund}, {Spilker},
  {Stark}, {Sulzenauer}, {Vieira}, {Vizgan}, \& {Wei{\ss}}}]{LK23}
---. 2023, \apj, 949, 87, \dodoi{10.3847/1538-4357/acc93a}

\bibitem[{{Liu} {et~al.}(2019){Liu}, {Lang}, {Magnelli}, {Schinnerer},
  {Leslie}, {Fudamoto}, {Bondi}, {Groves}, {Jim{\'e}nez-Andrade}, {Harrington},
  {Karim}, {Oesch}, {Sargent}, {Vardoulaki}, {B{\v{a}}descu}, {Moser},
  {Bertoldi}, {Battisti}, {da Cunha}, {Zavala}, {Vaccari}, {Davidzon},
  {Riechers}, \& {Aravena}}]{LD19}
{Liu}, D., {Lang}, P., {Magnelli}, B., {et~al.} 2019, \apjs, 244, 40,
  \dodoi{10.3847/1538-4365/ab42da}

\bibitem[{{Liu} {et~al.}(2024){Liu}, {Fan}, {Yang}, {Ba{\~n}ados}, {Wang},
  {Wolf}, {Barth}, {Costa}, {Decarli}, {Eilers}, {Loiacono}, {Shen}, {Farina},
  {Jin}, {Jun}, {Li}, {Lupi}, {Marshall}, {Pan}, {Pudoka}, {Zhuang},
  {Champagne}, {Li}, {Sun}, {Tee}, {Vayner}, \& {Zhang}}]{LW24}
{Liu}, W., {Fan}, X., {Yang}, J., {et~al.} 2024, \apj, 976, 33,
  \dodoi{10.3847/1538-4357/ad7de4}

\bibitem[{{Livermore} {et~al.}(2015){Livermore}, {Jones}, {Richard}, {Bower},
  {Swinbank}, {Yuan}, {Edge}, {Ellis}, {Kewley}, {Smail}, {Coppin}, \&
  {Ebeling}}]{LR15}
{Livermore}, R.~C., {Jones}, T.~A., {Richard}, J., {et~al.} 2015, \mnras, 450,
  1812, \dodoi{10.1093/mnras/stv686}

\bibitem[{{Loiacono} {et~al.}(2024){Loiacono}, {Decarli}, {Mignoli}, {Farina},
  {Ba{\~n}ados}, {Bosman}, {Eilers}, {Schindler}, {Strauss}, {Vestergaard},
  {Wang}, {Blecha}, {Carilli}, {Comastri}, {Connor}, {Costa}, {Dotti}, {Fan},
  {Gilli}, {Jun}, {Liu}, {Lupi}, {Marshall}, {Mazzucchelli}, {Meyer},
  {Neeleman}, {Overzier}, {Pensabene}, {Riechers}, {Trakhtenbrot}, {Trebitsch},
  {Venemans}, {Walter}, \& {Yang}}]{LF24}
{Loiacono}, F., {Decarli}, R., {Mignoli}, M., {et~al.} 2024, \aap, 685, A121,
  \dodoi{10.1051/0004-6361/202348535}

\bibitem[{{Lu} {et~al.}(2017{\natexlab{a}}){Lu}, {Zhao}, {D{\'\i}az-Santos},
  {Xu}, {Gao}, {Armus}, {Isaak}, {Mazzarella}, {van der Werf}, {Appleton},
  {Charmandaris}, {Evans}, {Howell}, {Iwasawa}, {Leech}, {Lord}, {Petric},
  {Privon}, {Sanders}, {Schulz}, \& {Surace}}]{L17a}
{Lu}, N., {Zhao}, Y., {D{\'\i}az-Santos}, T., {et~al.} 2017{\natexlab{a}},
  \apjs, 230, 1, \dodoi{10.3847/1538-4365/aa6476}

\bibitem[{{Lu} {et~al.}(2017{\natexlab{b}}){Lu}, {Zhao}, {D{\'\i}az-Santos},
  {Xu}, {Charmandaris}, {Gao}, {van der Werf}, {Privon}, {Inami}, {Rigopoulou},
  {Sanders}, \& {Zhu}}]{LN17a}
---. 2017{\natexlab{b}}, \apjl, 842, L16, \dodoi{10.3847/2041-8213/aa77fc}

\bibitem[{{Lu} {et~al.}(2018){Lu}, {Cao}, {D{\'\i}az-Santos}, {Zhao}, {Privon},
  {Cheng}, {Gao}, {Xu}, {Charmandaris}, {Rigopoulou}, {van der Werf}, {Huang},
  {Wang}, {Evans}, \& {Sanders}}]{LN18}
{Lu}, N., {Cao}, T., {D{\'\i}az-Santos}, T., {et~al.} 2018, \apj, 864, 38,
  \dodoi{10.3847/1538-4357/aad3c9}

\bibitem[{{Luridiana} {et~al.}(2015){Luridiana}, {Morisset}, \&
  {Shaw}}]{Luridiana15}
{Luridiana}, V., {Morisset}, C., \& {Shaw}, R.~A. 2015, \aap, 573, A42,
  \dodoi{10.1051/0004-6361/201323152}

\bibitem[{{Lutz} {et~al.}(2007){Lutz}, {Sturm}, {Tacconi}, {Valiante},
  {Schweitzer}, {Netzer}, {Maiolino}, {Andreani}, {Shemmer}, \&
  {Veilleux}}]{LD07}
{Lutz}, D., {Sturm}, E., {Tacconi}, L.~J., {et~al.} 2007, \apjl, 661, L25,
  \dodoi{10.1086/518537}

\bibitem[{{Ma} {et~al.}(2015){Ma}, {Gonzalez}, {Spilker}, {Strandet}, {Ashby},
  {Aravena}, {B{\'e}thermin}, {Bothwell}, {de Breuck}, {Brodwin}, {Chapman},
  {Fassnacht}, {Greve}, {Gullberg}, {Hezaveh}, {Malkan}, {Marrone},
  {Saliwanchik}, {Vieira}, {Weiss}, \& {Welikala}}]{MJ15}
{Ma}, J., {Gonzalez}, A.~H., {Spilker}, J.~S., {et~al.} 2015, \apj, 812, 88,
  \dodoi{10.1088/0004-637X/812/1/88}

\bibitem[{{Ma} {et~al.}(2024){Ma}, {Sun}, {Cheng}, {Yan}, {Ling}, {Sun}, {Foo},
  {Egami}, {Diego}, {Cohen}, {Jansen}, {Summers}, {Windhorst}, {D'Silva},
  {Koekemoer}, {Coe}, {Conselice}, {Driver}, {Frye}, {Grogin}, {Marshall},
  {Nonino}, {Ortiz}, {Pirzkal}, {Robotham}, {Ryan}, {Willmer}, {Adams},
  {Hathi}, {Dole}, {Willner}, {Espada}, {Furtak}, {Hsiao}, {Li}, {Chen},
  {Jolly}, \& {Chen}}]{MZ24}
{Ma}, Z., {Sun}, B., {Cheng}, C., {et~al.} 2024, \apj, 975, 87,
  \dodoi{10.3847/1538-4357/ad7b32}

\bibitem[{{Madau} \& {Dickinson}(2014)}]{madau14}
{Madau}, P., \& {Dickinson}, M. 2014, \araa, 52, 415,
  \dodoi{10.1146/annurev-astro-081811-125615}

\bibitem[{{Madden} {et~al.}(2013){Madden}, {R{\'e}my-Ruyer}, {Galametz},
  {Cormier}, {Lebouteiller}, {Galliano}, {Hony}, {Bendo}, {Smith}, {Pohlen},
  {Roussel}, {Sauvage}, {Wu}, {Sturm}, {Poglitsch}, {Contursi}, {Doublier},
  {Baes}, {Barlow}, {Boselli}, {Boquien}, {Carlson}, {Ciesla}, {Cooray},
  {Cortese}, {de Looze}, {Irwin}, {Isaak}, {Kamenetzky}, {Karczewski}, {Lu},
  {MacHattie}, {O'Halloran}, {Parkin}, {Rangwala}, {Schirm}, {Schulz},
  {Spinoglio}, {Vaccari}, {Wilson}, \& {Wozniak}}]{M13}
{Madden}, S.~C., {R{\'e}my-Ruyer}, A., {Galametz}, M., {et~al.} 2013, \pasp,
  125, 600, \dodoi{10.1086/671138}

\bibitem[{{Madden} {et~al.}(2020){Madden}, {Cormier}, {Hony}, {Lebouteiller},
  {Abel}, {Galametz}, {De Looze}, {Chevance}, {Polles}, {Lee}, {Galliano},
  {Lambert-Huyghe}, {Hu}, \& {Ramambason}}]{madden20}
{Madden}, S.~C., {Cormier}, D., {Hony}, S., {et~al.} 2020, \aap, 643, A141,
  \dodoi{10.1051/0004-6361/202038860}

\bibitem[{{Magdis} {et~al.}(2011){Magdis}, {Daddi}, {Elbaz}, {Sargent},
  {Dickinson}, {Dannerbauer}, {Aussel}, {Walter}, {Hwang}, {Charmandaris},
  {Hodge}, {Riechers}, {Rigopoulou}, {Carilli}, {Pannella}, {Mullaney},
  {Leiton}, \& {Scott}}]{MG11}
{Magdis}, G.~E., {Daddi}, E., {Elbaz}, D., {et~al.} 2011, \apjl, 740, L15,
  \dodoi{10.1088/2041-8205/740/1/L15}

\bibitem[{{Magdis} {et~al.}(2012){Magdis}, {Daddi}, {B{\'e}thermin}, {Sargent},
  {Elbaz}, {Pannella}, {Dickinson}, {Dannerbauer}, {da Cunha}, {Walter},
  {Rigopoulou}, {Charmandaris}, {Hwang}, \& {Kartaltepe}}]{MG12}
{Magdis}, G.~E., {Daddi}, E., {B{\'e}thermin}, M., {et~al.} 2012, \apj, 760, 6,
  \dodoi{10.1088/0004-637X/760/1/6}

\bibitem[{{Magdis} {et~al.}(2014){Magdis}, {Rigopoulou}, {Hopwood}, {Huang},
  {Farrah}, {Pearson}, {Alonso-Herrero}, {Bock}, {Clements}, {Cooray},
  {Griffin}, {Oliver}, {Perez Fournon}, {Riechers}, {Swinyard}, {Scott},
  {Thatte}, {Valtchanov}, \& {Vaccari}}]{MG14}
{Magdis}, G.~E., {Rigopoulou}, D., {Hopwood}, R., {et~al.} 2014, \apj, 796, 63,
  \dodoi{10.1088/0004-637X/796/1/63}

\bibitem[{{Maiolino} {et~al.}(2009){Maiolino}, {Caselli}, {Nagao}, {Walmsley},
  {De Breuck}, \& {Meneghetti}}]{MR09}
{Maiolino}, R., {Caselli}, P., {Nagao}, T., {et~al.} 2009, \aap, 500, L1,
  \dodoi{10.1051/0004-6361/200912265}

\bibitem[{{Maiolino} {et~al.}(2005){Maiolino}, {Cox}, {Caselli}, {Beelen},
  {Bertoldi}, {Carilli}, {Kaufman}, {Menten}, {Nagao}, {Omont}, {Wei{\ss}},
  {Walmsley}, \& {Walter}}]{MR05}
{Maiolino}, R., {Cox}, P., {Caselli}, P., {et~al.} 2005, \aap, 440, L51,
  \dodoi{10.1051/0004-6361:200500165}

\bibitem[{{Maiolino} {et~al.}(2012){Maiolino}, {Gallerani}, {Neri}, {Cicone},
  {Ferrara}, {Genzel}, {Lutz}, {Sturm}, {Tacconi}, {Walter}, {Feruglio},
  {Fiore}, \& {Piconcelli}}]{MR12}
{Maiolino}, R., {Gallerani}, S., {Neri}, R., {et~al.} 2012, \mnras, 425, L66,
  \dodoi{10.1111/j.1745-3933.2012.01303.x}

\bibitem[{{Maiolino} {et~al.}(2015){Maiolino}, {Carniani}, {Fontana},
  {Vallini}, {Pentericci}, {Ferrara}, {Vanzella}, {Grazian}, {Gallerani},
  {Castellano}, {Cristiani}, {Brammer}, {Santini}, {Wagg}, \&
  {Williams}}]{MR15}
{Maiolino}, R., {Carniani}, S., {Fontana}, A., {et~al.} 2015, \mnras, 452, 54,
  \dodoi{10.1093/mnras/stv1194}

\bibitem[{{Malhotra} {et~al.}(2017){Malhotra}, {Rhoads}, {Finkelstein}, {Yang},
  {Carilli}, {Combes}, {Dassas}, {Finkelstein}, {Frye}, {Gerin}, {Guillard},
  {Nesvadba}, {Rigby}, {Shin}, {Spaans}, {Strauss}, \& {Papovich}}]{MS17}
{Malhotra}, S., {Rhoads}, J.~E., {Finkelstein}, K., {et~al.} 2017, \apj, 835,
  110, \dodoi{10.3847/1538-4357/835/1/110}

\bibitem[{{Marconcini} {et~al.}(2024{\natexlab{a}}){Marconcini}, {D'Eugenio},
  {Maiolino}, {Arribas}, {Bunker}, {Carniani}, {Charlot}, {Perna},
  {Rodr{\'\i}guez Del Pino}, {{\"U}bler}, {Willott}, {B{\"o}ker}, {Cresci},
  {Curti}, {Jones}, {Lamperti}, {Parlanti}, \& {Venturi}}]{MC24a}
{Marconcini}, C., {D'Eugenio}, F., {Maiolino}, R., {et~al.} 2024{\natexlab{a}},
  \mnras, 533, 2488, \dodoi{10.1093/mnras/stae1971}

\bibitem[{{Marconcini} {et~al.}(2024{\natexlab{b}}){Marconcini}, {D'Eugenio},
  {Maiolino}, {Arribas}, {Bunker}, {Carniani}, {Charlot}, {Perna},
  {Rodr{\'\i}guez Del Pino}, {{\"U}bler}, {P{\'e}rez-Gonz{\'a}lez}, {Willott},
  {B{\"o}ker}, {Cresci}, {Curti}, {Lamperti}, {Scholtz}, {Parlanti}, \&
  {Venturi}}]{MC24b}
---. 2024{\natexlab{b}}, arXiv e-prints, arXiv:2411.08627,
  \dodoi{10.48550/arXiv.2411.08627}

\bibitem[{{Markov} {et~al.}(2022){Markov}, {Carniani}, {Vallini}, {Ferrara},
  {Pallottini}, {Maiolino}, {Gallerani}, \& {Pentericci}}]{MV22}
{Markov}, V., {Carniani}, S., {Vallini}, L., {et~al.} 2022, \aap, 663, A172,
  \dodoi{10.1051/0004-6361/202243336}

\bibitem[{{Marrone} {et~al.}(2018){Marrone}, {Spilker}, {Hayward}, {Vieira},
  {Aravena}, {Ashby}, {Bayliss}, {B{\'e}thermin}, {Brodwin}, {Bothwell},
  {Carlstrom}, {Chapman}, {Chen}, {Crawford}, {Cunningham}, {De Breuck},
  {Fassnacht}, {Gonzalez}, {Greve}, {Hezaveh}, {Lacaille}, {Litke}, {Lower},
  {Ma}, {Malkan}, {Miller}, {Morningstar}, {Murphy}, {Narayanan}, {Phadke},
  {Rotermund}, {Sreevani}, {Stalder}, {Stark}, {Strandet}, {Tang}, \&
  {Wei{\ss}}}]{MD18}
{Marrone}, D.~P., {Spilker}, J.~S., {Hayward}, C.~C., {et~al.} 2018, \nat, 553,
  51, \dodoi{10.1038/nature24629}

\bibitem[{{Marshall} {et~al.}(2023){Marshall}, {Perna}, {Willott}, {Maiolino},
  {Scholtz}, {{\"U}bler}, {Carniani}, {Arribas}, {L{\"u}tzgendorf}, {Bunker},
  {Charlot}, {Ferruit}, {Jakobsen}, {Rix}, {Rodr{\'\i}guez Del Pino},
  {B{\"o}ker}, {Cameron}, {Cresci}, {Curtis-Lake}, {Jones}, {Kumari},
  {P{\'e}rez-Gonz{\'a}lez}, \& {Reed}}]{MM23}
{Marshall}, M.~A., {Perna}, M., {Willott}, C.~J., {et~al.} 2023, \aap, 678,
  A191, \dodoi{10.1051/0004-6361/202346113}

\bibitem[{{Marshall} {et~al.}(2024){Marshall}, {Yue}, {Eilers}, {Scholtz},
  {Perna}, {Willott}, {Maiolino}, {{\"U}bler}, {Arribas}, {Bunker}, {Charlot},
  {Rodr{\'\i}guez Del Pino}, {B{\"o}ker}, {Carniani}, {Circosta}, {Cresci},
  {D'Eugenio}, {Jones}, {Venturi}, {Bordoloi}, {Kashino}, {Mackenzie},
  {Matthee}, {Naidu}, \& {Simcoe}}]{MM24}
{Marshall}, M.~A., {Yue}, M., {Eilers}, A.-C., {et~al.} 2024, arXiv e-prints,
  arXiv:2410.11035, \dodoi{10.48550/arXiv.2410.11035}

\bibitem[{{Matsuoka} {et~al.}(2016){Matsuoka}, {Onoue}, {Kashikawa}, {Iwasawa},
  {Strauss}, {Nagao}, {Imanishi}, {Niida}, {Toba}, {Akiyama}, {Asami}, {Bosch},
  {Foucaud}, {Furusawa}, {Goto}, {Gunn}, {Harikane}, {Ikeda}, {Kawaguchi},
  {Kikuta}, {Komiyama}, {Lupton}, {Minezaki}, {Miyazaki}, {Morokuma},
  {Murayama}, {Nishizawa}, {Ono}, {Ouchi}, {Price}, {Sameshima}, {Silverman},
  {Sugiyama}, {Tait}, {Takada}, {Takata}, {Tanaka}, {Tang}, \& {Utsumi}}]{MY16}
{Matsuoka}, Y., {Onoue}, M., {Kashikawa}, N., {et~al.} 2016, \apj, 828, 26,
  \dodoi{10.3847/0004-637X/828/1/26}

\bibitem[{{Matsuoka} {et~al.}(2018{\natexlab{a}}){Matsuoka}, {Onoue},
  {Kashikawa}, {Iwasawa}, {Strauss}, {Nagao}, {Imanishi}, {Lee}, {Akiyama},
  {Asami}, {Bosch}, {Foucaud}, {Furusawa}, {Goto}, {Gunn}, {Harikane}, {Ikeda},
  {Izumi}, {Kawaguchi}, {Kikuta}, {Kohno}, {Komiyama}, {Lupton}, {Minezaki},
  {Miyazaki}, {Morokuma}, {Murayama}, {Niida}, {Nishizawa}, {Oguri}, {Ono},
  {Ouchi}, {Price}, {Sameshima}, {Schulze}, {Shirakata}, {Silverman},
  {Sugiyama}, {Tait}, {Takada}, {Takata}, {Tanaka}, {Tang}, {Toba}, {Utsumi},
  \& {Wang}}]{MY18a}
---. 2018{\natexlab{a}}, \pasj, 70, S35, \dodoi{10.1093/pasj/psx046}

\bibitem[{{Matsuoka} {et~al.}(2018{\natexlab{b}}){Matsuoka}, {Iwasawa},
  {Onoue}, {Kashikawa}, {Strauss}, {Lee}, {Imanishi}, {Nagao}, {Akiyama},
  {Asami}, {Bosch}, {Furusawa}, {Goto}, {Gunn}, {Harikane}, {Ikeda}, {Izumi},
  {Kawaguchi}, {Kato}, {Kikuta}, {Kohno}, {Komiyama}, {Lupton}, {Minezaki},
  {Miyazaki}, {Morokuma}, {Murayama}, {Niida}, {Nishizawa}, {Oguri}, {Ono},
  {Ouchi}, {Price}, {Sameshima}, {Schulze}, {Shirakata}, {Silverman},
  {Sugiyama}, {Tait}, {Takada}, {Takata}, {Tanaka}, {Tang}, {Toba}, {Utsumi},
  {Wang}, \& {Yamashita}}]{MY18b}
{Matsuoka}, Y., {Iwasawa}, K., {Onoue}, M., {et~al.} 2018{\natexlab{b}}, \apjs,
  237, 5, \dodoi{10.3847/1538-4365/aac724}

\bibitem[{{Matsuoka} {et~al.}(2019){Matsuoka}, {Onoue}, {Kashikawa}, {Strauss},
  {Iwasawa}, {Lee}, {Imanishi}, {Nagao}, {Akiyama}, {Asami}, {Bosch},
  {Furusawa}, {Goto}, {Gunn}, {Harikane}, {Ikeda}, {Izumi}, {Kawaguchi},
  {Kato}, {Kikuta}, {Kohno}, {Komiyama}, {Koyama}, {Lupton}, {Minezaki},
  {Miyazaki}, {Murayama}, {Niida}, {Nishizawa}, {Noboriguchi}, {Oguri}, {Ono},
  {Ouchi}, {Price}, {Sameshima}, {Schulze}, {Shirakata}, {Silverman},
  {Sugiyama}, {Tait}, {Takada}, {Takata}, {Tanaka}, {Tang}, {Toba}, {Utsumi},
  {Wang}, \& {Yamashita}}]{MY19}
{Matsuoka}, Y., {Onoue}, M., {Kashikawa}, N., {et~al.} 2019, \apjl, 872, L2,
  \dodoi{10.3847/2041-8213/ab0216}

\bibitem[{{Matthee} {et~al.}(2017){Matthee}, {Sobral}, {Boone},
  {R{\"o}ttgering}, {Schaerer}, {Girard}, {Pallottini}, {Vallini}, {Ferrara},
  {Darvish}, \& {Mobasher}}]{MJ17}
{Matthee}, J., {Sobral}, D., {Boone}, F., {et~al.} 2017, \apj, 851, 145,
  \dodoi{10.3847/1538-4357/aa9931}

\bibitem[{{Matthee} {et~al.}(2019){Matthee}, {Sobral}, {Boogaard},
  {R{\"o}ttgering}, {Vallini}, {Ferrara}, {Paulino-Afonso}, {Boone},
  {Schaerer}, \& {Mobasher}}]{MJ19}
{Matthee}, J., {Sobral}, D., {Boogaard}, L.~A., {et~al.} 2019, \apj, 881, 124,
  \dodoi{10.3847/1538-4357/ab2f81}

\bibitem[{{Mazzucchelli} {et~al.}(2017){Mazzucchelli}, {Ba{\~n}ados},
  {Venemans}, {Decarli}, {Farina}, {Walter}, {Eilers}, {Rix}, {Simcoe},
  {Stern}, {Fan}, {Schlafly}, {De Rosa}, {Hennawi}, {Chambers}, {Greiner},
  {Burgett}, {Draper}, {Kaiser}, {Kudritzki}, {Magnier}, {Metcalfe}, {Waters},
  \& {Wainscoat}}]{MC17}
{Mazzucchelli}, C., {Ba{\~n}ados}, E., {Venemans}, B.~P., {et~al.} 2017, \apj,
  849, 91, \dodoi{10.3847/1538-4357/aa9185}

\bibitem[{{Mazzucchelli} {et~al.}(2025){Mazzucchelli}, {Decarli}, {Belladitta},
  {Ba{\~n}ados}, {Meyer}, {Connor}, {Momjian}, {Rojas-Ruiz}, {Eilers},
  {Khusanova}, {Farina}, {Drake}, {Walter}, {Wang}, {Onoue}, \&
  {Venemans}}]{MC25}
{Mazzucchelli}, C., {Decarli}, R., {Belladitta}, S., {et~al.} 2025, \aap, 694,
  A171, \dodoi{10.1051/0004-6361/202451290}

\bibitem[{{McGaugh}(1991)}]{mcgaugh91}
{McGaugh}, S.~S. 1991, \apj, 380, 140, \dodoi{10.1086/170569}

\bibitem[{{Meijerink} {et~al.}(2007){Meijerink}, {Spaans}, \&
  {Israel}}]{meijerink07}
{Meijerink}, R., {Spaans}, M., \& {Israel}, F.~P. 2007, \aap, 461, 793,
  \dodoi{10.1051/0004-6361:20066130}

\bibitem[{{Mellier} {et~al.}(1991){Mellier}, {Fort}, {Soucail}, {Mathez}, \&
  {Cailloux}}]{MY91}
{Mellier}, Y., {Fort}, B., {Soucail}, G., {Mathez}, G., \& {Cailloux}, M. 1991,
  \apj, 380, 334, \dodoi{10.1086/170592}

\bibitem[{{M{\'e}ndez-Delgado} {et~al.}(2023){M{\'e}ndez-Delgado}, {Esteban},
  {Garc{\'\i}a-Rojas}, {Kreckel}, \& {Peimbert}}]{mendez23}
{M{\'e}ndez-Delgado}, J.~E., {Esteban}, C., {Garc{\'\i}a-Rojas}, J., {Kreckel},
  K., \& {Peimbert}, M. 2023, \nat, 618, 249,
  \dodoi{10.1038/s41586-023-05956-2}

\bibitem[{{Messias} {et~al.}(2014){Messias}, {Dye}, {Nagar}, {Orellana},
  {Bussmann}, {Calanog}, {Dannerbauer}, {Fu}, {Ibar}, {Inohara}, {Ivison},
  {Negrello}, {Riechers}, {Sheen}, {Aguirre}, {Amber}, {Birkinshaw}, {Bourne},
  {Bradford}, {Clements}, {Cooray}, {De Zotti}, {Demarco}, {Dunne}, {Eales},
  {Fleuren}, {Kamenetzky}, {Lupu}, {Maddox}, {Marrone}, {Micha{\l}owski},
  {Murphy}, {Nguyen}, {Omont}, {Rowlands}, {Smith}, {Smith}, {Valiante}, \&
  {Vieira}}]{MH14}
{Messias}, H., {Dye}, S., {Nagar}, N., {et~al.} 2014, \aap, 568, A92,
  \dodoi{10.1051/0004-6361/201424410}

\bibitem[{{Meyer} {et~al.}(2025){Meyer}, {Venemans}, {Neeleman}, {Decarli}, \&
  {Walter}}]{MR25}
{Meyer}, R.~A., {Venemans}, B., {Neeleman}, M., {Decarli}, R., \& {Walter}, F.
  2025, \apj, 980, 20, \dodoi{10.3847/1538-4357/ada351}

\bibitem[{{Meyer} {et~al.}(2022){Meyer}, {Walter}, {Cicone}, {Cox}, {Decarli},
  {Neri}, {Novak}, {Pensabene}, {Riechers}, \& {Weiss}}]{MR22}
{Meyer}, R.~A., {Walter}, F., {Cicone}, C., {et~al.} 2022, \apj, 927, 152,
  \dodoi{10.3847/1538-4357/ac4e94}

\bibitem[{{Miley} {et~al.}(1984){Miley}, {Neugebauer}, {Clegg}, {Harris},
  {Rowan-Robinson}, {Soifer}, \& {Young}}]{miley84}
{Miley}, G., {Neugebauer}, G., {Clegg}, P.~E., {et~al.} 1984, \apjl, 278, L79,
  \dodoi{10.1086/184228}

\bibitem[{{Miley} {et~al.}(1985){Miley}, {Neugebauer}, \& {Soifer}}]{miley85}
{Miley}, G.~K., {Neugebauer}, G., \& {Soifer}, B.~T. 1985, \apjl, 293, L11,
  \dodoi{10.1086/184481}

\bibitem[{{Miller} {et~al.}(2020){Miller}, {Chapman}, {Hayward}, {Behroozi},
  {Bradford}, {Willott}, \& {Wagg}}]{MT20}
{Miller}, T.~B., {Chapman}, S.~C., {Hayward}, C.~C., {et~al.} 2020, \apj, 889,
  98, \dodoi{10.3847/1538-4357/ab63dd}

\bibitem[{{Miller} {et~al.}(2018){Miller}, {Chapman}, {Aravena}, {Ashby},
  {Hayward}, {Vieira}, {Wei{\ss}}, {Babul}, {B{\'e}thermin}, {Bradford},
  {Brodwin}, {Carlstrom}, {Chen}, {Cunningham}, {De Breuck}, {Gonzalez},
  {Greve}, {Harnett}, {Hezaveh}, {Lacaille}, {Litke}, {Ma}, {Malkan},
  {Marrone}, {Morningstar}, {Murphy}, {Narayanan}, {Pass}, {Perry}, {Phadke},
  {Rennehan}, {Rotermund}, {Simpson}, {Spilker}, {Sreevani}, {Stark},
  {Strandet}, \& {Strom}}]{MT18}
{Miller}, T.~B., {Chapman}, S.~C., {Aravena}, M., {et~al.} 2018, \nat, 556,
  469, \dodoi{10.1038/s41586-018-0025-2}

\bibitem[{{Mitsuhashi} {et~al.}(2021){Mitsuhashi}, {Matsuda}, {Smail},
  {Hayatsu}, {Simpson}, {Swinbank}, {Umehata},
  {Dudzevi{\v{c}}i{\={u}}t{\.{e}}}, {Birkin}, {Ikarashi}, {Chen}, {Tadaki},
  {Yajima}, {Harikane}, {Inami}, {Chapman}, {Hatsukade}, {Iono}, {Bunker},
  {Ao}, {Saito}, {Ueda}, \& {Sakamoto}}]{MI21}
{Mitsuhashi}, I., {Matsuda}, Y., {Smail}, I., {et~al.} 2021, \apj, 907, 122,
  \dodoi{10.3847/1538-4357/abcc72}

\bibitem[{{Mitsuhashi} {et~al.}(2024){Mitsuhashi}, {Tadaki}, {Ikeda},
  {Herrera-Camus}, {Aravena}, {De Looze}, {F{\"o}rster Schreiber},
  {Gonz{\'a}lez-L{\'o}pez}, {Spilker}, {Assef}, {Bouwens}, {Barcos-Munoz},
  {Birkin}, {Bowler}, {Calistro Rivera}, {Davies}, {Da Cunha},
  {D{\'\i}az-Santos}, {Ferrara}, {Fisher}, {Lee}, {Li}, {Lutz}, {Rela{\~n}o},
  {Naab}, {Palla}, {Posses}, {Solimano}, {Tacconi}, {{\"U}bler}, {van der
  Giessen}, \& {Veilleux}}]{MI24}
{Mitsuhashi}, I., {Tadaki}, K.-i., {Ikeda}, R., {et~al.} 2024, \aap, 690, A197,
  \dodoi{10.1051/0004-6361/202348782}

\bibitem[{{Mitsuhashi} {et~al.}(2025){Mitsuhashi}, {Zavala}, {Bakx}, {Inoue},
  {Castellano}, {Calabro}, {Casey}, {Franco}, {Hatsukade}, {Hathi}, {Ikeda},
  {Koekemoer}, {Kartaltepe}, {Knudsen}, {Santini}, {Saito}, {Terlevich}, \&
  {Terlevich}}]{MI25}
{Mitsuhashi}, I., {Zavala}, J.~A., {Bakx}, T. J.~L.~C., {et~al.} 2025, arXiv
  e-prints, arXiv:2501.19384, \dodoi{10.48550/arXiv.2501.19384}

\bibitem[{{Molyneux} {et~al.}(2022){Molyneux}, {Smit}, {Schaerer}, {Bouwens},
  {Bradley}, {Hodge}, {Longmore}, {Schouws}, {van der Werf}, {Zitrin}, \&
  {Phillips}}]{MS22}
{Molyneux}, S.~J., {Smit}, R., {Schaerer}, D., {et~al.} 2022, \mnras, 512, 535,
  \dodoi{10.1093/mnras/stac557}

\bibitem[{{Morishita} {et~al.}(2023){Morishita}, {Roberts-Borsani}, {Treu},
  {Brammer}, {Mason}, {Trenti}, {Vulcani}, {Wang}, {Acebron}, {Bah{\'e}},
  {Bergamini}, {Boyett}, {Bradac}, {Calabr{\`o}}, {Castellano}, {Chen}, {De
  Lucia}, {Filippenko}, {Fontana}, {Glazebrook}, {Grillo}, {Henry}, {Jones},
  {Kelly}, {Koekemoer}, {Leethochawalit}, {Lu}, {Marchesini}, {Mascia},
  {Mercurio}, {Merlin}, {Metha}, {Nanayakkara}, {Nonino}, {Paris},
  {Pentericci}, {Rosati}, {Santini}, {Strait}, {Vanzella}, {Windhorst}, \&
  {Xie}}]{MT23}
{Morishita}, T., {Roberts-Borsani}, G., {Treu}, T., {et~al.} 2023, \apjl, 947,
  L24, \dodoi{10.3847/2041-8213/acb99e}

\bibitem[{{Morrissey} {et~al.}(2007){Morrissey}, {Conrow}, {Barlow}, {Small},
  {Seibert}, {Wyder}, {Budav{\'a}ri}, {Arnouts}, {Friedman}, {Forster},
  {Martin}, {Neff}, {Schiminovich}, {Bianchi}, {Donas}, {Heckman}, {Lee},
  {Madore}, {Milliard}, {Rich}, {Szalay}, {Welsh}, \& {Yi}}]{morrissey07}
{Morrissey}, P., {Conrow}, T., {Barlow}, T.~A., {et~al.} 2007, \apjs, 173, 682,
  \dodoi{10.1086/520512}

\bibitem[{{Moustakas} \& {Kennicutt}(2006)}]{M06}
{Moustakas}, J., \& {Kennicutt}, Jr., R.~C. 2006, \apjs, 164, 81,
  \dodoi{10.1086/500971}

\bibitem[{{Nagao} {et~al.}(2012){Nagao}, {Maiolino}, {De Breuck}, {Caselli},
  {Hatsukade}, \& {Saigo}}]{NT12}
{Nagao}, T., {Maiolino}, R., {De Breuck}, C., {et~al.} 2012, \aap, 542, L34,
  \dodoi{10.1051/0004-6361/201219518}

\bibitem[{{Nakajima} {et~al.}(2023){Nakajima}, {Ouchi}, {Isobe}, {Harikane},
  {Zhang}, {Ono}, {Umeda}, \& {Oguri}}]{NK23}
{Nakajima}, K., {Ouchi}, M., {Isobe}, Y., {et~al.} 2023, \apjs, 269, 33,
  \dodoi{10.3847/1538-4365/acd556}

\bibitem[{{Neeleman} {et~al.}(2017){Neeleman}, {Kanekar}, {Prochaska},
  {Rafelski}, {Carilli}, \& {Wolfe}}]{NM17}
{Neeleman}, M., {Kanekar}, N., {Prochaska}, J.~X., {et~al.} 2017, Science, 355,
  1285, \dodoi{10.1126/science.aal1737}

\bibitem[{{Neeleman} {et~al.}(2019{\natexlab{a}}){Neeleman}, {Kanekar},
  {Prochaska}, {Rafelski}, \& {Carilli}}]{NM19b}
{Neeleman}, M., {Kanekar}, N., {Prochaska}, J.~X., {Rafelski}, M.~A., \&
  {Carilli}, C.~L. 2019{\natexlab{a}}, \apjl, 870, L19,
  \dodoi{10.3847/2041-8213/aaf871}

\bibitem[{{Neeleman} {et~al.}(2020){Neeleman}, {Prochaska}, {Kanekar}, \&
  {Rafelski}}]{NM20}
{Neeleman}, M., {Prochaska}, J.~X., {Kanekar}, N., \& {Rafelski}, M. 2020,
  \nat, 581, 269, \dodoi{10.1038/s41586-020-2276-y}

\bibitem[{{Neeleman} {et~al.}(2019{\natexlab{b}}){Neeleman}, {Ba{\~n}ados},
  {Walter}, {Decarli}, {Venemans}, {Carilli}, {Fan}, {Farina}, {Mazzucchelli},
  {Novak}, {Riechers}, {Rix}, \& {Wang}}]{NM19a}
{Neeleman}, M., {Ba{\~n}ados}, E., {Walter}, F., {et~al.} 2019{\natexlab{b}},
  \apj, 882, 10, \dodoi{10.3847/1538-4357/ab2ed3}

\bibitem[{{Negrello} {et~al.}(2010){Negrello}, {Hopwood}, {De Zotti}, {Cooray},
  {Verma}, {Bock}, {Frayer}, {Gurwell}, {Omont}, {Neri}, {Dannerbauer},
  {Leeuw}, {Barton}, {Cooke}, {Kim}, {da Cunha}, {Rodighiero}, {Cox},
  {Bonfield}, {Jarvis}, {Serjeant}, {Ivison}, {Dye}, {Aretxaga}, {Hughes},
  {Ibar}, {Bertoldi}, {Valtchanov}, {Eales}, {Dunne}, {Driver}, {Auld},
  {Buttiglione}, {Cava}, {Grady}, {Clements}, {Dariush}, {Fritz}, {Hill},
  {Hornbeck}, {Kelvin}, {Lagache}, {Lopez-Caniego}, {Gonzalez-Nuevo}, {Maddox},
  {Pascale}, {Pohlen}, {Rigby}, {Robotham}, {Simpson}, {Smith}, {Temi},
  {Thompson}, {Woodgate}, {York}, {Aguirre}, {Beelen}, {Blain}, {Baker},
  {Birkinshaw}, {Blundell}, {Bradford}, {Burgarella}, {Danese}, {Dunlop},
  {Fleuren}, {Glenn}, {Harris}, {Kamenetzky}, {Lupu}, {Maddalena}, {Madore},
  {Maloney}, {Matsuhara}, {Micha{\l}owski}, {Murphy}, {Naylor}, {Nguyen},
  {Popescu}, {Rawlings}, {Rigopoulou}, {Scott}, {Scott}, {Seibert}, {Smail},
  {Tuffs}, {Vieira}, {van der Werf}, \& {Zmuidzinas}}]{NM10}
{Negrello}, M., {Hopwood}, R., {De Zotti}, G., {et~al.} 2010, Science, 330,
  800, \dodoi{10.1126/science.1193420}

\bibitem[{{Neri} {et~al.}(2014){Neri}, {Downes}, {Cox}, \& {Walter}}]{NR14}
{Neri}, R., {Downes}, D., {Cox}, P., \& {Walter}, F. 2014, \aap, 562, A35,
  \dodoi{10.1051/0004-6361/201322528}

\bibitem[{{Neugebauer} {et~al.}(1984){Neugebauer}, {Habing}, {van Duinen},
  {Aumann}, {Baud}, {Beichman}, {Beintema}, {Boggess}, {Clegg}, {de Jong},
  {Emerson}, {Gautier}, {Gillett}, {Harris}, {Hauser}, {Houck}, {Jennings},
  {Low}, {Marsden}, {Miley}, {Olnon}, {Pottasch}, {Raimond}, {Rowan-Robinson},
  {Soifer}, {Walker}, {Wesselius}, \& {Young}}]{neugebauer84}
{Neugebauer}, G., {Habing}, H.~J., {van Duinen}, R., {et~al.} 1984, \apjl, 278,
  L1, \dodoi{10.1086/184209}

\bibitem[{{Nicholls} {et~al.}(2014){Nicholls}, {Dopita}, {Sutherland},
  {Jerjen}, \& {Kewley}}]{nicholls14}
{Nicholls}, D.~C., {Dopita}, M.~A., {Sutherland}, R.~S., {Jerjen}, H., \&
  {Kewley}, L.~J. 2014, \apj, 790, 75, \dodoi{10.1088/0004-637X/790/1/75}

\bibitem[{{Novak} {et~al.}(2019){Novak}, {Ba{\~n}ados}, {Decarli}, {Walter},
  {Venemans}, {Neeleman}, {Farina}, {Mazzucchelli}, {Carilli}, {Fan}, {Rix}, \&
  {Wang}}]{NM19c}
{Novak}, M., {Ba{\~n}ados}, E., {Decarli}, R., {et~al.} 2019, \apj, 881, 63,
  \dodoi{10.3847/1538-4357/ab2beb}

\bibitem[{{Oesch} {et~al.}(2016){Oesch}, {Brammer}, {van Dokkum},
  {Illingworth}, {Bouwens}, {Labb{\'e}}, {Franx}, {Momcheva}, {Ashby}, {Fazio},
  {Gonzalez}, {Holden}, {Magee}, {Skelton}, {Smit}, {Spitler}, {Trenti}, \&
  {Willner}}]{OP16}
{Oesch}, P.~A., {Brammer}, G., {van Dokkum}, P.~G., {et~al.} 2016, \apj, 819,
  129, \dodoi{10.3847/0004-637X/819/2/129}

\bibitem[{{Oliver} {et~al.}(2012){Oliver}, {Bock}, {Altieri}, {Amblard},
  {Arumugam}, {Aussel}, {Babbedge}, {Beelen}, {B{\'e}thermin}, {Blain},
  {Boselli}, {Bridge}, {Brisbin}, {Buat}, {Burgarella},
  {Castro-Rodr{\'\i}guez}, {Cava}, {Chanial}, {Cirasuolo}, {Clements},
  {Conley}, {Conversi}, {Cooray}, {Dowell}, {Dubois}, {Dwek}, {Dye}, {Eales},
  {Elbaz}, {Farrah}, {Feltre}, {Ferrero}, {Fiolet}, {Fox}, {Franceschini},
  {Gear}, {Giovannoli}, {Glenn}, {Gong}, {Gonz{\'a}lez Solares}, {Griffin},
  {Halpern}, {Harwit}, {Hatziminaoglou}, {Heinis}, {Hurley}, {Hwang}, {Hyde},
  {Ibar}, {Ilbert}, {Isaak}, {Ivison}, {Lagache}, {Le Floc'h}, {Levenson},
  {Faro}, {Lu}, {Madden}, {Maffei}, {Magdis}, {Mainetti}, {Marchetti},
  {Marsden}, {Marshall}, {Mortier}, {Nguyen}, {O'Halloran}, {Omont}, {Page},
  {Panuzzo}, {Papageorgiou}, {Patel}, {Pearson}, {P{\'e}rez-Fournon}, {Pohlen},
  {Rawlings}, {Raymond}, {Rigopoulou}, {Riguccini}, {Rizzo}, {Rodighiero},
  {Roseboom}, {Rowan-Robinson}, {S{\'a}nchez Portal}, {Schulz}, {Scott},
  {Seymour}, {Shupe}, {Smith}, {Stevens}, {Symeonidis}, {Trichas}, {Tugwell},
  {Vaccari}, {Valtchanov}, {Vieira}, {Viero}, {Vigroux}, {Wang}, {Ward},
  {Wardlow}, {Wright}, {Xu}, \& {Zemcov}}]{OS12}
{Oliver}, S.~J., {Bock}, J., {Altieri}, B., {et~al.} 2012, \mnras, 424, 1614,
  \dodoi{10.1111/j.1365-2966.2012.20912.x}

\bibitem[{{Omont} {et~al.}(2001){Omont}, {Cox}, {Bertoldi}, {McMahon},
  {Carilli}, \& {Isaak}}]{OA01}
{Omont}, A., {Cox}, P., {Bertoldi}, F., {et~al.} 2001, \aap, 374, 371,
  \dodoi{10.1051/0004-6361:20010721}

\bibitem[{{Ono} {et~al.}(2012){Ono}, {Ouchi}, {Mobasher}, {Dickinson},
  {Penner}, {Shimasaku}, {Weiner}, {Kartaltepe}, {Nakajima}, {Nayyeri},
  {Stern}, {Kashikawa}, \& {Spinrad}}]{OY12}
{Ono}, Y., {Ouchi}, M., {Mobasher}, B., {et~al.} 2012, \apj, 744, 83,
  \dodoi{10.1088/0004-637X/744/2/83}

\bibitem[{{Osterbrock}(1989)}]{osterbrock89}
{Osterbrock}, D.~E. 1989, {Astrophysics of gaseous nebulae and active galactic
  nuclei}

\bibitem[{{Ota} {et~al.}(2014){Ota}, {Walter}, {Ohta}, {Hatsukade}, {Carilli},
  {da Cunha}, {Gonz{\'a}lez-L{\'o}pez}, {Decarli}, {Hodge}, {Nagai}, {Egami},
  {Jiang}, {Iye}, {Kashikawa}, {Riechers}, {Bertoldi}, {Cox}, {Neri}, \&
  {Weiss}}]{OK14}
{Ota}, K., {Walter}, F., {Ohta}, K., {et~al.} 2014, \apj, 792, 34,
  \dodoi{10.1088/0004-637X/792/1/34}

\bibitem[{{Oteo} {et~al.}(2016){Oteo}, {Ivison}, {Dunne}, {Smail}, {Swinbank},
  {Zhang}, {Lewis}, {Maddox}, {Riechers}, {Serjeant}, {Van der Werf}, {Biggs},
  {Bremer}, {Cigan}, {Clements}, {Cooray}, {Dannerbauer}, {Eales}, {Ibar},
  {Messias}, {Micha{\l}owski}, {P{\'e}rez-Fournon}, \& {van Kampen}}]{OI16}
{Oteo}, I., {Ivison}, R.~J., {Dunne}, L., {et~al.} 2016, \apj, 827, 34,
  \dodoi{10.3847/0004-637X/827/1/34}

\bibitem[{{Ouchi} {et~al.}(2013){Ouchi}, {Ellis}, {Ono}, {Nakanishi}, {Kohno},
  {Momose}, {Kurono}, {Ashby}, {Shimasaku}, {Willner}, {Fazio}, {Tamura}, \&
  {Iono}}]{OM13}
{Ouchi}, M., {Ellis}, R., {Ono}, Y., {et~al.} 2013, \apj, 778, 102,
  \dodoi{10.1088/0004-637X/778/2/102}

\bibitem[{{Oyabu} {et~al.}(2009){Oyabu}, {Kawara}, {Tsuzuki}, {Matsuoka},
  {Sameshima}, {Asami}, \& {Ohyama}}]{OS09}
{Oyabu}, S., {Kawara}, K., {Tsuzuki}, Y., {et~al.} 2009, \apj, 697, 452,
  \dodoi{10.1088/0004-637X/697/1/452}

\bibitem[{{Parlanti} {et~al.}(2024){Parlanti}, {Carniani}, {{\"U}bler},
  {Venturi}, {Circosta}, {D'Eugenio}, {Arribas}, {Bunker}, {Charlot},
  {L{\"u}tzgendorf}, {Maiolino}, {Perna}, {Rodr{\'\i}guez Del Pino}, {Willott},
  {B{\"o}ker}, {Cameron}, {Chevallard}, {Cresci}, {Jones}, {Kumari},
  {Lamperti}, \& {Scholtz}}]{PE24}
{Parlanti}, E., {Carniani}, S., {{\"U}bler}, H., {et~al.} 2024, \aap, 684, A24,
  \dodoi{10.1051/0004-6361/202347914}

\bibitem[{{Parlanti} {et~al.}(2025){Parlanti}, {Carniani}, {Venturi},
  {Herrera-Camus}, {Arribas}, {Bunker}, {Charlot}, {D'Eugenio}, {Maiolino},
  {Perna}, {{\"U}bler}, {B{\"o}ker}, {Cresci}, {Curti}, {Jones}, {Lamperti},
  {P{\'e}rez-Gonz{\'a}lez}, {Del Pino}, \& {Zamora}}]{PE25}
{Parlanti}, E., {Carniani}, S., {Venturi}, G., {et~al.} 2025, \aap, 695, A6,
  \dodoi{10.1051/0004-6361/202451692}

\bibitem[{{Pavesi} {et~al.}(2019){Pavesi}, {Riechers}, {Faisst}, {Stacey}, \&
  {Capak}}]{PR19}
{Pavesi}, R., {Riechers}, D.~A., {Faisst}, A.~L., {Stacey}, G.~J., \& {Capak},
  P.~L. 2019, \apj, 882, 168, \dodoi{10.3847/1538-4357/ab3a46}

\bibitem[{{Pavesi} {et~al.}(2016){Pavesi}, {Riechers}, {Capak}, {Carilli},
  {Sharon}, {Stacey}, {Karim}, {Scoville}, \& {Smol{\v{c}}i{\'c}}}]{PR16}
{Pavesi}, R., {Riechers}, D.~A., {Capak}, P.~L., {et~al.} 2016, \apj, 832, 151,
  \dodoi{10.3847/0004-637X/832/2/151}

\bibitem[{{Pavesi} {et~al.}(2018){Pavesi}, {Riechers}, {Sharon},
  {Smol{\v{c}}i{\'c}}, {Faisst}, {Schinnerer}, {Carilli}, {Capak}, {Scoville},
  \& {Stacey}}]{PR18}
{Pavesi}, R., {Riechers}, D.~A., {Sharon}, C.~E., {et~al.} 2018, \apj, 861, 43,
  \dodoi{10.3847/1538-4357/aac6b6}

\bibitem[{{Peng} {et~al.}(2021){Peng}, {Lamarche}, {Stacey}, {Nikola},
  {Vishwas}, {Ferkinhoff}, {Rooney}, {Ball}, {Brisbin}, {Higdon}, \&
  {Higdon}}]{P21}
{Peng}, B., {Lamarche}, C., {Stacey}, G.~J., {et~al.} 2021, \apj, 908, 166,
  \dodoi{10.3847/1538-4357/abd4e2}

\bibitem[{{Peng} {et~al.}(2023){Peng}, {Vishwas}, {Stacey}, {Nikola},
  {Lamarche}, {Rooney}, {Ball}, {Ferkinhoff}, \& {Spoon}}]{PB23}
{Peng}, B., {Vishwas}, A., {Stacey}, G., {et~al.} 2023, \apjl, 944, L36,
  \dodoi{10.3847/2041-8213/acb59c}

\bibitem[{{Pensabene} {et~al.}(2021){Pensabene}, {Decarli}, {Ba{\~n}ados},
  {Venemans}, {Walter}, {Bertoldi}, {Fan}, {Farina}, {Li}, {Mazzucchelli},
  {Novak}, {Riechers}, {Rix}, {Strauss}, {Wang}, {Wei{\ss}}, {Yang}, \&
  {Yang}}]{PA21}
{Pensabene}, A., {Decarli}, R., {Ba{\~n}ados}, E., {et~al.} 2021, \aap, 652,
  A66, \dodoi{10.1051/0004-6361/202039696}

\bibitem[{{Pentericci} {et~al.}(2011){Pentericci}, {Fontana}, {Vanzella},
  {Castellano}, {Grazian}, {Dijkstra}, {Boutsia}, {Cristiani}, {Dickinson},
  {Giallongo}, {Giavalisco}, {Maiolino}, {Moorwood}, {Paris}, \&
  {Santini}}]{PL11}
{Pentericci}, L., {Fontana}, A., {Vanzella}, E., {et~al.} 2011, \apj, 743, 132,
  \dodoi{10.1088/0004-637X/743/2/132}

\bibitem[{{Pentericci} {et~al.}(2016){Pentericci}, {Carniani}, {Castellano},
  {Fontana}, {Maiolino}, {Guaita}, {Vanzella}, {Grazian}, {Santini}, {Yan},
  {Cristiani}, {Conselice}, {Giavalisco}, {Hathi}, \& {Koekemoer}}]{PL16}
{Pentericci}, L., {Carniani}, S., {Castellano}, M., {et~al.} 2016, \apjl, 829,
  L11, \dodoi{10.3847/2041-8205/829/1/L11}

\bibitem[{{Pereira-Santaella} {et~al.}(2010){Pereira-Santaella},
  {Diamond-Stanic}, {Alonso-Herrero}, \& {Rieke}}]{P10}
{Pereira-Santaella}, M., {Diamond-Stanic}, A.~M., {Alonso-Herrero}, A., \&
  {Rieke}, G.~H. 2010, \apj, 725, 2270, \dodoi{10.1088/0004-637X/725/2/2270}

\bibitem[{{P{\'e}rez-Beaupuits} {et~al.}(2011){P{\'e}rez-Beaupuits}, {Spoon},
  {Spaans}, \& {Smith}}]{perez11}
{P{\'e}rez-Beaupuits}, J.~P., {Spoon}, H.~W.~W., {Spaans}, M., \& {Smith},
  J.~D. 2011, \aap, 533, A56, \dodoi{10.1051/0004-6361/201117153}

\bibitem[{{P{\'e}rez-Montero} \& {Contini}(2009)}]{perez09}
{P{\'e}rez-Montero}, E., \& {Contini}, T. 2009, \mnras, 398, 949,
  \dodoi{10.1111/j.1365-2966.2009.15145.x}

\bibitem[{{Pettini} \& {Pagel}(2004)}]{pettini04}
{Pettini}, M., \& {Pagel}, B. E.~J. 2004, \mnras, 348, L59,
  \dodoi{10.1111/j.1365-2966.2004.07591.x}

\bibitem[{{Pety} {et~al.}(2004){Pety}, {Beelen}, {Cox}, {Downes}, {Omont},
  {Bertoldi}, \& {Carilli}}]{PJ04}
{Pety}, J., {Beelen}, A., {Cox}, P., {et~al.} 2004, \aap, 428, L21,
  \dodoi{10.1051/0004-6361:200400096}

\bibitem[{{Pilbratt} {et~al.}(2010){Pilbratt}, {Riedinger}, {Passvogel},
  {Crone}, {Doyle}, {Gageur}, {Heras}, {Jewell}, {Metcalfe}, {Ott}, \&
  {Schmidt}}]{pilbratt10}
{Pilbratt}, G.~L., {Riedinger}, J.~R., {Passvogel}, T., {et~al.} 2010, \aap,
  518, L1, \dodoi{10.1051/0004-6361/201014759}

\bibitem[{{Pilyugin} {et~al.}(2004){Pilyugin}, {Contini}, \&
  {V{\'\i}lchez}}]{pilyugin04}
{Pilyugin}, L.~S., {Contini}, T., \& {V{\'\i}lchez}, J.~M. 2004, \aap, 423,
  427, \dodoi{10.1051/0004-6361:20035745}

\bibitem[{{Pilyugin} \& {Grebel}(2016)}]{pilyugin16}
{Pilyugin}, L.~S., \& {Grebel}, E.~K. 2016, \mnras, 457, 3678,
  \dodoi{10.1093/mnras/stw238}

\bibitem[{{Podigachoski} {et~al.}(2015){Podigachoski}, {Barthel}, {Haas},
  {Leipski}, {Wilkes}, {Kuraszkiewicz}, {Westhues}, {Willner}, {Ashby},
  {Chini}, {Clements}, {Fazio}, {Labiano}, {Lawrence}, {Meisenheimer},
  {Peletier}, {Siebenmorgen}, \& {Verdoes Kleijn}}]{PP15}
{Podigachoski}, P., {Barthel}, P.~D., {Haas}, M., {et~al.} 2015, \aap, 575,
  A80, \dodoi{10.1051/0004-6361/201425137}

\bibitem[{{Poglitsch} {et~al.}(2010){Poglitsch}, {Waelkens}, {Geis},
  {Feuchtgruber}, {Vandenbussche}, {Rodriguez}, {Krause}, {Renotte}, {van
  Hoof}, {Saraceno}, {Cepa}, {Kerschbaum}, {Agn{\`e}se}, {Ali}, {Altieri},
  {Andreani}, {Augueres}, {Balog}, {Barl}, {Bauer}, {Belbachir}, {Benedettini},
  {Billot}, {Boulade}, {Bischof}, {Blommaert}, {Callut}, {Cara}, {Cerulli},
  {Cesarsky}, {Contursi}, {Creten}, {De Meester}, {Doublier}, {Doumayrou},
  {Duband}, {Exter}, {Genzel}, {Gillis}, {Gr{\"o}zinger}, {Henning},
  {Herreros}, {Huygen}, {Inguscio}, {Jakob}, {Jamar}, {Jean}, {de Jong},
  {Katterloher}, {Kiss}, {Klaas}, {Lemke}, {Lutz}, {Madden}, {Marquet},
  {Martignac}, {Mazy}, {Merken}, {Montfort}, {Morbidelli}, {M{\"u}ller},
  {Nielbock}, {Okumura}, {Orfei}, {Ottensamer}, {Pezzuto}, {Popesso},
  {Putzeys}, {Regibo}, {Reveret}, {Royer}, {Sauvage}, {Schreiber}, {Stegmaier},
  {Schmitt}, {Schubert}, {Sturm}, {Thiel}, {Tofani}, {Vavrek}, {Wetzstein},
  {Wieprecht}, \& {Wiezorrek}}]{poglitsch10}
{Poglitsch}, A., {Waelkens}, C., {Geis}, N., {et~al.} 2010, \aap, 518, L2,
  \dodoi{10.1051/0004-6361/201014535}

\bibitem[{{Pope} {et~al.}(2006){Pope}, {Scott}, {Dickinson}, {Chary},
  {Morrison}, {Borys}, {Sajina}, {Alexander}, {Daddi}, {Frayer}, {MacDonald},
  \& {Stern}}]{PA06}
{Pope}, A., {Scott}, D., {Dickinson}, M., {et~al.} 2006, \mnras, 370, 1185,
  \dodoi{10.1111/j.1365-2966.2006.10575.x}

\bibitem[{{Pope} {et~al.}(2017){Pope}, {Monta{\~n}a}, {Battisti}, {Limousin},
  {Marchesini}, {Wilson}, {Alberts}, {Aretxaga}, {Avila-Reese}, {Ram{\'o}n
  Bermejo-Climent}, {Brammer}, {Bravo-Alfaro}, {Calzetti}, {Chary}, {Cybulski},
  {Giavalisco}, {Hughes}, {Kado-Fong}, {Keller}, {Kirkpatrick}, {Labbe},
  {Lange-Vagle}, {Lowenthal}, {Murphy}, {Oesch}, {Rosa Gonzalez},
  {S{\'a}nchez-Arg{\"u}elles}, {Shipley}, {Stefanon}, {Vega}, {Whitaker},
  {Williams}, {Yun}, {Zavala}, \& {Zeballos}}]{PA17}
{Pope}, A., {Monta{\~n}a}, A., {Battisti}, A., {et~al.} 2017, \apj, 838, 137,
  \dodoi{10.3847/1538-4357/aa6573}

\bibitem[{{Pope} {et~al.}(2023){Pope}, {McKinney}, {Kamieneski}, {Battisti},
  {Aretxaga}, {Brammer}, {Diego}, {Hughes}, {Keller}, {Marchesini}, {Mizener},
  {Monta{\~n}a}, {Murphy}, {Whitaker}, {Wilson}, \& {Yun}}]{PA23a}
{Pope}, A., {McKinney}, J., {Kamieneski}, P., {et~al.} 2023, \apjl, 951, L46,
  \dodoi{10.3847/2041-8213/acdf5a}

\bibitem[{{Popping}(2023)}]{PG23}
{Popping}, G. 2023, \aap, 669, L8, \dodoi{10.1051/0004-6361/202244831}

\bibitem[{{Posses} {et~al.}(2024){Posses}, {Aravena}, {Gonz{\'a}lez-L{\'o}pez},
  {F{\"o}rster Schreiber}, {Liu}, {Lee}, {Solimano}, {D{\'\i}az-Santos},
  {Assef}, {Barcos-Mu{\~n}oz}, {Bovino}, {Bowler}, {Calistro Rivera}, {da
  Cunha}, {Davies}, {Killi}, {De Looze}, {Ferrara}, {Fisher}, {Herrera-Camus},
  {Ikeda}, {Lambert}, {Li}, {Lutz}, {Mitsuhashi}, {Palla}, {Rela{\~n}o},
  {Spilker}, {Naab}, {Tadaki}, {Telikova}, {{\"U}bler}, {van der Giessen}, \&
  {Villanueva}}]{PA24}
{Posses}, A., {Aravena}, M., {Gonz{\'a}lez-L{\'o}pez}, J., {et~al.} 2024, arXiv
  e-prints, arXiv:2403.03379, \dodoi{10.48550/arXiv.2403.03379}

\bibitem[{{Posses} {et~al.}(2023){Posses}, {Aravena}, {Gonz{\'a}lez-L{\'o}pez},
  {Assef}, {Lambert}, {Jones}, {Bouwens}, {Brisbin}, {D{\'\i}az-Santos},
  {Herrera-Camus}, {Ricci}, \& {Smit}}]{PA23b}
{Posses}, A.~C., {Aravena}, M., {Gonz{\'a}lez-L{\'o}pez}, J., {et~al.} 2023,
  \aap, 669, A46, \dodoi{10.1051/0004-6361/202243399}

\bibitem[{{Rawle} {et~al.}(2014){Rawle}, {Egami}, {Bussmann}, {Gurwell},
  {Ivison}, {Boone}, {Combes}, {Danielson}, {Rex}, {Richard}, {Smail},
  {Swinbank}, {Altieri}, {Blain}, {Clement}, {Dessauges-Zavadsky}, {Edge},
  {Fazio}, {Jones}, {Kneib}, {Omont}, {P{\'e}rez-Gonz{\'a}lez}, {Schaerer},
  {Valtchanov}, {van der Werf}, {Walth}, {Zamojski}, \& {Zemcov}}]{RT14}
{Rawle}, T.~D., {Egami}, E., {Bussmann}, R.~S., {et~al.} 2014, \apj, 783, 59,
  \dodoi{10.1088/0004-637X/783/1/59}

\bibitem[{{Ren} {et~al.}(2023){Ren}, {Fudamoto}, {Inoue}, {Sugahara},
  {Tokuoka}, {Tamura}, {Matsuo}, {Kohno}, {Umehata}, {Hashimoto}, {Bouwens},
  {Smit}, {Kashikawa}, {Okamoto}, {Shibuya}, \& {Shimizu}}]{RY23}
{Ren}, Y.~W., {Fudamoto}, Y., {Inoue}, A.~K., {et~al.} 2023, \apj, 945, 69,
  \dodoi{10.3847/1538-4357/acb8ab}

\bibitem[{{Reuter} {et~al.}(2020){Reuter}, {Vieira}, {Spilker}, {Weiss},
  {Aravena}, {Archipley}, {B{\'e}thermin}, {Chapman}, {De Breuck}, {Dong},
  {Everett}, {Fu}, {Greve}, {Hayward}, {Hill}, {Hezaveh}, {Jarugula}, {Litke},
  {Malkan}, {Marrone}, {Narayanan}, {Phadke}, {Stark}, \& {Strandet}}]{RC20}
{Reuter}, C., {Vieira}, J.~D., {Spilker}, J.~S., {et~al.} 2020, \apj, 902, 78,
  \dodoi{10.3847/1538-4357/abb599}

\bibitem[{{Richard} {et~al.}(2011{\natexlab{a}}){Richard}, {Jones}, {Ellis},
  {Stark}, {Livermore}, \& {Swinbank}}]{RJ11b}
{Richard}, J., {Jones}, T., {Ellis}, R., {et~al.} 2011{\natexlab{a}}, \mnras,
  413, 643, \dodoi{10.1111/j.1365-2966.2010.18161.x}

\bibitem[{{Richard} {et~al.}(2011{\natexlab{b}}){Richard}, {Kneib}, {Ebeling},
  {Stark}, {Egami}, \& {Fiedler}}]{RJ11a}
{Richard}, J., {Kneib}, J.-P., {Ebeling}, H., {et~al.} 2011{\natexlab{b}},
  \mnras, 414, L31, \dodoi{10.1111/j.1745-3933.2011.01050.x}

\bibitem[{{Riechers} {et~al.}(2008){Riechers}, {Walter}, {Brewer}, {Carilli},
  {Lewis}, {Bertoldi}, \& {Cox}}]{RD08}
{Riechers}, D.~A., {Walter}, F., {Brewer}, B.~J., {et~al.} 2008, \apj, 686,
  851, \dodoi{10.1086/591434}

\bibitem[{{Riechers} {et~al.}(2009){Riechers}, {Walter}, {Carilli}, \&
  {Lewis}}]{RD09}
{Riechers}, D.~A., {Walter}, F., {Carilli}, C.~L., \& {Lewis}, G.~F. 2009,
  \apj, 690, 463, \dodoi{10.1088/0004-637X/690/1/463}

\bibitem[{{Riechers} {et~al.}(2013){Riechers}, {Bradford}, {Clements},
  {Dowell}, {P{\'e}rez-Fournon}, {Ivison}, {Bridge}, {Conley}, {Fu}, {Vieira},
  {Wardlow}, {Calanog}, {Cooray}, {Hurley}, {Neri}, {Kamenetzky}, {Aguirre},
  {Altieri}, {Arumugam}, {Benford}, {B{\'e}thermin}, {Bock}, {Burgarella},
  {Cabrera-Lavers}, {Chapman}, {Cox}, {Dunlop}, {Earle}, {Farrah}, {Ferrero},
  {Franceschini}, {Gavazzi}, {Glenn}, {Solares}, {Gurwell}, {Halpern},
  {Hatziminaoglou}, {Hyde}, {Ibar}, {Kov{\'a}cs}, {Krips}, {Lupu}, {Maloney},
  {Martinez-Navajas}, {Matsuhara}, {Murphy}, {Naylor}, {Nguyen}, {Oliver},
  {Omont}, {Page}, {Petitpas}, {Rangwala}, {Roseboom}, {Scott}, {Smith},
  {Staguhn}, {Streblyanska}, {Thomson}, {Valtchanov}, {Viero}, {Wang},
  {Zemcov}, \& {Zmuidzinas}}]{RD13}
{Riechers}, D.~A., {Bradford}, C.~M., {Clements}, D.~L., {et~al.} 2013, \nat,
  496, 329, \dodoi{10.1038/nature12050}

\bibitem[{{Riechers} {et~al.}(2014){Riechers}, {Carilli}, {Capak}, {Scoville},
  {Smol{\v{c}}i{\'c}}, {Schinnerer}, {Yun}, {Cox}, {Bertoldi}, {Karim}, \&
  {Yan}}]{RD14}
{Riechers}, D.~A., {Carilli}, C.~L., {Capak}, P.~L., {et~al.} 2014, \apj, 796,
  84, \dodoi{10.1088/0004-637X/796/2/84}

\bibitem[{{Riechers} {et~al.}(2020){Riechers}, {Hodge}, {Pavesi}, {Daddi},
  {Decarli}, {Ivison}, {Sharon}, {Smail}, {Walter}, {Aravena}, {Capak},
  {Carilli}, {Cox}, {Cunha}, {Dannerbauer}, {Dickinson}, {Neri}, \&
  {Wagg}}]{RD20}
{Riechers}, D.~A., {Hodge}, J.~A., {Pavesi}, R., {et~al.} 2020, \apj, 895, 81,
  \dodoi{10.3847/1538-4357/ab8c48}

\bibitem[{{Rieke} {et~al.}(2004){Rieke}, {Young}, {Engelbracht}, {Kelly},
  {Low}, {Haller}, {Beeman}, {Gordon}, {Stansberry}, {Misselt}, {Cadien},
  {Morrison}, {Rivlis}, {Latter}, {Noriega-Crespo}, {Padgett}, {Stapelfeldt},
  {Hines}, {Egami}, {Muzerolle}, {Alonso-Herrero}, {Blaylock}, {Dole}, {Hinz},
  {Le Floc'h}, {Papovich}, {P{\'e}rez-Gonz{\'a}lez}, {Smith}, {Su}, {Bennett},
  {Frayer}, {Henderson}, {Lu}, {Masci}, {Pesenson}, {Rebull}, {Rho}, {Keene},
  {Stolovy}, {Wachter}, {Wheaton}, {Werner}, \& {Richards}}]{rieke04}
{Rieke}, G.~H., {Young}, E.~T., {Engelbracht}, C.~W., {et~al.} 2004, \apjs,
  154, 25, \dodoi{10.1086/422717}

\bibitem[{{Rigby} {et~al.}(2008){Rigby}, {Marcillac}, {Egami}, {Rieke},
  {Richard}, {Kneib}, {Fadda}, {Willmer}, {Borys}, {van der Werf},
  {P{\'e}rez-Gonz{\'a}lez}, {Knudsen}, \& {Papovich}}]{RJ08}
{Rigby}, J.~R., {Marcillac}, D., {Egami}, E., {et~al.} 2008, \apj, 675, 262,
  \dodoi{10.1086/525273}

\bibitem[{{Rigopoulou} {et~al.}(2018){Rigopoulou}, {Pereira-Santaella},
  {Magdis}, {Cooray}, {Farrah}, {Marques-Chaves}, {Perez-Fournon}, \&
  {Riechers}}]{RD18}
{Rigopoulou}, D., {Pereira-Santaella}, M., {Magdis}, G.~E., {et~al.} 2018,
  \mnras, 473, 20, \dodoi{10.1093/mnras/stx2311}

\bibitem[{{Rizzo} {et~al.}(2021){Rizzo}, {Vegetti}, {Fraternali}, {Stacey}, \&
  {Powell}}]{RF21}
{Rizzo}, F., {Vegetti}, S., {Fraternali}, F., {Stacey}, H.~R., \& {Powell}, D.
  2021, \mnras, 507, 3952, \dodoi{10.1093/mnras/stab2295}

\bibitem[{{Rizzo} {et~al.}(2020){Rizzo}, {Vegetti}, {Powell}, {Fraternali},
  {McKean}, {Stacey}, \& {White}}]{RF20}
{Rizzo}, F., {Vegetti}, S., {Powell}, D., {et~al.} 2020, \nat, 584, 201,
  \dodoi{10.1038/s41586-020-2572-6}

\bibitem[{{Rojas-Ruiz} {et~al.}(2021){Rojas-Ruiz}, {Ba{\~n}ados}, {Neeleman},
  {Connor}, {Eilers}, {Venemans}, {Khusanova}, {Carilli}, {Mazzucchelli},
  {Decarli}, {Momjian}, \& {Novak}}]{RS21}
{Rojas-Ruiz}, S., {Ba{\~n}ados}, E., {Neeleman}, M., {et~al.} 2021, \apj, 920,
  150, \dodoi{10.3847/1538-4357/ac1a13}

\bibitem[{{Roman-Oliveira} {et~al.}(2023){Roman-Oliveira}, {Fraternali}, \&
  {Rizzo}}]{RF23}
{Roman-Oliveira}, F., {Fraternali}, F., \& {Rizzo}, F. 2023, \mnras, 521, 1045,
  \dodoi{10.1093/mnras/stad530}

\bibitem[{{Rooney} {et~al.}(2025){Rooney}, {Peng}, {Vishwas}, {Stacey},
  {Nikola}, {Lamarche}, {Ball}, {Ferkinhoff}, {Brisbin}, \&
  {Hailey-Dunsheath}}]{RC25}
{Rooney}, C., {Peng}, B., {Vishwas}, A., {et~al.} 2025, \apj, 987, 61,
  \dodoi{10.3847/1538-4357/add9a1}

\bibitem[{{Rosenberg} {et~al.}(2015){Rosenberg}, {van der Werf}, {Aalto},
  {Armus}, {Charmandaris}, {D{\'\i}az-Santos}, {Evans}, {Fischer}, {Gao},
  {Gonz{\'a}lez-Alfonso}, {Greve}, {Harris}, {Henkel}, {Israel}, {Isaak},
  {Kramer}, {Meijerink}, {Naylor}, {Sanders}, {Smith}, {Spaans}, {Spinoglio},
  {Stacey}, {Veenendaal}, {Veilleux}, {Walter}, {Wei{\ss}}, {Wiedner}, {van der
  Wiel}, \& {Xilouris}}]{R15}
{Rosenberg}, M.~J.~F., {van der Werf}, P.~P., {Aalto}, S., {et~al.} 2015, \apj,
  801, 72, \dodoi{10.1088/0004-637X/801/2/72}

\bibitem[{{Rowland} {et~al.}(2024){Rowland}, {Hodge}, {Bouwens}, {Mancera
  Pi{\~n}a}, {Hygate}, {Algera}, {Aravena}, {Bowler}, {da Cunha}, {Dayal},
  {Ferrara}, {Herard-Demanche}, {Inami}, {van Leeuwen}, {de Looze}, {Oesch},
  {Pallottini}, {Phillips}, {Rybak}, {Schouws}, {Smit}, {Sommovigo},
  {Stefanon}, \& {van der Werf}}]{RL24}
{Rowland}, L.~E., {Hodge}, J., {Bouwens}, R., {et~al.} 2024, \mnras, 535, 2068,
  \dodoi{10.1093/mnras/stae2217}

\bibitem[{{Rowland} {et~al.}(2025){Rowland}, {Stefanon}, {Bouwens}, {Hodge},
  {Algera}, {Fisher}, {Dayal}, {Pallottini}, {Stark}, {Heintz}, {Aravena},
  {Bowler}, {Cescon}, {Endsley}, {Ferrara}, {Gonzalez}, {Graziani}, {Gulis},
  {Herard-Demanche}, {Inami}, {Laza-Ramos}, {van Leeuwen}, {de Looze},
  {Nanayakkara}, {Oesch}, {Ormerod}, {Sartorio}, {Schouws}, {Smit},
  {Sommovigo}, {Toft}, {Weaver}, \& {van der Werf}}]{RL25}
{Rowland}, L.~E., {Stefanon}, M., {Bouwens}, R., {et~al.} 2025, arXiv e-prints,
  arXiv:2501.10559, \dodoi{10.48550/arXiv.2501.10559}

\bibitem[{{Roy} {et~al.}(2024){Roy}, {Heckman}, {Overzier}, {Saxena}, {Duncan},
  {Miley}, {Villar Mart{\'\i}n}, {Gab{\'a}nyi}, {Aydar}, {Bosman},
  {Rottgering}, {Pentericci}, {Onoue}, \& {Reynaldi}}]{RN24}
{Roy}, N., {Heckman}, T., {Overzier}, R., {et~al.} 2024, \apj, 970, 69,
  \dodoi{10.3847/1538-4357/ad4bda}

\bibitem[{{Rujopakarn} {et~al.}(2012){Rujopakarn}, {Rieke}, {Papovich},
  {Weiner}, {Rigby}, {Rex}, {Bian}, {Kuhn}, \& {Thompson}}]{RW12}
{Rujopakarn}, W., {Rieke}, G.~H., {Papovich}, C.~J., {et~al.} 2012, \apj, 755,
  168, \dodoi{10.1088/0004-637X/755/2/168}

\bibitem[{{Rybak} {et~al.}(2020{\natexlab{a}}){Rybak}, {Hodge}, {Vegetti}, {van
  der Werf}, {Andreani}, {Graziani}, \& {McKean}}]{RM20a}
{Rybak}, M., {Hodge}, J.~A., {Vegetti}, S., {et~al.} 2020{\natexlab{a}},
  \mnras, 494, 5542, \dodoi{10.1093/mnras/staa879}

\bibitem[{{Rybak} {et~al.}(2015{\natexlab{a}}){Rybak}, {McKean}, {Vegetti},
  {Andreani}, \& {White}}]{RM15a}
{Rybak}, M., {McKean}, J.~P., {Vegetti}, S., {Andreani}, P., \& {White},
  S.~D.~M. 2015{\natexlab{a}}, \mnras, 451, L40, \dodoi{10.1093/mnrasl/slv058}

\bibitem[{{Rybak} {et~al.}(2015{\natexlab{b}}){Rybak}, {Vegetti}, {McKean},
  {Andreani}, \& {White}}]{RM15b}
{Rybak}, M., {Vegetti}, S., {McKean}, J.~P., {Andreani}, P., \& {White},
  S.~D.~M. 2015{\natexlab{b}}, \mnras, 453, L26, \dodoi{10.1093/mnrasl/slv092}

\bibitem[{{Rybak} {et~al.}(2020{\natexlab{b}}){Rybak}, {Zavala}, {Hodge},
  {Casey}, \& {Werf}}]{RM20b}
{Rybak}, M., {Zavala}, J.~A., {Hodge}, J.~A., {Casey}, C.~M., \& {Werf}, P.
  v.~d. 2020{\natexlab{b}}, \apjl, 889, L11, \dodoi{10.3847/2041-8213/ab63de}

\bibitem[{{Rybak} {et~al.}(2019){Rybak}, {Calistro Rivera}, {Hodge}, {Smail},
  {Walter}, {van der Werf}, {da Cunha}, {Chen}, {Dannerbauer}, {Ivison},
  {Karim}, {Simpson}, {Swinbank}, \& {Wardlow}}]{RM19}
{Rybak}, M., {Calistro Rivera}, G., {Hodge}, J.~A., {et~al.} 2019, \apj, 876,
  112, \dodoi{10.3847/1538-4357/ab0e0f}

\bibitem[{{Rybak} {et~al.}(2021){Rybak}, {da Cunha}, {Groves}, {Hodge},
  {Aravena}, {Maseda}, {Boogaard}, {Berg}, {Charlot}, {Decarli}, {Erb},
  {Nelson}, {Pacifici}, {Schmidt}, {Walter}, \& {van der Wel}}]{RM21}
{Rybak}, M., {da Cunha}, E., {Groves}, B., {et~al.} 2021, \apj, 909, 130,
  \dodoi{10.3847/1538-4357/abd946}

\bibitem[{{Rybak} {et~al.}(2023){Rybak}, {Lemsom}, {Lundgren}, {Zavala},
  {Hodge}, {de Breuck}, {Casey}, {Decarli}, {Torstensson}, {Wardlow}, \& {van
  der Werf}}]{RM23}
{Rybak}, M., {Lemsom}, L., {Lundgren}, A., {et~al.} 2023, Research Notes of the
  American Astronomical Society, 7, 188, \dodoi{10.3847/2515-5172/acf579}

\bibitem[{{Saintonge} {et~al.}(2013){Saintonge}, {Lutz}, {Genzel}, {Magnelli},
  {Nordon}, {Tacconi}, {Baker}, {Bandara}, {Berta}, {F{\"o}rster Schreiber},
  {Poglitsch}, {Sturm}, {Wuyts}, \& {Wuyts}}]{SA13}
{Saintonge}, A., {Lutz}, D., {Genzel}, R., {et~al.} 2013, \apj, 778, 2,
  \dodoi{10.1088/0004-637X/778/1/2}

\bibitem[{{Salak} {et~al.}(2024){Salak}, {Hashimoto}, {Inoue}, {Bakx},
  {Donevski}, {Tamura}, {Sugahara}, {Kuno}, {Miyamoto}, {Fujimoto}, \&
  {Suphapolthaworn}}]{SD24}
{Salak}, D., {Hashimoto}, T., {Inoue}, A.~K., {et~al.} 2024, \apj, 962, 1,
  \dodoi{10.3847/1538-4357/ad0df5}

\bibitem[{{Salim} \& {Narayanan}(2020)}]{salim20}
{Salim}, S., \& {Narayanan}, D. 2020, \araa, 58, 529,
  \dodoi{10.1146/annurev-astro-032620-021933}

\bibitem[{{Samsonyan} {et~al.}(2016){Samsonyan}, {Weedman}, {Lebouteiller},
  {Barry}, \& {Sargsyan}}]{S16}
{Samsonyan}, A., {Weedman}, D., {Lebouteiller}, V., {Barry}, D., \& {Sargsyan},
  L. 2016, \apjs, 226, 11, \dodoi{10.3847/0067-0049/226/1/11}

\bibitem[{{Sanders} {et~al.}(2003){Sanders}, {Mazzarella}, {Kim}, {Surace}, \&
  {Soifer}}]{S03}
{Sanders}, D.~B., {Mazzarella}, J.~M., {Kim}, D.~C., {Surace}, J.~A., \&
  {Soifer}, B.~T. 2003, \aj, 126, 1607, \dodoi{10.1086/376841}

\bibitem[{{Sanders} \& {Mirabel}(1996)}]{sanders96}
{Sanders}, D.~B., \& {Mirabel}, I.~F. 1996, \araa, 34, 749,
  \dodoi{10.1146/annurev.astro.34.1.749}

\bibitem[{{Sawamura} {et~al.}(2025){Sawamura}, {Izumi}, {Nakanishi}, {Okuda},
  {Strauss}, {Imanishi}, {Matsuoka}, {Toba}, {Umehata}, {Hashimoto}, {Baba},
  {Goto}, {Kawaguchi}, {Kohno}, {Salak}, {Kawamuro}, {Iwasawa}, {Onoue}, {Lee},
  \& {Lee}}]{SM25a}
{Sawamura}, M., {Izumi}, T., {Nakanishi}, K., {et~al.} 2025, arXiv e-prints,
  arXiv:2502.16858, \dodoi{10.48550/arXiv.2502.16858}

\bibitem[{{Saxena} {et~al.}(2024){Saxena}, {Overzier}, {Villar-Mart{\'\i}n},
  {Heckman}, {Roy}, {Duncan}, {R{\"o}ttgering}, {Miley}, {Aydar}, {Best},
  {Bosman}, {Cameron}, {Gab{\'a}nyi}, {Humphrey}, {Morais}, {Onoue},
  {Pentericci}, {Reynaldi}, \& {Venemans}}]{SA24}
{Saxena}, A., {Overzier}, R.~A., {Villar-Mart{\'\i}n}, M., {et~al.} 2024,
  \mnras, 531, 4391, \dodoi{10.1093/mnras/stae1406}

\bibitem[{{Schaerer} {et~al.}(2015{\natexlab{a}}){Schaerer}, {Boone},
  {Zamojski}, {Staguhn}, {Dessauges-Zavadsky}, {Finkelstein}, \&
  {Combes}}]{SD15b}
{Schaerer}, D., {Boone}, F., {Zamojski}, M., {et~al.} 2015{\natexlab{a}}, \aap,
  574, A19, \dodoi{10.1051/0004-6361/201424649}

\bibitem[{{Schaerer} {et~al.}(2015{\natexlab{b}}){Schaerer}, {Boone}, {Jones},
  {Dessauges-Zavadsky}, {Sklias}, {Zamojski}, {Cava}, {Richard}, {Ellis},
  {Rawle}, {Egami}, \& {Combes}}]{SD15a}
{Schaerer}, D., {Boone}, F., {Jones}, T., {et~al.} 2015{\natexlab{b}}, \aap,
  576, L2, \dodoi{10.1051/0004-6361/201425542}

\bibitem[{{Schenker} {et~al.}(2012){Schenker}, {Stark}, {Ellis}, {Robertson},
  {Dunlop}, {McLure}, {Kneib}, \& {Richard}}]{SM12}
{Schenker}, M.~A., {Stark}, D.~P., {Ellis}, R.~S., {et~al.} 2012, \apj, 744,
  179, \dodoi{10.1088/0004-637X/744/2/179}

\bibitem[{{Scholtz} {et~al.}(2025){Scholtz}, {Curti}, {D'Eugenio}, {{\"U}bler},
  {Maiolino}, {Marconcini}, {Smit}, {Perna}, {Witstok}, {Arribas}, {B{\"o}ker},
  {Bunker}, {Carniani}, {Charlot}, {Cresci}, {Lamperti}, {Parlanti},
  {P{\'e}rez-Gonz{\'a}lez}, {Rodr{\'\i}guez Del Pino}, \& {Venturi}}]{SJ25}
{Scholtz}, J., {Curti}, M., {D'Eugenio}, F., {et~al.} 2025, \mnras, 539, 2463,
  \dodoi{10.1093/mnras/staf518}

\bibitem[{{Schouws} {et~al.}(2022){Schouws}, {Stefanon}, {Bouwens}, {Smit},
  {Hodge}, {Labb{\'e}}, {Algera}, {Boogaard}, {Carniani}, {Fudamoto},
  {Holwerda}, {Illingworth}, {Maiolino}, {Maseda}, {Oesch}, \& {van der
  Werf}}]{SS22}
{Schouws}, S., {Stefanon}, M., {Bouwens}, R., {et~al.} 2022, \apj, 928, 31,
  \dodoi{10.3847/1538-4357/ac4605}

\bibitem[{{Schouws} {et~al.}(2023){Schouws}, {Bouwens}, {Smit}, {Hodge},
  {Stefanon}, {Witstok}, {Hilhorst}, {Labb{\'e}}, {Algera}, {Boogaard},
  {Maseda}, {Oesch}, {R{\"o}ttgering}, \& {van der Werf}}]{SS23}
{Schouws}, S., {Bouwens}, R., {Smit}, R., {et~al.} 2023, \apj, 954, 103,
  \dodoi{10.3847/1538-4357/ace10c}

\bibitem[{{Schouws} {et~al.}(2024){Schouws}, {Bouwens}, {Ormerod}, {Smit},
  {Algera}, {Sommovigo}, {Hodge}, {Ferrara}, {Oesch}, {Rowland}, {van Leeuwen},
  {Stefanon}, {Herard-Demanche}, {Fudamoto}, {R{\"o}ttgering}, \& {van der
  Werf}}]{SS24}
{Schouws}, S., {Bouwens}, R.~J., {Ormerod}, K., {et~al.} 2024, arXiv e-prints,
  arXiv:2409.20549, \dodoi{10.48550/arXiv.2409.20549}

\bibitem[{{Schouws} {et~al.}(2025){Schouws}, {Bouwens}, {Algera}, {Smit},
  {Kumari}, {Rowland}, {van Leeuwen}, {Sommovigo}, {Ferrara}, {Oesch},
  {Ormerod}, {Stefanon}, {Herard-Demanche}, {Hodge}, {Fudamoto},
  {R{\"o}ttgering}, \& {van der Werf}}]{SS25}
{Schouws}, S., {Bouwens}, R.~J., {Algera}, H., {et~al.} 2025, arXiv e-prints,
  arXiv:2502.01610, \dodoi{10.48550/arXiv.2502.01610}

\bibitem[{{Schreiber} {et~al.}(2018){Schreiber}, {Labb{\'e}}, {Glazebrook},
  {Bekiaris}, {Papovich}, {Costa}, {Elbaz}, {Kacprzak}, {Nanayakkara}, {Oesch},
  {Pannella}, {Spitler}, {Straatman}, {Tran}, \& {Wang}}]{SC18}
{Schreiber}, C., {Labb{\'e}}, I., {Glazebrook}, K., {et~al.} 2018, \aap, 611,
  A22, \dodoi{10.1051/0004-6361/201731917}

\bibitem[{{Schreiber} {et~al.}(2021){Schreiber}, {Glazebrook}, {Papovich},
  {D{\'\i}az-Santos}, {Verma}, {Elbaz}, {Kacprzak}, {Nanayakkara}, {Oesch},
  {Pannella}, {Spitler}, {Straatman}, {Tran}, \& {Wang}}]{SC21}
{Schreiber}, C., {Glazebrook}, K., {Papovich}, C., {et~al.} 2021, \aap, 646,
  A68, \dodoi{10.1051/0004-6361/201936460}

\bibitem[{{Seitz} {et~al.}(1998){Seitz}, {Saglia}, {Bender}, {Hopp}, {Belloni},
  \& {Ziegler}}]{SS98}
{Seitz}, S., {Saglia}, R.~P., {Bender}, R., {et~al.} 1998, \mnras, 298, 945,
  \dodoi{10.1046/j.1365-8711.1998.01443.x}

\bibitem[{{Shao} {et~al.}(2017){Shao}, {Wang}, {Jones}, {Carilli}, {Walter},
  {Fan}, {Riechers}, {Bertoldi}, {Wagg}, {Strauss}, {Omont}, {Cox}, {Jiang},
  {Narayanan}, \& {Menten}}]{SY17}
{Shao}, Y., {Wang}, R., {Jones}, G.~C., {et~al.} 2017, \apj, 845, 138,
  \dodoi{10.3847/1538-4357/aa826c}

\bibitem[{{Shao} {et~al.}(2019){Shao}, {Wang}, {Carilli}, {Wagg}, {Walter},
  {Li}, {Fan}, {Jiang}, {Riechers}, {Bertoldi}, {Strauss}, {Cox}, {Omont}, \&
  {Menten}}]{SY19}
{Shao}, Y., {Wang}, R., {Carilli}, C.~L., {et~al.} 2019, \apj, 876, 99,
  \dodoi{10.3847/1538-4357/ab133d}

\bibitem[{{Shao} {et~al.}(2022){Shao}, {Wang}, {Weiss}, {Wagg}, {Carilli},
  {Strauss}, {Walter}, {Cox}, {Fan}, {Menten}, {Narayanan}, {Riechers},
  {Bertoldi}, {Omont}, \& {Jiang}}]{SY22}
{Shao}, Y., {Wang}, R., {Weiss}, A., {et~al.} 2022, \aap, 668, A121,
  \dodoi{10.1051/0004-6361/202244610}

\bibitem[{{Sharda} {et~al.}(2018){Sharda}, {Federrath}, {da Cunha}, {Swinbank},
  \& {Dye}}]{SP18}
{Sharda}, P., {Federrath}, C., {da Cunha}, E., {Swinbank}, A.~M., \& {Dye}, S.
  2018, \mnras, 477, 4380, \dodoi{10.1093/mnras/sty886}

\bibitem[{{Sharon} {et~al.}(2019){Sharon}, {Tagore}, {Baker}, {Rivera},
  {Keeton}, {Lutz}, {Genzel}, {Wilner}, {Hicks}, {Allam}, \& {Tucker}}]{SC19}
{Sharon}, C.~E., {Tagore}, A.~S., {Baker}, A.~J., {et~al.} 2019, \apj, 879, 52,
  \dodoi{10.3847/1538-4357/ab22b9}

\bibitem[{{Simpson} {et~al.}(2020){Simpson}, {Smail},
  {Dudzevi{\v{c}}i{\={u}}t{\.{e}}}, {Matsuda}, {Hsieh}, {Wang}, {Swinbank},
  {Stach}, {An}, {Birkin}, {Ao}, {Bunker}, {Chapman}, {Chen}, {Coppin},
  {Ikarashi}, {Ivison}, {Mitsuhashi}, {Saito}, {Umehata}, {Wang}, \&
  {Zhao}}]{SJ20}
{Simpson}, J.~M., {Smail}, I., {Dudzevi{\v{c}}i{\={u}}t{\.{e}}}, U., {et~al.}
  2020, \mnras, 495, 3409, \dodoi{10.1093/mnras/staa1345}

\bibitem[{{Simpson}(1975)}]{simpson75}
{Simpson}, J.~P. 1975, \aap, 39, 43

\bibitem[{{Smail} {et~al.}(1997){Smail}, {Ivison}, \& {Blain}}]{smail97}
{Smail}, I., {Ivison}, R.~J., \& {Blain}, A.~W. 1997, \apjl, 490, L5,
  \dodoi{10.1086/311017}

\bibitem[{{Smail} {et~al.}(2005){Smail}, {Smith}, \& {Ivison}}]{SI05}
{Smail}, I., {Smith}, G.~P., \& {Ivison}, R.~J. 2005, \apj, 631, 121,
  \dodoi{10.1086/432641}

\bibitem[{{Smail} {et~al.}(2023){Smail}, {Dudzevi{\v{c}}i{\={u}}t{\.{e}}},
  {Gurwell}, {Fazio}, {Willner}, {Swinbank}, {Arumugam}, {Summers}, {Cohen},
  {Jansen}, {Windhorst}, {Meena}, {Zitrin}, {Keel}, {Cheng}, {Coe},
  {Conselice}, {D'Silva}, {Driver}, {Frye}, {Grogin}, {Koekemoer}, {Marshall},
  {Nonino}, {Pirzkal}, {Robotham}, {Rutkowski}, {Ryan}, {Tompkins}, {Willmer},
  {Yan}, {Broadhurst}, {Diego}, {Kamieneski}, \& {Yun}}]{SI23}
{Smail}, I., {Dudzevi{\v{c}}i{\={u}}t{\.{e}}}, U., {Gurwell}, M., {et~al.}
  2023, \apj, 958, 36, \dodoi{10.3847/1538-4357/acf931}

\bibitem[{{Smit} {et~al.}(2018){Smit}, {Bouwens}, {Carniani}, {Oesch},
  {Labb{\'e}}, {Illingworth}, {van der Werf}, {Bradley}, {Gonzalez}, {Hodge},
  {Holwerda}, {Maiolino}, \& {Zheng}}]{SR18}
{Smit}, R., {Bouwens}, R.~J., {Carniani}, S., {et~al.} 2018, \nat, 553, 178,
  \dodoi{10.1038/nature24631}

\bibitem[{{Smol{\v{c}}i{\'c}} {et~al.}(2015){Smol{\v{c}}i{\'c}}, {Karim},
  {Miettinen}, {Novak}, {Magnelli}, {Riechers}, {Schinnerer}, {Capak}, {Bondi},
  {Ciliegi}, {Aravena}, {Bertoldi}, {Bourke}, {Banfield}, {Carilli}, {Civano},
  {Ilbert}, {Intema}, {Le F{\`e}vre}, {Finoguenov}, {Hallinan}, {Kl{\"o}ckner},
  {Koekemoer}, {Laigle}, {Masters}, {McCracken}, {Mooley}, {Murphy},
  {Navarette}, {Salvato}, {Sargent}, {Sheth}, {Toft}, \& {Zamorani}}]{SV15}
{Smol{\v{c}}i{\'c}}, V., {Karim}, A., {Miettinen}, O., {et~al.} 2015, \aap,
  576, A127, \dodoi{10.1051/0004-6361/201424996}

\bibitem[{{Solimano} {et~al.}(2024){Solimano}, {Gonz{\'a}lez-L{\'o}pez},
  {Aravena}, {Herrera-Camus}, {De Looze}, {F{\"o}rster Schreiber}, {Spilker},
  {Tadaki}, {Assef}, {Barcos-Mu{\~n}oz}, {Davies}, {D{\'\i}az-Santos},
  {Ferrara}, {Fisher}, {Guaita}, {Ikeda}, {Johnston}, {Lutz}, {Mitsuhashi},
  {Moya-Sierralta}, {Rela{\~n}o}, {Naab}, {Posses}, {Telikova}, {{\"U}bler},
  {van der Giessen}, {Veilleux}, \& {Villanueva}}]{SM24}
{Solimano}, M., {Gonz{\'a}lez-L{\'o}pez}, J., {Aravena}, M., {et~al.} 2024,
  \aap, 689, A145, \dodoi{10.1051/0004-6361/202449192}

\bibitem[{{Solimano} {et~al.}(2025){Solimano}, {Gonz{\'a}lez-L{\'o}pez},
  {Aravena}, {Alcalde Pampliega}, {Assef}, {B{\'e}thermin}, {Boquien},
  {Bovino}, {Casey}, {Cassata}, {da Cunha}, {Davies}, {De Looze}, {Ding},
  {D{\'\i}az-Santos}, {Faisst}, {Ferrara}, {Fisher}, {F{\"o}rster-Schreiber},
  {Fujimoto}, {Ginolfi}, {Gruppioni}, {Guaita}, {Hathi}, {Herrera-Camus},
  {Ibar}, {Inami}, {Jones}, {Koekemoer}, {Lee}, {Li}, {Liu}, {Liu}, {Molina},
  {Ogle}, {Posses}, {Pozzi}, {Rela{\~n}o}, {Riechers}, {Romano}, {Spilker},
  {Sulzenauer}, {Telikova}, {Vallini}, {Vasan}, {Veilleux}, {Vergani},
  {Villanueva}, {Wang}, {Yan}, \& {Zamorani}}]{SM25b}
---. 2025, \aap, 693, A70, \dodoi{10.1051/0004-6361/202451551}

\bibitem[{{Sommariva} {et~al.}(2012){Sommariva}, {Mannucci}, {Cresci},
  {Maiolino}, {Marconi}, {Nagao}, {Baroni}, \& {Grazian}}]{SV12}
{Sommariva}, V., {Mannucci}, F., {Cresci}, G., {et~al.} 2012, \aap, 539, A136,
  \dodoi{10.1051/0004-6361/201118134}

\bibitem[{{Sommovigo} {et~al.}(2022){Sommovigo}, {Ferrara}, {Pallottini},
  {Dayal}, {Bouwens}, {Smit}, {da Cunha}, {De Looze}, {Bowler}, {Hodge},
  {Inami}, {Oesch}, {Endsley}, {Gonzalez}, {Schouws}, {Stark}, {Stefanon},
  {Aravena}, {Graziani}, {Riechers}, {Schneider}, {van der Werf}, {Algera},
  {Barrufet}, {Fudamoto}, {Hygate}, {Labb{\'e}}, {Li}, {Nanayakkara}, \&
  {Topping}}]{SL22}
{Sommovigo}, L., {Ferrara}, A., {Pallottini}, A., {et~al.} 2022, \mnras, 513,
  3122, \dodoi{10.1093/mnras/stac302}

\bibitem[{{Spilker} {et~al.}(2016){Spilker}, {Marrone}, {Aravena},
  {B{\'e}thermin}, {Bothwell}, {Carlstrom}, {Chapman}, {Crawford}, {de Breuck},
  {Fassnacht}, {Gonzalez}, {Greve}, {Hezaveh}, {Litke}, {Ma}, {Malkan},
  {Rotermund}, {Strandet}, {Vieira}, {Weiss}, \& {Welikala}}]{SJ16}
{Spilker}, J.~S., {Marrone}, D.~P., {Aravena}, M., {et~al.} 2016, \apj, 826,
  112, \dodoi{10.3847/0004-637X/826/2/112}

\bibitem[{{Spinoglio} {et~al.}(2015){Spinoglio}, {Pereira-Santaella}, {Dasyra},
  {Calzoletti}, {Malkan}, {Tommasin}, \& {Busquet}}]{S15}
{Spinoglio}, L., {Pereira-Santaella}, M., {Dasyra}, K.~M., {et~al.} 2015, \apj,
  799, 21, \dodoi{10.1088/0004-637X/799/1/21}

\bibitem[{{Spinoglio} {et~al.}(2022){Spinoglio}, {Fern{\'a}ndez-Ontiveros},
  {Malkan}, {Kumar}, {Pereira-Santaella}, {P{\'e}rez-D{\'\i}az},
  {P{\'e}rez-Montero}, {Krabbe}, {Vacca}, {Colditz}, \& {Fischer}}]{S22}
{Spinoglio}, L., {Fern{\'a}ndez-Ontiveros}, J.~A., {Malkan}, M.~A., {et~al.}
  2022, \apj, 926, 55, \dodoi{10.3847/1538-4357/ac37b7}

\bibitem[{{Spoon} {et~al.}(2022){Spoon}, {Hern{\'a}n-Caballero}, {Rupke},
  {Waters}, {Lebouteiller}, {Tielens}, {Loredo}, {Su}, \& {Viola}}]{spoon22}
{Spoon}, H.~W.~W., {Hern{\'a}n-Caballero}, A., {Rupke}, D., {et~al.} 2022,
  \apjs, 259, 37, \dodoi{10.3847/1538-4365/ac4989}

\bibitem[{{Springob} {et~al.}(2005){Springob}, {Haynes}, {Giovanelli}, \&
  {Kent}}]{S05}
{Springob}, C.~M., {Haynes}, M.~P., {Giovanelli}, R., \& {Kent}, B.~R. 2005,
  \apjs, 160, 149, \dodoi{10.1086/431550}

\bibitem[{{Stacey}(2011)}]{stacey11}
{Stacey}, G.~J. 2011, IEEE Transactions on Terahertz Science and Technology, 1,
  241, \dodoi{10.1109/TTHZ.2011.2159649}

\bibitem[{{Stacey} {et~al.}(1991){Stacey}, {Geis}, {Genzel}, {Lugten},
  {Poglitsch}, {Sternberg}, \& {Townes}}]{stacey91}
{Stacey}, G.~J., {Geis}, N., {Genzel}, R., {et~al.} 1991, \apj, 373, 423,
  \dodoi{10.1086/170062}

\bibitem[{{Stacey} {et~al.}(2010){Stacey}, {Hailey-Dunsheath}, {Ferkinhoff},
  {Nikola}, {Parshley}, {Benford}, {Staguhn}, \& {Fiolet}}]{SG10}
{Stacey}, G.~J., {Hailey-Dunsheath}, S., {Ferkinhoff}, C., {et~al.} 2010, \apj,
  724, 957, \dodoi{10.1088/0004-637X/724/2/957}

\bibitem[{{Stacey} {et~al.}(1983){Stacey}, {Smyers}, {Kurtz}, \&
  {Harwit}}]{stacey83}
{Stacey}, G.~J., {Smyers}, S.~D., {Kurtz}, N.~T., \& {Harwit}, M. 1983, \apjl,
  265, L7, \dodoi{10.1086/183948}

\bibitem[{{Stacey} {et~al.}(2018){Stacey}, {McKean}, {Robertson}, {Ivison},
  {Isaak}, {Schleicher}, {van der Werf}, {Baan}, {Berciano Alba}, {Garrett}, \&
  {Loenen}}]{SH18}
{Stacey}, H.~R., {McKean}, J.~P., {Robertson}, N.~C., {et~al.} 2018, \mnras,
  476, 5075, \dodoi{10.1093/mnras/sty458}

\bibitem[{{Stark} {et~al.}(2015){Stark}, {Richard}, {Charlot}, {Cl{\'e}ment},
  {Ellis}, {Siana}, {Robertson}, {Schenker}, {Gutkin}, \& {Wofford}}]{SD15c}
{Stark}, D.~P., {Richard}, J., {Charlot}, S., {et~al.} 2015, \mnras, 450, 1846,
  \dodoi{10.1093/mnras/stv688}

\bibitem[{{Stasinska}(2019)}]{stasinska19}
{Stasinska}, G. 2019, arXiv e-prints, arXiv:1906.04520,
  \dodoi{10.48550/arXiv.1906.04520}

\bibitem[{{Stiavelli} {et~al.}(2023){Stiavelli}, {Morishita}, {Chiaberge},
  {Grillo}, {Leethochawalit}, {Rosati}, {Schuldt}, {Trenti}, \& {Treu}}]{SM23}
{Stiavelli}, M., {Morishita}, T., {Chiaberge}, M., {et~al.} 2023, \apjl, 957,
  L18, \dodoi{10.3847/2041-8213/ad0159}

\bibitem[{{Strandet} {et~al.}(2016){Strandet}, {Weiss}, {Vieira}, {de Breuck},
  {Aguirre}, {Aravena}, {Ashby}, {B{\'e}thermin}, {Bradford}, {Carlstrom},
  {Chapman}, {Crawford}, {Everett}, {Fassnacht}, {Furstenau}, {Gonzalez},
  {Greve}, {Gullberg}, {Hezaveh}, {Kamenetzky}, {Litke}, {Ma}, {Malkan},
  {Marrone}, {Menten}, {Murphy}, {Nadolski}, {Rotermund}, {Spilker}, {Stark},
  \& {Welikala}}]{SM16}
{Strandet}, M.~L., {Weiss}, A., {Vieira}, J.~D., {et~al.} 2016, \apj, 822, 80,
  \dodoi{10.3847/0004-637X/822/2/80}

\bibitem[{{Strandet} {et~al.}(2017){Strandet}, {Weiss}, {De Breuck}, {Marrone},
  {Vieira}, {Aravena}, {Ashby}, {B{\'e}thermin}, {Bothwell}, {Bradford},
  {Carlstrom}, {Chapman}, {Cunningham}, {Chen}, {Fassnacht}, {Gonzalez},
  {Greve}, {Gullberg}, {Hayward}, {Hezaveh}, {Litke}, {Ma}, {Malkan}, {Menten},
  {Miller}, {Murphy}, {Narayanan}, {Phadke}, {Rotermund}, {Spilker}, \&
  {Sreevani}}]{SM17}
{Strandet}, M.~L., {Weiss}, A., {De Breuck}, C., {et~al.} 2017, \apjl, 842,
  L15, \dodoi{10.3847/2041-8213/aa74b0}

\bibitem[{{Sturm} {et~al.}(2010){Sturm}, {Verma}, {Graci{\'a}-Carpio},
  {Hailey-Dunsheath}, {Contursi}, {Fischer}, {Gonz{\'a}lez-Alfonso},
  {Poglitsch}, {Sternberg}, {Genzel}, {Lutz}, {Tacconi}, {Christopher}, \& {de
  Jong}}]{SE10}
{Sturm}, E., {Verma}, A., {Graci{\'a}-Carpio}, J., {et~al.} 2010, \aap, 518,
  L36, \dodoi{10.1051/0004-6361/201014560}

\bibitem[{{Stutzki} {et~al.}(1988){Stutzki}, {Stacey}, {Genzel}, {Harris},
  {Jaffe}, \& {Lugten}}]{stuzki88}
{Stutzki}, J., {Stacey}, G.~J., {Genzel}, R., {et~al.} 1988, \apj, 332, 379,
  \dodoi{10.1086/166663}

\bibitem[{{Sugahara} {et~al.}(2021){Sugahara}, {Inoue}, {Hashimoto},
  {Yamanaka}, {Fujimoto}, {Tamura}, {Matsuo}, {Binggeli}, \&
  {Zackrisson}}]{SY21}
{Sugahara}, Y., {Inoue}, A.~K., {Hashimoto}, T., {et~al.} 2021, \apj, 923, 5,
  \dodoi{10.3847/1538-4357/ac2a36}

\bibitem[{{Sugahara} {et~al.}(2025){Sugahara}, {{\'A}lvarez-M{\'a}rquez},
  {Hashimoto}, {Colina}, {Inoue}, {Costantin}, {Fudamoto}, {Mawatari}, {Ren},
  {Arribas}, {Bakx}, {Blanco-Prieto}, {Ceverino}, {Crespo G{\'o}mez},
  {Hagimoto}, {Hashigaya}, {Marques-Chaves}, {Matsuo}, {Nakazato},
  {Pereira-Santaella}, {Tamura}, {Usui}, \& {Yoshida}}]{SY25}
{Sugahara}, Y., {{\'A}lvarez-M{\'a}rquez}, J., {Hashimoto}, T., {et~al.} 2025,
  \apj, 981, 135, \dodoi{10.3847/1538-4357/adb02a}

\bibitem[{{Sun} {et~al.}(2024){Sun}, {Helton}, {Egami}, {Hainline}, {Rieke},
  {Willmer}, {Eisenstein}, {Johnson}, {Rieke}, {Robertson}, {Tacchella},
  {Alberts}, {Baker}, {Bhatawdekar}, {Boyett}, {Bunker}, {Charlot}, {Chen},
  {Chevallard}, {Curtis-Lake}, {Danhaive}, {DeCoursey}, {Ji}, {Lyu},
  {Maiolino}, {Rujopakarn}, {Sandles}, {Shivaei}, {{\"U}bler}, {Willott}, \&
  {Witstok}}]{SF24}
{Sun}, F., {Helton}, J.~M., {Egami}, E., {et~al.} 2024, \apj, 961, 69,
  \dodoi{10.3847/1538-4357/ad07e3}

\bibitem[{{Sun} {et~al.}(2025){Sun}, {Yang}, {Wang}, {Eisenstein}, {Decarli},
  {Fan}, {Rieke}, {Ba{\~n}ados}, {Bosman}, {Cai}, {Champagne}, {Colina},
  {D'Eugenio}, {Fudamoto}, {Li}, {Lin}, {Liu}, {Lyu}, {Mazzucchelli}, {Jin},
  {Jun}, {Wu}, \& {Zhang}}]{SF25}
{Sun}, F., {Yang}, J., {Wang}, F., {et~al.} 2025, arXiv e-prints,
  arXiv:2506.06418, \dodoi{10.48550/arXiv.2506.06418}

\bibitem[{{Sutherland} {et~al.}(1993){Sutherland}, {Bicknell}, \&
  {Dopita}}]{sutherland93}
{Sutherland}, R.~S., {Bicknell}, G.~V., \& {Dopita}, M.~A. 1993, \apj, 414,
  510, \dodoi{10.1086/173099}

\bibitem[{{Sutter} {et~al.}(2019){Sutter}, {Dale}, {Croxall}, {Pelligrini},
  {Smith}, {Appleton}, {Beir{\~a}o}, {Bolatto}, {Calzetti}, {Crocker}, {De
  Looze}, {Draine}, {Galametz}, {Groves}, {Helou}, {Herrera-Camus}, {Hunt},
  {Kennicutt}, {Roussel}, \& {Wolfire}}]{S19}
{Sutter}, J., {Dale}, D.~A., {Croxall}, K.~V., {et~al.} 2019, \apj, 886, 60,
  \dodoi{10.3847/1538-4357/ab4da5}

\bibitem[{{Swinbank} {et~al.}(2010){Swinbank}, {Smail}, {Longmore}, {Harris},
  {Baker}, {De Breuck}, {Richard}, {Edge}, {Ivison}, {Blundell}, {Coppin},
  {Cox}, {Gurwell}, {Hainline}, {Krips}, {Lundgren}, {Neri}, {Siana},
  {Siringo}, {Stark}, {Wilner}, \& {Younger}}]{SA10}
{Swinbank}, A.~M., {Smail}, I., {Longmore}, S., {et~al.} 2010, \nat, 464, 733,
  \dodoi{10.1038/nature08880}

\bibitem[{{Swinbank} {et~al.}(2011){Swinbank}, {Papadopoulos}, {Cox}, {Krips},
  {Ivison}, {Smail}, {Thomson}, {Neri}, {Richard}, \& {Ebeling}}]{SA11}
{Swinbank}, A.~M., {Papadopoulos}, P.~P., {Cox}, P., {et~al.} 2011, \apj, 742,
  11, \dodoi{10.1088/0004-637X/742/1/11}

\bibitem[{{Swinbank} {et~al.}(2012){Swinbank}, {Karim}, {Smail}, {Hodge},
  {Walter}, {Bertoldi}, {Biggs}, {de Breuck}, {Chapman}, {Coppin}, {Cox},
  {Danielson}, {Dannerbauer}, {Ivison}, {Greve}, {Knudsen}, {Menten},
  {Simpson}, {Schinnerer}, {Wardlow}, {Wei{\ss}}, \& {van der Werf}}]{SA12}
{Swinbank}, A.~M., {Karim}, A., {Smail}, I., {et~al.} 2012, \mnras, 427, 1066,
  \dodoi{10.1111/j.1365-2966.2012.22048.x}

\bibitem[{{Tacchella} {et~al.}(2023){Tacchella}, {Eisenstein}, {Hainline},
  {Johnson}, {Baker}, {Helton}, {Robertson}, {Suess}, {Chen}, {Nelson},
  {Pusk{\'a}s}, {Sun}, {Alberts}, {Egami}, {Hausen}, {Rieke}, {Rieke},
  {Shivaei}, {Williams}, {Willmer}, {Bunker}, {Cameron}, {Carniani}, {Charlot},
  {Curti}, {Curtis-Lake}, {Looser}, {Maiolino}, {Maseda}, {Rawle}, {Rix},
  {Smit}, {{\"U}bler}, {Willott}, {Witstok}, {Baum}, {Bhatawdekar}, {Boyett},
  {Danhaive}, {de Graaff}, {Endsley}, {Ji}, {Lyu}, {Sandles}, {Saxena},
  {Scholtz}, {Topping}, \& {Whitler}}]{TS23}
{Tacchella}, S., {Eisenstein}, D.~J., {Hainline}, K., {et~al.} 2023, \apj, 952,
  74, \dodoi{10.3847/1538-4357/acdbc6}

\bibitem[{{Tadaki} {et~al.}(2019){Tadaki}, {Iono}, {Hatsukade}, {Kohno}, {Lee},
  {Matsuda}, {Michiyama}, {Nakanishi}, {Nagao}, {Saito}, {Tamura}, {Ueda}, \&
  {Umehata}}]{TK19}
{Tadaki}, K.-i., {Iono}, D., {Hatsukade}, B., {et~al.} 2019, \apj, 876, 1,
  \dodoi{10.3847/1538-4357/ab1415}

\bibitem[{{Tadaki} {et~al.}(2022){Tadaki}, {Tsujita}, {Tamura}, {Kohno},
  {Hatsukade}, {Iono}, {Lee}, {Matsuda}, {Michiyama}, {Nagao}, {Nakanishi},
  {Nishimura}, {Saito}, {Umehata}, \& {Zavala}}]{TK22}
{Tadaki}, K.-i., {Tsujita}, A., {Tamura}, Y., {et~al.} 2022, \pasj, 74, L9,
  \dodoi{10.1093/pasj/psac018}

\bibitem[{{Tadhunter} {et~al.}(1989){Tadhunter}, {Robinson}, \&
  {Morganti}}]{tadhunter89}
{Tadhunter}, C.~N., {Robinson}, A., \& {Morganti}, R. 1989, in European
  Southern Observatory Conference and Workshop Proceedings, Vol.~32, European
  Southern Observatory Conference and Workshop Proceedings, 293

\bibitem[{{Tamura} {et~al.}(2020){Tamura}, {Mawatari}, {Hashimoto}, {Inoue},
  {Zackrissonm}, {Christensen}, {Binggeli}, {Matsuda}, {Matsuo}, {Takeuchi},
  {Asano}, {Sunaga}, {Shimizu}, {Okamoto}, {Yoshida}, {Lee}, {Shibuya},
  {Taniguchi}, {Umehata}, {Hatsukade}, {Kohno}, \& {Ota}}]{TY20}
{Tamura}, Y., {Mawatari}, K., {Hashimoto}, T., {et~al.} 2020, in IAU Symposium,
  Vol. 341, Panchromatic Modelling with Next Generation Facilities, ed.
  M.~{Boquien}, E.~{Lusso}, C.~{Gruppioni}, \& P.~{Tissera}, 211--215,
  \dodoi{10.1017/S1743921319002436}

\bibitem[{{Tamura} {et~al.}(2023){Tamura}, {C. Bakx}, {Inoue}, {Hashimoto},
  {Tokuoka}, {Imamura}, {Hatsukade}, {Lee}, {Moriwaki}, {Okamoto}, {Ota},
  {Umehata}, {Yoshida}, {Zackrisson}, {Hagimoto}, {Matsuo}, {Shimizu},
  {Sugahara}, \& {Takeuchi}}]{TY23}
{Tamura}, Y., {C. Bakx}, T. J.~L., {Inoue}, A.~K., {et~al.} 2023, \apj, 952, 9,
  \dodoi{10.3847/1538-4357/acd637}

\bibitem[{{Telikova} {et~al.}(2025){Telikova}, {Gonz{\'a}lez-L{\'o}pez},
  {Aravena}, {Posses}, {Villanueva}, {Baeza-Garay}, {Jones}, {Solimano}, {Lee},
  {De Looze}, {F{\"o}rster Schreiber}, {Herrera-Camus}, {Tadaki}, {Assef},
  {Diaz Santos}, {Ferrara}, {Ikeda}, {Lamperti}, {Mitsuhashi}, {Perna},
  {Relano}, \& {{\"U}bler}}]{TK25}
{Telikova}, K., {Gonz{\'a}lez-L{\'o}pez}, J., {Aravena}, M., {et~al.} 2025,
  \aap, 699, A5, \dodoi{10.1051/0004-6361/202452990}

\bibitem[{{Temi} {et~al.}(2018){Temi}, {Hoffman}, {Ennico}, \& {Le}}]{temi18}
{Temi}, P., {Hoffman}, D., {Ennico}, K., \& {Le}, J. 2018, Journal of
  Astronomical Instrumentation, 7, 1840011, \dodoi{10.1142/S2251171718400111}

\bibitem[{{Teplitz} {et~al.}(2000){Teplitz}, {McLean}, {Becklin}, {Figer},
  {Gilbert}, {Graham}, {Larkin}, {Levenson}, \& {Wilcox}}]{TH00}
{Teplitz}, H.~I., {McLean}, I.~S., {Becklin}, E.~E., {et~al.} 2000, \apjl, 533,
  L65, \dodoi{10.1086/312595}

\bibitem[{{Thomson} {et~al.}(2012){Thomson}, {Ivison}, {Smail}, {Swinbank},
  {Weiss}, {Kneib}, {Papadopoulos}, {Baker}, {Sharon}, \& {van Moorsel}}]{TA12}
{Thomson}, A.~P., {Ivison}, R.~J., {Smail}, I., {et~al.} 2012, \mnras, 425,
  2203, \dodoi{10.1111/j.1365-2966.2012.21584.x}

\bibitem[{{Tielens} \& {Hollenbach}(1985)}]{tielens85}
{Tielens}, A.~G.~G.~M., \& {Hollenbach}, D. 1985, \apj, 291, 722,
  \dodoi{10.1086/163111}

\bibitem[{{Tokuoka} {et~al.}(2022){Tokuoka}, {Inoue}, {Hashimoto}, {Ellis},
  {Laporte}, {Sugahara}, {Matsuo}, {Tamura}, {Fudamoto}, {Moriwaki},
  {Roberts-Borsani}, {Shimizu}, {Yamanaka}, {Yoshida}, {Zackrisson}, \&
  {Zheng}}]{TT22}
{Tokuoka}, T., {Inoue}, A.~K., {Hashimoto}, T., {et~al.} 2022, \apjl, 933, L19,
  \dodoi{10.3847/2041-8213/ac7447}

\bibitem[{{Tripodi} {et~al.}(2023{\natexlab{a}}){Tripodi}, {Lelli}, {Feruglio},
  {Fiore}, {Fontanot}, {Bischetti}, \& {Maiolino}}]{TR23a}
{Tripodi}, R., {Lelli}, F., {Feruglio}, C., {et~al.} 2023{\natexlab{a}}, \aap,
  671, A44, \dodoi{10.1051/0004-6361/202245202}

\bibitem[{{Tripodi} {et~al.}(2022){Tripodi}, {Feruglio}, {Fiore}, {Bischetti},
  {D'Odorico}, {Carniani}, {Cristiani}, {Gallerani}, {Maiolino}, {Marconi},
  {Pallottini}, {Piconcelli}, {Vallini}, \& {Zana}}]{TR22}
{Tripodi}, R., {Feruglio}, C., {Fiore}, F., {et~al.} 2022, \aap, 665, A107,
  \dodoi{10.1051/0004-6361/202243920}

\bibitem[{{Tripodi} {et~al.}(2023{\natexlab{b}}){Tripodi}, {Feruglio},
  {Kemper}, {Civano}, {Costa}, {Elvis}, {Bischetti}, {Carniani}, {Di Mascia},
  {D'Odorico}, {Fiore}, {Gallerani}, {Ginolfi}, {Maiolino}, {Piconcelli},
  {Valiante}, \& {Zappacosta}}]{TR23b}
{Tripodi}, R., {Feruglio}, C., {Kemper}, F., {et~al.} 2023{\natexlab{b}},
  \apjl, 946, L45, \dodoi{10.3847/2041-8213/acc58d}

\bibitem[{{Tripodi} {et~al.}(2024){Tripodi}, {Feruglio}, {Fiore}, {Zappacosta},
  {Piconcelli}, {Bischetti}, {Bongiorno}, {Carniani}, {Civano}, {Chen},
  {Cristiani}, {Cupani}, {Di Mascia}, {D'Odorico}, {Fan}, {Ferrara},
  {Gallerani}, {Ginolfi}, {Maiolino}, {Mainieri}, {Marconi}, {Saccheo},
  {Salvestrini}, {Tortosa}, \& {Valiante}}]{TR24}
{Tripodi}, R., {Feruglio}, C., {Fiore}, F., {et~al.} 2024, \aap, 689, A220,
  \dodoi{10.1051/0004-6361/202349054}

\bibitem[{{Tsukui} {et~al.}(2023){Tsukui}, {Wisnioski}, {Krumholz}, \&
  {Battisti}}]{TT23}
{Tsukui}, T., {Wisnioski}, E., {Krumholz}, M.~R., \& {Battisti}, A. 2023,
  \mnras, 523, 4654, \dodoi{10.1093/mnras/stad1464}

\bibitem[{{{\"U}bler} {et~al.}(2023){{\"U}bler}, {Maiolino}, {Curtis-Lake},
  {P{\'e}rez-Gonz{\'a}lez}, {Curti}, {Perna}, {Arribas}, {Charlot}, {Marshall},
  {D'Eugenio}, {Scholtz}, {Bunker}, {Carniani}, {Ferruit}, {Jakobsen}, {Rix},
  {Rodr{\'\i}guez Del Pino}, {Willott}, {Boeker}, {Cresci}, {Jones}, {Kumari},
  \& {Rawle}}]{UH23}
{{\"U}bler}, H., {Maiolino}, R., {Curtis-Lake}, E., {et~al.} 2023, \aap, 677,
  A145, \dodoi{10.1051/0004-6361/202346137}

\bibitem[{{{\"U}bler} {et~al.}(2024{\natexlab{a}}){{\"U}bler}, {Maiolino},
  {P{\'e}rez-Gonz{\'a}lez}, {D'Eugenio}, {Perna}, {Curti}, {Arribas}, {Bunker},
  {Carniani}, {Charlot}, {Rodr{\'\i}guez Del Pino}, {Baker}, {B{\"o}ker},
  {Cresci}, {Dunlop}, {Grogin}, {Jones}, {Kumari}, {Lamperti}, {Laporte},
  {Marshall}, {Mazzolari}, {Parlanti}, {Rawle}, {Scholtz}, {Venturi}, \&
  {Witstok}}]{UH24a}
{{\"U}bler}, H., {Maiolino}, R., {P{\'e}rez-Gonz{\'a}lez}, P.~G., {et~al.}
  2024{\natexlab{a}}, \mnras, 531, 355, \dodoi{10.1093/mnras/stae943}

\bibitem[{{{\"U}bler} {et~al.}(2024{\natexlab{b}}){{\"U}bler}, {D'Eugenio},
  {Perna}, {Arribas}, {Jones}, {Bunker}, {Carniani}, {Charlot}, {Maiolino},
  {Rodr{\'\i}guez del Pino}, {Willott}, {B{\"o}ker}, {Cresci}, {Kumari},
  {Lamperti}, {Parlanti}, {Scholtz}, \& {Venturi}}]{UH24b}
{{\"U}bler}, H., {D'Eugenio}, F., {Perna}, M., {et~al.} 2024{\natexlab{b}},
  \mnras, 533, 4287, \dodoi{10.1093/mnras/stae1993}

\bibitem[{{Umehata} {et~al.}(2015){Umehata}, {Tamura}, {Kohno}, {Ivison},
  {Alexander}, {Geach}, {Hatsukade}, {Hughes}, {Ikarashi}, {Kato}, {Izumi},
  {Kawabe}, {Kubo}, {Lee}, {Lehmer}, {Makiya}, {Matsuda}, {Nakanishi}, {Saito},
  {Smail}, {Yamada}, {Yamaguchi}, \& {Yun}}]{UH15}
{Umehata}, H., {Tamura}, Y., {Kohno}, K., {et~al.} 2015, \apjl, 815, L8,
  \dodoi{10.1088/2041-8205/815/1/L8}

\bibitem[{{Umehata} {et~al.}(2017){Umehata}, {Matsuda}, {Tamura}, {Kohno},
  {Smail}, {Ivison}, {Steidel}, {Chapman}, {Geach}, {Hayes}, {Nagao}, {Ao},
  {Kawabe}, {Yun}, {Hatsukade}, {Kubo}, {Kato}, {Saito}, {Ikarashi},
  {Nakanishi}, {Lee}, {Izumi}, {Mori}, \& {Ouchi}}]{UH17}
{Umehata}, H., {Matsuda}, Y., {Tamura}, Y., {et~al.} 2017, \apjl, 834, L16,
  \dodoi{10.3847/2041-8213/834/2/L16}

\bibitem[{{Umehata} {et~al.}(2021){Umehata}, {Smail}, {Steidel}, {Hayes},
  {Scott}, {Swinbank}, {Ivison}, {Nagao}, {Kubo}, {Nakanishi}, {Matsuda},
  {Ikarashi}, {Tamura}, \& {Geach}}]{UH21}
{Umehata}, H., {Smail}, I., {Steidel}, C.~C., {et~al.} 2021, \apj, 918, 69,
  \dodoi{10.3847/1538-4357/ac1106}

\bibitem[{{Umehata} {et~al.}(2025){Umehata}, {Steidel}, {Smail}, {Swinbank},
  {Monson}, {Rosario}, {Lehmer}, {Nakanishi}, {Kubo}, {Iono}, {Alexander},
  {Kohno}, {Tamura}, {Ivison}, {Saito}, {Mitsuhashi}, {Huang}, \&
  {Matsuda}}]{UH25}
{Umehata}, H., {Steidel}, C.~C., {Smail}, I., {et~al.} 2025, \pasj, 77, 432,
  \dodoi{10.1093/pasj/psaf010}

\bibitem[{{Uzgil} {et~al.}(2016){Uzgil}, {Bradford}, {Hailey-Dunsheath},
  {Maloney}, \& {Aguirre}}]{UB16}
{Uzgil}, B.~D., {Bradford}, C.~M., {Hailey-Dunsheath}, S., {Maloney}, P.~R., \&
  {Aguirre}, J.~E. 2016, \apj, 832, 209, \dodoi{10.3847/0004-637X/832/2/209}

\bibitem[{{Valentino} {et~al.}(2022){Valentino}, {Brammer}, {Fujimoto},
  {Heintz}, {Weaver}, {Strait}, {Gould}, {Mason}, {Watson}, {Laursen}, \&
  {Toft}}]{VF22}
{Valentino}, F., {Brammer}, G., {Fujimoto}, S., {et~al.} 2022, \apjl, 929, L9,
  \dodoi{10.3847/2041-8213/ac62cc}

\bibitem[{{Valtchanov} {et~al.}(2011){Valtchanov}, {Virdee}, {Ivison},
  {Swinyard}, {van der Werf}, {Rigopoulou}, {da Cunha}, {Lupu}, {Benford},
  {Riechers}, {Smail}, {Jarvis}, {Pearson}, {Gomez}, {Hopwood}, {Altieri},
  {Birkinshaw}, {Coia}, {Conversi}, {Cooray}, {de Zotti}, {Dunne}, {Frayer},
  {Leeuw}, {Marston}, {Negrello}, {Portal}, {Scott}, {Thompson}, {Vaccari},
  {Baes}, {Clements}, {Micha{\l}owski}, {Dannerbauer}, {Serjeant}, {Auld},
  {Buttiglione}, {Cava}, {Dariush}, {Dye}, {Eales}, {Fritz}, {Ibar}, {Maddox},
  {Pascale}, {Pohlen}, {Rigby}, {Rodighiero}, {Smith}, {Temi}, {Carpenter},
  {Bolatto}, {Gurwell}, \& {Vieira}}]{VI11}
{Valtchanov}, I., {Virdee}, J., {Ivison}, R.~J., {et~al.} 2011, \mnras, 415,
  3473, \dodoi{10.1111/j.1365-2966.2011.18959.x}

\bibitem[{{van Leeuwen} {et~al.}(2024){van Leeuwen}, {Bouwens}, {van der Werf},
  {Hodge}, {Schouws}, {Stefanon}, {Algera}, {Aravena}, {Boogaard}, {Bowler},
  {da Cunha}, {Dayal}, {Decarli}, {Gonzalez}, {Inami}, {de Looze}, {Sommovigo},
  {Venemans}, {Walter}, {Barrufet}, {Ferrara}, {Graziani}, {Hygate}, {Oesch},
  {Palla}, {Rowland}, \& {Schneider}}]{vI24}
{van Leeuwen}, I.~F., {Bouwens}, R.~J., {van der Werf}, P.~P., {et~al.} 2024,
  \mnras, 534, 2062, \dodoi{10.1093/mnras/stae2171}

\bibitem[{{Vanzella} {et~al.}(2011){Vanzella}, {Pentericci}, {Fontana},
  {Grazian}, {Castellano}, {Boutsia}, {Cristiani}, {Dickinson}, {Gallozzi},
  {Giallongo}, {Giavalisco}, {Maiolino}, {Moorwood}, {Paris}, \&
  {Santini}}]{VE11}
{Vanzella}, E., {Pentericci}, L., {Fontana}, A., {et~al.} 2011, \apjl, 730,
  L35, \dodoi{10.1088/2041-8205/730/2/L35}

\bibitem[{{Venemans} {et~al.}(2019){Venemans}, {Neeleman}, {Walter}, {Novak},
  {Decarli}, {Hennawi}, \& {Rix}}]{VB19}
{Venemans}, B.~P., {Neeleman}, M., {Walter}, F., {et~al.} 2019, \apjl, 874,
  L30, \dodoi{10.3847/2041-8213/ab11cc}

\bibitem[{{Venemans} {et~al.}(2016){Venemans}, {Walter}, {Zschaechner},
  {Decarli}, {De Rosa}, {Findlay}, {McMahon}, \& {Sutherland}}]{VB16}
{Venemans}, B.~P., {Walter}, F., {Zschaechner}, L., {et~al.} 2016, \apj, 816,
  37, \dodoi{10.3847/0004-637X/816/1/37}

\bibitem[{{Venemans} {et~al.}(2012){Venemans}, {McMahon}, {Walter}, {Decarli},
  {Cox}, {Neri}, {Hewett}, {Mortlock}, {Simpson}, \& {Warren}}]{VB12}
{Venemans}, B.~P., {McMahon}, R.~G., {Walter}, F., {et~al.} 2012, \apjl, 751,
  L25, \dodoi{10.1088/2041-8205/751/2/L25}

\bibitem[{{Venemans} {et~al.}(2017{\natexlab{a}}){Venemans}, {Walter},
  {Decarli}, {Ba{\~n}ados}, {Hodge}, {Hewett}, {McMahon}, {Mortlock}, \&
  {Simpson}}]{VB17a}
{Venemans}, B.~P., {Walter}, F., {Decarli}, R., {et~al.} 2017{\natexlab{a}},
  \apj, 837, 146, \dodoi{10.3847/1538-4357/aa62ac}

\bibitem[{{Venemans} {et~al.}(2017{\natexlab{b}}){Venemans}, {Walter},
  {Decarli}, {Ba{\~n}ados}, {Carilli}, {Winters}, {Schuster}, {da Cunha},
  {Fan}, {Farina}, {Mazzucchelli}, {Rix}, \& {Weiss}}]{VB17b}
---. 2017{\natexlab{b}}, \apjl, 851, L8, \dodoi{10.3847/2041-8213/aa943a}

\bibitem[{{Venemans} {et~al.}(2020){Venemans}, {Walter}, {Neeleman}, {Novak},
  {Otter}, {Decarli}, {Ba{\~n}ados}, {Drake}, {Farina}, {Kaasinen},
  {Mazzucchelli}, {Carilli}, {Fan}, {Rix}, \& {Wang}}]{VB20}
{Venemans}, B.~P., {Walter}, F., {Neeleman}, M., {et~al.} 2020, \apj, 904, 130,
  \dodoi{10.3847/1538-4357/abc563}

\bibitem[{{Venturi} {et~al.}(2024){Venturi}, {Carniani}, {Parlanti},
  {Kohandel}, {Curti}, {Pallottini}, {Vallini}, {Arribas}, {Bunker}, {Cameron},
  {Castellano}, {Ferrara}, {Fontana}, {Gallerani}, {Gelli}, {Maiolino},
  {Ntormousi}, {Pacifici}, {Pentericci}, {Salvadori}, \& {Vanzella}}]{VG24}
{Venturi}, G., {Carniani}, S., {Parlanti}, E., {et~al.} 2024, \aap, 691, A19,
  \dodoi{10.1051/0004-6361/202449855}

\bibitem[{{Venturini} \& {Solomon}(2003)}]{VS03}
{Venturini}, S., \& {Solomon}, P.~M. 2003, \apj, 590, 740,
  \dodoi{10.1086/375050}

\bibitem[{{Villanueva} {et~al.}(2024){Villanueva}, {Herrera-Camus},
  {Gonz{\'a}lez-L{\'o}pez}, {Aravena}, {Assef}, {Baeza-Garay},
  {Barcos-Mu{\~n}oz}, {Bovino}, {Bowler}, {da Cunha}, {De Looze},
  {Diaz-Santos}, {Ferrara}, {F{\"o}rster Schreiber}, {Algera}, {Ikeda},
  {Killi}, {Mitsuhashi}, {Naab}, {Relano}, {Spilker}, {Solimano}, {Palla},
  {Price}, {Posses}, {Tadaki}, {Telikova}, \& {{\"U}bler}}]{VV24}
{Villanueva}, V., {Herrera-Camus}, R., {Gonz{\'a}lez-L{\'o}pez}, J., {et~al.}
  2024, \aap, 691, A133, \dodoi{10.1051/0004-6361/202451490}

\bibitem[{{Wagg} {et~al.}(2010){Wagg}, {Carilli}, {Wilner}, {Cox}, {De Breuck},
  {Menten}, {Riechers}, \& {Walter}}]{WJ10}
{Wagg}, J., {Carilli}, C.~L., {Wilner}, D.~J., {et~al.} 2010, \aap, 519, L1,
  \dodoi{10.1051/0004-6361/201015424}

\bibitem[{{Wagg} {et~al.}(2012){Wagg}, {Wiklind}, {Carilli}, {Espada}, {Peck},
  {Riechers}, {Walter}, {Wootten}, {Aravena}, {Barkats}, {Cortes}, {Hills},
  {Hodge}, {Impellizzeri}, {Iono}, {Leroy}, {Mart{\'\i}n}, {Rawlings},
  {Maiolino}, {McMahon}, {Scott}, {Villard}, \& {Vlahakis}}]{WJ12}
{Wagg}, J., {Wiklind}, T., {Carilli}, C.~L., {et~al.} 2012, \apjl, 752, L30,
  \dodoi{10.1088/2041-8205/752/2/L30}

\bibitem[{{Wagg} {et~al.}(2020){Wagg}, {Aravena}, {Brisbin}, {Valtchanov},
  {Carilli}, {Daddi}, {Dannerbauer}, {Decarli}, {D{\'\i}az-Santos}, {Riechers},
  {Sargent}, \& {Walter}}]{WJ20}
{Wagg}, J., {Aravena}, M., {Brisbin}, D., {et~al.} 2020, \mnras, 499, 1788,
  \dodoi{10.1093/mnras/staa2884}

\bibitem[{{Walter} {et~al.}(2009){Walter}, {Riechers}, {Cox}, {Neri},
  {Carilli}, {Bertoldi}, {Weiss}, \& {Maiolino}}]{WF09}
{Walter}, F., {Riechers}, D., {Cox}, P., {et~al.} 2009, \nat, 457, 699,
  \dodoi{10.1038/nature07681}

\bibitem[{{Walter} {et~al.}(2012){Walter}, {Decarli}, {Carilli}, {Bertoldi},
  {Cox}, {da Cunha}, {Daddi}, {Dickinson}, {Downes}, {Elbaz}, {Ellis}, {Hodge},
  {Neri}, {Riechers}, {Weiss}, {Bell}, {Dannerbauer}, {Krips}, {Krumholz},
  {Lentati}, {Maiolino}, {Menten}, {Rix}, {Robertson}, {Spinrad}, {Stark}, \&
  {Stern}}]{WF12}
{Walter}, F., {Decarli}, R., {Carilli}, C., {et~al.} 2012, \nat, 486, 233,
  \dodoi{10.1038/nature11073}

\bibitem[{{Walter} {et~al.}(2018){Walter}, {Riechers}, {Novak}, {Decarli},
  {Ferkinhoff}, {Venemans}, {Ba{\~n}ados}, {Bertoldi}, {Carilli}, {Fan},
  {Farina}, {Mazzucchelli}, {Neeleman}, {Rix}, {Strauss}, {Uzgil}, \&
  {Wang}}]{WF18}
{Walter}, F., {Riechers}, D., {Novak}, M., {et~al.} 2018, \apjl, 869, L22,
  \dodoi{10.3847/2041-8213/aaf4fa}

\bibitem[{{Walter} {et~al.}(2022){Walter}, {Neeleman}, {Decarli}, {Venemans},
  {Meyer}, {Weiss}, {Ba{\~n}ados}, {Bosman}, {Carilli}, {Fan}, {Riechers},
  {Rix}, \& {Thompson}}]{WF22}
{Walter}, F., {Neeleman}, M., {Decarli}, R., {et~al.} 2022, \apj, 927, 21,
  \dodoi{10.3847/1538-4357/ac49e8}

\bibitem[{{Wang} {et~al.}(2019{\natexlab{a}}){Wang}, {Wang}, {Fan}, {Wu},
  {Yang}, {Neri}, \& {Yue}}]{WF19a}
{Wang}, F., {Wang}, R., {Fan}, X., {et~al.} 2019{\natexlab{a}}, \apj, 880, 2,
  \dodoi{10.3847/1538-4357/ab2717}

\bibitem[{{Wang} {et~al.}(2019{\natexlab{b}}){Wang}, {Yang}, {Fan}, {Wu},
  {Yue}, {Li}, {Bian}, {Jiang}, {Ba{\~n}ados}, {Schindler}, {Findlay},
  {Davies}, {Decarli}, {Farina}, {Green}, {Hennawi}, {Huang}, {Mazzuccheli},
  {McGreer}, {Venemans}, {Walter}, {Dye}, {Lyke}, {Myers}, \& {Nunez}}]{WF19b}
{Wang}, F., {Yang}, J., {Fan}, X., {et~al.} 2019{\natexlab{b}}, \apj, 884, 30,
  \dodoi{10.3847/1538-4357/ab2be5}

\bibitem[{{Wang} {et~al.}(2021{\natexlab{a}}){Wang}, {Yang}, {Fan}, {Hennawi},
  {Barth}, {Banados}, {Bian}, {Boutsia}, {Connor}, {Davies}, {Decarli},
  {Eilers}, {Farina}, {Green}, {Jiang}, {Li}, {Mazzucchelli}, {Nanni},
  {Schindler}, {Venemans}, {Walter}, {Wu}, \& {Yue}}]{WF21a}
---. 2021{\natexlab{a}}, \apjl, 907, L1, \dodoi{10.3847/2041-8213/abd8c6}

\bibitem[{{Wang} {et~al.}(2021{\natexlab{b}}){Wang}, {Fan}, {Yang},
  {Mazzucchelli}, {Wu}, {Li}, {Ba{\~n}ados}, {Farina}, {Nanni}, {Ai}, {Bian},
  {Davies}, {Decarli}, {Hennawi}, {Schindler}, {Venemans}, \& {Walter}}]{WF21b}
{Wang}, F., {Fan}, X., {Yang}, J., {et~al.} 2021{\natexlab{b}}, \apj, 908, 53,
  \dodoi{10.3847/1538-4357/abcc5e}

\bibitem[{{Wang} {et~al.}(2024{\natexlab{a}}){Wang}, {Yang}, {Fan}, {Venemans},
  {Decarli}, {Ba{\~n}ados}, {Walter}, {Barth}, {Bian}, {Davies}, {Eilers},
  {Farina}, {Hennawi}, {Li}, {Mazzucchelli}, {Wang}, {Wu}, \& {Yue}}]{WF24a}
{Wang}, F., {Yang}, J., {Fan}, X., {et~al.} 2024{\natexlab{a}}, \apj, 968, 9,
  \dodoi{10.3847/1538-4357/ad3fb4}

\bibitem[{{Wang} {et~al.}(2024{\natexlab{b}}){Wang}, {Yang}, {Hennawi}, {Fan},
  {Yue}, {Ba{\~n}ados}, {Bechtel}, {Bian}, {Bosman}, {Champagne}, {Davies},
  {Decarli}, {Farina}, {Mazzucchelli}, {Venemans}, \& {Walter}}]{WF24b}
{Wang}, F., {Yang}, J., {Hennawi}, J.~F., {et~al.} 2024{\natexlab{b}}, \apjl,
  962, L11, \dodoi{10.3847/2041-8213/ad20ef}

\bibitem[{{Wang} {et~al.}(2021{\natexlab{c}}){Wang}, {Hill}, {Chapman},
  {Wei{\ss}}, {Scott}, {Apostolovski}, {Aravena}, {Archipley}, {B{\'e}thermin},
  {Canning}, {De Breuck}, {Dong}, {Everett}, {Gonzalez}, {Greve}, {Hayward},
  {Hezaveh}, {Jarugula}, {Marrone}, {Phadke}, {Reuter}, {Rotermund}, {Spilker},
  \& {Vieira}}]{WG21}
{Wang}, G. C.~P., {Hill}, R., {Chapman}, S.~C., {et~al.} 2021{\natexlab{c}},
  \mnras, 508, 3754, \dodoi{10.1093/mnras/stab2800}

\bibitem[{{Wang} {et~al.}(2013){Wang}, {Wagg}, {Carilli}, {Walter}, {Lentati},
  {Fan}, {Riechers}, {Bertoldi}, {Narayanan}, {Strauss}, {Cox}, {Omont},
  {Menten}, {Knudsen}, {Neri}, \& {Jiang}}]{WR13}
{Wang}, R., {Wagg}, J., {Carilli}, C.~L., {et~al.} 2013, \apj, 773, 44,
  \dodoi{10.1088/0004-637X/773/1/44}

\bibitem[{{Wang} {et~al.}(2016){Wang}, {Wu}, {Neri}, {Fan}, {Walter},
  {Carilli}, {Momjian}, {Bertoldi}, {Strauss}, {Li}, {Wang}, {Riechers},
  {Jiang}, {Omont}, {Wagg}, \& {Cox}}]{WR16}
{Wang}, R., {Wu}, X.-B., {Neri}, R., {et~al.} 2016, \apj, 830, 53,
  \dodoi{10.3847/0004-637X/830/1/53}

\bibitem[{{Wang} {et~al.}(2019{\natexlab{c}}){Wang}, {Shao}, {Carilli},
  {Jones}, {Walter}, {Fan}, {Riechers}, {Decarli}, {Bertoldi}, {Wagg},
  {Strauss}, {Omont}, {Cox}, {Jiang}, {Narayanan}, {Menten}, \&
  {Venemans}}]{WR19}
{Wang}, R., {Shao}, Y., {Carilli}, C.~L., {et~al.} 2019{\natexlab{c}}, \apj,
  887, 40, \dodoi{10.3847/1538-4357/ab4d4b}

\bibitem[{{Wang} {et~al.}(2024{\natexlab{c}}){Wang}, {Wylezalek}, {De Breuck},
  {Vernet}, {Rupke}, {Zakamska}, {Vayner}, {Lehnert}, {Nesvadba}, \&
  {Stern}}]{WW24}
{Wang}, W., {Wylezalek}, D., {De Breuck}, C., {et~al.} 2024{\natexlab{c}},
  \aap, 683, A169, \dodoi{10.1051/0004-6361/202348531}

\bibitem[{{Wang} {et~al.}(2025){Wang}, {De Breuck}, {Wylezalek}, {Vernet},
  {Lehnert}, {Stern}, {Rupke}, {Nesvadba}, {Vayner}, {Zakamska}, {Lin},
  {Kukreti}, {Dall'Agnol de Oliveira}, \& {Groth}}]{WW25}
{Wang}, W., {De Breuck}, C., {Wylezalek}, D., {et~al.} 2025, \aap, 696, A88,
  \dodoi{10.1051/0004-6361/202553668}

\bibitem[{{Ward} {et~al.}(2022){Ward}, {Eales}, {Pons}, {Smith}, {McMahon},
  {Dunne}, {Ivison}, {Maddox}, \& {Negrello}}]{WB22}
{Ward}, B.~A., {Eales}, S.~A., {Pons}, E., {et~al.} 2022, \mnras, 510, 2261,
  \dodoi{10.1093/mnras/stab3300}

\bibitem[{{Wardlow} {et~al.}(2013){Wardlow}, {Cooray}, {De Bernardis},
  {Amblard}, {Arumugam}, {Aussel}, {Baker}, {B{\'e}thermin}, {Blundell},
  {Bock}, {Boselli}, {Bridge}, {Buat}, {Burgarella}, {Bussmann},
  {Cabrera-Lavers}, {Calanog}, {Carpenter}, {Casey}, {Castro-Rodr{\'\i}guez},
  {Cava}, {Chanial}, {Chapin}, {Chapman}, {Clements}, {Conley}, {Cox},
  {Dowell}, {Dye}, {Eales}, {Farrah}, {Ferrero}, {Franceschini}, {Frayer},
  {Frazer}, {Fu}, {Gavazzi}, {Glenn}, {Gonz{\'a}lez Solares}, {Griffin},
  {Gurwell}, {Harris}, {Hatziminaoglou}, {Hopwood}, {Hyde}, {Ibar}, {Ivison},
  {Kim}, {Lagache}, {Levenson}, {Marchetti}, {Marsden}, {Martinez-Navajas},
  {Negrello}, {Neri}, {Nguyen}, {O'Halloran}, {Oliver}, {Omont}, {Page},
  {Panuzzo}, {Papageorgiou}, {Pearson}, {P{\'e}rez-Fournon}, {Pohlen},
  {Riechers}, {Rigopoulou}, {Roseboom}, {Rowan-Robinson}, {Schulz}, {Scott},
  {Scoville}, {Seymour}, {Shupe}, {Smith}, {Streblyanska}, {Strom},
  {Symeonidis}, {Trichas}, {Vaccari}, {Vieira}, {Viero}, {Wang}, {Xu}, {Yan},
  \& {Zemcov}}]{WJ13}
{Wardlow}, J.~L., {Cooray}, A., {De Bernardis}, F., {et~al.} 2013, \apj, 762,
  59, \dodoi{10.1088/0004-637X/762/1/59}

\bibitem[{{Wardlow} {et~al.}(2017){Wardlow}, {Cooray}, {Osage}, {Bourne},
  {Clements}, {Dannerbauer}, {Dunne}, {Dye}, {Eales}, {Farrah}, {Furlanetto},
  {Ibar}, {Ivison}, {Maddox}, {Micha{\l}owski}, {Riechers}, {Rigopoulou},
  {Scott}, {Smith}, {Wang}, {van der Werf}, {Valiante}, {Valtchanov}, \&
  {Verma}}]{WJ17}
{Wardlow}, J.~L., {Cooray}, A., {Osage}, W., {et~al.} 2017, \apj, 837, 12,
  \dodoi{10.3847/1538-4357/837/1/12}

\bibitem[{{Watson} {et~al.}(2015){Watson}, {Christensen}, {Knudsen}, {Richard},
  {Gallazzi}, \& {Micha{\l}owski}}]{WD15}
{Watson}, D., {Christensen}, L., {Knudsen}, K.~K., {et~al.} 2015, \nat, 519,
  327, \dodoi{10.1038/nature14164}

\bibitem[{{Wei{\ss}} {et~al.}(2007){Wei{\ss}}, {Downes}, {Neri}, {Walter},
  {Henkel}, {Wilner}, {Wagg}, \& {Wiklind}}]{WA07}
{Wei{\ss}}, A., {Downes}, D., {Neri}, R., {et~al.} 2007, \aap, 467, 955,
  \dodoi{10.1051/0004-6361:20066117}

\bibitem[{{Wei{\ss}} {et~al.}(2003){Wei{\ss}}, {Henkel}, {Downes}, \&
  {Walter}}]{WA03}
{Wei{\ss}}, A., {Henkel}, C., {Downes}, D., \& {Walter}, F. 2003, \aap, 409,
  L41, \dodoi{10.1051/0004-6361:20031337}

\bibitem[{{Wei{\ss}} {et~al.}(2013){Wei{\ss}}, {De Breuck}, {Marrone},
  {Vieira}, {Aguirre}, {Aird}, {Aravena}, {Ashby}, {Bayliss}, {Benson},
  {B{\'e}thermin}, {Biggs}, {Bleem}, {Bock}, {Bothwell}, {Bradford}, {Brodwin},
  {Carlstrom}, {Chang}, {Chapman}, {Crawford}, {Crites}, {de Haan}, {Dobbs},
  {Downes}, {Fassnacht}, {George}, {Gladders}, {Gonzalez}, {Greve},
  {Halverson}, {Hezaveh}, {High}, {Holder}, {Holzapfel}, {Hoover}, {Hrubes},
  {Husband}, {Keisler}, {Lee}, {Leitch}, {Lueker}, {Luong-Van}, {Malkan},
  {McIntyre}, {McMahon}, {Mehl}, {Menten}, {Meyer}, {Murphy}, {Padin},
  {Plagge}, {Reichardt}, {Rest}, {Rosenman}, {Ruel}, {Ruhl}, {Schaffer},
  {Shirokoff}, {Spilker}, {Stalder}, {Staniszewski}, {Stark}, {Story},
  {Vanderlinde}, {Welikala}, \& {Williamson}}]{WA13}
{Wei{\ss}}, A., {De Breuck}, C., {Marrone}, D.~P., {et~al.} 2013, \apj, 767,
  88, \dodoi{10.1088/0004-637X/767/1/88}

\bibitem[{{Welch} {et~al.}(2023){Welch}, {Coe}, {Zitrin}, {Diego}, {Windhorst},
  {Mandelker}, {Vanzella}, {Ravindranath}, {Zackrisson}, {Florian}, {Bradley},
  {Sharon}, {Brada{\v{c}}}, {Rigby}, {Frye}, \& {Fujimoto}}]{WB23}
{Welch}, B., {Coe}, D., {Zitrin}, A., {et~al.} 2023, \apj, 943, 2,
  \dodoi{10.3847/1538-4357/aca8a8}

\bibitem[{{Werner} {et~al.}(2004){Werner}, {Roellig}, {Low}, {Rieke}, {Rieke},
  {Hoffmann}, {Young}, {Houck}, {Brandl}, {Fazio}, {Hora}, {Gehrz}, {Helou},
  {Soifer}, {Stauffer}, {Keene}, {Eisenhardt}, {Gallagher}, {Gautier}, {Irace},
  {Lawrence}, {Simmons}, {Van Cleve}, {Jura}, {Wright}, \&
  {Cruikshank}}]{werner04}
{Werner}, M.~W., {Roellig}, T.~L., {Low}, F.~J., {et~al.} 2004, \apjs, 154, 1,
  \dodoi{10.1086/422992}

\bibitem[{{Willott} {et~al.}(2015{\natexlab{a}}){Willott}, {Bergeron}, \&
  {Omont}}]{WC15b}
{Willott}, C.~J., {Bergeron}, J., \& {Omont}, A. 2015{\natexlab{a}}, \apj, 801,
  123, \dodoi{10.1088/0004-637X/801/2/123}

\bibitem[{{Willott} {et~al.}(2017){Willott}, {Bergeron}, \& {Omont}}]{WC17}
---. 2017, \apj, 850, 108, \dodoi{10.3847/1538-4357/aa921b}

\bibitem[{{Willott} {et~al.}(2015{\natexlab{b}}){Willott}, {Carilli}, {Wagg},
  \& {Wang}}]{WC15a}
{Willott}, C.~J., {Carilli}, C.~L., {Wagg}, J., \& {Wang}, R.
  2015{\natexlab{b}}, \apj, 807, 180, \dodoi{10.1088/0004-637X/807/2/180}

\bibitem[{{Willott} {et~al.}(2013){Willott}, {Omont}, \& {Bergeron}}]{WC13}
{Willott}, C.~J., {Omont}, A., \& {Bergeron}, J. 2013, \apj, 770, 13,
  \dodoi{10.1088/0004-637X/770/1/13}

\bibitem[{{Willott} {et~al.}(2007){Willott}, {Delorme}, {Omont}, {Bergeron},
  {Delfosse}, {Forveille}, {Albert}, {Reyl{\'e}}, {Hill}, {Gully-Santiago},
  {Vinten}, {Crampton}, {Hutchings}, {Schade}, {Simard}, {Sawicki}, {Beelen},
  \& {Cox}}]{WC07}
{Willott}, C.~J., {Delorme}, P., {Omont}, A., {et~al.} 2007, \aj, 134, 2435,
  \dodoi{10.1086/522962}

\bibitem[{{Witstok} {et~al.}(2022){Witstok}, {Smit}, {Maiolino}, {Kumari},
  {Aravena}, {Boogaard}, {Bouwens}, {Carniani}, {Hodge}, {Jones}, {Stefanon},
  {van der Werf}, \& {Schouws}}]{WJ22}
{Witstok}, J., {Smit}, R., {Maiolino}, R., {et~al.} 2022, \mnras, 515, 1751,
  \dodoi{10.1093/mnras/stac1905}

\bibitem[{{Wolfire} {et~al.}(2022){Wolfire}, {Vallini}, \&
  {Chevance}}]{wolfire22}
{Wolfire}, M.~G., {Vallini}, L., \& {Chevance}, M. 2022, \araa, 60, 247,
  \dodoi{10.1146/annurev-astro-052920-010254}

\bibitem[{{Wong} {et~al.}(2022{\natexlab{a}}){Wong}, {Wang}, {Hashimoto},
  {Takagi}, {Goto}, {Kim}, {Wu}, {On}, {Santos}, {Lu}, {Kilerci-Eser}, {Ho}, \&
  {Hsiao}}]{WY22a}
{Wong}, Y. H.~V., {Wang}, P., {Hashimoto}, T., {et~al.} 2022{\natexlab{a}},
  arXiv e-prints, arXiv:2202.13613, \dodoi{10.48550/arXiv.2202.13613}

\bibitem[{{Wong} {et~al.}(2022{\natexlab{b}}){Wong}, {Wang}, {Hashimoto},
  {Takagi}, {Goto}, {Kim}, {Wu}, {On}, {Santos}, {Lu}, {Kilerci-Eser}, {Ho}, \&
  {Hsiao}}]{WY22b}
---. 2022{\natexlab{b}}, \apj, 929, 161, \dodoi{10.3847/1538-4357/ac5cc7}

\bibitem[{{Wright} {et~al.}(2010){Wright}, {Eisenhardt}, {Mainzer}, {Ressler},
  {Cutri}, {Jarrett}, {Kirkpatrick}, {Padgett}, {McMillan}, {Skrutskie},
  {Stanford}, {Cohen}, {Walker}, {Mather}, {Leisawitz}, {Gautier}, {McLean},
  {Benford}, {Lonsdale}, {Blain}, {Mendez}, {Irace}, {Duval}, {Liu}, {Royer},
  {Heinrichsen}, {Howard}, {Shannon}, {Kendall}, {Walsh}, {Larsen}, {Cardon},
  {Schick}, {Schwalm}, {Abid}, {Fabinsky}, {Naes}, \& {Tsai}}]{wright10}
{Wright}, E.~L., {Eisenhardt}, P. R.~M., {Mainzer}, A.~K., {et~al.} 2010, \aj,
  140, 1868, \dodoi{10.1088/0004-6256/140/6/1868}

\bibitem[{{Wu} {et~al.}(2021){Wu}, {Cai}, {Neeleman}, {Finlator}, {Zhang},
  {Prochaska}, {Wang}, {Emonts}, {Fan}, {Keating}, {Wang}, {Yang}, {Hennawi},
  \& {Wang}}]{WY21}
{Wu}, Y., {Cai}, Z., {Neeleman}, M., {et~al.} 2021, Nature Astronomy, 5, 1110,
  \dodoi{10.1038/s41550-021-01471-4}

\bibitem[{{Wuyts} {et~al.}(2012{\natexlab{a}}){Wuyts}, {Rigby}, {Gladders},
  {Gilbank}, {Sharon}, {Gralla}, \& {Bayliss}}]{WE12b}
{Wuyts}, E., {Rigby}, J.~R., {Gladders}, M.~D., {et~al.} 2012{\natexlab{a}},
  \apj, 745, 86, \dodoi{10.1088/0004-637X/745/1/86}

\bibitem[{{Wuyts} {et~al.}(2012{\natexlab{b}}){Wuyts}, {Rigby}, {Sharon}, \&
  {Gladders}}]{WE12a}
{Wuyts}, E., {Rigby}, J.~R., {Sharon}, K., \& {Gladders}, M.~D.
  2012{\natexlab{b}}, \apj, 755, 73, \dodoi{10.1088/0004-637X/755/1/73}

\bibitem[{{Xiao} {et~al.}(2024){Xiao}, {Oesch}, {Elbaz}, {Bing}, {Nelson},
  {Weibel}, {Illingworth}, {van Dokkum}, {Naidu}, {Daddi}, {Bouwens},
  {Matthee}, {Wuyts}, {Chisholm}, {Brammer}, {Dickinson}, {Magnelli}, {Leroy},
  {Schaerer}, {Herard-Demanche}, {Lim}, {Barrufet}, {Endsley}, {Fudamoto},
  {G{\'o}mez-Guijarro}, {Gottumukkala}, {Labb{\'e}}, {Magee}, {Marchesini},
  {Maseda}, {Qin}, {Reddy}, {Shapley}, {Shivaei}, {Shuntov}, {Stefanon},
  {Whitaker}, \& {Wyithe}}]{XM24}
{Xiao}, M., {Oesch}, P.~A., {Elbaz}, D., {et~al.} 2024, \nat, 635, 311,
  \dodoi{10.1038/s41586-024-08094-5}

\bibitem[{{Yang} {et~al.}(2017){Yang}, {Fan}, {Wu}, {Wang}, {Bian}, {Yang},
  {McGreer}, {Yi}, {Jiang}, {Green}, {Yue}, {Wang}, {Li}, {Ding}, {Dye}, \&
  {Lawrence}}]{YJ17}
{Yang}, J., {Fan}, X., {Wu}, X.-B., {et~al.} 2017, \aj, 153, 184,
  \dodoi{10.3847/1538-3881/aa6577}

\bibitem[{{Yang} {et~al.}(2019{\natexlab{a}}){Yang}, {Venemans}, {Wang}, {Fan},
  {Novak}, {Decarli}, {Walter}, {Yue}, {Momjian}, {Keeton}, {Wang},
  {Zabludoff}, {Wu}, \& {Bian}}]{YJ19a}
{Yang}, J., {Venemans}, B., {Wang}, F., {et~al.} 2019{\natexlab{a}}, \apj, 880,
  153, \dodoi{10.3847/1538-4357/ab2a02}

\bibitem[{{Yang} {et~al.}(2019{\natexlab{b}}){Yang}, {Wang}, {Fan}, {Yue},
  {Wu}, {Li}, {Bian}, {Jiang}, {Ba{\~n}ados}, \& {Beletsky}}]{YJ19b}
{Yang}, J., {Wang}, F., {Fan}, X., {et~al.} 2019{\natexlab{b}}, \aj, 157, 236,
  \dodoi{10.3847/1538-3881/ab1be1}

\bibitem[{{Yang} {et~al.}(2020){Yang}, {Wang}, {Fan}, {Hennawi}, {Davies},
  {Yue}, {Banados}, {Wu}, {Venemans}, {Barth}, {Bian}, {Boutsia}, {Decarli},
  {Farina}, {Green}, {Jiang}, {Li}, {Mazzucchelli}, \& {Walter}}]{YJ20}
---. 2020, \apjl, 897, L14, \dodoi{10.3847/2041-8213/ab9c26}

\bibitem[{{Yang} {et~al.}(2021){Yang}, {Wang}, {Fan}, {Barth}, {Hennawi},
  {Nanni}, {Bian}, {Davies}, {Farina}, {Schindler}, {Ba{\~n}ados}, {Decarli},
  {Eilers}, {Green}, {Guo}, {Jiang}, {Li}, {Venemans}, {Walter}, {Wu}, \&
  {Yue}}]{YJ21}
---. 2021, \apj, 923, 262, \dodoi{10.3847/1538-4357/ac2b32}

\bibitem[{{Yi} {et~al.}(2014){Yi}, {Wang}, {Wu}, {Yang}, {Bai}, {Fan},
  {Brandt}, {Ho}, {Zuo}, {Kim}, {Wang}, {Yang}, {Zhang}, {Wang}, {Wang}, {Ai},
  {Fan}, {Chang}, {Wang}, {Lun}, \& {Xin}}]{YW14}
{Yi}, W.-M., {Wang}, F., {Wu}, X.-B., {et~al.} 2014, \apjl, 795, L29,
  \dodoi{10.1088/2041-8205/795/2/L29}

\bibitem[{{Yue} {et~al.}(2021){Yue}, {Yang}, {Fan}, {Wang}, {Spilker},
  {Georgiev}, {Keeton}, {Litke}, {Marrone}, {Walter}, {Wang}, {Wu}, {Venemans},
  \& {Zabludoff}}]{YM21}
{Yue}, M., {Yang}, J., {Fan}, X., {et~al.} 2021, \apj, 917, 99,
  \dodoi{10.3847/1538-4357/ac0af4}

\bibitem[{{Yue} {et~al.}(2024){Yue}, {Eilers}, {Simcoe}, {Mackenzie},
  {Matthee}, {Kashino}, {Bordoloi}, {Lilly}, \& {Naidu}}]{YM24}
{Yue}, M., {Eilers}, A.-C., {Simcoe}, R.~A., {et~al.} 2024, \apj, 966, 176,
  \dodoi{10.3847/1538-4357/ad3914}

\bibitem[{{Yun} {et~al.}(2015){Yun}, {Aretxaga}, {Gurwell}, {Hughes},
  {Monta{\~n}a}, {Narayanan}, {Rosa-Gonz{\'a}lez}, {S{\'a}nchez-Arg{\"u}elles},
  {Schloerb}, {Snell}, {Vega}, {Wilson}, {Zeballos}, {Chavez}, {Cybulski},
  {D{\'\i}az-Santos}, {De La Luz}, {Erickson}, {Ferrusca}, {Gim}, {Heyer},
  {Iono}, {Pope}, {Rogstad}, {Scott}, {Souccar}, {Terlevich}, {Terlevich},
  {Wilner}, \& {Zavala}}]{YM15}
{Yun}, M.~S., {Aretxaga}, I., {Gurwell}, M.~A., {et~al.} 2015, \mnras, 454,
  3485, \dodoi{10.1093/mnras/stv1963}

\bibitem[{{Zamora} {et~al.}(2024){Zamora}, {Venturi}, {Carniani}, {Bertola},
  {Parlanti}, {Perna}, {Arribas}, {B{\"o}ker}, {Bunker}, {Charlot},
  {D'Eugenio}, {Maiolino}, {Rodr{\'\i}guez Del Pino}, {{\"U}bler}, {Cresci},
  {Jones}, \& {Lamperti}}]{ZS24}
{Zamora}, S., {Venturi}, G., {Carniani}, S., {et~al.} 2024, arXiv e-prints,
  arXiv:2412.02751, \dodoi{10.48550/arXiv.2412.02751}

\bibitem[{{Zanella} {et~al.}(2018){Zanella}, {Daddi}, {Magdis}, {Diaz Santos},
  {Cormier}, {Liu}, {Cibinel}, {Gobat}, {Dickinson}, {Sargent}, {Popping},
  {Madden}, {Bethermin}, {Hughes}, {Valentino}, {Rujopakarn}, {Pannella},
  {Bournaud}, {Walter}, {Wang}, {Elbaz}, \& {Coogan}}]{ZA18}
{Zanella}, A., {Daddi}, E., {Magdis}, G., {et~al.} 2018, \mnras, 481, 1976,
  \dodoi{10.1093/mnras/sty2394}

\bibitem[{{Zanella} {et~al.}(2024){Zanella}, {Iani}, {Dessauges-Zavadsky},
  {Richard}, {De Breuck}, {Vernet}, {Kohandel}, {Arrigoni Battaia},
  {Bolamperti}, {Calura}, {Chen}, {Devereaux}, {Ferrara}, {Mainieri},
  {Pallottini}, {Rodighiero}, {Vallini}, \& {Vanzella}}]{ZA24}
{Zanella}, A., {Iani}, E., {Dessauges-Zavadsky}, M., {et~al.} 2024, \aap, 685,
  A80, \dodoi{10.1051/0004-6361/202349074}

\bibitem[{{Zavala} {et~al.}(2015){Zavala}, {Yun}, {Aretxaga}, {Hughes},
  {Wilson}, {Geach}, {Egami}, {Gurwell}, {Wilner}, {Smail}, {Blain}, {Chapman},
  {Coppin}, {Dessauges-Zavadsky}, {Edge}, {Monta{\~n}a}, {Nakajima}, {Rawle},
  {S{\'a}nchez-Arg{\"u}elles}, {Swinbank}, {Webb}, \& {Zeballos}}]{ZJ15}
{Zavala}, J.~A., {Yun}, M.~S., {Aretxaga}, I., {et~al.} 2015, \mnras, 452,
  1140, \dodoi{10.1093/mnras/stv1351}

\bibitem[{{Zavala} {et~al.}(2018){Zavala}, {Monta{\~n}a}, {Hughes}, {Yun},
  {Ivison}, {Valiante}, {Wilner}, {Spilker}, {Aretxaga}, {Eales},
  {Avila-Reese}, {Ch{\'a}vez}, {Cooray}, {Dannerbauer}, {Dunlop}, {Dunne},
  {G{\'o}mez-Ruiz}, {Micha{\l}owski}, {Narayanan}, {Nayyeri}, {Oteo}, {Rosa
  Gonz{\'a}lez}, {S{\'a}nchez-Arg{\"u}elles}, {Schloerb}, {Serjeant}, {Smith},
  {Terlevich}, {Vega}, {Villalba}, {van der Werf}, {Wilson}, \&
  {Zeballos}}]{ZJ18}
{Zavala}, J.~A., {Monta{\~n}a}, A., {Hughes}, D.~H., {et~al.} 2018, Nature
  Astronomy, 2, 56, \dodoi{10.1038/s41550-017-0297-8}

\bibitem[{{Zavala} {et~al.}(2024){Zavala}, {Bakx}, {Mitsuhashi}, {Castellano},
  {Calabro}, {Akins}, {Buat}, {Casey}, {Fernandez-Arenas}, {Franco}, {Fontana},
  {Hatsukade}, {Ho}, {Ikeda}, {Kartaltepe}, {Koekemoer}, {McKinney},
  {Napolitano}, {P{\'e}rez-Gonz{\'a}lez}, {Santini}, {Serjeant}, {Terlevich},
  {Terlevich}, \& {Yung}}]{ZJ24}
{Zavala}, J.~A., {Bakx}, T., {Mitsuhashi}, I., {et~al.} 2024, \apjl, 977, L9,
  \dodoi{10.3847/2041-8213/ad8f38}

\bibitem[{{Zavala} {et~al.}(2025){Zavala}, {Castellano}, {Akins}, {Bakx},
  {Burgarella}, {Casey}, {Ch{\'a}vez Ortiz}, {Dickinson}, {Finkelstein},
  {Mitsuhashi}, {Nakajima}, {P{\'e}rez-Gonz{\'a}lez}, {Arrabal Haro},
  {Bergamini}, {Buat}, {Backhaus}, {Calabr{\`o}}, {Cleri},
  {Fern{\'a}ndez-Arenas}, {Fontana}, {Franco}, {Grillo}, {Giavalisco},
  {Grogin}, {Hathi}, {Hirschmann}, {Ikeda}, {Jung}, {Kartaltepe}, {Koekemoer},
  {Larson}, {McKinney}, {Papovich}, {Rosati}, {Saito}, {Santini}, {Terlevich},
  {Terlevich}, {Treu}, \& {Yung}}]{ZJ25}
{Zavala}, J.~A., {Castellano}, M., {Akins}, H.~B., {et~al.} 2025, Nature
  Astronomy, 9, 155, \dodoi{10.1038/s41550-024-02397-3}

\bibitem[{{Zhang} {et~al.}(2018){Zhang}, {Ivison}, {George}, {Zhao}, {Dunne},
  {Herrera-Camus}, {Lewis}, {Liu}, {Naylor}, {Oteo}, {Riechers}, {Smail},
  {Yang}, {Eales}, {Hopwood}, {Maddox}, {Omont}, \& {van der Werf}}]{ZZ18}
{Zhang}, Z.-Y., {Ivison}, R.~J., {George}, R.~D., {et~al.} 2018, \mnras, 481,
  59, \dodoi{10.1093/mnras/sty2082}

\bibitem[{{Zhao} {et~al.}(2016){Zhao}, {Lu}, {Xu}, {Gao}, {Lord},
  {Charmandaris}, {Diaz-Santos}, {Evans}, {Howell}, {Petric}, {van der Werf},
  \& {Sanders}}]{Z16}
{Zhao}, Y., {Lu}, N., {Xu}, C.~K., {et~al.} 2016, \apj, 819, 69,
  \dodoi{10.3847/0004-637X/819/1/69}

\bibitem[{{Zhu} {et~al.}(2024){Zhu}, {Bakx}, {Ikeda}, {Umehata}, {Becker},
  {Cain}, {Champagne}, {Fan}, {Fudamoto}, {Jin}, {Ma}, {Sun}, {Takeuchi}, \&
  {Tee}}]{ZY24}
{Zhu}, Y., {Bakx}, T. J.~L.~C., {Ikeda}, R., {et~al.} 2024, Research Notes of
  the American Astronomical Society, 8, 284, \dodoi{10.3847/2515-5172/ad91ad}

\end{thebibliography}


\end{document}